%% file: FRG1.tex
\documentclass[groupeaddress,aps,article,nofootinbib,10pt]{revtex4-2}

\usepackage{array}
\usepackage{titlesec}
\usepackage{amsbsy}
\titleformat{\paragraph}
{\normalfont\normalsize\bfseries}{\theparagraph}{1em}{}
\titlespacing*{\paragraph}
{0pt}{3.25ex plus 1ex minus .2ex}{1.5ex plus .2ex}
\renewcommand*{\theparagraph}{\roman{paragraph})}
\renewcommand*{\thesubparagraph}{\roman{subparagraph}}

\usepackage{amssymb, amsmath,amsthm,bm,bbm}    
\usepackage{graphicx}   
\usepackage{verbatim}   
\usepackage{color}      
\usepackage[labelformat=simple]{subcaption}
\usepackage{mathrsfs} 
\usepackage{dsfont}
\usepackage[linktocpage = true]{hyperref}  
\usepackage{cancel}
\usepackage{graphicx}
\usepackage{caption} 
\usepackage{tcolorbox}
\usepackage{yfonts}
\usepackage{multirow}
\definecolor{green}{RGB}{35,142,35}
\usepackage{float}
\usepackage{upgreek}

\usepackage{empheq} 
 
\usepackage{ulem}

\usepackage[framemethod=tikz]{mdframed}

 \usepackage{physics}
 
 \usepackage{bigints}

 \usepackage{braket}

\hypersetup{
    colorlinks   = true,
    linkcolor = blue,
    citecolor    = red
}  

\usepackage{cleveref}
\usepackage{multirow}

\allowdisplaybreaks

\newcommand{\eq}[1]{Eq. \eqref{#1}}
\newcommand{\fig}[1]{Fig. \ref{#1}}

\newcommand{\dslash}{\not{\hbox{\kern-2pt $\partial$}}}
\newcommand{\bqa}{\begin{eqnarray}} 
\newcommand{\eqa}{\end{eqnarray}}
\newcommand{\nn}{\nonumber \\}

\newcommand{\beq}{\begin{equation}}
\newcommand{\eeq}{\end{equation}}

\def\be{\begin{eqnarray}}
\def\ee{\end{eqnarray}}

\newcommand\HD{{\bf H}_d}
\newcommand\DD{\Delta_d^{(\nu)}}
\newcommand\KFthetadim{{\bf K}_{F,\theta}}
\newcommand\KFdim{{\bf K}_{F}}
\newcommand\KFtheta{K_{F,\theta}}
\newcommand\KFAV{{\bf k}_{F}}
\newcommand\kFAV{k_{F}}

\newcommand\thetaq{\Theta (\delta,\theta,\vec q)}
\newcommand\thetasq{\Theta (\theta,\vec q)}
\newcommand\deltaq{\Delta (\delta,\theta,\vec q)}

\newcommand\thetasqp{\Theta (\theta',\vec q)}
\newcommand\deltaqp{\Delta (\delta',\theta',\vec q)}
\newcommand\OLU{OLU(1) }

\newcommand\edim{{\pmb e}}
\newcommand\lambdadim{{\pmb \lambda}}
\newcommand\udim{{\pmb u}}

\newcommand\dsco{d_{SC}^{(0)}}
\newcommand\snu{\mathfrak{s}^{(\nu)}}

\newcommand{\Lc}{{}}

\newcommand{\mtheta}{ \theta_{\overline{12}} }
\newcommand{\btheta}{\bar \theta_{\overline{12}} }
\newcolumntype{P}[1]{>{\centering\arraybackslash}p{#1}}

\begin{document}

\title{\textbf{
Space of 
non-Fermi liquids
}}

        \author{Shubham Kukreja$^{1,2}$ }
        \author{Afshin Besharat$^{1,2}$ }
        \author{Sung-Sik Lee$^{1,2}$}
        \affiliation{$^{1}$Department of Physics \& Astronomy, McMaster University, Hamilton ON L8S 4M1, Canada}
        \affiliation{$^{2}$Perimeter Institute for Theoretical Physics, Waterloo ON N2L 2Y5, Canada}
    
        \date{\today}


\begin{abstract}


In metals, low-energy effective theories are characterized by a set of coupling functions.
Among them, the angle-dependent Fermi momentum specifies the size and shape of Fermi surface.
Since the size of Fermi momentum 
grows incessantly
relative to an energy scale that is lowered
under the renormalization group (RG) flow,
a metallic fixed point is defined only modulo a rescaling of Fermi momentum.
In this paper, we discuss 
 the physical consequences of  this projective nature of fixed points for non-Fermi liquids with hot Fermi surfaces.
The first is the absence of a unique dynamical critical exponent that dictates the relative scaling between energy and momentum across all low-energy observables.
The second is mismatches between the scaling dimensions of couplings and their relevancy.
Nonetheless, each projective fixed point is characterized by a few marginal and relevant coupling functions,
and the notion of universality survives. 
We illustrate our findings by charting the space of 
projective fixed points and extracting their universal properties 
for 
the Ising-nematic quantum critical metal beyond the patch theory.
To approach the interacting theory in two space dimensions from a controlled limit, 
we use the dimensional regularization scheme that tunes the co-dimension of Fermi surface as a control parameter.
Near the upper critical dimension,
two exactly marginal coupling functions span the space of stable projective fixed points:
one specifies the 
 shape of the Fermi surface 
and the other sets the angle-dependent Fermi velocity.
All other coupling functions,
including the Landau functions and the universal pairing interaction,
are fixed by those two marginal functions.
As the space dimension is lowered, 
vertex corrections alter the fate of the four-fermion coupling in a channel-dependent way.
To the leading order,
the forward scattering remains irrelevant 
while the pairing interaction
becomes relevant near two dimensions.
In two dimensions,
it is expected that the universal superconducting fluctuations lower the symmetry of the non-Fermi liquid realized above the superconducting transition temperatures 
from the loop U(1) group 
to a proper subgroup.

\end{abstract}

\maketitle
\newpage
\tableofcontents
\newpage

\input{mainbody.tex}

\input{appendix.tex}

\section*{Acknowledgement}

This research was supported by the Natural Sciences 
and Engineering Research Council of
Canada. Research at the Perimeter Institute is supported in part by the
Government of Canada through Industry Canada, and by the Province of
Ontario through the Ministry of Research and Information.

\bibliographystyle{apsrev4-1}

\bibliography{references}

\end{document}

%% file: mainbody.tex
\section*{Notation}

\begin{enumerate}

\item scales
\begin{itemize}
\item $\Lambda$ : UV energy cutoff
\item $\mu = \Lambda e^{-l}$ : renormalization group energy scale 
\item $\KFthetadim$ : the magnitude of Fermi momentum at angle $\theta$
\item 
$\KFAV = \frac{1}{2\pi} \int d \theta \KFthetadim$ : 
the averaged Fermi momentum 
\item  
$\KFtheta = \KFthetadim/\mu$,
$\kFAV =  \KFAV/\mu$ :
dimensionless Fermi momentum
\end{itemize}

\item momentum
\begin{itemize}
\item $\vec k = (\KFthetadim+\delta) (\cos \theta, \sin \theta)$ : two-dimensional momentum of fermion radially displaced by $\delta$ from the Fermi surface at angle $\theta$
\item $\vec q = q (\cos \varphi, \sin \varphi)$ : two-dimensional momentum of boson
\item 
${\bf K} = (k_0,.., k_{d-2})$,
${\bf Q} = (q_0,.., q_{d-2})$ : 
Matsubara frequency combined with 
 $(d-2)$ components of spatial momentum in general $d$
\item 
${\bf k} = ({\bf K}, \vec k)$,
${\bf q} = ({\bf Q}, \vec q)$ : 
$(d+1)$-dimensional frequency-momentum vector 
\end{itemize}

\item `coupling' functions
\begin{itemize}
\item
$\kappa_{F,\theta} = \KFthetadim/\KFAV$ :
the normalized Fermi surface shape function
\item$v_{F,\theta}$ : the angle-dependent Fermi velocity in the radial direction
\item $e_{\theta_1,\theta_2}$ : the angle-dependent fermion-boson coupling function
\item 
$\lambda_{ \left(\begin{smallmatrix} \theta_1 & \theta_2   \\ \theta_4 & \theta_3    \end{smallmatrix}\right)}$ :
the general four-fermion coupling written as a function of angles of the two incoming and two outgoing fermions
\begin{itemize}
\item
$
\lambda
^{(\nu,s)}_{\theta', \theta}(\vec q)$ 
for $\nu = F_{\pm}$ :
the four-fermion coupling in the forward scattering channel written as a function of the angles of two fermions and the momentum transfer $\vec q$ 
\item
$
\lambda
^{(\nu,s)}_{\theta', \theta}(\vec q)$ 
for $\nu = P$ :
the four-fermion coupling in the pairing channel written as a function of the angles and the total center of mass momentum $\vec q$ of Cooper pairs

\end{itemize}
\end{itemize}
\end{enumerate}

\newpage
\section{Introduction}

Universal low-energy physics of condensed matter systems is encoded in 
 infrared fixed points of the renormalization group flow\cite{WILSONRG1,WILSONRG2,WILSON1}.
Therefore, 
mapping the space of low-energy fixed points 
is of utmost importance 
for classifying and characterizing phases of matter.
In gapped phases, fixed points arise as isolated points in the space of couplings\cite{
10.1142/S0217979290000139,
10.1063/1.3149481,
2009AIPC.1134...22K,
PhysRevB.83.035107}.
This is usually the case for critical states with the conformal symmetry as well\cite{DIFRANCESCO}. 
For metallic systems, however, low-energy fixed points generally form extended spaces spanned by marginal parameters 
that can be tuned for the infinitely many gapless modes residing on the Fermi surfaces.
In Fermi liquids\cite{LANDAU,LANDAU2}, 
the shape of the Fermi surface, the angle-dependent Fermi velocity and the Landau function for the forward scatterings are the marginal parameters \cite{PhysRevB.42.9967,POLCHINSKI1,SHANKAR},
which span the space of Fermi liquid fixed points.
Notably, pairing interactions, which 
 generate another kinematically allowed scatterings at low energies, is negligible at the Fermi liquid fixed points.

The space of non-Fermi liquids is less well charted\cite{
HERTZ,
MILLIS,
VARMALI,
POLCHINSKI2,
PLEE1,
ALTSHULER,
YBKIM,
ABANOV2,
SSLEE,
MAX0,
MAX2,
NAYAK2,
STEWART,
SCHOFIELD,
DENNIS,
SENTHIL,
MROSS,
FITZPATRICK,
SHOUVIK2,
SCHLIEF,
PhysRevX.11.021005,
PhysRevB.107.165152,
annurev:/content/journals/10.1146/annurev-conmatphys-031218-013339}.
This is partly because the strong coupling, which is usually what makes non-Fermi liquids what they are, makes it difficult to access the low-energy physics in a controlled way\cite{
NAYAK2,
MROSS,
DENNIS,
SHOUVIK2,
SCHLIEF,
doi:10.1126/science.abq6011}.
The strong coupling problem is compounded by the fact that 
in metals the low-energy physics is fully defined only after certain short-distance information is specified. 
For example, couplings that are functions of momentum along the Fermi surface are parts of low-energy data that need to be specified over the Fermi surface.
Fermi momentum,
which sets the size of Fermi surface,
is not a high-energy but
large momentum scale that determines the phase space of the low-energy modes.
It acts as a relevant parameter because its size relative to 
the momentum scale perpendicular to the Fermi surface continues to grow 
as the low-energy limit is taken.

Being a large momentum scale that controls the low-energy physics,
Fermi momentum renders the notion of scale invariance and fixed point far from being trivial.
Since the large momentum scale can singularly affect the low-energy physics,
the low-energy limit does not generally commute with the large Fermi momentum limit\footnote{
This is a hallmark of ultraviolet/infrared (UV/IR) mixing. 
However, there is an important difference from the UV/IR mixing that arises from a non-commutative structure of space or an extended nature of physical objects such as strings,
where high-energy scales is intertwined with low-energy physics \cite{Minwalla:2000um}.  The UV/IR mixing that arises in metals
is rooted to the fact that large-momentum but low-energy scale singularly affects the low-energy physics\cite{IPSITA,PhysRevLett.128.106402}
(this includes the UV/IR mixing that arises  in `Bose' metals\cite{ PhysRevB.100.024519, PhysRevB.104.014517, PhysRevB.104.235116}).
The former is a `vertical' UV/IR mixing while the latter is a `horizontal' one.
In principle, the horizontal UV/IR mixing can make the low-energy physics become more sensitive to the high-energy physics if large-angle scatterings between low-energy fermions are mediated by high-energy bosons\cite{BORGES2023169221}.}.
As the finiteness of Fermi momentum can not be `forgotten' for the low-energy physics,
metals do not have a scale invariance in the strict sense.
Fermi momentum also alters the relevancy/irrelevancy of couplings as the degree of infrared singularity a coupling generates can be shifted from what is expected from its scaling dimension.
For example, couplings that carry negative scaling dimensions can still give rise to infrared singularities 
assisted by the volume of the low-energy phase space controlled by the Fermi momentum.
Conversely, seemingly relevant couplings can be suppressed by the damping created by the extensive particle-hole excitations.
To correctly capture such 
phase-space-induced enhancement/suppression of infrared singularities, 
one has to include the effects of Fermi momentum non-perturbatively and
generalize the notion of renormalizable field theories for metals\cite{BORGES2023169221}.
Here, a renormalizable field theory is the minimal theory that captures the low-energy physics of a universality class.
Generalized notion of renormalizable theory also has an implication for the emergent symmetry of metals.
For example, 
large-angle scatterings 
enhanced by the extensive phase space
can prevent an emergent conservation of fermion number in each patch of Fermi surface
\cite{2005cond.mat..5529H,PhysRevX.11.021005} 
even if the corresponding four-fermion coupling has a negative 
 scaling dimension.

The best way to study the entirety of metals is to include all low-energy degrees of freedom within an
effective theory.
In such frameworks,
momentum-dependent interactions of fermions are captured through couplings that are functions of angles around the Fermi surface.
Naturally, universality classes are encoded in the functional renormalization group (RG) flow of those coupling functions.
One can in principle use the exact RG equation\cite{POLCHINSKIFRG,WETTERICH}
that describes the flow of the fully momentum-dependent vertex functions\cite{
MORRIS,
REUTER,
ROSA,
HOFLING,
HONERKAMP,
GIES,
GIES2,
BRAUN,
METZNER,
doi:10.1080/00018732.2013.862020,
PhysRevB.61.13609,
PhysRevLett.102.047005,
PhysRevB.61.7364,
SCHERER,
JANSSEN2,
MESTERHAZY,
PhysRevB.87.045104,
EBERLEINMETZNER,
PLATT,
WANGEBERLEIN,
JANSSEN,
MAIEREBERLEIN,
EBERLEIN2,
EBERLEIN3,
JAKUBCZYK,
MAIER,
TORRES}.
However, the exact beta functionals are usually too difficult to solve for interacting theories.
Alternatively, one can discretize the momentum space to trade coupling functions with a discrete set of coupling constants.
However, a discrete mesh fails to capture singularities that angle-dependent coupling functions can dynamically develop at low energies in quantum critical metals.
In order to maximize the tractability of a theory without losing the essential low-energy physics,
what is needed is  
 a functional renormalization group for renormalizable field theories of metals
 that {\it fully} keep
the low-energy data but only {\it minimally}.
The field-theoretic functional renormalization group formalism\cite{BORGES2023169221,PhysRevB.109.045143} provides a route to this goal.

A recent application of the formalism to the antiferromagnetic quantum critical metal\cite{
ABANOV3,
ABANOV2,
HARTNOLL,
ABRAHAMS,
PhysRevB.87.045104,
VANUILDO,
DECARVALHO,
PATEL,
PATEL2,
VARMA2,
MAIER,
VARMA3,
MAX2,
SHOUVIK,
SHOUVIK3,
SUNGSIKREVIEW,
MAX1,
LIHAI2,
SCHATTNER2,
GERLACH,
LIHAI,
WANG2} 
has revealed a non-perturbative fixed point in the full space of coupling functions\cite{BORGES2023169221}.
It further shows that superconducting fluctuations generated from the critical spin fluctuations induce strong inter-patch scatterings,
which makes the use of the patch theory questionable near the fixed point\cite{PhysRevB.95.174520}.
However, the antiferromagnetic quantum critical metal
is a rather special example of non-Fermi liquids 
in that only a discrete set of points on the Fermi surface stay incoherent in the low-energy limit.
While the coupling functions exhibit universal singularities in the vicinity of the hot spots, electrons far away from the hot spots are coherent in the low-energy limit.
On the other hand, 
non-Fermi liquids with hot Fermi surfaces
are expected to exhibit qualitatively different structures of infrared fixed points.
The goal of this paper is to chart the space of non-Fermi liquids and characterize their universal low-energy physics in a quantum critical with hot Fermi surface.

A prototypical example of non-Fermi liquids with hot Fermi surface is the Ising-nematic quantum critical metal
in which the majority of Fermi surface remains strongly coupled with critical nematic fluctuations\cite{OGANESYAN,
MAX0,
DENNIS,
SCHATTNER,
LEDERER,
LEDERERREV}.
Its low-energy effective field theory has been extensively studied, 
but most of the previous theoretical works are based on the patch theory 
whose validity is in priori unclear in the presence of large-angle scatterings.
In particular, in order to capture superconducting fluctuations 
one has to include inter-patch couplings\cite{PhysRevD.59.094019,PhysRevB.91.115111}.
In this paper, we study the low-energy effective field theory of the full Fermi surface
at the Ising-nematic quantum critical point,
and identify the space of low-energy fixed points 
through the field-theoretic functional renormalization group formalism\cite{BORGES2023169221}.
For a controlled access to the low-energy physics, we use the dimensional regularization scheme\cite{DENNIS} 
in which the co-dimension of Fermi surface is tuned as a control parameter.
The theory defined in general space dimension $d$ continuously interpolates between
the nematic quantum critical point of a semi-metal with Fermi ring in $d=3$ and the non-Fermi liquid theory in $d=2$.
The coupling between the critical nematic fluctuation and electron becomes marginal at the upper critical dimension $d_c=5/2$,
and the low-energy theories become interacting in $d<d_c$.

\begin{figure}[t]
\centering
  \includegraphics[width=0.6\linewidth]{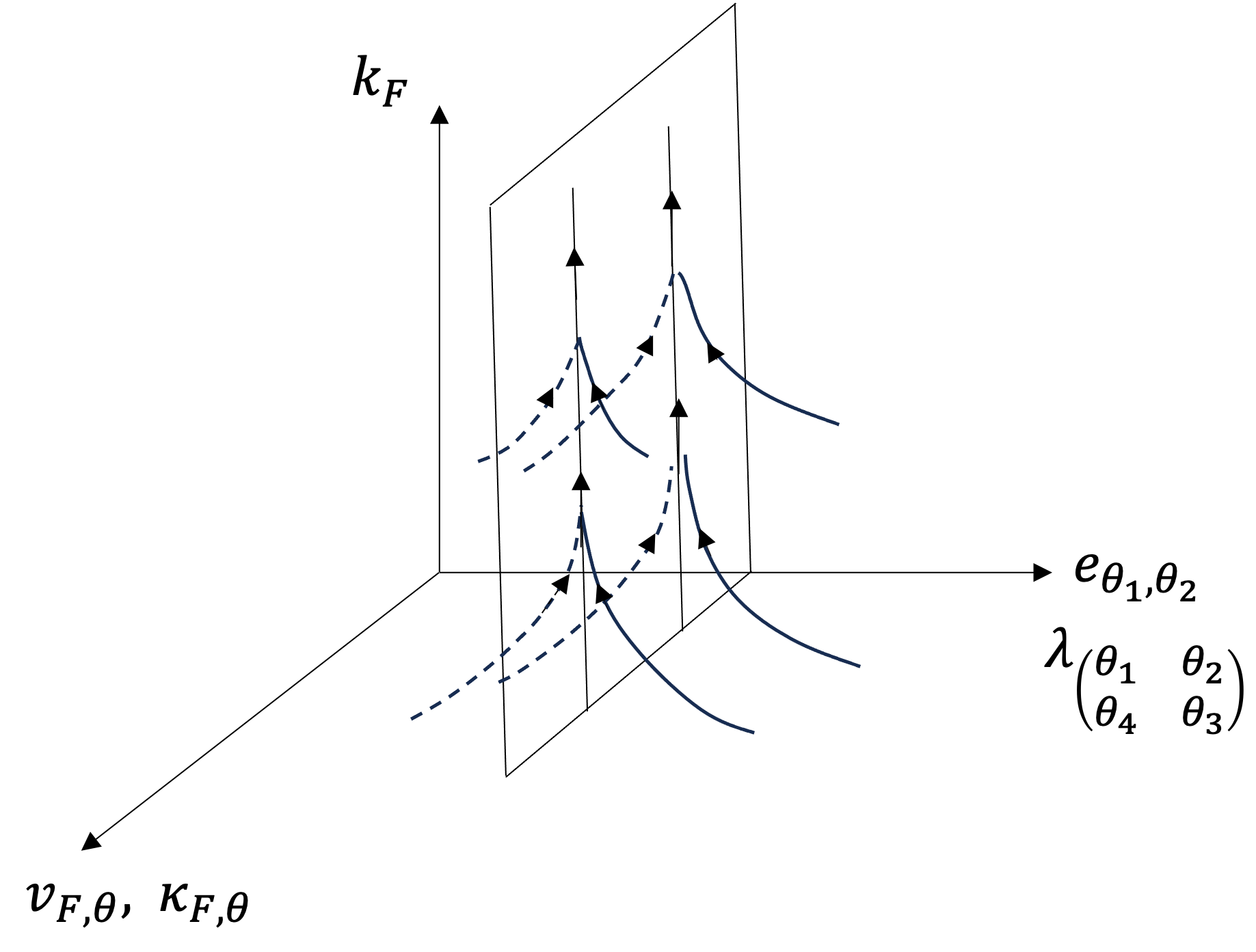}
\caption{
The low-energy effective theory of the Ising-nematic quantum critical metal is characterized by
an overall size of Fermi surface
($\kFAV$), 
a function that determines the shape of the Fermi surface
($\kappa_{F,\theta}$),
a Fermi velocity
 function ($v_{F,\theta}$),
a Yukawa coupling function
($e_{\theta_1,\theta_2}$)
and a four-fermion coupling function
$\left(\lambda_{ \left(\begin{smallmatrix} \theta_1 & \theta_2   \\ \theta_4 & \theta_3    \end{smallmatrix}\right)}\right)$.
Under the renormalization group flow, a microscopic theory is attracted to a one-dimensional manifold along which $\kFAV$ keeps increasing toward the infinity.
The space of non-Fermi liquid fixed points is characterized by marginal and relevant couplings defined in the space that is transverse to $\kFAV$.
Near the upper critical dimension, $\kappa_{F,\theta}$ and $v_{F,\theta}$ are exactly marginal, and they fix the remaining coupling functions.
 }
\label{fig:projective_fp}
\end{figure}

The low-energy theory is characterized by 
the Fermi momentum $\KFthetadim$ that specifies the size and shape of the Fermi surface,
the Fermi velocity $v_{F,\theta}$,
the Yukawa coupling function $e_{\theta_1,\theta_2}$
and the four-fermion coupling functions
$\lambda_{ \left(\begin{smallmatrix} \theta_1 & \theta_2   \\ \theta_4 & \theta_3    \end{smallmatrix}\right)}$.
They are functions of angles around the Fermi surface.
The Yukawa 
and four-fermion
coupling functions 
 generally depend on multiple angles for the in and out electrons involved in scatterings.
Among all `coupling' functions, the most prominent one is the Fermi momentum.
Under the RG flow, its size grows indefinitely
because the Fermi momentum keep increasing relative to the RG energy scale that is sent to zero in the low-energy limit.
Because of the non-stop flow of the Fermi momentum,
metals can not have fixed points in the usual sense\cite{BORGES2023169221}. 
Nonetheless, the notion of universality classes remains well defined 
because the low-energy physics is still characterized by a small set of coupling functions.

For the purpose of 
charting the set of distinct metallic universality classes,
it is convenient to decompose $\KFthetadim$ into one scale that represents the overall size of Fermi surface
and a dimensionless function that captures its shape as
\bqa
\KFAV \equiv \frac{1}{2\pi} \int d \theta \KFthetadim, 
~~~
\kappa_{F,\theta} \equiv \KFthetadim/\KFAV,
\label{eq:KFkappaF}
\eqa
where
$\KFAV$ denotes the Fermi momentum averaged over the Fermi surface
and
$\kappa_{F,\theta}$ 
 is an angle-dependent function 
with normalization 
$\frac{1}{2\pi}  \int d \theta ~ \kappa_{F,\theta}=1$.
The full functional RG takes place in 
the  space of  $\left\{ \KFAV, \kappa_{F,\theta}, v_{F,\theta}, e_{\theta_1,\theta_2}, \lambda_{ \left(\begin{smallmatrix} \theta_1 & \theta_2   \\ \theta_4 & \theta_3    \end{smallmatrix}\right)} \right\}$.
As the RG energy scale $\mu$ is lowered,
the size of Fermi surface measured in the unit of $\mu$, $\kFAV\equiv \KFAV/\mu$, keeps increasing because the manifold of gapless modes does not shrink under the coarse graining.
The RG flow can be still attracted toward a universal profile 
within the subspace of 
$\left\{  \kappa_{F,\theta}, v_{F,\theta}, e_{\theta_1,\theta_2}, \lambda_{ \left(\begin{smallmatrix} \theta_1 & \theta_2   \\ \theta_4 & \theta_3    \end{smallmatrix}\right)} \right\}$.
However, the profile of
$\left\{  \kappa_{F,\theta}, v_{F,\theta}, e_{\theta_1,\theta_2}, \lambda_{ \left(\begin{smallmatrix} \theta_1 & \theta_2   \\ \theta_4 & \theta_3    \end{smallmatrix}\right)} \right\}$ that emerges in the low-energy limit continues to depend on $\kFAV$,
which makes it impossible to take the
$\kFAV \rightarrow \infty$ limit up front.
This is illustrated in \fig{fig:projective_fp}.
The incessant growth of $\kFAV$ implies that metallic universality classes are characterized by {\it projective fixed points} defined modulo the rescaling of $\kFAV$.
Due to the projective nature of fixed points, low-energy observables do not take simple power-law forms even in the low-energy limit.
In particular, there is no unique dynamical critical exponent that sets the relative scaling between energy and momentum across all low-energy observables.
Furthermore, whether a coupling is relevant or not can not be in general determined based on its scaling dimension.
Instead, it should be determined from its contribution to low-energy observables.
If a coupling singularly affects any low-energy observable,
it should be deemed relevant irrespective of its scaling dimension.
In this paper, we will illustrate these points through the example of the Ising-nematic quantum critical metal.
However, the general consequences of the projective nature of fixed points do not depend on the details of the theory.

\begin{figure}[t]
\centering
\begin{subfigure}{.5\textwidth}
  \centering
  \includegraphics[width=1.0\linewidth]{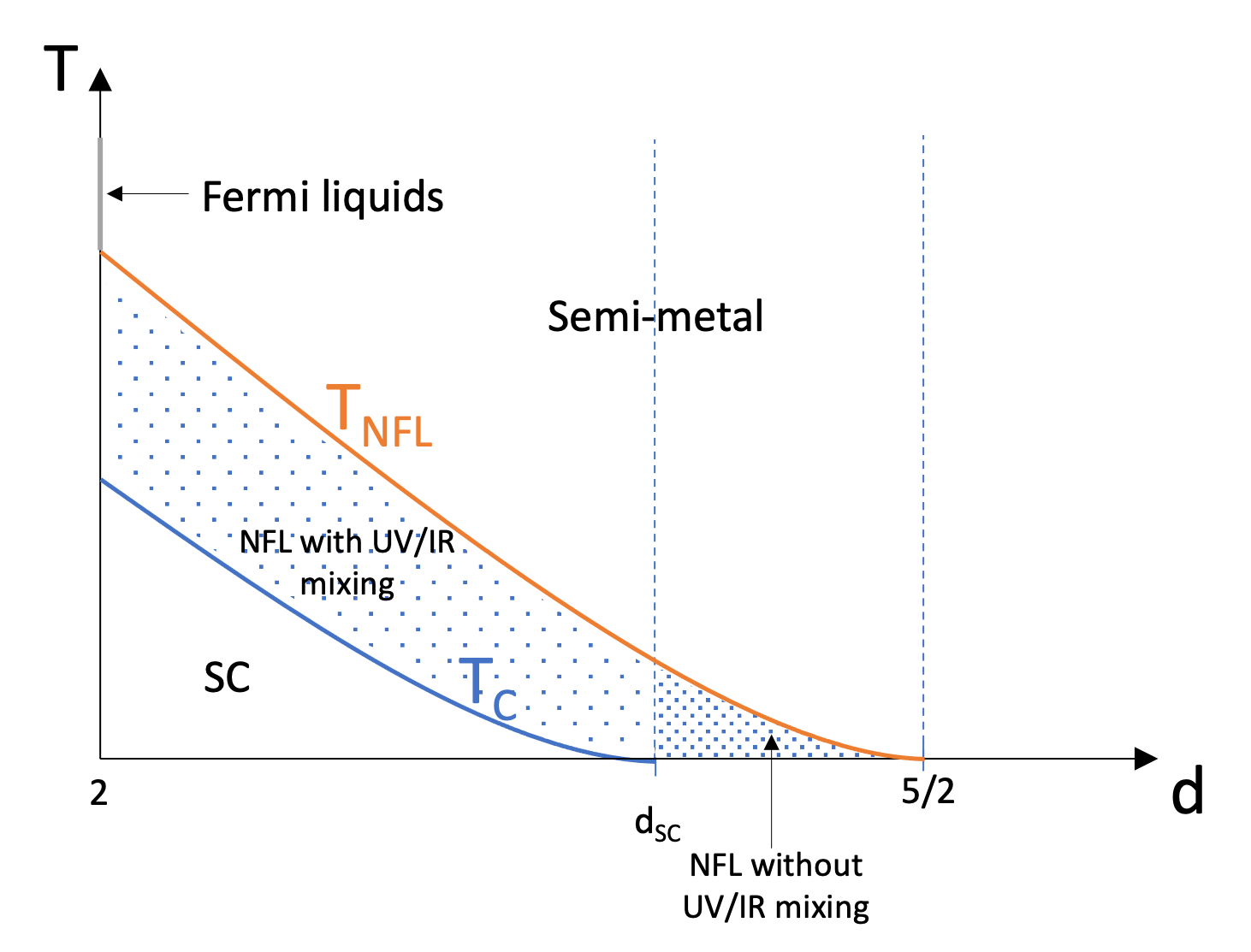}
  \caption{}
  \label{fig:overview}
\end{subfigure}%
\begin{subfigure}{.5\textwidth}
  \centering
  \includegraphics[width=1.0\linewidth]{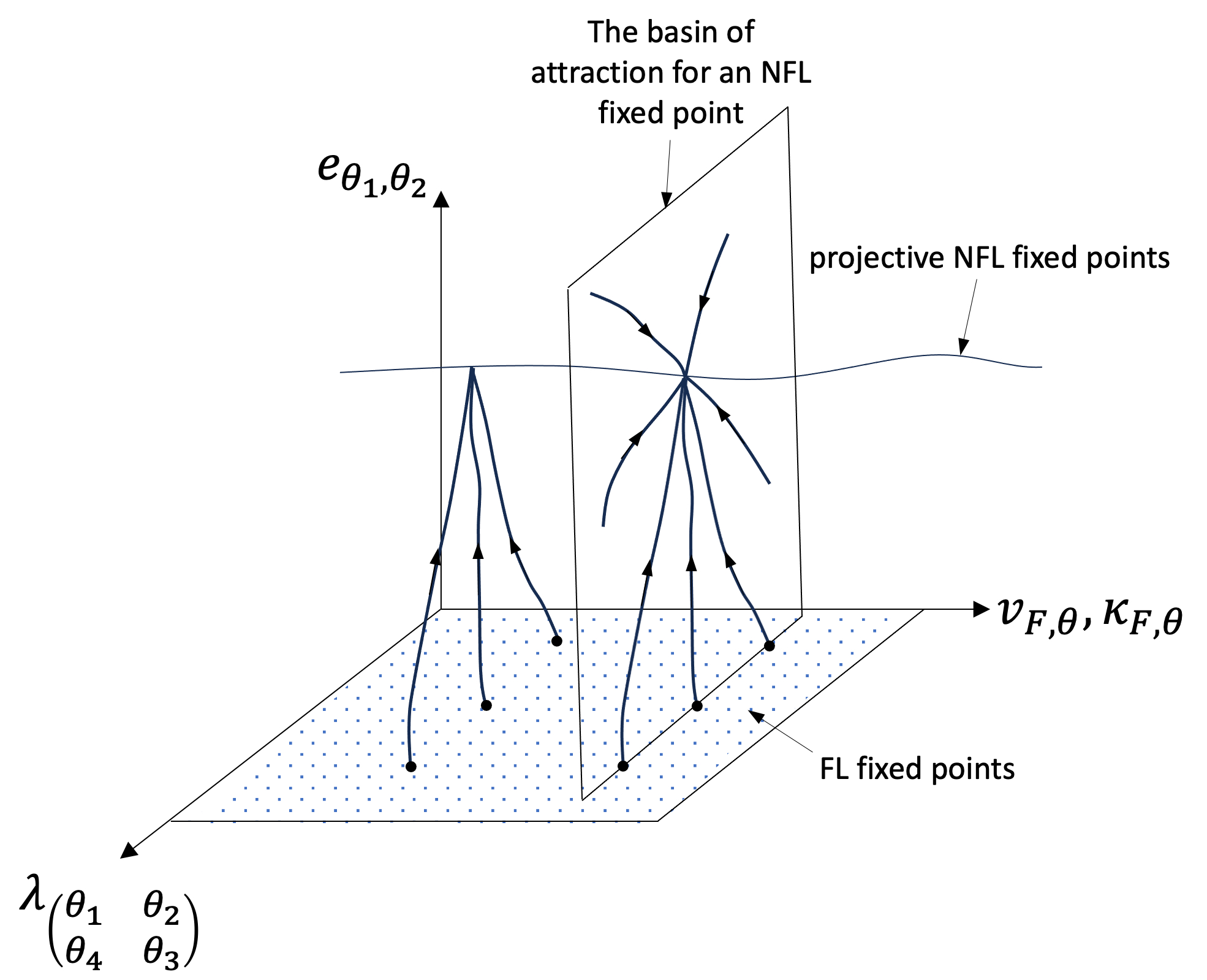}
  \caption{
}
  \label{fig:spaceNFL}
\end{subfigure}
\caption{
(a) 
The `phase diagram' of the Ising-nematic quantum critical metal 
with a one-dimensional Fermi surface
embedded in the $d$-dimensional momentum space.
In $d_{SC}<d<5/2$, the low-energy theories flow to stable non-Fermi liquid fixed points with negligible inter-patch couplings.
Therefore, the short-distance data is unimportant for physical observables that are local in momentum space (no UV/IR mixing).
In $d< d_{SC}$, the ground state is likely to be a superconducting state.
But, there is a window of energy scale controlled by non-Fermi liquid quasi-fixed points, 
which is  characterized by a strong inter-patch coupling. 
The full description of the low-energy physics requires short-distance data 
 beyond what is included in the patch theory (UV/IR mixing).
(b) The schematic functional renormalization group flow of the coupling functions 
in $d>d_{SC}$.
The space of the Ising-nematic non-Fermi liquid (NFL) fixed points is smaller than that of Fermi liquids in that the Landau function is completely  fixed by the shape of the Fermi surface ($\kappa_{F,\theta}$) and the angle-dependent Fermi velocity ($v_{F,\theta}$).
}
\label{fig:1}
\end{figure}

Charting the universality classes of non-Fermi liquids amounts to identifying projective fixed points of the functional RG flow for the coupling functions.
\fig{fig:overview} is a schematic phase diagram for the Ising-nematic quantum critical metal drawn in the plane of space dimension $d$ and temperature $T$.
Between the upper critical dimension $d_{c}=2.5$ and another critical dimension $d_{SC}$, the theories flow to stable projective non-Fermi liquid fixed points in the low-energy limit.
The space of stable non-Fermi liquids are 
 spanned by two marginal coupling functions,
$\kappa_{F,\theta}$  and  $v_{F,\theta}$.
At the fixed points,
$e_{\theta_1,\theta_2}$
and 
$\lambda_{
\left(\begin{smallmatrix} \theta_1 & \theta_2   \\
    \theta_4 & \theta_3   
\end{smallmatrix}\right)}$
are completely fixed by those marginal couplings and $\kFAV$.
Accordingly, deformations made in the Yukawa coupling and the four-fermion coupling functions away from the fixed points are irrelevant. 
The schematic functional renormalization group flow and the space of infrared fixed points are illustrated in   \fig{fig:spaceNFL}.
Apart from the fact that there exist no well-defined quasiparticles in non-Fermi liquids,
two features that are qualitatively different from Fermi liquids arise
:
(1) the Landau function is no longer marginal as it is fixed by 
$\kappa_{F,\theta}$  and  $v_{F,\theta}$
and 
(2) the pairing interaction is non-vanishing due to universal superconducting fluctuations generated from quantum criticality\cite{PhysRevB.92.205104}.
In short,
there are generally fewer non-Fermi liquids than Fermi liquids,
and superconducting fluctuations are an intrinsic part of non-Fermi liquids.

\begin{figure}[t]
\centering
  \includegraphics[width=0.4\linewidth]{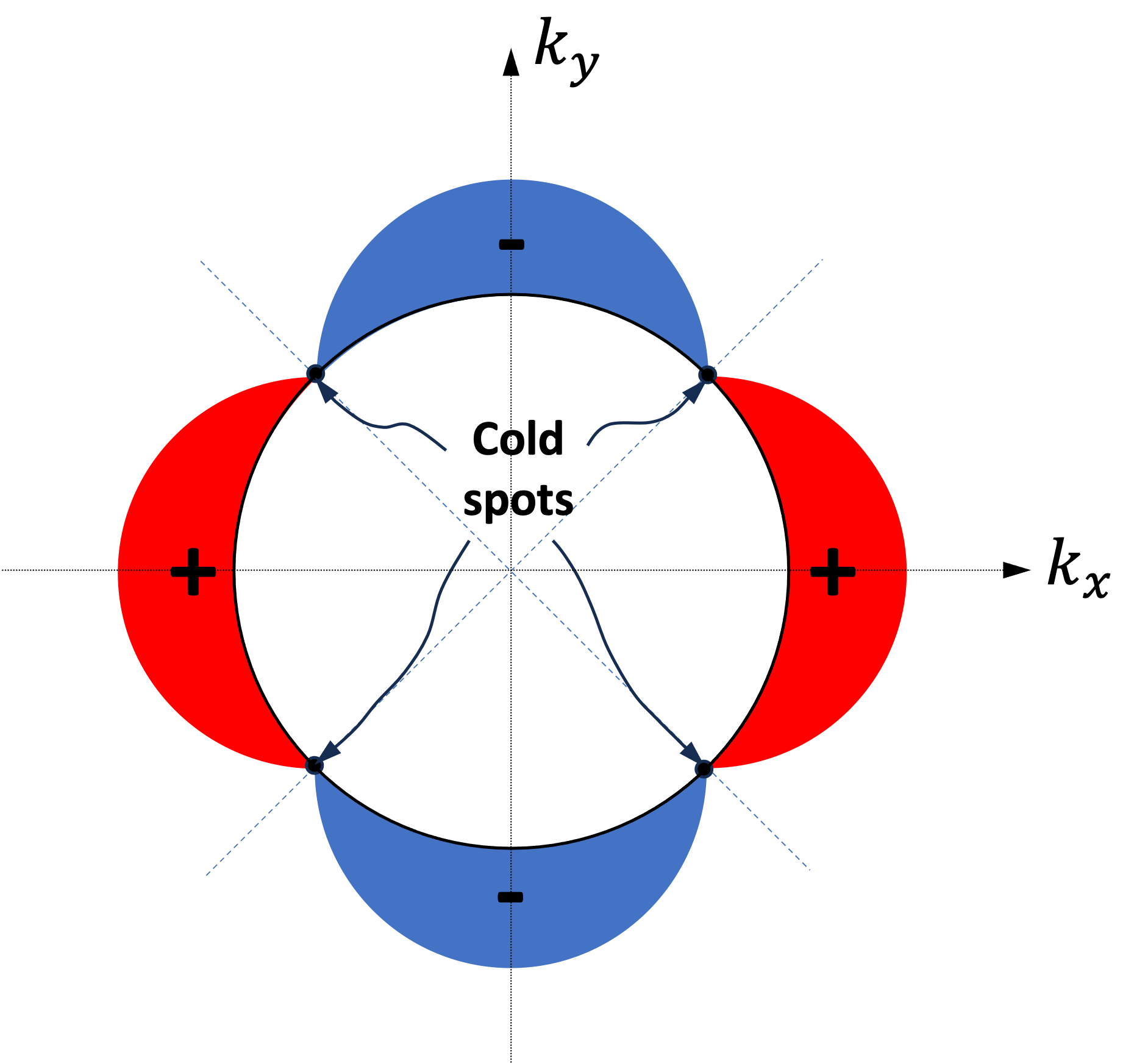}
\caption{
The bare Yukawa coupling function $e_{\theta,\theta}$ has the d-wave form factor as a function of the angle around the Fermi surface because the critical mode describes the nematic fluctuations in the angular momentum two channel.
The anisotropic bare coupling makes electrons move faster near the nodal directions compared to those in the anti-nodal directions.
Remarkably, electrons near (but not exactly at) the nodal directions end up being subject to a stronger renormalized Yukawa coupling at low energies due to a weakened screening effect.
The angular profiles of the Fermi velocity and the renormalized Yukawa coupling conspire to generate the angle-independent anomalous dimension except at the cold spots.
}
\label{fig:dwave_form}
\end{figure}

The nature of the stable projective fixed points is encoded in the momentum-dependent coupling functions that emerge in the low-energy limit.
At the low-energy fixed points,
the Yukawa coupling function acquires a momentum profile 
in such a way that 
the anomalous dimension of electron,
which measures the degree of incoherence,
becomes independent of angle around the Fermi surface
everywhere except for the exact cold spots.
While this emergent sliding symmetry is expected from the patch theory\cite{MAX0,DENNIS},
it has an interesting consequence from the global perspective.
Suppose that at a UV scale
the Fermi velocity is angle-independent,
and the Yukawa coupling has the d-wave form factor with nodes along the diagonal directions at which the cold spots are located 
(see \fig{fig:dwave_form}).
This form factor arises from the fact that density fluctuations of angular momentum two become critical at the Ising-nematic critical point.
Under the renormalization group flow, the Fermi velocity around the nodal directions
becomes bigger compared to that in the anti-nodal directions
because electrons in the nodal directions become less heavier thanks to the weaker interaction.
The flip side of this is that the faster electrons near the nodal directions  become less effective in screening the interaction.
As a result, the Yukawa coupling near the nodal directions 
(but away from the exact 
$\pm 45^\circ$ directions) 
eventually becomes stronger than the coupling in the anti-nodal direction 
in the low-energy limit
even if the bare coupling is weaker near the nodal direction.
In the end, electrons near the cold spots 
become faster 
and subject to stronger interactions
relative to those in the anti-nodal directions. 
Consequently, the anomalous dimension, which is determined by a ratio between the coupling and the Fermi velocity, becomes angle independent in the low-energy limit. 
The functional RG flow of the Yukawa coupling function and the Fermi velocity is summarized in \fig{fig:flowoffv}.

For the four-fermion coupling function,
its scaling dimension controls the functional RG flow of the coupling function both in energy and 
 momentum.
The former determines how the coupling function  evolves (vertically) under the renormalization group flow
while the latter determines the profile of the fixed-point coupling (horizontally) as a function of the angle around the Fermi surface.
%
Near the upper critical dimension, both the Landau function and the pairing interaction are irrelevant.
We emphasize that this does not mean that the four-fermion couplings are zero at the fixed point.
The four-fermion interaction is non-vanishing in both channels but they are fixed by the Fermi momentum and Fermi velocity.
Therefore, small perturbations added to the UV coupling decay out under the renormalization group flow:
the four-fermion coupling is irrelevant `vertically'.
Furthermore, the universal coupling function that arises in the IR decays quickly as a function of the angular separation between the fermions so that
the large-angle scatterings are also irrelevant `horizontally'.

With the negligible inter-patch coupling, 
the patch theory is sufficient for describing low-energy observables that are local in momentum space 
such as the single-particle spectral function.
However, the patch theory is not the full description 
 even in this case because there exist low-energy observables that are not local in momentum space.
Examples include the thermodynamic observables that probe the entire low-energy modes and multi-point vertex functions where low-energy fermions are probed at different locations on the Fermi surface.
Those `non-local' observables generally depend on $\kFAV$ differently than `local' observables do. 
As a result, there exists no single power-law scaling under which all low-energy observables are scale invariant.
For example, the momentum of particle-hole and particle-particle pairs in the four-fermion vertex acquire different scaling dimensions depending on whether the momentum is tangential or perpendicular to the Fermi surface.
Consequently, the fixed point lacks a universal dynamical critical exponent.
Another peculiar aspect of metallic fixed points is that scaling dimensions alone do not determine the relevancy of couplings.
A coupling can become relevant with the help of $\kFAV$ even if it has a negative scaling dimension.
This $\kFAV$-assisted infrared singularities is a crucial ingredient for understanding the physics that emerges as the dimension is lowered toward the target dimension $d=2$.

With decreasing $d$, the four-fermion coupling is renormalized by the vertex correction more strongly.
To the leading order, the Landau functions remain irrelevant down to two space dimensions due to a kinematic constraint that limits the phase space for the forward scatterings.
On the other hand, the pairing interaction becomes relevant below a critical dimension $d_{SC}$ which is between $d=2$ and the upper critical dimension $d_c=2.5$. 
Interestingly, it turns relevant while the scaling dimension of the coupling remains strictly negative:
the scaling dimension of the four-fermion coupling becomes $-1/2$ at that critical dimension.
Despite its negative scaling dimension, 
which would naively imply that the coupling is irrelevant at low energies,
the pairing interaction becomes relevant due to an enhancement by the Fermi momentum that provides an extensive phase space for Cooper pairs.
Consequently, the normal state becomes unstable against superconductivity in $d<d_{SC}$. 
Although the ground state is likely to be a superconducting state in $d<d_{SC}$,
there exists a window of energy scale above the superconducting transition temperature 
in which the physics is controlled by the 
non-Fermi liquid scaling.
Below but near $d_{SC}$,
the size of the energy window for the non-Fermi liquid normal state is large due to a separation between the scale below which the non-Fermi liquid physics sets in and the superconducting transition temperature.
The non-Fermi liquid state realized in $d<d_{SC}$ is characterized by strong inter-patch superconducting fluctuations.
In particular, the critical superconducting fluctuations generate large-angle scatterings that violate the conservation of the number of low-energy electrons within each patch.
Such scatterings lower the emergent symmetry of the normal state from the loop U(1) group of Fermi liquids\cite{2005cond.mat..5529H,PhysRevX.11.021005} to its proper subgroup, the loop U(1) group with odd parity.

Here is the outline of the rest of this paper.
In Sec. \ref{sec:ii}, we begin with the low-energy effective field theory that includes the entire Fermi surface beyond the patch theory
for the Ising-nematic quantum critical metal in two space dimensions.
The theory is then extended to general space dimensions between $2$ and $3$ through the dimensional regularization scheme that tunes the co-dimension of the Fermi surface.
In Sec. \ref{sec:iii}, 
the functional renormalization group formalism is discussed.
While this is a standard material for relativistic field theories, here we emphasize the difference that arises from the 
 additional scale, $\kFAV$.
In Sec. \ref{sec:beta}, we derive the beta functionals for all coupling functions to the leading order in $\epsilon = 5/2-d$.
In Sec. \ref{sec:iv}, 
we solve the beta functionals to identify the low-energy fixed points. 
The solution allows us to chart the space of projective fixed points parameterized by the marginal couplings.
In Sec. \ref{sec:NFLfp},
we extract the universal profiles of the four-fermion coupling functions that are fixed by the marginal coupling functions near the upper critical dimension.
In Sec. \ref{sec:SC_locality}, we discuss how the Landau function and the pairing interaction evolve as the dimension is lowered toward the physical dimension.
In particular, we discuss the role of large-angle scatterings that are enhanced by vertex corrections,
and its implications on the symmetry and instability of the normal state. 
In Sec. 
\ref{sec:conclusion},
we conclude with a summary and open questions.


\section{
Low-energy effective field theories for hot Fermi surfaces
}
\label{sec:ii}


\begin{figure}[th]
  \centering
  \includegraphics[width=0.3\linewidth]{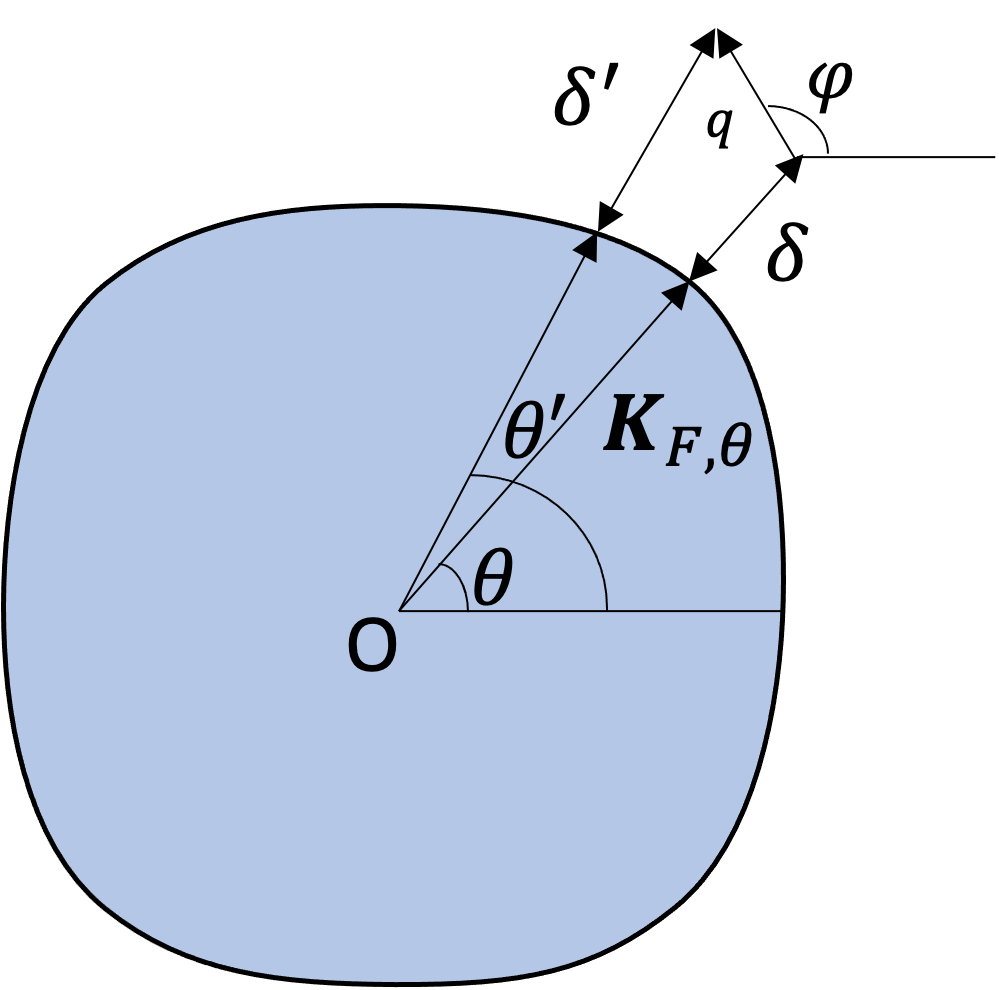}
  \caption{
In the Fermi polar coordinate system, the two-dimensional momentum 
 of a fermion is written as $(\theta, \delta)$,
where $\theta$ is the polar angle
and $\delta$ is the radial displacement of fermion away from the angle-dependent Fermi momentum $\KFthetadim$ at angle $\theta$.
For momentum of boson $\vec q$, we use the usual polar coordinate.
A fermion at angle $\theta$ and radial displacement $\delta$
absorbs a boson with momentum $\vec q=q(\cos \varphi, \sin \varphi)$ to change its angle and radial displacement to
$\theta' = 
\thetasq$
and
 $\delta' = \deltaq$.
}
\label{Fig:Fermi_Surface_Angles}
\end{figure}

\subsection{Action 
and dimensional regularization}

\begin{figure}[th]
\centering
\begin{subfigure}{.3\textwidth}
  \centering
  \includegraphics[width=1.0\linewidth]{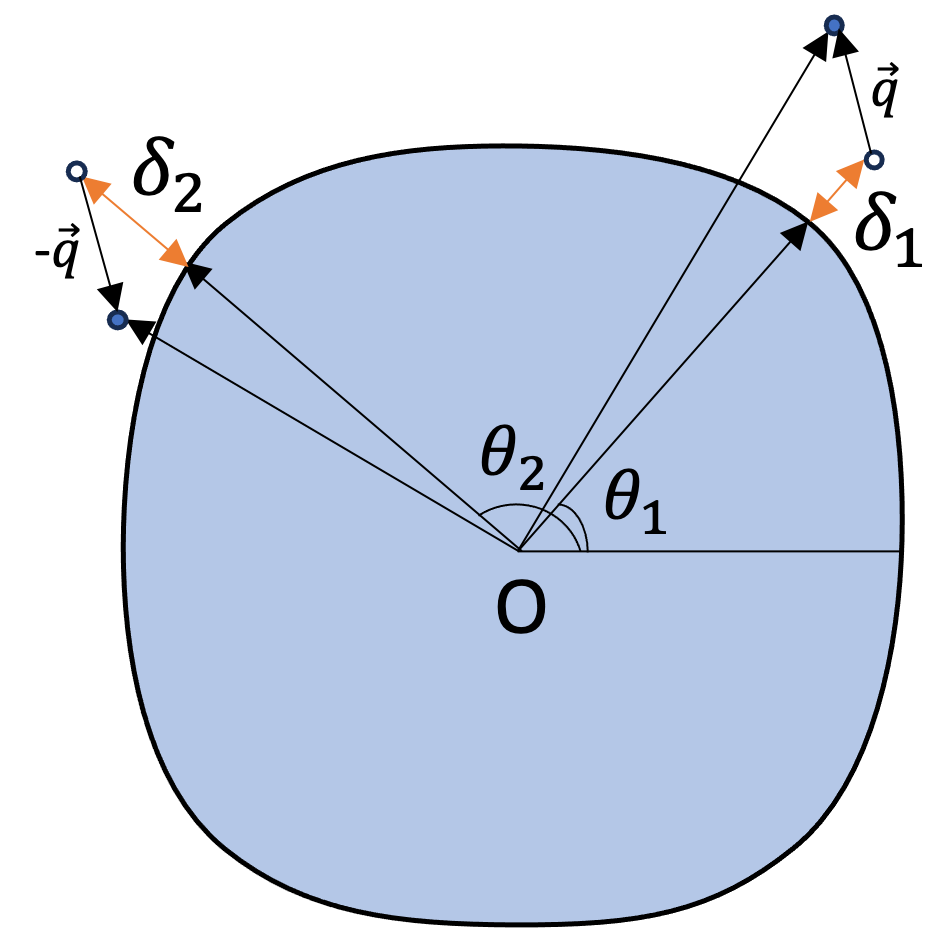}
  \caption{}
  \label{fig:forward_scattering}
\end{subfigure}%
\begin{subfigure}{.3\textwidth}
  \centering
  \includegraphics[width=1.0\linewidth]{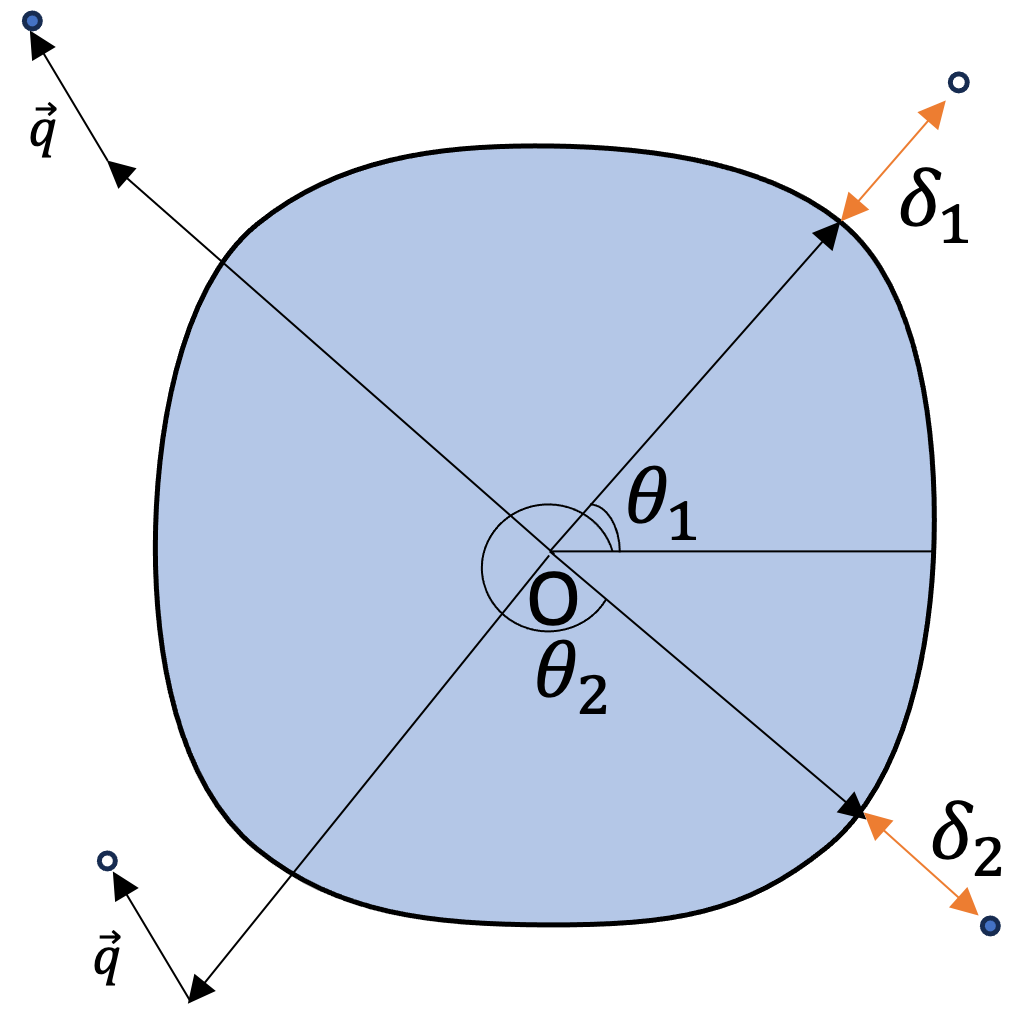}
  \caption{}
  \label{fig:bcs_scattering}
\end{subfigure}
\caption{
Low-energy two-body interactions of fermions.
(a) Particles at angle $\theta_1$ and $\theta_2$ exchange momentum $\vec q$ through a near forward scattering.
(b) A Cooper pair with total center of mass momentum $\vec q$ at 
angles $\theta_1$ and $\theta_1+\pi$
is scattered to angles $\theta_2$ and $\theta_2+\pi$.
At low energies,
$\vec q$ is small but generally non-zero.
}
\label{fig:4f_scattering}
\end{figure}

We consider a Fermi surface 
that undergoes
the Ising-nematic quantum phase transition 
associated with a spontaneous symmetry breaking of the four-fold rotational symmetry to the two-fold subgroup in (2+1)-dimensions.
At the quantum critical point, a real scalar mode that captures nematic fluctuations becomes gapless.
The effective theory for the nematic collective mode and low-energy fermions  can be written as 
\begin{equation}
\begin{aligned}
S & =    
\int d_f^3 {\bf k} ~
\psi_{j}^\dagger (k_0,\delta,\theta )
\Bigl[ 
i k_0   +v_{F,\theta  }\delta   \Bigr] \psi_{j}(k_0,\delta,\theta ) 
+ \frac{1}{2} 
\int d_b^3 {\bf q}~
|{\bf q}|^2 
\phi^*(\mathbf{q})
\phi({\bf q}) \\ &
+ \frac{1}{\sqrt{N}}
\int d_f^3 {\bf k} ~ d_b^3 {\bf q}~
\edim_{\thetasq,\theta  }
\phi(\mathbf{q})
\psi^\dagger_{j}\left(k_0+q_0,
 \deltaq,
  \thetasq 
\right)\psi_{j}(k_0,\delta,\theta ) 
+ \int d_f^3 {\bf k} ~ d_f^3 {\bf k}' ~ d_b^3 {\bf q}~
\lambdadim_{\left(\begin{smallmatrix}
    \thetasq  & \theta^{\prime} 
    \\ \theta      & \thetasqp   
    \end{smallmatrix}\right)}
~  
\times \\ & \hspace{1cm}
\psi^{\dagger}_{j_1}\left(k_0+q_0,
 \deltaq,
  \thetasq 
\right)
\psi_{j_1}(k_0,\delta,\theta)
\psi^\dagger_{j_2}
(k^{\prime}_0,\delta^{\prime},\theta^{\prime} )
\psi_{j_2}
\left(k^{\prime}_0+q_0, \deltaqp, \thetasqp  \right) \\&
+ 
\udim  \int 
d_b^3 {\bf q}_1~
d_b^3 {\bf q}_2~
d_b^3 {\bf q}_3
~
\phi({\bf q}_1)  \phi({\bf q}_2)
\phi({\bf q}_3)  \phi(
-{\bf q}_1
-{\bf q}_2
-{\bf q}_3
). 
\label{eq:action_2d}
\end{aligned}
\end{equation}
Here,
$\psi_{j} (k_0,\delta,\theta )$ represents the fermionic field  
with frequency $k_0$
and spatial momentum 
$\vec k = 
(\KFthetadim+\delta) \left( 
\cos \theta, \sin \theta \right)$,
where $\KFthetadim$ is the 
 magnitude of Fermi momentum at angle $\theta$,
and $\delta$ is the radial displacement away from the Fermi surface
as is illustrated in 
  \fig{Fig:Fermi_Surface_Angles}.
We use $(\delta, \theta)$ to denote spatial momentum of fermion 
because $\delta$ not $\vec k$ is what is sent to zero in the low-energy limit.
We call this Fermi-polar coordinate.
Repeated flavour indices are  understood to be summed over $1 \leq j \leq N$.
$\phi({\bf q})$ with ${\bf q}=(q_0,\vec q)$ represents the critical boson with frequency $q_0$ and spatial momentum $\vec q = q ( \cos \varphi, \sin \varphi)$. 
The momentum of the boson is measured relative to the origin as the energy 
 of boson vanishes at zero momentum unlike fermions.
$\int d_f^3 {\bf k} \equiv   \int\frac{dk_0}{2\pi} \int  \frac{d\delta}{2\pi}   \int_{-\pi}^{\pi} \frac{d\theta }{2\pi}   \KFthetadim$
and
$\int d_b^3 {\bf q}  \equiv \int \frac{dq_0}{2\pi}   \int \frac{dqq}{2\pi}   \int_{-\pi}^{\pi} \frac{d\varphi }{2\pi}$
represent the Cartesian measure of fermionic and bosonic three-momenta, respectively
\footnote{The exact Cartesian measure for fermion is $\int\frac{dk_0}{2\pi} \int  \frac{d\delta}{2\pi}   \int_{-\pi}^{\pi} \frac{d\theta }{2\pi}   (\KFthetadim+\delta)$, but $(\KFthetadim+\delta)$ can be replaced with $\KFthetadim$ at low energies.}.
$v_{F,\theta}$ is the radial component of the Fermi velocity. 
The size and shape of the Fermi surface, and the dispersion of low-energy fermions are completely specified by two functions $\KFthetadim$
and $v_{F,\theta}$.
$\thetasq $
and
$\deltaq$
denote the angle and the radial displacement of $\vec k + \vec q$.
For $q,\delta \ll \KFthetadim$, they are given by
(see Appendix \ref{app:Momentum_conservation} for derivation),
\bqa
\thetasq 
= \theta+\mathscr{A}_{\varphi,\theta}\frac{q}{\KFthetadim}
\label{angle_scattered}, ~~~~~
\deltaq = \delta+\mathscr{F}_{\varphi,\theta}q+\mathscr{G}_{\varphi,\theta}\frac{q^2}{\KFthetadim}+\frac{q\delta}{\KFthetadim}\mathscr{I}_{\varphi,\theta},
\label{eq:ThetaDelta}
\eqa
where
\begin{equation}
    \begin{aligned}
        &\mathscr{A}_{\varphi,\theta} = 
       \sin(\varphi-\theta),
       ~~\mathscr{F}_{\varphi,\theta} =  \left(
        \cos(\varphi -\theta )-
        \sin(\varphi -\theta )
        \frac{{\bf K}^{\prime}_{F,\theta  }}{\KFthetadim}\right),\\
      &\mathscr{G}_{\varphi,\theta} =  
      \frac{1}{2}\left( 
        \sin^2(\varphi -\theta )\left[1-\frac{{\bf K}^{\prime\prime}_{F,\theta}}{\KFthetadim}\right]+\sin\left(2\left(\varphi-\theta\right)\right)\frac{{\bf K}^{\prime}_{F,\theta}}{\KFthetadim}\right),~~
        \mathscr{I}_{\varphi,\theta} = \frac{\KFthetadim^\prime}{\KFthetadim}\sin\left(\varphi-\theta\right)
\label{eq:scattering_functions}
    \end{aligned}
\end{equation}
 with
${\bf K}^{\prime}_{F,\theta}$
and
${\bf K}^{\prime\prime}_{F,\theta}$
representing the first 
and second derivatives of 
$\KFthetadim$ with respect to $\theta$.
For low-energy bosons with $q/\KFthetadim \ll 1$,
we only need the diagonal coupling function  $\edim_{\theta,\theta}$.
At a UV scale,
it takes the form of
\bqa
\edim_{\theta,\theta}  
\sim \cos(2 \theta),
\eqa
where the form factor represents the nematic nature of the boson in the d-wave channel
as is shown in \fig{fig:dwave_form}. 
Accordingly, the coupling between the critical nematic mode and fermions vanishes along the nodal directions at
$\theta =  (2n+1) \pi/4$ for integer $n$.
$\lambdadim_{\left(\begin{smallmatrix} \theta_1  & \theta_2 \\ \theta_4     & \theta_3   \\ \end{smallmatrix}\right)}$ 
denotes the four-fermion coupling that scatters low-energy fermions from angles 
$\theta_4$ and $\theta_3$
to 
$\theta_1$ and $\theta_2$.
In the four-fermion coupling, $\theta_i$ are not necessarily close to each other, and $\lambdadim$ include both the forward and pairing interactions
as is shown in
\fig{fig:4f_scattering}.
$\udim$ represents the quartic boson coupling.
%
%

In two space dimensions, 
the theory becomes strongly coupled at low energies.
In order to access the interacting low-energy theory in a controlled manner, we extend the spatial dimension by embedding the one-dimensional Fermi surface in a $d$-dimensional momentum space.
To understand the construction, it is instructive to first consider the case in three space dimensions in which 
the usual Fermi surface is two-dimensional.
One can gap out the electronic mode except for the ring located at $k_z=0$ by turning on a $p_z$-wave pairing amplitude,
and consider the Ising-nematic quantum phase transition for the nodal line.
In general $2 \leq d \leq 3$, one adds $(d-2)$ extra spatial dimensions $(k_1,..,k_{d-2})$,
which provides one-parameter family of theories that interpolates between the original theory in $d=2$ 
and the $p_z$-wave superconducting state with a nodal ring in $d=3$\cite{DENNIS}.
One can combine the frequency and the extra components of momentum to form a $(d-1)$-dimensional vector ${\bf K}=(k_0,..,k_{d-2})$.
Imposing the $O(d-1)$ symmetry in ${\bf K}$ fixes the form of the action in general $d$. 
Due to the pairing term, the $U(1)$ charge conservation symmetry is broken down to $Z_2$ in $d>2$.
In this construction, it is convenient to interpret the low-energy modes near the Fermi surface as a collection of (1+1)-dimensional Dirac fermion,
where a two-component spinor is constructed out of fermion fields at angle $\theta$ and $\theta+\pi$ as
$
\Psi_j(k_0,\delta,\theta ) = \left( 
\begin{array}{c}
	\psi_{j}(k_0,\delta ,\theta  ) \\
\psi_{j}^\dagger(-k_0,\delta ,\theta  +\pi) 
\end{array}
\right)$.
The theory in spatial dimensions $d$ is written as 
\begin{equation}\begin{aligned}
   S & =   
  \int^{'} d_{f}^{d+1} {\bf k}~
\bar \Psi_j(\mathbf{K},\delta,\theta )
\Bigl[ 
i {\bf \Gamma} \cdot{\bf K} + i 
\gamma_{d-1} 
v_{F,\theta  }
\delta 
\Bigr]
\Psi_{j}(\mathbf{K},\delta,\theta ) 
+  \frac{1}{2}
\int d_{b}^{d+1} {\bf q}~
|{\bf q}|^2
\phi^*(\mathbf{q})\phi(\mathbf{q})
\\
&+ \frac{i}{\sqrt{N}} 
\int^{'}  d_{f}^{d+1} {\bf k} ~
\int d_{b}^{d+1} {\bf q}~
\edim_{
\thetasq ,
\theta  }
~
\phi(\mathbf{q})
\bar{\Psi}_{j}\left(\mathbf{K}+\mathbf{Q},\deltaq,\thetasq \right)\gamma_{d-1} \Psi_{j}(\mathbf{K},\delta,\theta )\\
&+ 
\sum_{\nu=F_\pm,P}
~
\sum_{s=d,e}
\int^{'}
d_{f}^{d+1} {\bf k} ~
d_{f}^{d+1} {\bf k}' ~
\int
d_{b}^{d+1} {\bf q}~~
\lambdadim^{(\nu,s)}_{\theta', \theta}(\vec q)
~~
{\bf O}^{(\nu,s)} ( \mathbf{K}^{\prime},  \delta^\prime, \theta'; \mathbf{K}, \delta, \theta; \mathbf{Q}, \vec q) \\&
+ 
\udim  \int 
d_b^{d+1} {\bf q}_1~
d_b^{d+1} {\bf q}_2~
d_b^{d+1} {\bf q}_3
~
\phi({\bf q}_1)  \phi({\bf q}_2)
\phi({\bf q}_3)  \phi(
-{\bf q}_1
-{\bf q}_2
-{\bf q}_3
). 
\label{action_generald}
\end{aligned}
\end{equation}
The Fermi surface is folded into the half as one spinor covers a pair of anti-podal points.
The measure of the fermionic $(d+1)$-momentum integration becomes $\int^{'} d_{f}^{d+1} {\bf k} \equiv   \int\frac{d^{d-1}{\bf K}}{2\pi} \int  \frac{d\delta}{2\pi}   \int_{-\pi/2}^{\pi/2} \frac{d\theta }{2\pi}   \KFthetadim$.
$\mathbf{\Gamma}$'s are the gamma matrices associated with \textbf{K}. 
Similarly, the $(d+1)$-dimensional boson frequency-momentum vector is written as
${\bf q} = ({\bf Q}, \vec q)$ with $\vec q = q ( \cos \varphi, \sin \varphi)$
and
$\int d_{b}^{d+1} {\bf q} \equiv   \int\frac{d^{d-1}{\bf Q}}{2\pi} \int  \frac{dq  q}{2\pi}   \int_{-\pi}^{\pi} \frac{d\varphi }{2\pi}   $.
$\lambdadim^{(\nu,s)}_{\theta', \theta}(\vec q)$ denotes the coupling function 
in channel $(\nu,s)$.
$\nu=F_+, F_-$ represent the forward scattering in the even and odd angular momentum channels, respectively,
and
$\nu=P$ denotes the pairing channel.
In the pairing channel, one does not need to consider even and odd angular momentum channels 
separately because 
the anti-symmetric nature of fermion pair wavefunction fixes
the parity of orbital angular momentum in terms of the flavour (spin) wavefunction that is to be specified below.
$s=d, e$ denote the direct and exchange channels for the flavour quantum number, respectively.
In each channel,
the four-fermion operator is written as
\bqa
&& {\bf O}^{(\nu,s)} 
( \mathbf{K}^{\prime},  \delta^\prime, \theta'; \mathbf{K}, \delta, \theta; \mathbf{Q}, \vec q) 
  \equiv 
\nn && \hspace{0.5cm} 
T^{(\nu,s)}_{\left(\begin{smallmatrix}     j_1  & j_2      \\ j_4      & j_3       \end{smallmatrix}\right)}
 \bar{\Psi}_{j_1}
(\mathbf{K}^{\prime},\delta^{\prime},\theta^{\prime} )
 I^{(\nu)}_m
 \Psi_{j_4}
\left(\mathbf{K}'+\mathbf{Q}, \deltaqp, \thetasqp  \right)  
~
\bar{\Psi}_{j_2}\left(\mathbf{K}+\mathbf{Q},
 \deltaq, \thetasq 
\right)
I^{(\nu)}_m
\Psi_{j_3}(\mathbf{K},\delta,\theta).
\label{eq:Onus}
\eqa
For each $\nu=F_+,F_-,P$, 
$m$ is summed over 
$1$, $(d-1)$, $(4-d)$ matrices, respectively,
where their elements are given by
\begin{equation}\begin{aligned}
   I^{(F_+)} = i\gamma_{d-1}, ~~~
   I^{(F_-)} = 
   (\gamma_0,..,\gamma_{d-2}), ~~~
I^{(P)} = ( \mathbbm{1}, i\gamma_d,...,i\gamma_2).
   \label{4f_channels}
\end{aligned}\end{equation}
$\{ \mathbbm{1}, \gamma_0,..,\gamma_2 \}$ form the complete set of $2 \times 2$ matrices.
It is noted that ${\bf \Gamma}$ used in the kinetic term is nothing but
$I^{(F_-)}$.
From now on, all repeated indices are assumed to be summed over.
In general $d$, we choose $\gamma_0$ = $\sigma_y$
and 
$\gamma_{d-1}$ = $\sigma_x$.
The fermion bi-linear $ \bar \Psi({\bf k}+{\bf q}) I^{(\nu)}_m\Psi({\bf k}) $ represents a particle-hole pair with momentum ${\bf q}$ for $\nu=F_\pm$ and a particle-particle pair with the center of mass momentum ${\bf q}$ for $\nu=P$\footnote{ 
Fluctuations of Fermi surface can be described in terms of these bosonic variables\cite{ PhysRevResearch.4.033131, 
10.21468/SciPostPhys.16.3.069,
2023arXiv230614955H}
}.
$s=d$ or $e$ controls how the flavour indices are contracted through
\bqa
T^{(\nu,s)}_{\left(\begin{smallmatrix}     j_1  & j_2      \\ j_4      & j_3       \end{smallmatrix}\right)} =
\left\{
\begin{array}{cl}
\delta_{j_1,j_4} \delta_{j_2,j_3}, 
&   \nu=F_\pm,~  s=d     \\
\delta_{j_1,j_3} \delta_{j_2,j_4},
&  \nu=F_\pm, ~  s=e 
~~~\mbox{or}~~~ \nu=P, ~ s=d
\\
\delta_{j_1,j_2} \delta_{j_3,j_4},
&  \nu=P, ~ s=e     
\end{array}
\label{Flavour_Tensor}
\right..
\eqa%

One advantage of the spinor representation lies in the fact that the forward scattering and the pairing interaction are 
 kinematically `unified'.
Since a particle-hole pair and a Cooper pair
are both represented by $\bar \Psi(k+q) I_m \Psi(k)$
for some $I_m$, 
the same kinematic constraints enter into quantum corrections in both channels.
Since we include the direct and exchange channels separately,
we only need to consider fermion bi-linears with small ${\bf q}$ 
in the four-fermion operators
at low energies.
In particular, we don't need to separately include four-fermion operators that involve bi-linears with large momenta such as 
\bqa
 \bar{\Psi}_{j_1}
(\mathbf{K}^{\prime},\delta^{\prime},\theta^{\prime} )
 I^{(\nu)}_m
\Psi_{j_3}(\mathbf{K},\delta,\theta)
\bar{\Psi}_{j_2}\left(\mathbf{K}+\mathbf{Q},
 \deltaq, \thetasq 
\right)
I^{(\nu)}_m
 \Psi_{j_4} \left(\mathbf{K}'+\mathbf{Q}, \deltaqp, \thetasqp  \right)
\label{eq:Onus2}
\eqa
because it can be cast into
the form of \eq{eq:Onus}
through the Fierz transformation
(see Appendix \ref{sec:Fierz_Transformation} for detail).

In the $d \rightarrow 2$ limit, 
Eq. \eqref{action_generald}
is reduced to the low-energy limit of the original action
provided that 
$\gamma_0$ = $\sigma_y$,
$\gamma_{1}$ = $\sigma_x$,
$\gamma_2 = \sigma_z$.
In order to see this, 
it is useful to explicitly write \eq{eq:Onus} 
in $d=2$\footnote{
For simplicity, Eq. (\ref{4f_Operator}) is written for $\vec q=0$ with the dependence on frequency and $\delta$ suppressed.
However, it is crucial to allow small but non-zero $\vec q$ in the full action for locality.
},
\begin{equation}
    \begin{aligned}
    {\bf O}^{(F_+,s)}_{\theta', \theta}
        &=
        T^{(F_+,s)}_{\left(\begin{smallmatrix}     j_1  & j_2      \\ j_4      & j_3       \end{smallmatrix}\right)}\left[\psi^{\dagger}_{j_1}\left(\theta^{\prime}\right)\psi_{j_4}\left(\theta'\right)\psi^{\dagger}_{j_2}\left(\theta\right)\psi_{j_3}\left(\theta\right)+\psi^{\dagger}_{j_1}\left(\theta^{\prime}\right)\psi_{j_4}\left( {\theta'}\right)\psi^{\dagger}_{j_3}\left( \theta+\pi\right)\psi_{j_2}\left(\theta+\pi\right)\right.\\&\left.+
        \psi^{\dagger}_{j_4}\left( \theta^{\prime}+\pi\right)\psi_{j_1}\left(\theta^{\prime}+\pi\right)\psi_{j_2}^{\dagger}\left(\theta\right)\psi_{j_3}\left(\theta\right)+
        \psi^{\dagger}_{j_4}\left(\theta'+\pi\right)\psi_{j_1}\left(\theta^{\prime}+\pi\right)\psi^{\dagger}_{j_3}\left( \theta+\pi\right)\psi_{j_2}\left(\theta+\pi\right)
        \right], \\[10pt]
        {\bf O}^{(F_-,s)}_{\theta', \theta}
        &=
        T^{(F_-,s)}_{\left(\begin{smallmatrix}     j_1  & j_2      \\ j_4      & j_3       \end{smallmatrix}\right)}\left[\psi^{\dagger}_{j_1}\left(\theta^{\prime}\right)\psi_{j_4}\left(\theta'\right)\psi^{\dagger}_{j_2}\left(\theta\right)\psi_{j_3}\left(\theta\right)-\psi^{\dagger}_{j_1}\left(\theta^{\prime}\right)\psi_{j_4}\left( {\theta'}\right)\psi^{\dagger}_{j_3}\left( \theta+\pi\right)\psi_{j_2}\left(\theta+\pi\right)\right.\\&\left.-
        \psi^{\dagger}_{j_4}\left( \theta^{\prime}+\pi\right)\psi_{j_1}\left(\theta^{\prime}+\pi\right)\psi_{j_2}^{\dagger}\left(\theta\right)\psi_{j_3}\left(\theta\right)+
        \psi^{\dagger}_{j_4}\left(\theta'+\pi\right)\psi_{j_1}\left(\theta^{\prime}+\pi\right)\psi^{\dagger}_{j_3}\left( \theta+\pi\right)\psi_{j_2}\left(\theta+\pi\right)
        \right], \\[10pt]
        {\bf O}^{(P,s)}_{\theta', \theta}
        &=
                2T^{(P,s)}_{\left(\begin{smallmatrix}     j_1  & j_2      \\ j_4      & j_3       \end{smallmatrix}\right)}\left[\psi^{\dagger}_{j_1}\left(\theta^{\prime}\right)\psi^{\dagger}_{j_4}\left(\theta'+\pi \right)\psi_{j_2}\left(\theta+\pi\right)\psi_{j_3}\left(\theta\right)+\psi_{j_2}^{\dagger}\left(\theta\right)\psi^{\dagger}_{j_3}\left( \theta+\pi\right)\psi_{j_1}\left( \theta^{\prime}+\pi\right)\psi_{j_4}\left(\theta^{\prime}\right)
        \right].
\label{4f_Operator}
    \end{aligned}
\end{equation}
The full low-energy theory is specified by
$\Bigl\{ 
v_{F,\theta},
\KFthetadim,
\edim_{\theta,\theta'}, 
\lambdadim^{(\nu,s)}_{\theta,\theta'}(\vec q),
\udim
\Bigr\}$.
In relativistic field theories, it is sufficient to expand momentum-dependent coupling functions near the origin and keep a few coupling constants associated with low powers of momentum.
In metals, one has to consider general coupling functions of the angle 
 around the Fermi surface 
 because gapless degrees of freedom are parameterized by the angle.

\subsection{Symmetry}
\label{sec:symmetry}

\begin{table}[h]
\centering
 \begin{tabular}{cccc}
 \hline
 \ttfamily  & &\ttfamily $d=2$ & \ttfamily $d>2$ \\                \hline
  ~~~(a) UV symmetry ~~ & & \ttfamily ~~$U(1) \times SU(N)$~~ & \ttfamily ~~$Z_2 \times SO(N)$~~ \\                \hline
 \multirow{2}{*}{~~(b) IR symmetry~~} & 
 \multicolumn{1}{l}{~i. ~
 $\lambda_{\theta,\theta'}  \sim \delta(\theta-\theta')$
 ~~}  
 & \multicolumn{1}{l}{~~$LU(1) \times ELSU(N)$~~} 
 & \multicolumn{1}{l}{~~$OLU(1) \times ELSO(N)$~~} \\\cline{2-4}
 & \multicolumn{1}{l}{~ii. ~ 
 \mbox{general} $\lambda_{\theta,\theta'}$
 ~~ }  & \multicolumn{1}{l}{~~$OLU(1) \times SU(N)$~~}  & \multicolumn{1}{l}{~~$OLU(1) \times SO(N)$~~} \\
    \hline
    \end{tabular}
\caption{
(a) The UV symmetry of the theory.  
(b)-i. 
The IR symmetry that emerges if the inter-patch four-fermion couplings outside the pairs of anti-podal patches are irrelevant.  
(b)-ii.  The IR symmetry that emerges if general inter-patch couplings are relevant.
Under each group, fermion angle at $\theta$ is transformed as follows. 
U(1) :  
$\psi_j(\theta) \rightarrow e^{i \gamma} \psi_j(\theta)$, 
SU(N) :  
$\psi_j(\theta) \rightarrow U_{ji} \psi_i(\theta)$ with 
$U \in SU(N)$, 
$Z_2$ :  $\psi_j(\theta) \rightarrow \pm \psi_j(\theta)$, 
SO(N) :  $\psi_j(\theta) \rightarrow U_{ji} \psi_i(\theta)$ with $U \in SO(N)$, 
loop U(1) [LU(1)]  :  $\psi_j(\theta) \rightarrow e^{i \gamma(\theta)} \psi_j(\theta)$, even loop SU(N) [ELSU(N)] : $\psi_j(\theta) \rightarrow U_{ji}(\theta) \psi_i(\theta)$ with $U(\theta) \in SU(N)$ and $U(\theta+\pi)=U(\theta)$, odd loop U(1) [OLU(1)] :  $\psi_j(\theta) \rightarrow e^{i \gamma(\theta)} \psi_j(\theta)$ with  $\gamma(\theta+\pi)= - \gamma(\theta)$, even loop SO(N) [ELSO(N)] : $\psi_j(\theta) \rightarrow U_{ji}(\theta) \psi_i(\theta)$ with  $U(\theta) \in SO(N)$ and  $U(\theta+\pi)=U(\theta)$. For the loop groups, group elements are smooth functions of angle.
}
\label{tab:symmetry}
\end{table}

In this section, we discuss the symmetry of the theory.

First, we consider the $SO(3)$ group under 
which $\Psi$ forms the spinor representation.
Here, it is simpler to consider $\Psi$ and $\bar \Psi$ as independent fields which are tranformed as
$\Psi \rightarrow e^{-\frac{i}{4} \Sigma_{\alpha \beta} \omega_{\alpha \beta}} \Psi$
and
$\bar \Psi \rightarrow 
\bar \Psi 
\gamma_{d-1}
e^{\frac{i}{4} \Sigma_{\alpha \beta} \omega_{\alpha \beta}}
\gamma_{d-1}$
with $\Sigma_{\alpha \beta}=\frac{i}{2}[\gamma_\alpha,\gamma_\beta]$.
Under this $SO(3)$,
$\bar \Psi \gamma_{d-1} \Psi$ is a scalar
and
$\bar \Psi 
( 
\gamma_0,...,\gamma_{d-2},
\mathbbm{1}, 
\gamma_d,...,\gamma_2) \Psi$
is a vector.
The action in \eq{action_generald} 
respects its subgroup, $SO(d-1) \times SO(4-d)$.
Under this unbroken group,
$\bar \Psi I^{(F_+)}_m \Psi$,
$\bar \Psi I^{(F_-)}_m \Psi$
and
$\bar \Psi I^{(P)}_m \Psi$
in \eq{4f_channels}
behave as  
${\bf 1} \times {\bf 1}$,
${\bf (d-1)}\times {\bf 1}$
and
${\bf 1}\times {\bf (4-d)}$, respectively.
Namely,
$\bar \Psi I^{(F_-)}_m \Psi$
and 
$\bar \Psi I^{(P)}_m \Psi$
are vectors under $SO(d-1)$ and $SO(4-d)$, respectively, whereas
$\bar \Psi I^{(F_+)}_m \Psi$
is a scalar. 
In $d=2$, the non-trivial $SO(2)$ corresponds to the global U(1) symmetry.
In $d=3$, the non-trivial $SO(2)$ is a spacetime symmetry acting on time and the extra co-dimension.
As $d$ decreases from $3$ to $2$, the unbroken symmetry evolves from one $SO(2)$ to the other.

Next, we consider the internal symmetry.
In $d=2$, the UV theory has the 
$SU(N)$ flavour symmetry as well as
the $U(1)$ charge symmetry.
In $d>2$, $U(1) \times SU(N)$ 
 is broken to its real subgroup, $Z_2 \times SO(N)$ due to the pairing term that 
 is introduced to keep the dimension of Fermi surface fixed.
In the kinetic term, 
$\sum_{\mu=1}^{d-2} \bar \Psi \Gamma_\mu K_\mu \Psi$
is proportional to
$\{ \psi_j(\theta)
\psi_j(\theta+\pi) 
+ h.c. \}$ in $d=3$.
Similarly, the four-fermion operator 
 in the pairing channel includes terms that break the global U(1) symmetry :
in $d=3$,  ${\bf O}^{(P,s)}_{\theta', \theta}$ becomes
\bqa
{\bf O}^{(P,s)}_{\theta', \theta} 
= T^{(P,s)}_{\left(\begin{smallmatrix}     j_1  & j_2      \\ j_4      & j_3       \end{smallmatrix}\right)}\left[-\psi^{\dagger}_{j_1}\left(\theta^{\prime}\right)\psi^{\dagger}_{j_4}\left(\theta'+\pi \right)\psi^{\dagger}_{j_2}\left (\theta\right)\psi^{\dagger}_{j_3}\left(\theta+\pi\right) +\psi^{\dagger}_{j_1}\left(\theta^{\prime}\right)\psi^{\dagger}_{j_4}\left(\theta'+\pi\right)\psi_{j_2}\left(\theta+\pi\right)\psi_{j_3}\left(\theta\right)
\right. \\
\left.
+ \psi_{j_2}^{\dagger}\left(\theta\right)\psi^{\dagger}_{j_3}\left( \theta+\pi\right)\psi_{j_1}\left(\theta^{\prime}+\pi\right)\psi_{j_4}\left(\theta^{\prime}\right)-\psi_{j_1}\left(\theta^{\prime}+\pi\right)\psi_{j_4}\left(\theta'\right)\psi_{j_2}\left(\theta+\pi\right)\psi_{j_3}\left(\theta\right) \right].
\eqa
These symmetry breaking effects are artifacts of the  dimensional regularization  scheme which go away in the $d \rightarrow 2$ limit.

In the low-energy limit,
the energetic constraint that restricts fermions to be on the Fermi surface enlarges
the symmetries.
At low energies,
momentum carried by particle-hole or particle-particle pairs becomes small.
Accordingly, the four-fermion coupling function 
 with non-zero $q$ is suppressed\footnote{
Later, we will show that the four-fermion coupling at energy scale $\mu$ scales as
\bqa
\lambdadim^{(\nu,s)}_{\theta', \theta}(\vec q; \mu)
 \sim \left( \frac{ q^*(\theta',\theta,\mu)}{q} \right)^{\Delta_q}
\lambdadim^{(\nu,s)}_{\theta', \theta}(\vec 0; \mu)
\eqa
for $q$ that is much larger than a characteristic momentum scale  $q^*(\theta',\theta,\mu)$ that goes to zero in the small $\mu$ limit,
where 
$\Delta_q$
is a positive exponent.}.
%
The IR symmetry is then determined by the coupling functions  in the small $q$ limit.

To understand possible emergent low-energy symmetries, we begin by considering the limit in which the couplings between different patches of Fermi surface are negligible except for those between anti-podal patches.
In this case, 
the global symmetry is enlarged to a loop group as the symmetry rotation can depend on angle around the Fermi surface\cite{2005cond.mat..5529H,PhysRevX.11.021005}.
In $d=2$,
the U(1) and SU(N) groups are enlarged to the loop U(1) [LU(1)] and the even loop SU(N) [ELSU(N)] groups, respectively.
Under this symmetry group, the fermion field is transformed as
$\psi_j(\theta) \rightarrow e^{i \gamma(\theta)} 
U_{ji}(\theta) 
\psi_i(\theta)$ 
with smooth angle-dependent U(1) phase $\gamma(\theta)$
and SU(N) element $U(\theta)$
with 
$U(\theta+\pi)=U(\theta)$.
For the loop SU(N) group, 
only the transformations in even angular momentum channels are symmetry because 
the $2k_F$ scatterings mix flavour quantum number between anti-podal patches.
There is no such constraint for $\gamma(\theta)$ because the particle number remains conserved within each patch due to the Abelian nature for the 
 U(1) group.
In $d>2$, the pairing term lowers this loop group to
$OLU(1) \times ELSO(N)$,
where under the 
odd loop U(1) [OLU(1)]  
the fermion is transformed as
$\psi_j(\theta) \rightarrow e^{i \gamma(\theta)} \psi_j(\theta)$ with  $\gamma(\theta+\pi)= - \gamma(\theta)$, 
and the even loop SO(N) [ELSO(N)] 
is the SO(N) counterpart of $ELSU(N)$.

%

%
%
In the presence of general inter-patch couplings, the non-Abelian loop group is broken down to the global subgroup.
On the other hand, the Abelian loop group is less fragile.
In $d>2$, the OLU(1) remains intact because general inter-patch couplings still preserve the number of fermions in the odd angular momentum channels.
What is less clear is the fate of LU(1) in $d=2$.
To figure out the emergent symmetry in $d=2$, it is useful to understand how the LU(1) symmetry breaking terms evolve as a function of $d$.
In $d>2$,
there are two sources 
 that break LU(1) to \OLU.
The first is the pairing term introduced in the regularization scheme, which explicitly breaks the global U(1) symmetry in the action.
This symmetry breaking effect disappears in the $d \rightarrow 2$ limit.
The second is the one that keeps the global U(1) but breaks LU(1) through large-angle scatterings.
In particular, the pairing interaction in \eq{4f_Operator}
scatters a Cooper pair 
at angles $\theta$ and $\theta+\pi$ to $\theta'$ and $\theta'+\pi$.
While this respects the global U(1) symmetry, it breaks LU(1) to \OLU if the interaction is non-negligible for large $|\theta' - \theta|$.
Whether \OLU 
 is enhanced to LU(1) or not in the $d \rightarrow 2$ limit
hinges on the profile of 
$
\lambdadim^{(P,s)}_{\theta', \theta}(\vec q)
$ as a function of $\theta'$ and $\theta$.
If it decays `quickly' at large 
$|\theta'- \theta|$,
inter-patch 
 scatterings are suppressed,
and
the pairing interaction remains more or less local in the momentum space.
In this case,
\OLU is enhanced to LU(1) in the $d \rightarrow 2$ limit
due to the emergent momentum-space locality\cite{PhysRevX.11.021005}.
On the other hand,  
if large-angle Cooper pair scatterings are strong,
the low-energy symmetry remains to be \OLU in $d=2$.
In 
Secs. 
\ref{sec:NFLfp} 
and
\ref{sec:SC_locality},
we will discuss the precise criterion for 
the relevance/irrelevance of the large-angle scatterings.
The leading-order calculation suggests that the low-energy symmetry of the Ising-nematic quantum critical metal remains to be \OLU in $d=2$ due to 
universal superconducting fluctuations that create large-angle scatterings.

\section{
Field-theoretic functional renormalization group 
}
\label{sec:iii}

\subsection{Tree-level scaling}

We begin our renormalization group analysis by setting a tree-level scale transformation.
For a patch theory that only considers a small portion of Fermi surface,
one can employ an anisotropic scale transformation that assigns different scaling dimensions to the momentum components perpendicular and tangential to the patch.
Under the patch scaling,
a small patch of the full Fermi surface is `magnified' into a scale-invariant parabola\cite{DENNIS}.
However, parts of the Fermi surface that are far away from the  center of the patch are distorted in complicated ways.
This is fine as far as 
one only cares about observables defined within one patch {\it and}
the inter-patch coupling is weak.
However, the patch scaling is not convenient for describing the entire Fermi surface,
which is necessary in the presence of strong inter-patch couplings\footnote{
Even if the inter-patch coupling is weak,
a global description of Fermi surface is needed for observables that are non-local in momentum space.
}.

The fact that the patch scaling does not leave the whole Fermi surface invariant is due to a general property of metals:
there exists no scale transformation that keeps the local action of  a compact Fermi surface invariant.
This is because 
(1) in local theories, 
particle-hole or particle-particle pairs are allowed to carry small but non-zero momenta,
and 
(2) the Fermi momentum measured in the unit of the typical momentum of the bosonic modes
grows as the low-energy limit is taken.
Given that there exists no scale transformation that leaves the theory invariant even at the tree-level,
we choose the simplest scale transformation 
in which angles are dimensionless
while frequency and all components of momentum have scaling dimension $1$.
Under the scale transformation,
\begin{equation}\begin{aligned}
    \mathbf{K} = b^{-1}\tilde{\mathbf{K}},~~~\delta   = b^{-1}\tilde{\delta},~~~ 
   q=b^{-1} \tilde{q}, ~~~   
    \Psi(\mathbf{K},{\delta},{\theta})
    =b^{(\frac{d+2}{2})}\tilde{\Psi}(\tilde{\mathbf{K}}, \tilde{\delta}  ,{\theta } ), ~~~
\phi({\mathbf{Q}},{q},{\varphi} )=b^{(\frac{d+3}{2})}\tilde{\phi}(\tilde{\mathbf{Q}}, \tilde{q},{\varphi} )
\label{rescale_momenta_fields}
\end{aligned}
\end{equation}
with $b>1$,
the coupling functions 
are transformed into 
\begin{equation}
 \tilde{{\bf K}}_{F,{\theta }  }=
 b \KFthetadim,
 ~~~~~
\tilde{v}_{F,{\theta } }= v_{F,{\theta } },  ~~~~~
 \tilde\edim_{
 \theta_1 ,\theta_2 } =
 b^{\frac{3-d}{2}}
	\edim_{\theta_1 ,
 \theta_2},
 ~~~~~
\tilde \lambdadim_{ \theta, \theta'}(\vec q)
=
b^{1-d}
\lambdadim_{ \theta, 
\theta'}
(
b^{-1}
\vec q
), 
~~~~~
\tilde \udim
 =
b^{3-d}
\udim.
\label{eq:rescaledcouplings}
\end{equation}
%
With increasing $b$,
the momentum of fermion is sent toward the Fermi surface. 
One may think that the Fermi momentum should be left intact under such scale transformation.
However, this is not the case.
In \eq{eq:rescaledcouplings},
the Fermi momentum is transformed with dimension one and it becomes a relevant parameter of the theory.
This is because even low-energy scatterings that occur near the Fermi surface are kinematically controlled by the Fermi momentum\footnote{
One can in principle sense the size of the earth by observing how far our feet go above the ground under a linear displacement tangential to the surface of the earth.
}.
For example, a fermion on the Fermi surface that absorb a small but non-zero momentum $\vec q$ that is tangential to the Fermi surface changes its angle and energy by
$\Delta \theta \sim q/\KFthetadim$
and $\Delta \delta \sim q^2/\KFthetadim$.
In the rescaled variables,
these changes become
$\Delta \theta \sim \tilde q/(b \KFthetadim)$
and $\Delta \tilde \delta \sim \tilde q^2/(b \KFthetadim)$.
It captures the fact that
the Fermi momentum measured in the unit of the momentum $q$ and $\delta$ increases as $q$ and $\delta$
become smaller at low energies.
The fact that the Fermi momentum grows under the scale transformation does not contract Luttinger's theorem\cite{LUTTINGERFLT}.
While the volume of Fermi sea,
$V_{FS} = 
\frac{1}{(2\pi)^2}
\int_{ k< \KFthetadim }  
d\theta
dk  k
$ has scaling dimension $2$ in two space dimensions, the unit cell area $a^2$ has scaling dimension $-2$.
The dimensionless filling fraction 
$\nu = a^2 V_{FS}$,
which is what is protected from renormalization,
does not run under the scaling transformation.
As expected, the Fermi velocity is dimensionless. 
The Yukawa coupling, 
the four-fermion coupling 
and the quartic boson coupling
have dimensions 
$\frac{3-d}{2}$,
$1-d$
and
$3-d$,
respectively\footnote{
One can in principle choose a scale transformation in which the 
 Fermi momentum is kept invariant at the expense of endowing the angle around the Fermi surface a positive scaling dimension.
Under such a scale transformation, however, the kinematic constraints that govern the momentum-conservation in the fermion-boson interaction `run'.
In the end, the physics does not depend on how we choose our tree-level scale transformation.}.
%
%
However, infrared divergences that arise from the interactions are not controlled by these scaling dimensions 
because
the Fermi momentum is a relevant parameter 
and
the actual degree of infrared divergence is non-perturbatively modified by 
 it.
The mismatch between the scaling dimension and the degree of infrared singularity can go either way.
Couplings with negative scaling dimensions can exhibit infrared singularities,
 while some couplings with positive scaling dimensions are actually irrelevant at low energies.
The four-fermion interaction belongs to the first type.
Even if its scaling dimension is negative in $d>1$, it can give rise to infrared singularities at and above $d=2$ with the help of the extensive phase space of low-energy fermions. 
For example, Cooper pairs can be scattered across the entire Fermi surface
and the pairing interaction is enhanced by $\kFAV$.
As a result, the four-fermion coupling gives rise to the logarithmic singularity in metals with co-dimension one.
In non-Fermi liquids, even stronger infrared singularities can arise as the four-fermion couplings further acquire anomalous dimensions through critical collective modes as will be seen later.
This is why we keep the four-fermion coupling in the effective field theory.
On the other hand, the quartic boson coupling belongs to the second type.
Although it has a positive scaling dimension in $d<3$, it is rendered irrelevant everywhere in $2 \leq d \leq 3$ due to the Landau damping enhanced by Fermi momentum.

\begin{figure}[h]
 \centering
 \begin{subfigure}{.25\textwidth}
  \includegraphics[width=1\linewidth]{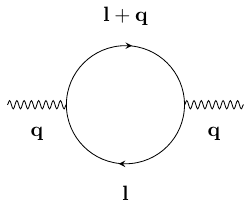}
  \caption{}
  \label{Fig:BSE}
\end{subfigure}
\hspace{5 mm}
\begin{subfigure}{.35\textwidth}
  \includegraphics[width=1\linewidth]{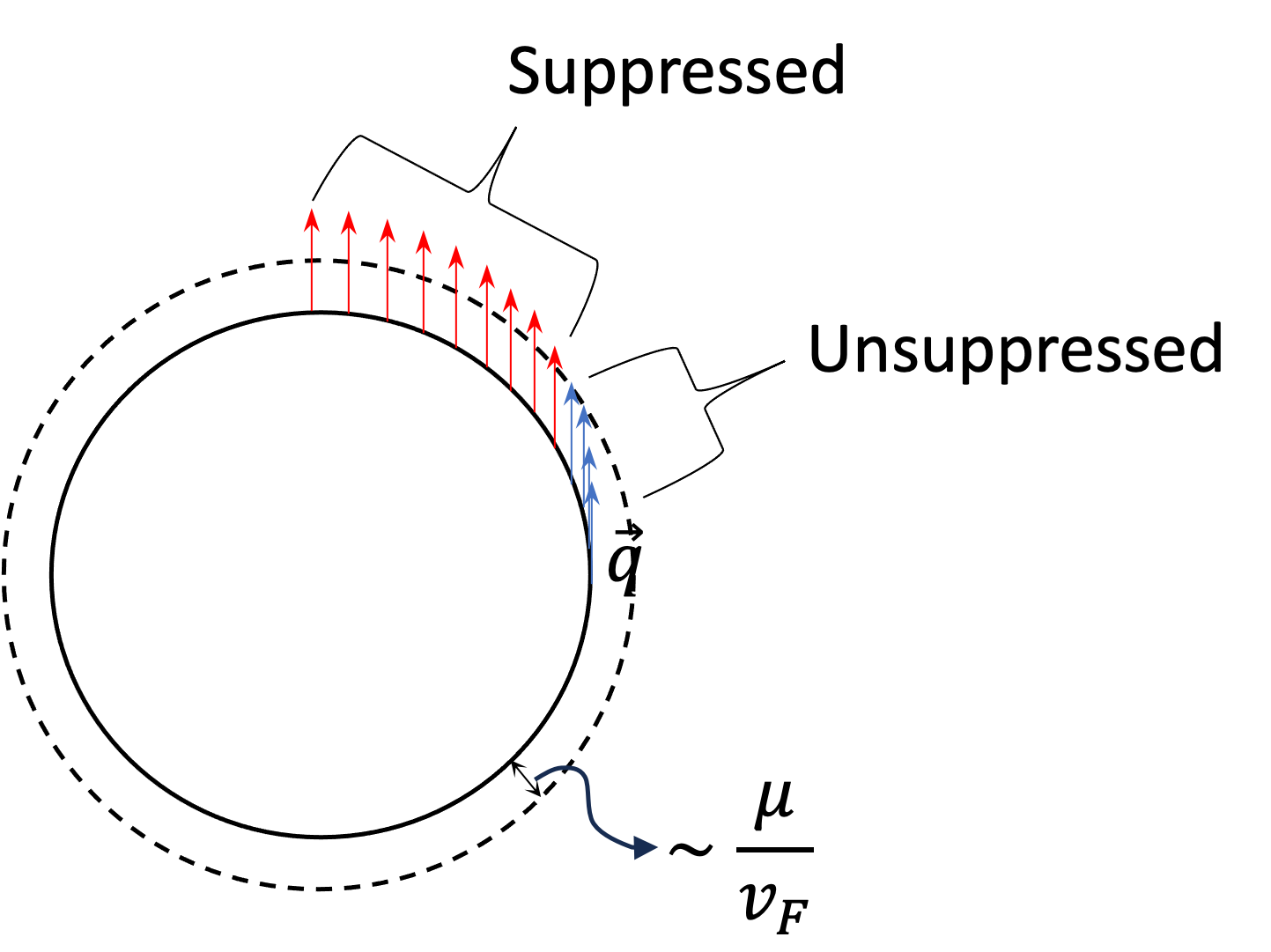}
  \caption{}
  \label{Fig:Lthetaq}
\end{subfigure}
\caption{
(a) One-loop boson self-energy.
(b) 
The energy of particle-hole pair with momentum $\vec q$ 
($L_\theta(\vec q)$) 
is small in the region where $\vec q$ is tangential to the Fermi surface.
At energy scale $\mu$, 
virtual excitations with 
$L_\theta(\vec q) \gg \mu$ 
are suppressed.
}
\end{figure}

The quartic boson coupling becomes irrelevant because the boson kinetic energy is non-perturbatively dressed by the Landau damping.
The one-loop boson self-energy in Fig. \ref{Fig:BSE} reads
(see Appendix \ref{sec:boson_self_energy})
\begin{equation}
    \begin{aligned}
        \Pi_1(\mathbf{q}) = v_d  \int_{-\frac{\pi}{2}}^{\frac{\pi}{2}}\frac{d\theta}{2\pi}\frac{\KFthetadim}{v_{F,\theta}}\boldsymbol{e}_{\theta}^2\frac{Q^2}{\left(Q^2+ \left(L_{\theta}(\vec q)\right)^2\right)^{\frac{4-d}{2}}}.
        \label{General_BSE}
    \end{aligned}
\end{equation}
Here, 
$\edim_\theta=\edim_{\theta,\theta}$
and
$v_d = \frac{1}{2}\frac{\Omega_{d-1}\Gamma\left(\frac{d-1}{2}\right)\Gamma\left(\frac{4-d}{2}\right)\Gamma^2\left(\frac{d}{2}\right)}{\Gamma\left(\frac{3}{2}\right)(2\pi)^{d-1}\Gamma(d)} $.
$L_{\theta}(\vec q)$ is the energy of the a particle-hole pair with momentum $\vec{q}$ created near the Fermi surface at angle $\theta$,
\begin{equation}
    \begin{aligned}
        L_{\theta}(\vec q) 
        = v_{F,\theta}\left(\mathscr{F}_{\varphi,\theta}q+
        \frac{\mathscr{G}_{\varphi,\theta}}{\KFthetadim}q^2\right).
        \label{eq:Lqvarphi}
    \end{aligned}
\end{equation}
Depending on the value of  $\frac{|\mathbf{Q}|}{q}$,
the self-energy takes different forms,
\begin{equation}\begin{aligned}
    \Pi_1(\mathbf{q})  =  
   \begin{cases}
   \boldsymbol{f}_{d,\varphi }\frac{|\mathbf{Q}|^{d-1}}{q}
    & \mbox{for} ~~~~~
q\gg|\mathbf{Q}|, \\
       \boldsymbol{p}_{d}|\mathbf{Q}|^{d-2} 
    &  \mbox{for} ~~~~~
q\ll|\mathbf{Q}|,
   \end{cases} 
   \label{eq:Pi1}
\end{aligned}\end{equation}
where
$         \boldsymbol{f}_{d,\varphi } = \beta_d\frac{\mathbf{K}_{F,\vartheta(\varphi )}\boldsymbol{e}_{\vartheta(\varphi )}^2}{|\chi_{\varphi}|v^2_{F,\vartheta(\varphi )}}$
with
$\beta_d = \frac{\Gamma^{2}(\frac{d}{2})}{2^{d-1}\pi^{\frac{d-1}{2}}\Gamma(d)\Gamma(\frac{d-1}{2})|cos(\frac{\pi d}{2})|}$
and
$\boldsymbol{p}_{d} = v_d \int_{-\frac{\pi}{2}}^{\frac{\pi}{2}}\frac{d\theta}{2\pi}\frac{\KFthetadim}{v_{F,\theta}}\boldsymbol{e}_{\theta}^2 $.
$\vartheta(\varphi) = \varphi -\arctan\left(\frac{\mathbf{K}_{F,\vartheta(\varphi )}}{{\bf K}^{\prime}_{F,\vartheta(\varphi )}}\right)$
is the angular position at which $\vec q = q (\cos \varphi, \sin \varphi)$ is tangential to the Fermi surface.
In this paper, we focus on the case in which the Fermi surface has no inflection points and 
$\vartheta(\varphi)$ is single-valued within $-\pi/2 \leq \theta < \pi/2$\footnote{For Fermi surfaces with inflection points,
$\vartheta(\varphi)$ is multi-valued. 
}.
$\chi_{\varphi} = \sin  \left(\varphi -\vartheta(\varphi )\right)\left[
1
-\frac{{\bf K}^{\prime\prime}_{F,\vartheta(\varphi )}}{\mathbf{K}_{F,\vartheta(\varphi )}}
+\left(\frac{{\bf K}^{\prime}_{F,\vartheta(\varphi )}}{\mathbf{K}_{F,\vartheta(\varphi )}}\right)^2
\right]
+\cos  \left(\varphi -\vartheta(\varphi )\right)\frac{{\bf K}^{\prime}_{F,\vartheta(\varphi )}}{\mathbf{K}_{F,\vartheta(\varphi )}}$.  
The self-energy is proportional to $\KFthetadim$ because the phase space for particle-hole excitations increases with increasing Fermi momentum.
For 
$q\gg|\mathbf{Q}|$,
the boson self-energy at 
$\vec q$
is determined by the Yukawa coupling at angle 
$\vartheta(\varphi)$
because 
particle-hole excitations with momentum $\vec q$ 
but with energies much smaller than $q$ 
can be created only near that angle at which the Fermi surface is tangential to $\vec q$.
The curvature radius of the local patch of the Fermi surface, which is proportional to $\mathbf{K}_{F,\vartheta(\varphi)}$, 
controls the phase space of the particle-hole excitations and the Landau damping.
%
In the other limit of 
$q \ll |\mathbf{Q}|$, the entire Fermi surface contributes to the self-energy.

The Landau damping, which is proportional to the Fermi momentum, is strongly relevant compared to the bare kinetic term of the boson.
This allows us to simplify the effective action in two ways.
First, $|{\bf Q}|^2$ term in the boson kinetic energy is strictly sub-leading  compared to the self-energy in $d \leq 3$.
This allows us to drop 
$|{\bf Q}|^2$ in the boson kinetic energy  $|{\bf q}|^2 = |{\bf Q}|^2  + q^2$.
The flip side of this is that the self-energy of the boson should be included up front in loops that include internal bosons. 
Second, the quartic interaction of boson can be dropped because the Landau damping 
suppresses the infrared singularity associated with the coupling.
For example, the one-loop quantum correction to $\udim$ given by
\bqa
\udim^2\int d_{b}^{d+1} {\bf q}~
\left( 
\frac{1}{
|{\bf q}|^2 + \Pi({\bf q})
}
\right)^2
\eqa
is IR finite in $d>1$ 
because the boson self-energy
cuts off the IR singularity.
%
It is noted that the typical magnitude of the boson momentum is
\bqa
q \sim
\left( \edim^2 \KFthetadim |{\bf Q}|^{d-1} \right)^{1/3},
\label{eq:typicalq}
\eqa
which is 
much bigger than $|{\bf Q}|$
in the loop.
Therefore, the first expression in \eq{eq:Pi1} is the relevant one inside the loop.

Dropping the $\KFdim$-rendered irrelevant terms 
but keeping terms that have potential to become relevant from an enhancement by $\KFdim$,
we write down the low-energy action as
\begin{equation}\begin{aligned}
   S & =   \int^{'}   d_{f}^{d+1} {\bf k}~
\bar \Psi_j(\mathbf{K},\delta,\theta )
\Bigl[ 
i {\bf \Gamma} \cdot{\bf K} + i v_{F,\theta  }\delta \gamma_{d-1} \Bigr]
\Psi_{j}(\mathbf{K},\delta,\theta ) 
+ 
\frac{1}{2} \int d_{b}^{d+1} {\bf q}~ 
 q^2
\phi^*(\mathbf{q})\phi(\mathbf{q}) \\
&+ i \frac{\mu^\frac{3-d}{2}}{\sqrt{N}} 
\int^{'} 
d_{f}^{d+1} {\bf k}~
\int
d_{b}^{d+1} {\bf q}~
e_{
\thetasq ,
\theta  }
\phi(\mathbf{q})   
\bar{\Psi}_{j}\left(\mathbf{K}+\mathbf{Q},\deltaq,
\thetasq 
\right)\gamma_{d-1} \Psi_{j}(\mathbf{K},\delta,\theta )\\
&+ \mu^{1-d}
\sum_{\nu}
\sum_{s=d,e}
\int^{'}
d_{f}^{d+1} {\bf k}~
d_{f}^{d+1} {\bf k}'~
\int
d_{b}^{d+1} {\bf q}~~
\lambda^{(\nu,s)}_{ \theta',\theta}(\vec q)
~
{\bf O}^{(\nu,s)}_{\Lc} ( \mathbf{K}^{\prime},  \delta^\prime, \theta'; \mathbf{K}, \delta, \theta; \mathbf{Q}, \vec q).
\label{action_low_energy}
\end{aligned}\end{equation}
Here, $\mu$ is a floating energy scale.
The coupling functions will be fixed in terms of physical observables evaluated at frequency $\mu$.
In the action, powers of $\mu$ have been factored out to make the couplings dimensionless.
It is noted that the low-energy theory is fully specified once
the set of dimensionless coupling functions
$\Bigl\{  v_{F,\theta}, 
\KFtheta, 
e_{\theta,\theta'},  \lambda^{(\nu,s)}_{\theta,\theta'}(\vec q)
 \Bigr\}$
is specified,
where 
\bqa
\KFtheta \equiv \frac{\KFthetadim}{\mu}
\label{eq:kFtheta}
\eqa
denotes the dimensionless Fermi momentum measured in the unit of $\mu$\footnote{
It is  noted that  $\KFtheta$ can be decomposed as 
$\KFtheta=
\left( \frac{\KFAV}{\mu} 
 \right) 
 \kappa_{F,\theta}$,
where $\KFAV$ and $\kappa_{F,\theta}$ are defined in \eq{eq:KFkappaF}.
$\KFAV/\mu$
is the overall size of Fermi surface that runs to infinity in the low-energy limit
and $\kappa_{F,\theta}$ is a function that specifies the shape of the Fermi surface.}.


\subsection{
Functional renormalization group equation
}
\label{RG scheme}

In order to extract the dynamical information of the theory, we compute the quantum effective action 
as a functional of 
$\Bigl\{  v_{F,\theta}, \KFtheta, e_{\theta,\theta'},  \lambda^{(\nu,s)}_{\theta,\theta'}(\vec q) \Bigr\}$.
Counter terms are added to \eq{action_low_energy} to impose renormalization group conditions on physical observables.
Adding counter terms amounts to calibrating the couplings of the theory in terms of physical observables.
The set of physical observables that are needed to specify the  entire vertex function within errors that vanish as positive powers of energy scale at which the vertex function is evaluated comprises the complete low-energy data\cite{BORGES2023169221}.
{\it Renormalizable theory} is the theory that includes the minimal set of couplings required to specify the full low-energy data. 
In metals, low-energy modes can carry large momenta and the vertex functions need to be specified as functions of angles around the Fermi surface.
This makes it necessary to consider couplings that are general functions of angles.
For systems without extensive gapless modes, renormalizable theories only include those couplings with non-negative scaling dimensions.
However, this is not enough for metals because  couplings with negative scaling dimensions can still give rise to infrared singularities.
In particular, the four-fermion coupling should be included within the renormalizable theory at least near $d=2$\footnote{
$\Bigl\{ 
v_{F,\theta},
\KFthetadim,
\edim_{\theta,\theta'}, 
\lambdadim^{(\nu,s)}_{\theta,\theta'}(\vec q), \udim
\Bigr\}$
is more than what is strictly needed to specify the low-energy physics.
Some information included in these coupling functions are not low-energy data.
For example, $\edim_{\theta,\theta'}$  
and
$\lambdadim^{(\nu,s)}_{\theta,\theta'}(\vec q)$ 
for 
$\KFdim (\theta-\theta')$ and $\vec q$ 
larger than the probing energy scale
encodes the high-energy physics because
scatterings with large 
 $|\theta-\theta'|$ 
and $|\vec q|$ necessarily involve high-energy excitations. 
Therefore, one can expand 
$\edim_{\theta,\theta'}$  and  $\lambdadim^{(\nu,s)}_{\theta,\theta'}(\vec q)$ in powers of
 $\theta-\theta'$ and $\vec q$ to keep only a few low order terms to specify the low-energy physics.
 Indeed, we will only need the diagonal Yukawa coupling
$\edim_{\theta,\theta}$ at low energies. 
For the four-fermion coupling function, however, we consider the 
 general $\vec q$ dependent coupling function.
This is not because $\vec q$-dependence is necessarily a relevant data 
but because it is convenient to keep track of the full momentum dependence of the fermion four-point function through the coupling function.
}.

The full one-particle irreducible (1PI) vertex function is a functional of the coupling functions.
Let $\varGamma^{(2m,n)}(k_i)$ denote the 1PI vertex function with $2m$ external fermions and $n$ external bosons. 
The complete low-energy observables are comprised of the vertex functions evaluated at frequency $\mu \ll \Lambda$ ($\Lambda$ is a UV cutoff) and momenta chosen anywhere close to the Fermi surface for fermions and near the origin for bosons.
The coupling functions at energy scale $\mu$ are defined through the following renormalization conditions,
\bqa
&&
\Re 
\left[\tr{ -i\gamma_{d-1} \varGamma^{(2,0)}( {\bf k})}_{{\bf k}=(\mu \hat 1_{d-1}, \delta=0, \theta)}\right] = 
0,
\label{eq:RG1}\\
&&-\frac{i}{2(d-1)}
\tr{ {\bf \Gamma} \cdot \nabla_{{\bf K}}
  \varGamma^{(2,0)}({\bf k})}_{{\bf k}=(\mu \hat 1_{d-1},\delta=0, \theta )} = 1  + {\bf \mathcal{F}}_{1; \theta},\label{eq:RG2}\\
 && \frac{1}{2}
 \Re \left[ \tr{ -i\gamma_{d-1} 
 \frac{\partial }{\partial \delta}  
 \varGamma^{(2,0)}({\bf k}) }_{{\bf k}=(\mu \hat 1_{d-1},\delta=0, \theta )} 
 \right]
 = v_{F,\theta}  +\mathcal{F}_{2; \theta},\label{eq:RG3}\\
&&\frac{\partial}{\partial q^2}\varGamma^{(0,2)}\left(\mathbf{q}\right)\bigg|_{{\bf q}=
  \left(\mu \hat{1}_{d-1},\vec{q} = 0\right)} = 1+\mathcal{F}_{3},
 \label{eq:RG32} 
  \\
 &&-\frac{i \sqrt{N}}
 {2  \mu^{\frac{3-d}{2}}}
 \tr{\gamma_{d-1} \varGamma^{(2,1)}({\bf k}',{\bf k})}_{
 \scriptsize
   \begin{array}{l}
  {\bf k}' =(2 \mu \hat 1_{d-1}, \delta'=0, \theta' ) \\
  {\bf k}=( \mu \hat 1_{d-1}, \delta=0, \theta )  
 \end{array}
 } 
 = 
e_{\theta',\theta}  
+
\mathcal{F}_{4;\theta',\theta},
\label{eq:RG4}\\
&&\frac{1}{
\mu^{1-d}}
\varGamma^{(4,0)}
_{\{j_i\};\{a_i\}}
 (\{{\bf k}_i\})
 \bigg|_{{\bf k}_i=
  \mathbf{k}^{A}_i
  } = 
T^{(\nu,s)}_{\left(\begin{smallmatrix}     j_1  & j_2      \\ j_4      & j_3       \end{smallmatrix}\right)} 
\left(I_m^{(\nu)}\right)_{a_1 a_2}\left(I_m^{(\nu)}\right)_{a_3 a_4}
\left[
\lambda^{(\nu,s)}_{ \theta_1,\theta_2}(\vec q)
+ \mathcal{F}^{(\nu,s)}_{5; \theta_1,\theta_2}(\vec q)
\right].
 \label{eq:RG5}
\eqa
Here, 
$\varGamma^{(2,0)}_{ab}( k)$  
$\left(\varGamma^{(0,2)}(k)\right)$
is the two-point function of electron (boson).
$\varGamma^{(2,1)}_{ab}(k',k)$ is the electron-boson vertex function.
The traces in Eqs. \eqref{eq:RG1}-\eqref{eq:RG4} are over the spinor indices.
$\varGamma^{(4,0)}_{\{j_i\}, \{ a_i\}}(\{ k_i\})$
is the electron four-point function. 
%
$\hat 1_{d-1}$ is a unit vector defined in the $(d-1)$-dimensional space of ${\bf K}$.
Due to the $SO(d-1)$ symmetry,
any direction of $\hat 1_{d-1}$ can be used in the renormalization condition.
\eq{eq:RG1} is the defining equation for the angle-dependent Fermi momentum.
Since $\delta$ is measured with respect to the Fermi momentum\footnote{
It is noted that $\delta, \theta$ specify the two-dimensional momentum of fermion to be $(\KFthetadim+\delta) (\cos \theta, \sin \theta)$.},
\eq{eq:RG1} 
defines $\KFthetadim$ at energy scale $\mu$ to be the momentum at which the real part of the two-point function vanishes.
It is noted that the angle-dependent Fermi momentum can in general depend on the energy scale.
\eq{eq:RG2} 
and
\eq{eq:RG32} 
set normalization of the fields.
To set this condition for the electron two-point function, 
it is in general necessary to scale the electron field in an angle-dependent matter, which gives rise to an angle-dependent anomalous dimension of electron.
\eq{eq:RG3} defines the angle-dependent Fermi velocity in the radial direction in terms of the gradient of the renormalized energy. 
We use the freedom in how we scale frequency relative to momentum to set
\bqa
v_{F,0}=1.
\label{eq:VF0equalto1}
\eqa
This amounts to setting the unit of frequency/time such that the Fermi velocity at $\theta=0$ to be $1$ at all energy scales\footnote{
One could have chosen any point on the Fermi surface as a reference point to set a global clock.}.
This requires rescaling the frequency in an energy dependent manner, which gives rise to a non-trivial dynamical critical exponent.
With this choice of clock, Fermi velocity at $\theta \neq 0$ generally  becomes different from $1$ due to angle-dependent quantum corrections\cite{PhysRevB.108.245112}.
%
%
Eqs. \eqref{eq:RG4}
and
\eqref{eq:RG5}
define the Yukawa coupling function and the four-fermion coupling function
in terms of the cubic and quartic vertex functions, respectively.
For the four-fermion coupling function, the external momenta are chosen to be 
\begin{equation}
    \begin{aligned}
            \mathbf{k}^{A}_1 =\left(3\boldsymbol{\mu},
        0,\theta_1 \right),~~ 
        \mathbf{k}^{A}_2=\left(-\boldsymbol{\mu},\Delta(0,\theta_2,\vec q),\Theta(\theta_2,\vec q)\right),  ~~
        \mathbf{k}^{A}_3= \left( \boldsymbol{\mu}, 0,\theta_2 \right),   ~~
\mathbf{k}^{A}_4=\left(\boldsymbol{\mu},\Delta(0,\theta_1,\vec q),\Theta(\theta_1,\vec q)\right),
\label{eq:4fmomenta}
    \end{aligned}
\end{equation}
where
${\bf k}^{A}_1$  and ${\bf k}^{A}_3$ 
are on the Fermi surface at angle $\theta_1$ and $\theta_2$, respectively,
and 
${\bf k}^{A}_2$  and ${\bf k}^{A}_4$ are
displaced from the Fermi surface by momentum $\vec q$ (see Appendix \ref{app:lambda012} for details).
%
$\boldsymbol{\mu}$ is a $(d-1)$-dimensional vector with magnitude $\mu$.
At energy scale $\mu$,
$|\vec q|$ is at most order of $\sqrt{\KFdim \mu}$.
Frequencies are chosen such that the net frequency is non-vanishing in all $s,t,u$ channels to avoid infrared divergences.
$\mathcal{F}_{i;\{ \theta \}}$'s
 denote discrepancies between the vertex functions and the coupling functions, which depend on renormalization group scheme.
 
In order to guarantee that the coupling functions represent the actual physical observables 
 up to corrections that are non-divergent at low energies,
the discrepancies need to satisfy certain regularity conditions\cite{BORGES2023169221}.
First, 
$\mathcal{F}_{i;\{ \theta \}}$ needs to be regular at each angle in the small $\mu$ limit.
This is because what appear on the left hand sides of Eqs. \eqref{eq:RG1}-\eqref{eq:RG5}
are the dimensionless vertex functions measured in the unit of energy scale, which are physical observables themselves.
For relativistic field theories, this condition is sufficient.
However, in metals, we need an extra condition.
This is because there exist other dimensionless low-energy observables that can be constructed out of angular integrations of the four-fermion vertex function\cite{BORGES2023169221}.
For example, consider
\bqa
E^{(\nu,s)}_\theta 
\equiv 
\frac{1}{\mu}
\int d \theta' \mathbf{K}_{F,\theta'}
\lambda^{(\nu,s)}_{ \theta',\theta}(0).
\label{eq:int_theta_lambda}
\eqa
For $\nu = F_{\pm}$,
it represents the correction to the energy of electron at angle $\theta$,
measured in the unit of $\mu$,
which arises from the forward scattering with all other electrons 
 of energy less than $\mu$ 
 around the Fermi surface.
In Fermi liquids,
this corresponds to the change of the quasiparticle energy
generated from the Landau interaction with increasing chemical potential.
For $\nu = P$, it represents the pairing field 
that an s-wave pair condensate generates for a Cooper pair located at angle $\theta$ through the momentum-space Josephson coupling.
It is noted that  $\KFthetadim$ must appear in the measure of the angular integration
in \eq{eq:int_theta_lambda} to account for the phase space properly.
The dimensionless measure 
diverges in the small $\mu$ limit because $\KFtheta = \KFthetadim/\mu$ 
grows as the low-energy limit is taken.
For this reason, 
 its integration can be singular
 even if  $\lambda^{(\nu,s)}_{ \theta',\theta}(0)$ is regular at each $\theta'$ and $\theta$.
For this reason, we need to make sure that
 the difference between \eq{eq:int_theta_lambda} and the angular integration constructed from the actual vertex function is regular.
Therefore, we require that 
$\int
 d\theta' K_{F,\theta'}
~
{\mathcal F}_{5; \theta, \theta'}  
f_{\theta'}
$
is regular 
for any normalizable function 
$f_{\theta'}$
in the small $\mu$ limit\cite{BORGES2023169221}.
This is a more stringent condition than requiring $\mathcal{F}_{i;\{ \theta \}}$ to be regular at each angle.

To enforce the renormalization conditions
in Eqs. \eqref{eq:RG1}-\eqref{eq:RG5},
we add a local counter-term action,
\begin{equation}\begin{aligned}
   \hspace{-60pt}
   S_{CT} & =   
  \int^{'}
d_{f}^{d+1} {\bf k}~
\bar \Psi_j(\mathbf{K},\delta,\theta )
\Bigl[ 
iA_1(\theta ) {\bf \Gamma} \cdot{\bf K} + 
i
v_{F,\theta  }
\left\{
A_2(\theta ) 
\delta 
+
c(\theta ) 
\mu
\right\}
\gamma_{d-1} 
\Bigr]
\Psi_{j}(\mathbf{K},\delta,\theta )  \\ &
+ \frac{A_3}{2}
\int
d_{b}^{d+1} {\bf q}~
q^2
\phi^*(\mathbf{q})\phi(\mathbf{q}) 
+ i \frac{\mu^{\frac{3-d}{2}}}{\sqrt{N}}  
\int^{'}
d_{f}^{d+1} {\bf k}
\int
d_{b}^{d+1} {\bf q}~
A_4\left(
\thetasq ,
\theta  \right)~
e_{
\thetasq ,
\theta  }
\\ &~~~~\times
\phi(\mathbf{q}) 
\bar{\Psi}_{j}\left(\mathbf{K}+\mathbf{Q},\Delta(\delta,\theta,\vec q),\thetasq \right)\gamma_{d-1} \Psi_{j}(\mathbf{K},\delta,\theta )
\\ &
+ \mu^{1-d}
\sum_{\nu}
\sum_{s=d,e}
\int^{'}
d_{f}^{d+1} {\bf k}
d_{f}^{d+1} {\bf k}'
\int
d_{b}^{d+1} {\bf q}~
A^{(\nu,s)}_{5}\left(\theta',\theta,\vec{q}\right)
\lambda^{(\nu,s)}_{ \theta',\theta}(\vec q) 
{\bf O}^{(\nu,s)}_{\Lc} ( \mathbf{K}^{\prime},  \delta^\prime, \theta'; \mathbf{K}, \delta, \theta; \mathbf{Q}, \vec q).
\label{eq:SCT}
\end{aligned}
\end{equation}
Electronic quantum corrections in general depend on angles around the Fermi surface.
At any non-zero energy scale $\mu$, the angle-dependent counter terms must be smooth functions of angles to make sure that the theory maintains its locality in real space.
The counter term for the electronic kinetic energy is separated into a piece ($A_2(\theta)$) that renormalize the Fermi velocity
and an angle-dependent shift ($c(\theta) \mu$) 
of the Fermi momentum\footnote{
While the total volume encoded by the Fermi surface is protected from quantum correction,
the Fermi momentum can in principle be renormalized in an angle-dependent manner.
}.
%
The counter term added to the original action gives the bare action, 
\begin{equation}\begin{aligned}
 S_B  & =   
  \int^{'} d_{f}^{d+1} {\bf k}_B~
   \bar \Psi_{Bj}(\mathbf{K}_B,\delta_B,\theta_B)
\Bigl[ 
i {\bf \Gamma} \cdot{\bf Q} _B + i v_{FB,\theta_B}\delta_B\gamma_{d-1} \Bigr]
\Psi_{Bj}(\mathbf{K}_B,\delta_B,\theta_B) 
\\
&+ \frac{1}{2}
\int
d_{b}^{d+1} {\bf q}_B~
q_{B}^2
\phi^{*}_B(\mathbf{Q}_B,q_{B},\varphi_B)\phi_B(\mathbf{Q}_B,q_{B},\varphi_B)  
+ \frac{i}{\sqrt{N}}   
\int^{'}
d_{f}^{d+1} {\bf k}_B
\int
d_{b}^{d+1} {\bf q}_B~
\edim_{B,
\Theta(\theta_B, q_B,\varphi_B),
\theta_B}
\times
\\ &
~ ~~~~~ \phi_B(\mathbf{Q}_B,q_{B},\varphi_B) 
\bar{\Psi}_{Bj}\left(\mathbf{K}_B+\mathbf{Q}_B,
\Delta(\delta_B,\theta_B,q_B,\varphi_B),
\Theta(\theta_B, q_B,\varphi_B)
\right)\gamma_{d-1} \Psi_{Bj}(\mathbf{K}_B,\delta_B,\theta_B)\\
&+ 
\sum_{\nu}
\sum_{s=d,e}
\int^{'}
d_{f}^{d+1} {\bf k}_B
d_{f}^{d+1} {\bf k}_B'
\int
d_{b}^{d+1} {\bf q}_B ~
\lambdadim^{(\nu,s)}_{B, \theta_B',\theta_B}(q_B,\varphi_B) 
{\bf O}_{\Lc_B}^{(\nu,s)} ( \mathbf{K}_B^{\prime},  \delta_B^\prime, \theta_B'; \mathbf{K}_B, \delta_B, \theta_B; \mathbf{Q}_B, q_B, \varphi_B),
 \end{aligned}\end{equation}
where the bare quantities and the renormalized quantities are related through 
\begin{equation}\begin{aligned}
&	 
\mathbf{K}_B = Z_{\tau}\mathbf{K},
~~\delta_B = \delta+
\frac{c(\theta)}{Z_2(\theta)} \mu,
~~ \theta_B = \theta,
~~ \mathbf{Q}_B = Z_{\tau}\mathbf{Q},
~~q_{B}=q, 
~~\varphi_B = \varphi,\\ 
& \Psi_{jB}(\mathbf{K}_B, \delta_B,\theta_B)=\sqrt{Z_{\psi}(\theta )}\Psi_{j}(\mathbf{K}, \delta ,\theta  ),~~
\phi_B(\mathbf{Q}_B, q_{B},\varphi_B)=\sqrt{Z_{\phi}}\phi(\mathbf{Q},q,\varphi), \\
& v_{FB,\theta_B}=Z_{v_F}(\theta )v_{F,\theta  },~~
{\bf K}_{FB,\theta_B} = \mu
\left[
\KFtheta-
\frac{c(\theta)}{Z_2(\theta)}
\right],~~
\edim_{B,\theta_{1B},\theta_{2B}} = Z_{e}(\theta_1 ,\theta_2)
\mu^{\frac{3-d}{2}}
e_{\theta_1 ,\theta_2 }, \\
&\lambdadim^{(\nu,s)}_{B, \theta_B',\theta_B}(q_B,\varphi_B)=  Z^{(\nu,s)}_{\lambda}(\theta',\theta,\vec q) 
\mu^{1-d} \lambda^{(\nu,s)}_{\theta',\theta}(\vec q)
\label{eq:baretorenorm}
\end{aligned}\end{equation}
with\footnote{
Similarly, the bare four-fermion operator is given by
replacing the fermion fields with the bare fields in \eq{4f_Operator}.
%
}
\begin{equation}\begin{aligned}
  &  Z_{\tau} = \frac{Z_1(0)}{Z_2(0)}, ~~
    Z_{\psi}(\theta ) = Z_1(\theta )\left(\frac{Z_2(0)}{Z_1(0)}\right)^{d}, ~~
    Z_{\phi} = Z_3\left(\frac{Z_2(0)}{Z_1(0)}\right)^{d-1}, ~~
    Z_{v_F}(\theta ) = \frac{Z_2(\theta )}{Z_1(\theta )}\frac{Z_1(0)}{Z_2(0)},\\
    &
   Z_{e}(\theta_1 ,\theta_2) = \frac{Z_4(\theta_1 ,\theta_2)}{\sqrt{Z_1(\theta_1)Z_1(\theta_2  )Z_3}}\left(\frac{Z_1(0)}{Z_2(0)}\right)^{\frac{3-d}{2}}, ~~
Z^{(\nu,s)}_{\lambda} \left(\theta_1,\theta_2,\vec{q}\right)= Z^{(\nu,s)}_{5}\left(\theta_1,\theta_2,\vec{q}\right)\left(\frac{Z_1(0)}{Z_2(0)}\right)^{3-d} \prod_{i=1}^{4}\frac{1}{\sqrt{Z_1(\theta_i)}}
\label{eq:Z_multiplicative}
\end{aligned}\end{equation}
and $Z_i(\theta) = 1 + A_i(\theta)$.
Here, $Z_\tau$ is a factor with which the renormalized frequency is dilated relative to the bare frequency.
Because the frequency is dilated so that $v_{F,\theta=0}=1$, 
$Z_{v_F}(0)=1$
and $Z_\tau$ is determined from $Z_1(0)$ and $Z_2(0)$. The bare Fermi momentum is determined from ${\bf K}_{FB,\theta_B}+\delta_B = \KFthetadim+\delta$.

The bare vertex function and the renormalized one are related to each other through
\begin{equation}\begin{aligned}
  \varGamma_B ^{(2m,n)}\left( 	\{{\bf k}_{Bi}\}; [\edim_B,v_{FB},{\bf K}_{FB},\lambdadim_B]	\right)  &= 
  Z_{\tau}^{-(2m+n-1)(d-1)}
  Z_{\phi}^{-n/2}
  \left[\prod_{i=1}^{2m}Z^{-1/2}_{\psi}(\theta_i) \right]
  \varGamma^{(2m,n)}\left(
	\{ {\bf k}_i \}; [e,v_F,K_F,\lambda];\mu
	\right).
 \label{eq:GBG}
\end{aligned}\end{equation}
Taking the derivative on the logarithm of \eq{eq:GBG} with respect to $\ln \mu$ for a fixed bare theory, we obtain
\begin{equation}
\begin{aligned}
& \Bigg[ 
- (2m+n-1)(d-1) (z-1)
-n\eta_{\phi} 
-\sum_{i=1}^{2m}\eta_{\psi,\theta_i}
+\sum_{i=1}^{2m+n}(1-z)\mathbf{K_i}\cdot\nabla_{\mathbf{K_{i}}} 
- 
\sum_{i=1}^{2m}
\mu
\left[ \beta_{K_F}(\theta_i) 
+ K_{F,\theta_i} \right]
\frac{\partial}{\partial\delta_i} 
\\ & 
+\int d\theta  \beta_{K_F}(\theta )\frac{\delta}{\delta  \KFtheta}
+\int d\theta  \beta_{v_F}(\theta )\frac{\delta}{\delta v_{F,\theta }}
+ 
\int d\theta_1 d\theta_2  \beta_{e}(\theta_1 ,\theta_2)\frac{\delta}{\delta e_{\theta_1 ,\theta_2 }} 
  +  
  \sum_{\nu,s}\int d\theta_1d\theta_2 d \vec q ~\beta^{(\nu,s)}_{\lambda} \left(\theta_1,\theta_2,\vec{q}\right)\frac{\delta}{\delta \lambda^{(\nu,s)}_{\theta_1,\theta_2}(\vec q)}
 \\ &
 +\frac{\partial}{\partial~\mathrm{ln}~\mu}\Bigg] 
~~ 
\varGamma^{(2m,n)}
\left( \{ {\bf k}_i \}; [e,v_F,K_F,\lambda];\mu \right)
= 0,
\label{eq:logmuderivG}
\end{aligned}\end{equation}
where
\begin{equation}\begin{aligned}
z =  1+\frac{d~ \mathrm{ln}\left(Z_{\tau}\right)}{d ~\mathrm{ln}~\mu}, ~~~~
    \eta_{\psi,\theta} =
    \frac{d~ \mathrm{ln}\sqrt{Z_{\psi}(\theta)}}{d ~\mathrm{ln}~\mu},
    ~~~~~
   \eta_{\phi} = \frac{d~ \mathrm{ln}\sqrt{Z_{\phi}}}{d ~\mathrm{ln}~\mu}
   \label{dynam_crit_anomalous_dim}
\end{aligned}\end{equation}
are the dynamical critical exponent,
the anomalous dimensions of the fermion and boson, respectively. 
The anomalous dimension of fermion is a function of angle.
The following beta functionals describe the flow of the coupling functions at fixed angles with increasing energy scale,
\begin{equation}\begin{aligned}
&  \beta_{v_F}(\theta )= 
  \frac{d v_{F,\theta }}{d~\mathrm{ln}~\mu}, ~~~
   \beta_{K_F}(\theta ) = 
   \frac{d \KFtheta}{d~\mathrm{ln}~\mu}, ~~~
   \beta_{e}(\theta_1 ,\theta_2) =  \frac{de_{\theta_1 ,\theta_2 }}{d~\mathrm{ln}~\mu}, ~~~
   \beta^{(\nu,s)}_{\lambda} \left(\theta_1,\theta_2,\vec{q}\right)= 
\frac{d \lambda^{(\nu,s)}_{\theta_1,\theta_2}(\vec q)}{d~\mathrm{ln}~\mu}.
\label{eq:betafunctionals}
\end{aligned}\end{equation}

\begin{figure}[h]
 \centering
  \includegraphics[width=0.5\linewidth]{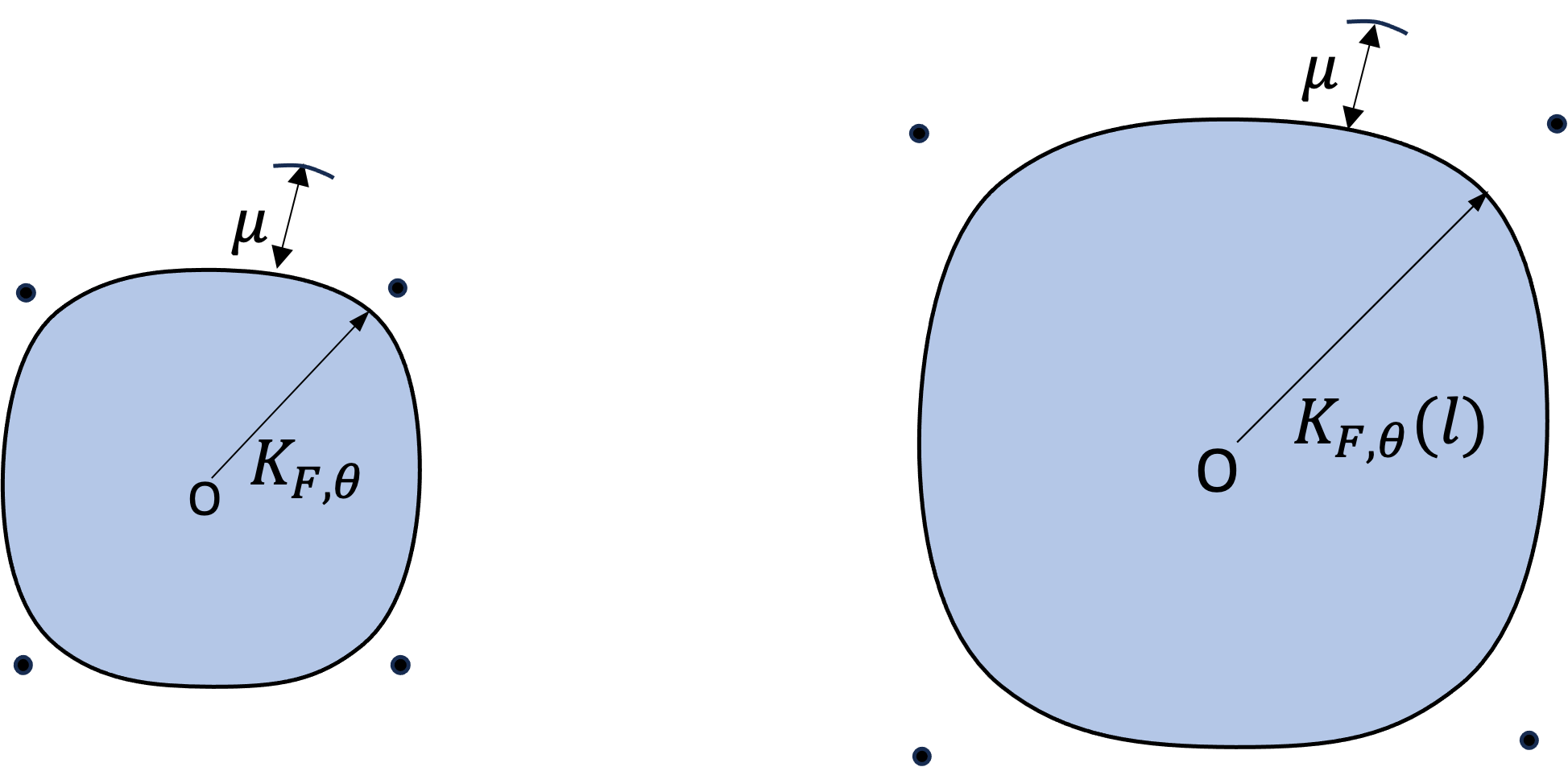}
\caption{
At a fixed point with the conformal symmetry, 
the $n$-point function of momenta $\{ k_i\}$ is related to 
that of momenta $\{ e^l k_i\}$ 
for the same theory.
In a metallic fixed point,
the $n$-point function 
of momenta $\{ \delta_i\}$ away from the Fermi surface
is related to 
that that of $\{ \delta_i(l) \}$ for a theory with a different Fermi momentum.
}
\label{Fig:scaling}
\end{figure}

Through the tree-level scale transformation in \eq{rescale_momenta_fields},
the vertex function for a theory with one set of coupling functions is related to that for a theory with another set of coupling functions as
\begin{equation}\begin{aligned}
\varGamma^{(2m,n)}\left(
	\{\tilde{{\bf k}}_i\}; 
 [\tilde{e},\tilde{v}_F,\tilde K_F,\tilde {\lambda}]; b \mu\right) = 
 b^{ D_{2m,n}  }
 \varGamma^{(2m,n)}\left(
	\{ {\bf k}_i \}; [e,v_F,K_F,\lambda];\mu\right),
\label{eq:Gtreelevelscaling}
\end{aligned}\end{equation}
where
$D_{2m,n} = -(2m+n-1)(d+1)+m (d+2) + n \frac{(d+3)}{2}$
is the tree-level scaling dimension of the vertex function,
and $\tilde {\bf k}$ and $\{ \tilde{e},\tilde{v}_F,\tilde K_F,\tilde{\lambda} \}$ denote the dilated momentum and coupling functions defined in  
Eqs. \eqref{rescale_momenta_fields}
and
\eqref{eq:rescaledcouplings}, respectively.
It is noted that 
only $\vec q$ inside the four-fermion coupling is dilated for dimensionless coupling functions. 
\eq{eq:Gtreelevelscaling} 
 can be written as a differential equation for the vertex function as
\begin{equation}
\begin{aligned}
\Bigg[ 
 &
\sum_{i=1}^{2m+n}\mathbf{K_i}\cdot\nabla_{\mathbf{K_{i}}}
+ \sum_{i=1}^{2m}
\delta_i\frac{\partial}{\partial\delta_i}
+  \sum_{j=2m+1}^{2m+n} 
{\vec k}_{j} \cdot \frac{\partial}{\partial {\vec k}_{j}}
-
\sum_{\nu,s}\int d\theta_1d\theta_2 
d \vec q~
\vec{q}\cdot \frac{\partial}{\partial \vec{q}}
~\lambda^{(\nu,s)}_{\theta_1,\theta_2}(\vec q)\frac{\delta}{\delta \lambda^{(\nu,s)}_{\theta_1,\theta_2}(\vec q)}
+\frac{\partial}{\partial~\mathrm{ln}~\mu}
-D_{2m,n}
\Bigg]  \times
\\ &
 \varGamma^{(2m,n)} 
\left( \{ {\bf k}_i \}; [e,v_F,K_F,\lambda];\mu \right)
=0.
\label{eq:resultoftreelevelscalingG}
\end{aligned}
\end{equation}
%
%
Combining Eqs.
\eqref{eq:logmuderivG}
and
\eqref{eq:resultoftreelevelscalingG},
we arrive at the functional renormalization group equation,
\begin{equation}\begin{aligned}
&
\Bigg[
\sum_{i=1}^{2m+n}z\mathbf{K_i}\cdot\nabla_{\mathbf{K_{i}}}+ \sum_{i=1}^{2m}
\left(\delta_i
+
\mu \left[ \beta_{K_F}(\theta_i) + K_{F,\theta_i} \right]
\right)\frac{\partial}{\partial\delta_i}
+
\sum_{j=2m+1}^{2m+n}
{\vec k}_{j} \cdot \frac{\partial}{\partial {\vec k}_{j}}
-
\int d\theta  
 \beta_{v_F}(\theta )
\frac{\delta}{\delta v_{F,\theta }} 
-
\int d\theta  
 \beta_{K_F}(\theta )
\frac{\delta}{\delta \KFtheta}
\\ & 
-
\int d\theta_1 d\theta_2   
\beta_{e}(\theta_1,\theta_2  )
\frac{\delta}{\delta e_{\theta_1 ,\theta_2 }}   
-
\sum_{\nu,s}\int d\theta_1d\theta_2 d \vec q ~
\left[
\beta^{(\nu,s)}_{\lambda} 
+ \vec{q}\cdot \frac{\partial}{\partial \vec{q}}~
\lambda^{(\nu,s)}_{\theta_1,\theta_2}(\vec q) 
\right]
\frac{\delta}{\delta \lambda^{(\nu,s)}_{\theta_1,\theta_2}(\vec q)}
-\tilde D_{2m,n}
\Bigg] 
\times \\ &
~~\varGamma^{(2m,n)}  \left( \{ {\bf k}_i \}; [e,v_F,K_F,\lambda];\mu \right) 
=0,
\label{G_loc}
\end{aligned}
\end{equation}
where
$\tilde D_{2m,n} =  D_{2m,n}
- (2m+n-1)(d-1) (z-1)
-n\eta_{\phi} 
-\sum_{i=1}^{2m}\eta_{\psi,\theta_i}$
is the full scaling dimension.
In \eq{G_loc}, the energy scale $\mu$ at which the coupling is defined is not varied,
and we can choose any scale to define the theory.
Here, let us set it to be $\Lambda$.
\eq{G_loc} 
summarizes the
scaling behaviors of physical observables in metals.
Its solution can be written as a scaling relation obeyed by the vertex function
\begin{equation}
    \begin{aligned}
\varGamma^{(2m,n)}  \left( \{ {\bf k}_i(l) \}; [e(l),v_F(l),K_F(l),\lambda(l)];
\Lambda 
\right) 
=&\varGamma ^{(2m,n)}  \left( \{ {\bf k}_i \}; [e,v_F,K_F,\lambda];
\Lambda 
\right)\\
&\hspace{-135pt}\times\exp\left\{\int_{0}^{l}dl'\left(-(2m+n-1)(d-1)z\left(l'\right)-n\eta_{\phi}\left(l'\right)-\sum_{i=1}^{2m}\eta_{\psi,\theta_i}\left(l'\right)+n\frac{\left(d-1\right)}{2}
+m(d-2)+2\right)\right\}
. 
\label{eq:scaling_relation_G}
    \end{aligned}
\end{equation}
Here, $l$ is a scale factor with which the frequencies and momenta are dilated as
${\bf K}_i(l) = e^{\int_0^l dl' z(l')} {\bf K}_i$,
and
${\vec k}_j(l) = e^{l} {\vec k}_j $.
$\delta_{i}(l)$ satisfies $\frac{d\delta_i(l)}{dl} = \delta_i(l)+\Lambda\left[\beta_{K_{F}}\left(\theta_i;l\right)+K_{F,\theta_i}(l)\right]$,
and
the scale-dependent coupling functions satisfy\footnote{
If one uses the momentum as a label for the Fermi surface instead of angle, 
the beta functionals include terms that dilate the momentum along the Fermi surface \cite{BORGES2023169221}.
}
\bqa
&&
\frac{d}{dl} \KFtheta(l) = - \beta_{K_F}(\theta), 
~~~~
\frac{d}{dl} v_{F,\theta}(l) = - \beta_{v_F}(\theta), 
~~~~
\frac{d}{dl} e_{\theta_1,\theta_2}(l) = - \beta_{e}(\theta_1,\theta_2),  \nn
&&
\frac{d}{dl} 
\lambda^{(\nu,s)}_{\theta_1,\theta_2}(\vec q;l) =
-
\beta^{(\nu,s)}_{\lambda}(\theta_1,\theta_2, \vec q)
- \vec{q}\cdot \frac{\partial}{\partial \vec{q}}~
\lambda^{(\nu,s)}_{\theta_1,\theta_2}(\vec q;l). 
\label{eq:full_betafunctionals}
\eqa
The beta functional for $\KFtheta$
in \eq{eq:full_betafunctionals}
can be decomposed into the ones for the overall Fermi momentum scale ($\kFAV$) and the Fermi surface shape function ($\kappa_{F,\theta}$) as
\bqa
\frac{d}{dl} \kFAV(l) = - \frac{1}{2\pi} \int d\theta \beta_{K_F}(\theta), ~~~
\frac{d}{dl} \kappa_{F,\theta}(l) = \kappa_{F,\theta} 
\left(
\frac{1}{\KFtheta} \frac{d
\KFtheta
}{dl} 
-
\frac{1}{\kFAV} \frac{d \kFAV}{dl}
\right),
\label{eq:betakFkappaF}
\eqa
where
$\kFAV \equiv \frac{1}{2\pi} \int d \theta \KFtheta$ 
and
$\kappa_{F,\theta} \equiv \KFtheta/\kFAV$.
\eq{eq:scaling_relation_G} relates the vertex function at one set of momenta with that at another set of momenta.
If all beta functions were zero, it would guarantee that the vertex function takes 
a power-law form as a function of the energy and momentum.
However, the Fermi momentum spoils the simple scaling.
Because the beta function for $\kFAV$ is
$\frac{d}{dl} \kFAV = \kFAV$,
it increases indefinitely in the large $l$ limit.
Because other beta functionals generally depend on $\kFAV$,
the incessant flow of $\kFAV$ also prevents other coupling functions from settling into scale invariant forms.
Fortunately, this does not lead to a loss of the notion of universality.
The low-energy physics still remains insensitive to most microscopic details, 
and the coupling functions flow to the profiles that are fixed by 
a small set of parameters (generally functions) that include $\kFAV$.


\subsection{ 
A lack of scale invariance
}

While the notion of universality is intact, the presence of relevant scale $\kFAV$ has non-trivial consequences.
Here, we discuss one of them on the general ground - a lack of unique scaling dimensions that can be endowed to energy and momentum for scale invariance of physical observables.
For simplicity, let us first suppose that there is no marginal or relevant parameter so that all coupling functions are entirely fixed by $\kFAV(l)$ in the large $l$ limit.
In this case, all physical observables are uniquely fixed at low energies.
This would correspond to an isolated `fixed' point.
However, it is not a fixed point in the usual sense because $\kFAV$ keeps growing.
Since the low-energy theory exhibits scale invariance only modulo a rescaling of $\kFAV$,  one can have a sense of fixed point only after theories with different $\kFAV$ are `identified'\cite{BORGES2023169221}.
To understand its implication for physical observables, let us consider the four-point function of fermions at external momenta given by \eq{eq:4fmomenta} with $\mu = \omega$.
It describes the scattering in which two fermions on the Fermi surface at angles $\theta_1$ and $\theta_2$ exchange momentum $\vec q$ either in the particle-hole or particle-particle channel.
In this case,
the form of the vertex function compatible with \eq{eq:scaling_relation_G}
is
\bqa
\varGamma^{(4,0)}( \omega,
q, \theta_1, \theta_2, \varphi; \KFdim; \Lambda ) =  
\omega^{\tilde D_{4,0}/z} 
f^{(4,0)}\left(  
\frac{q}{\omega^{1/z}}; \theta_1, \theta_2, \varphi; \frac{\KFdim}{\omega^{1/z}} \right),
\label{eq:fscaling}
\eqa
where we suppress the channel index because
the specifics of the channel does not matter for the current general discussion. 
The presence of a scale such as $\KFdim$ can lead to a correction to the exponent with which the observables depend on energy and momentum.
If the crossover function scales as
$\lim_{y \rightarrow \infty} 
f^{(4,0)}\left( 
x;
\theta_1, \theta_2, \varphi; y \right)
\sim y^{\tau}$ in the large $y$ limit,
$ \omega^{-  (\tilde D_{4,0} -\tau)/z } \varGamma^{(4,0)}( \omega, q; \theta_1, \theta_2, \varphi; \KFdim; \Lambda )$ 
would be scale invariant
in the low-energy limit taken with 
$\omega^{1/z}/q$ fixed
for a theory with a fixed $\KFdim$.
However, a complication arises because 
$\tau$ is not a smooth function of angle as will be shown later.
This can be understood from the fact that
$\KFdim$ controls physical observables differently depending on whether $\vec q$ is tangential to the Fermi surface or not.
For example, the energy of a fermion 
 at angle $\theta$ changes by
 $L_\theta(\vec q)$ in
\eq{eq:Lqvarphi} when it absorbs 
 momentum $\vec q$.
Since $L_\theta(\vec q)$ acts as an infrared cutoff,
the crossover function decays in powers of
$\omega^{1/z}/ L_{\theta_1}(\vec q)$
and
$\omega^{1/z}/ L_{\theta_2}(\vec q)$
at large $q$ as
(this will be derived in Sec. \ref{sec:NFLfp})
\bqa
f^{(4,0)} \sim
\left[
\frac{\omega^{2/z}}{q^2}
\frac{1}{ 
\left(\cos(\varphi - \theta_1)  + \frac{\sin^2(\varphi-\theta_1) q }{2 \KFdim}\right)  
\left(\cos(\varphi - \theta_2)  + \frac{\sin^2(\varphi-\theta_2) q }{2 \KFdim}\right)  }
\right]^{\eta
_d/
2
}.
\label{eq:LQ}
\eqa
Here, 
$\eta
_d
$ 
is a channel-dependent exponent whose
precise value is not important for the present discussion. 
The crossover function at small $\omega^{1/z}/L_\theta(\vec q)$ behaves differently, 
depending on whether $\vec q$ is tangential to the Fermi surface or not:
\bqa
f^{(4,0)} \sim
\begin{cases}
\left[ \frac{\omega^{2/z}}{q^2} \right]^{\eta
_d/
2
} 
~~~~~~~~\mbox{for}~~~
\cos (\varphi - \theta_1) \neq 0,
~\cos (\varphi - \theta_2) \neq 0,\\
\left[ \frac{\omega^{2/z} \KFdim}{q^3} \right]^{\eta
_d/
2
}
~~~\mbox{for}~~~
\cos (\varphi - \theta_1) =0, 
~\cos (\varphi - \theta_2) \neq 0,\\
\left[ \frac{\omega^{2/z} \KFdim^2}{q^4} \right]^{\eta
_d/
2
}
~~~\mbox{for}~~~
\cos (\varphi - \theta_1) = 
\cos (\varphi - \theta_2) = 0.
\end{cases}
\eqa
In these three different cases, 
$f^{(4,0)}$
becomes scale invariant 
if $q$ and $\omega$ are sent to zero 
with
$\omega^{1/z}/q$,
$\omega^{1/z}/q^{3/2}$
and $\omega^{1/z}/q^2$ 
fixed, respectively,
because $\KFthetadim$ is fixed 
in a given system.
The patch theory captures the scaling of the third kinematic region
where $\theta_1 \approx \theta_2$
and $\vec q$ is tangential to the Fermi surface 
at that angle.
However, the vertex functions in other kinematic regions are also low-energy observables, 
which is captured within the full low-energy theory.
This angle-dependent scaling arises because the local curvature of Fermi surface set by $\KFthetadim$ controls the low-energy physics in an angle-dependent manner.
Therefore, {\it there exists no single scaling dimension that can be assigned to $q$ and $\omega$ such that the vertex function is scale invariant at all angles.}

Above, we assumed that all coupling functions are uniquely fixed by $\kFAV$.
In general, there can be a set of marginal parameters that need to be specified along with $\kFAV$ to fix low-energy observables.
Each choice of those marginal parameters corresponds to a distinct projective fixed point.
The set of those projective fixed points forms the space of metallic universality classes.
Furthermore, there can be also relevant parameters (besides $\kFAV$) that require a fine tuning to stay within the space of projective fixed points\footnote{
If such fine tuning requires a breaking of  unitarity, 
Hermitian theories exhibit a run-away flow
\cite{
PhysRevD.80.125005,
VEYTSMAN1993315,
PhysRevLett.89.230401,
PhysRevB.69.020505,
PhysRevD.75.025005,
PhysRevLett.108.131601,
BORGES2023169221}.}.
The rest of the paper is devoted to identify the space of the projective non-Fermi liquid fixed points and characterize their universal physics for the Ising-nematic quantum critical metal.
The critical exponents and the crossover functions will be explicitly given in Sec. \ref{sec:NFLfp}.



\section{Beta Functionals}
\label{sec:beta}

In this section, 
the critical exponents
and the beta functionals are computed to the leading order in $\epsilon = 5/2-d$.
The details of the derivation can be found in 
Appendices \ref{app:deriv_beta}
and
\ref{sec:Quantum_Corrections}.
Here, we summarize the main results with emphasis on their physical implications.

\subsection{General expressions}

\begin{figure}[h]
\begin{subfigure}{.25\textwidth}
  \includegraphics[width=1\linewidth]{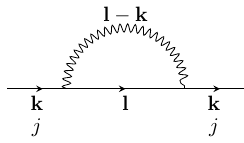}
  \caption{}
  \label{Fig:FSE}
\end{subfigure}
\begin{subfigure}{.20\textwidth}
  \includegraphics[width=1\linewidth]{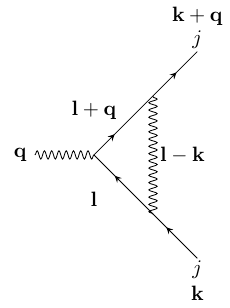}
  \caption{}
  \label{Fig:Yukawa_Vertex}
\end{subfigure}
%
\begin{subfigure}{.3\textwidth}
  \centering
  \includegraphics[width=0.70\linewidth]{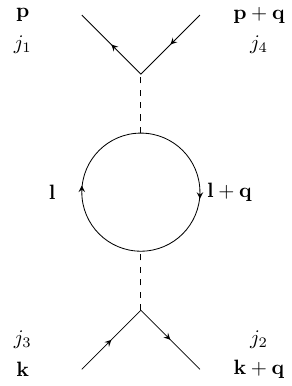}
  \caption{}
  \label{fish1}
\end{subfigure}
\begin{subfigure}{.3\textwidth}
  \centering
  \includegraphics[width=0.65\linewidth]{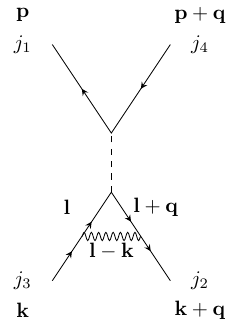}
  \caption{}
  \label{vertex3_1}
\end{subfigure}
\begin{subfigure}{.3\textwidth}
  \centering
  \includegraphics[width=0.65\linewidth]{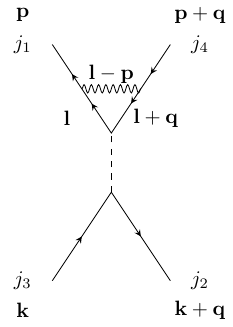}
  \caption{}
  \label{vertex3_2}
\end{subfigure}
\begin{subfigure}{.3\textwidth}
  \centering
  \includegraphics[width=1.0\linewidth]{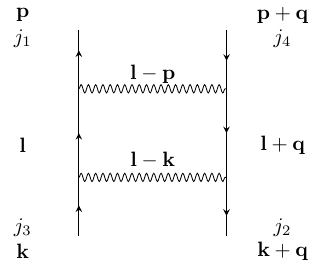}
  \caption{}
  \label{ladder1}
\end{subfigure}%
\caption{
The lowest order graphs that contribute to the
fermion self-energy and the vertex corrections.
The dashed line represents the short-range four-fermion interaction.
The wiggly line denotes the dressed boson propagator that includes the one-loop self-energy
in \fig{Fig:BSE}.
}
\label{fig:FSE_quarticfermion}
\end{figure}

The leading order quantum corrections are generated by  diagrams shown in 
\fig{fig:FSE_quarticfermion}.
Figs. \ref{Fig:FSE} 
and \ref{Fig:Yukawa_Vertex}
give the fermion self-energy 
and the cubic vertex correction, respectively.
Within the patch theory, these quantum corrections give rise to the non-Fermi liquid behaviour below the upper critical dimension $5/2$\cite{DENNIS}.
Here, we consider all patches simultaneously within one 
 unified theory,
which allows us to understand the angle-dependent  renormalization of low-energy fermions around the Fermi surface.
The remaining graphs, which contribute to the vertex correction of the four-fermion coupling, have not been considered in the patch theory as the four-fermion coupling is deemed to be `irrelevant'.
However, 
we will show that the four-fermion coupling becomes relevant near two dimensions due to large-angle scatterings and 
it qualitatively alters the nature of the non-Fermi liquid even above the superconducting transition temperature.
In the small $\vec q$ limit,
$\lambda_{\theta_1,\theta_2}(\vec q)$
at generic angles $\theta_1$ and $\theta_2$
receive a non-negligible vertex correction 
only if the following two conditions are satisfied.
First, all low-energy four-fermion operators can be composed of 
$\bar{\Psi}_{j}\left(
{\bf k} + {\bf q}\right)
I^{(\nu)}_m
\Psi_{j'}\left({\bf k} \right)$ with small 
${\bf q}$.
Other low-energy four-fermion operators can be always recast in that form through the Fierz transformation (see 
 \eq{eq:Onus2}).
Second, the loop should support nested fermions, that is, the internal fermion pairs should be able to stay close to the Fermi surface irrespective of the loop momentum.
Only those diagrams that involve nested fermion loops are enhanced by $\kFAV$ 
 due to the extensive phase space available for the virtual excitations.
\fig{fig:FSE_quarticfermion} includes the diagrams that satisfy both conditions\footnote{
It is noted that
non-1PI diagrams do not generate a counter term\cite{PhysRevB.94.115138} 
because the RG conditions are imposed on the 1PI vertex function in \eq{eq:RG5}. }.
Because we are using the spinor representation, each fermion loop represents virtual excitations both in the particle-hole and particle-particle channels.
For example, 
\fig{fish1} 
represents the RPA bubble 
that renormalizes the near-forward scattering\cite{PhysRevB.109.045143}
in the particle-hole channel 
and the loop of particle-particle pairs that renormalizes the pairing interaction.

All quantum corrections are functionals of the coupling functions,
$\Bigl\{ 
v_{F,\theta},
\KFthetadim,
\edim_{\theta,\theta'}, 
\lambdadim^{(\nu,s)}_{\theta,\theta'}(\vec q)
\Bigr\}$.
The angular dependence of the coupling functions are not in priori known because they are subject to renormalization.
Therefore, the critical exponents and the beta functionals are expressed as integrations of the coupling functions\cite{BORGES2023169221}.
We can write the dynamical critical exponent and the anomalous dimensions of the fields as (see Appendices \ref{app:deriv_beta}
and
\ref{sec:Quantum_Corrections}
for details)
\bqa
z  =  1+\bar{\omega}_{0}, ~~~
\eta_{\phi} =
    -\frac{d-1}{2}\bar{\omega}_{0}, ~~~
\eta_{\psi,\theta} =
     \frac{1}{2}
     \left(\bar{\omega}_{\theta}-d  \bar{\omega}_{0}\right).
     \label{eq:z_and_etas_in_convolution}
\eqa
$\bar{\omega}_{\theta}$ is the
quantum correction that renormalizes the fermion at angle $\theta$ (see Appendix \ref{sec:fermion_self_energy} for details),
\begin{equation}
    \begin{aligned}
        \bar{\omega}_{\theta} 
        = \int_{-\frac{\pi}{2}}^{\frac{\pi}{2}}\frac{d\theta'}{2\pi}\frac{K_{F,\theta'}}{v_{F,\theta'}} \omega_{d;\theta',\theta}(\mu),
        \label{Z1_logmu_convolution}
    \end{aligned}
\end{equation}
where $\KFtheta = \KFthetadim/\mu$ is the dimensionless Fermi momentum
and
\begin{equation}
    \begin{aligned}
    \omega_{d;\theta',\theta}(\mu) = -\frac{\mu \edim^2_{\theta,\theta'}}{4N(d-1)}\partial_{\log\mu}\left(\tr\left\{\mathbf{\Gamma}\cdot\nabla_{\mathbf{K}}\int\frac{d\mathbf{L}}{\left(2\pi\right)^{d-1}}
 D_{1;\mu}\left(\mathbf{L}-\mathbf{K},\theta^\prime,\theta\right)\frac{\mathbf{\Gamma}\cdot\mathbf{L}}{|\mathbf{L}|}\right\}\bigg|_{\mathbf{K} = \boldsymbol{\mu}}\right)
 \label{eq:omega12}
    \end{aligned}
\end{equation}
denotes the renormalization that
the fermion at $\theta$ receives from virtual particle-hole excitations created at angle $\theta'$. 
$        D_{1;\mu}\left(\mathbf{L},\theta_1,\theta_2\right) = \frac{1}{q(\theta_1,\theta_2)^{2}+ \boldsymbol{f}_{d,\vartheta^{-1}\left( 
\mtheta
\right)}
        \frac{|\mathbf{L}|
        ^{d-1}}{\sqrt{q(\theta_1,\theta_2)^{2}+\mu^2}}}
$
is the contribution of the critical boson,
where 
$\mtheta = \frac{\theta_1+\theta_2}{2}$
is the average angle between
$\theta_1$ and $\theta_2$\footnote{
For $|\theta_1-\theta_2| \ll 1$,
one can replace $\mtheta$ with any angle between $\theta_1$ and $\theta_2$.
}
and
$q\left(\theta_1,\theta_2\right)$
is the momentum that connects two points on the Fermi surface at angles $\theta_1$ and $\theta_2$,
\bqa
q\left(\theta_1,\theta_2\right) 
= \sqrt{\mathbf{K}^2_{F,\theta_1}+\mathbf{K}^2_{F,\theta_2}-2\mathbf{K}_{F,\theta_1}\mathbf{K}_{F,\theta_2}\cos\left(\theta_1-\theta_2\right)}.
\label{eq:qexp}
\eqa 
The net quantum correction 
in \eq{Z1_logmu_convolution}
is given by the angular integration over $\theta'$.
The dynamical critical exponent and the anomalous dimension of the boson 
depend on $\bar{\omega}_{0}$ because the frequency has been rescaled such that the Fermi velocity is $1$ at $\theta=0$ 
as is shown in  \eq{eq:VF0equalto1}.
There is no fundamental reason to choose $\theta=0$ as the reference.
This is one of many ways of choosing a reference clock.
What is fundamental though is the fact that the frequency-dependent kinetic term is renormalized  
in an angle-dependent manner on the Fermi surface through $A_1(\theta)$ in \eq{eq:SCT}.
Therefore, there is no clock in which Fermi velocity is kept uniform around the Fermi surface unless one introduces a curved momentum-spacetime metric\cite{PhysRevB.108.245112}. 
In this paper, we simply keep track of the effect of the momentum-dependent red shift in terms of the momentum-dependent Fermi velocity instead of using a curved momentum-spacetime picture with momentum independent Fermi velocity.
The anomalous dimension of fermion is generally angle-dependent :
$\eta_{\psi,\theta}$ 
depends both on $\bar{\omega}_{0}$ and $\bar{\omega}_{\theta}$,
where the dependence on $\bar{\omega}_{0}$ arises from $z$.

%

The beta functionals for the  
Fermi momentum,
Fermi velocity
and the Yukawa coupling function read
\bqa
 \frac{d \KFtheta}{dl}  &=& \KFtheta, \label{eq:betaKf_1} \\
\frac{ d v_{F,\theta}}{dl}  &=&
\left(
\bar{\omega}_{0}-\bar{\omega}_{\theta}\right)
v_{F,\theta }, 
\label{eq:betavf_general} \\
\frac{\mathrm{d}e_{\theta_1 ,\theta_2}}{\mathrm{d}l} &=& e_{\theta_1,\theta_2}\left[\frac{3-d}{2}+\frac{3-d}{2}\bar{\omega}_{0}-\frac{1}{2}\left(\bar{\omega}_{\theta_1}+\bar{\omega}_{\theta_2}\right)\right].
\label{eq:betae_general} 
\eqa
They describe the flow of coupling functions with increasing length scale at fixed angles
with $l= \ln \frac{\Lambda}{\mu}$ being the logarithmic length scale.
To the leading order, the beta functional for the Fermi momentum is completely fixed by its tree-level scaling\footnote{
However, the Fermi momentum can be in principle renormalized by higher order quantum corrections.
For example, the four-fermion coupling, 
which is of the higher order than the Yukawa coupling,
can create large-angle scatterings and renormalize the angle-dependent Fermi momentum.}.
The RG flow of the Fermi velocity at angle $\theta$ is determined by  
$\left(
\bar{\omega}_{0}-\bar{\omega}_{\theta}\right)$,
which sets the rate at which the frequency is dilated at angle $\theta$ relative to angle $0$.
The flow of the general Yukawa coupling function $e_{\theta_1,\theta_2}$ depends not only on the renormalization of fermions at angles $\theta_1$ and $\theta_2$ encoded in
$\bar{\omega}_{\theta_1}$ 
and $\bar{\omega}_{\theta_2}$,
but also on 
$\bar{\omega}_{0}$
through the dynamical critical exponent.

\begin{figure}[H]
 \centering
  \includegraphics[width=0.4\linewidth]{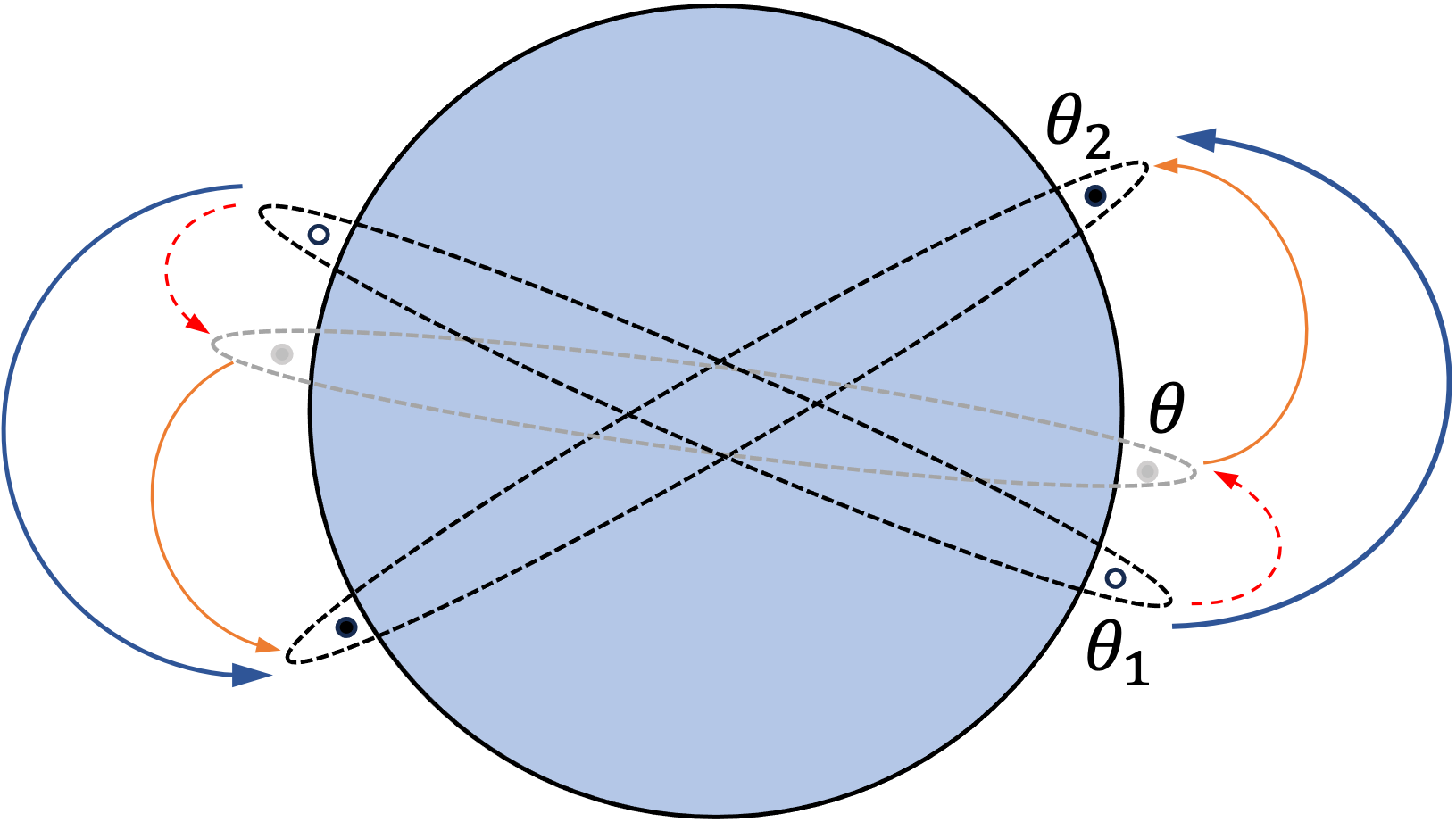}
\caption{
An interplay between small-angle and large-angle scatterings in the presence of critical collective modes.
A large-angle scattering that takes a Cooper pair from angle $\theta_1$  to angle $\theta_2$ (denoted as a solid arrow) can be created 
by bridging a small-angle scattering mediated by the critical boson from $\theta_1$ to $\theta$ (dashed line)
with another large-angle scattering from $\theta$ to $\theta_2$.
The critical mode creates a mixing between 
$\lambda^{(\nu,s)}_{\theta_1,\theta_2}$
and
$\lambda^{(\nu,s)}_{\theta,\theta_2}$,
where the mixing matrix is given by 
$\sum_{i = \pm}
\left( h_{i;d}
\cdot\frac{\mathscr{K}^{(\nu)}_{i;d}}{4} \right)_{\theta_1,\theta}
$
as is shown in 
\eq{eq:betalambda_convolution}.
Even if the critical boson mainly mediates small-angle scatterings,
the anomalous dimension generated from the mixing modifies the exponent with which the four-fermion coupling function decays 
at large angles.
}
\label{fig:small_large_angle}
\end{figure}

The beta functional for the four-fermion coupling function is
\bqa
&&
 \frac{\mathrm{d} \lambda^{(\nu,s)}}{dl} 
         =
        W_d\lambda^{(\nu,s)}
         -\mathcal{U}\cdot\lambda^{(\nu,s)}
        -\lambda^{(\nu,s)}\cdot \mathcal{U} 
        \nn
        &&
        -
        \sum_{i=\pm}
        \left\{
        M^{(\nu,s)}_{s_1,s_2}
        \lambda^{(\nu,s_1)}\cdot\mathscr{K}^{(\nu)}_{i;d}\cdot \lambda^{(\nu,s_2)}
        +
        \snu\left(
        h_{i;d}\cdot\frac{\mathscr{K}^{(\nu)}_{i;d}}{4}\cdot \lambda^{(\nu,s)}
        +\lambda^{(\nu,s)}\cdot\frac{\mathscr{K}^{(\nu)}_{i;d}}{4} \cdot h_{i;d}^{\dagger}\right)
      \right\}
        +S_d^{(\nu,s)},
\label{eq:betalambda_convolution}
\eqa
where $\lambda^{(\nu,s)}$ represents
the matrix defined in the space of angles with its matrix elements given by
$\lambda^{(\nu,s)}_{\theta_1,\theta_2}$
and the matrix multiplication  defined as
\begin{equation}
    \begin{aligned}
     \left(A\cdot B\right)_{\theta_1,\theta_2} = \int_{-\frac{\pi}{2}}^{\frac{\pi}{2}}\frac{d\theta}{2\pi}\frac{\KFtheta}{v_{F,\theta}} A_{\theta_1,\theta}B_{\theta,\theta_2}.
     \label{eq:matrixproduct}
    \end{aligned}
\end{equation}
Since the beta functional for the four-fermion coupling is of our central interest, we dissect each term one by one.
\begin{itemize}
\item 
$W_{d} = \left[ 1-d +(3-3d)(z-1)\right]$ 
is the tree-level scaling dimension of the four-fermion coupling corrected by the dynamical critical exponent.
\item 
$\mathcal{U}_{\theta_1,\theta_2}\left(\vec{q}\right) = 2\pi\left(\eta_{\psi,\theta_1}+\eta_{\psi,\Theta\left(\theta_1,\vec{q}\right)}\right)\frac{v_{F,\theta_1}}{K_{F,\theta_1}}\delta\left(\theta_1-\theta_2\right)$ 
is a diagonal matrix that represents the shift in the  dimension of the four-fermion 
 coupling generated by the anomalous dimension of the fermion field.
For small $|\vec q|$,
$\mathcal{U}_{\theta_1,\theta_2}\left(\vec{q}\right) \approx 4\pi \eta_{\psi,\theta_1}
\frac{v_{F,\theta_1}}{K_{F,\theta_1}}\delta\left(\theta_1-\theta_2\right)$.
\item 
$M^{(\nu,s)}_{s_1,s_2} \lambda^{(\nu,s_1)}\cdot
\left( \sum_{i=\pm} \mathscr{K}^{(\nu)}_{i;d} 
\right) \cdot \lambda^{(\nu,s_2)} $ 
is the Fermi liquid contribution
to the beta functional of the four-fermion coupling function (\fig{fish1}).
It includes the renormalization of the forward scattering 
and the pairing interaction created by the short-range four-fermion coupling.
$M^{(\nu,s)}_{s_1,s_2}$ is a matrix defined in the space of flavour channels $e$ and $d$,
\begin{equation}
    \begin{aligned}
       M^{(F_{\pm},d)}_{s_1,s_2} = 
        \begin{cases}
            N,&~
            (s_1, s_2) = (d,d)\\
            1,&~
            ~(s_1,s_2)=(e,d) ~~\text{or}~~(d,e)\\
            0,&~\text{otherwise}
        \end{cases}, & ~~~~~~
       M^{(F_{\pm},e)}_{s_1,s_2} = 
        \begin{cases}
            1,&~
            (s_1,s_2) = (e,e)\\
            0,&~\text{otherwise}
        \end{cases},\\
        M^{(P,d)}_{s_1,s_2} = 
        \begin{cases}
            1,&~
            (s_1,s_2)=(d,d) ~\text{or}~~(e,e)\\
            0,&~\text{otherwise}
        \end{cases}, & ~~~~~~
        M^{(P,e)}_{s_1,s_2} = 
        \begin{cases}
            1,&~
            (s_1,s_2)=(e,d) ~~\text{or}~~(d,e) \\
            0&~\text{otherwise}
        \end{cases}.
    \end{aligned}
\end{equation}
$\mathscr{K}_{+;d;\theta_1,\theta_2}^{(\nu)}(\vec q;\mu)
+
\mathscr{K}_{-;d;\theta_1,\theta_2}^{(\nu)}(\vec q;\mu)$ 
represents a matrix whose diagonal element captures the angle-dependent contribution of virtual fermion pairs within the loop\footnote{
Here, we keep the most general expression that is valid for any $\vec q$ and $\theta$ 
while in Fermi liquids one can ignore the measure zero set in which $\vec q$ becomes tangential to the Fermi surface.}
(see Appendix \ref{lambda_2_order}), 
\begin{equation}
    \begin{aligned}
        \mathscr{K}_{+;d;\theta_1,\theta_2}^{(\nu)}(\vec q;\mu) =2\pi
\frac{A_+^{(\nu)}(d) T_+(d)}{\left(\left(\mathscr{L}_{\mu,\theta_1}(\vec q)\right)^2+4\right)^{\frac{4-d}{2}}}\frac{v_{F,\theta_1}}{K_{F,\theta_1}}
\delta\left(\theta_1-\theta_2\right),
\label{main:FL_Plus_Kernel}
    \end{aligned}
\end{equation}
\begin{equation}
    \begin{aligned}
        \mathscr{K}_{-;d;\theta_1,\theta_2}^{(\nu)}(\vec q;\mu) = 2\pi\frac{A_-^{(\nu)}(d) T_-(d)\left(\left(\mathscr{L}_{\mu,\theta_1}(\vec q)\right)^2+2(d-2)\right)}{\left(\left(\mathscr{L}_{\mu,\theta}(\vec q)\right)^2+4\right)^{\frac{6-d}{2}}}
      \frac{v_{F,\theta_1}}{K_{F,\theta_1}}
      \delta\left(\theta_1-\theta_2\right).
      \label{main:FL_Minus_Kernel}
    \end{aligned}
\end{equation}
Here, the kinematically determined channel-dependent prefactors are given by
\bqa
  A_+^{(F_+)}(d) = 0,~~ A_-^{(F_+)}(d)  = 2,~~
        A_+^{(F_-)}(d) = \frac{2(2-d)}{d-1},~~ A_-^{(F_-)}(d)  = \frac{2}{d-1},~~
         A_+^{(P)}(d) = 2,~~ A_-^{(P)}(d)  = 0.
\eqa
$T_+(d)
= 
\frac{
4
\Omega_{d-1}
\Gamma\left(\frac{d+1}{2}\right)\Gamma\left(\frac{4-d}{2}\right)\Gamma^2\left(\frac{d}{2}\right)}{
\Gamma\left(\frac{3}{2}\right)(2\pi)^{d-1}\Gamma(d)}$
and
$
T_-(d)
= 
\frac{
4
\Omega_{d-1}
\Gamma\left(\frac{d-1}{2}\right)\Gamma\left(\frac{4-d}{2}\right)\Gamma^2\left(\frac{d}{2}\right)}{\Gamma\left(\frac{3}{2}\right)(2\pi)^{d-1}\Gamma(d)}$
are constants that are $O(1)$.
$\mathscr{L}_{\mu,\theta} \equiv L_{\theta}(\vec q) / \mu$, where $L_{\theta}(\vec q)$  denotes the energy of a particle-hole  or a particle-particle pair with net momentum $\vec q$ created at angle $\theta$ (see \eq{eq:Lqvarphi}).
%
%
%
$\mathscr{K}_{+;d;\theta_1,\theta_2}^{(\nu)}(\vec q;\mu)
+
\mathscr{K}_{-;d;\theta_1,\theta_2}^{(\nu)}(\vec q;\mu)$ 
has been separated into 
$\mathscr{K}_{i;d;\theta_1,\theta_2}^{(\nu)}(\vec q;\mu)$ with $i=\pm$ depending on its dependence on 
$\mathscr{L}_{\mu,\theta}$.
In two dimensions,
$\mathscr{K}_{+;d;\theta_1,\theta_2}^{(\nu)}(\vec q;\mu) \sim  A_+^{(\nu)}$ 
and
$\mathscr{K}_{-;d;\theta_1,\theta_2}^{(\nu)}(0;\mu) \sim A_-^{(\nu)} 
\mathscr{L}_{\mu,\theta}^2$ 
in the small $\vec q$ 
($\mathscr{L}_{\mu,\theta} \rightarrow 0$) limit.
In the forward scattering channel, 
$A_+^{(F_\pm)}=0$
and the only contribution is from
$\mathscr{K}_{-;d;\theta_1,\theta_2}^{(F_\pm)}(\vec q;\mu)$,
which becomes vanishingly small in the small 
$\mathscr{L}_{\mu,\theta}$ limit.
This reflects the fact that the phase space for virtual particle-hole excitation is severely limited in the forward scattering channel due to the chiral nature of low-energy fermions in metals with co-dimension one.
Consequently, the non-trivial renormalization arises mainly from 
$\mathscr{L}_{\mu,\theta} \sim 1$\cite{PhysRevB.109.045143}
and
the exact forward scattering is free from quantum correction in two dimensions\cite{SHANKAR,POLCHINSKI1}.
The pairing interaction is free from the kinematic constraint.
With $A_+^{(P)} \neq 0$,
the pairing interaction receives the non-vanishing quantum correction even in the small $\vec q$ limit, 
which is responsible for the BCS instability in Fermi liquids with an attractive interaction.
In the opposite limit with  $\mathscr{L}_{\mu,\theta} \gg 1$,
all quantum corrections become small
because virtual excitations with energies that exceed $\mu$ are suppressed.
\item
The second last term in \eq{eq:betalambda_convolution} is the vertex correction to the four-fermion coupling generated from the critical boson (Figs. \ref{vertex3_1}
and \ref{vertex3_2}).
$h_{+;d;\theta_1,\theta_2}\left(\vec{q};\mu\right)$ and $h_{-;d;\theta_1,\theta_2}\left(\vec{q};\mu\right)$ 
represent the mixing of  the four-fermion couplings with different angles.
As is the case for the Fermi liquid correction, 
the vertex correction has been also divided into
$
\left( h_{+;d}
\cdot\mathscr{K}^{(\nu)}_{+;d}\right)_{\theta_1,\theta_2}
$ and 
$
\left( h_{-;d}
\cdot\mathscr{K}^{(\nu)}_{-;d}\right)_{\theta_1,\theta_2}
$ 
which are proportional to 
$A_+^{(\nu)}$
and 
$A_-^{(\nu)}$, respectively.
Formally, 
those mixing matrices can be written as 
 (see Appendix \ref{lambda_1_order}),
\begin{equation}
    \begin{aligned}
h_{+;d;\theta_1,\theta_2}\left(\vec{q};\mu\right)
&= -\frac{\edim^2_{\theta_1,\theta_2}\mu}
{N}
\frac
{\left(\left(\mathscr{L}_{\mu,\theta_2}\left(\vec{q}\right)\right)^2+4\right)^{\frac{4-d}{2}}} 
{ T_+(d)}
\int\frac{d\mathbf{L}dE }{(2\pi)^{d}}
        \partial_{\log\mu}\left[
        D_{1;\mu}\left(\mathbf{L}-3\boldsymbol{\mu},\theta_1,\theta_2\right)\tilde{K}_+\left(\mathbf{L},\boldsymbol{\mu},E,\theta_2,\vec{q}\right)
        \right.\\&\left.+
        D_{1;\mu}\left(\mathbf{L}+\boldsymbol{\mu},\theta_1,\theta_2\right)\tilde{K}_+\left(\mathbf{L},-\boldsymbol{\mu},E,\theta_2,\vec{q}\right)
        \right], 
        \label{main:h1_def} 
    \end{aligned}
\end{equation}
\begin{equation}
    \begin{aligned}
h_{-;d;\theta_1,\theta_2}\left(\vec{q};\mu\right)
&= -\frac{\edim^2_{\theta_1,\theta_2}\mu}
{N}
\frac{\left(\left(\mathscr{L}_{\mu,\theta_2}\left(\vec{q}\right)\right)^2+4\right)^{\frac{6-d}{2}} }{T_-(d)\left(\left(\mathscr{L}_{\mu,\theta_2}\left(\vec{q}\right)\right)^2+2(d-2)\right)}
\int\frac{d\mathbf{L}dE }{(2\pi)^{d}}
\partial_{\log\mu}\left[
D_{1;\mu}\left(\mathbf{L}-3\boldsymbol{\mu},\theta_1,\theta_2\right)\tilde{K}_-\left(\mathbf{L},\boldsymbol{\mu},E,\theta_2,\vec{q}\right)
\right.\\&\left.
+
D_{1;\mu}\left(\mathbf{L}+\boldsymbol{\mu},\theta_1,\theta_2\right)\tilde{K}_-\left(\mathbf{L},-\boldsymbol{\mu},E,\theta_2,\vec{q}\right)
\right],
       \label{main:h2_def}
    \end{aligned}
\end{equation}
where $E = v_{F,\theta}\delta$ is the energy of virtual fermions,
    $\tilde{K}_\pm\left(\mathbf{L},\boldsymbol{\mu},E,\theta,\vec{q}\right) = \frac{\mathbf{L}\cdot(\mathbf{L}-2\boldsymbol{\mu})\pm E\left(E+L_{\theta}(\vec{q})\right)}{\left(|\mathbf{L}-2\boldsymbol{\mu}|^{2}+\left(E+L_{\theta}(\vec{q})\right)^2\right)\left(|\mathbf{L}|^{2}+E^2\right)}$
captures the contribution of virtual fermion pairs located at angle $\theta$.
%
The mixing matrices satisfy
$h^{\dagger}_{i;d;\theta_1,\theta_2}\left(\vec{q};\mu\right) = h_{i;d;\theta_2,\theta_1}\left(-\vec{q};\mu\right)$.
While the full expressions of 
$h_{+;d;\theta_1,\theta}\left(\vec{q};\mu\right)$ and 
$h_{-;d;\theta_1,\theta}\left(\vec{q};\mu\right)$ are complicated,
they become proportional to the boson propagator at large angles.
In the limit with $|\theta_1-\theta_2|\gg \KFtheta^{-\frac{1}{2}}$ and $\mathscr{L}_{\mu,\theta}\left(\vec{q}\right)\leq 1$, 
the matrices become
\begin{equation}
    \begin{aligned}
        h_{+;d;\theta_1,\theta_2}\left(\vec{q};\mu\right) = h_{-;d;\theta_1,\theta_2}\left(\vec{q};\mu\right) = \frac{e^2_{\theta_1,\theta_2}}{N}\left(\frac{\mu}{q\left(\theta_1,\theta_2\right)}\right)^2.
        \label{appendix:h1_h2_interpatch_def}
    \end{aligned}
\end{equation}
The mixing between four-fermion couplings with a large angular difference $|\theta_1-\theta_2|$ is suppressed as $1/q(\theta_1,\theta_2)^2$ 
because the critical boson has to carry the correspondingly large momentum to scatter fermions with widely different Fermi momenta.
Even if the boson-mediated mixing of the four-fermion vertices arises mainly from small-angle scatterings, it does not imply that large-angle scatterings are negligible for the four-fermion coupling itself.
Small-angle scatterings can enhance large-angle scatterings by generating an anomalous dimension for the large-angle four-fermion coupling.
This is illustrated in \fig{fig:small_large_angle}.
The sign that appears in the vertex correction is given by
\bqa       
\snu = 
        \begin{cases}
            1~~&\nu = F_{\pm}\\
            -1~~&\nu = P
\end{cases}.
\label{s_nu}
\eqa
It turns out that \eq{s_nu} does not lead to a difference in the sign of the anomalous dimension generated from the vertex correction as an extra sign that arises from 
$\mathscr{K}_{+;d;\theta_1,\theta_2}^{(\nu)}(\vec q;\mu)$ 
cancels \eq{s_nu}.
As a result, the anomalous dimension generated from the vertex correction in the small $\vec{q}$ limit is non-negative in all channels.
Instead, the channel dependent sign in \eq{s_nu} will pop up in the source term for the four-fermion coupling generated from the critical boson.
This will be shown explicitly in the next section.

\item
$S_{d;\theta_1,\theta_2}^{(\nu,s)}\left(\vec{q};\mu\right)$ 
is the source for the four-fermion coupling generated from the critical boson (\fig{ladder1}),
\begin{equation}
    \begin{aligned}
&S_{d;\theta_1,\theta_2}^{(\nu,s)}\left(\vec{q};\mu\right)
= \frac{\delta_{s,s_{\nu}}}{16}
\mu^{6-d}
       \sum_{i=\pm}  A_i^{(\nu)}(d)
\int \frac{ d\theta}
{2\pi}
\frac{\KFtheta}
{v_{F,\theta}}
   \frac{e^2_{\theta_1,\theta}
   e^2_{\theta,\theta_2}
   }{N^2}
        \int\frac{d\mathbf{L}dE }{(2\pi)^{d}}
        \partial_{\log\mu} \times \\ &
        \left\{ 
D_{1;\mu}\left(\mathbf{L}-3\boldsymbol{\mu},\theta_1,\theta\right)
\tilde{K}_i\left(\mathbf{L},\boldsymbol{\mu},E,\theta,\vec{q}\right) 
D_{1;\mu}\left(\mathbf{L}-\boldsymbol{\mu},\theta,\theta_2\right)
 +
D_{1;\mu}\left(\mathbf{L}+\boldsymbol{\mu},\theta_1,\theta\right)
\tilde{K}_i\left(\mathbf{L},-\boldsymbol{\mu},E,\theta,\vec{q}\right) 
D_{1;\mu}\left(\mathbf{L}-\boldsymbol{\mu},\theta,\theta_2\right)
        \right\},
        \label{main:(0)_general_beta_functional}
    \end{aligned}
\end{equation}
where $ s_{F_\pm}=e $ and $ s_{P}=d $ 
(see Appendix \ref{lambda_0_order} for details). 
The delta function $\delta_{s,s_\nu}$ shows the fact that the interaction is
generated in the 
$(F_{\pm},e)$ 
and 
$(P,d)$ 
channels only. 
This channel-dependence is due to the following facts.
First, the ladder diagram that generates the four-fermion coupling at generic ${\bf p}$ and ${\bf k}$ in \fig{ladder1} has to have small net momentum ${\bf q}$ in order for the fermion pairs in the loop to be able to stay close to the Fermi surface irrespective of the loop momentum ${\bf l}$.
Second, the critical boson is flavour neutral, and the flavour of an electron is preserved across the Yukawa interactions.
As a result, the interaction is generated only in the exchange channel for the forward scatterings 
and in the direct channel for the pairing interaction
(see \eq{Flavour_Tensor} for the definition of each channel).
\end{itemize}

For the four-fermion coupling,
the counter terms are chosen to cancel the average of the quantum corrections evaluated at two sets of external momenta with different frequencies but with same two-dimensional spatial momenta. 
This is responsible for the two terms with different factors of $\mu$ within the square brackets
of Eqs. (\ref{main:h1_def}),
(\ref{main:h2_def}) 
and 
\eqref{main:(0)_general_beta_functional}.
Such counter terms guarantee that the bare Hamiltonian stays Hermitian.

\subsection{Small-angle and large-angle scatterings}

In the previous section,
the critical exponents and the beta functionals are expressed as integrals over the angle around the Fermi surface.
Those integrations can not be done without the knowledge of the angular profiles of the coupling functions.
However, a simplification arises for quantum corrections  in which infrared singularities arise from virtual fermions that are confined within a small angular region. 
If a loop integration exhibits an infrared singularity at an isolated point in the space of loop momenta,
one can ignore the variation of the coupling functions around that point in the momentum space 
to the leading order in the infrared singularity\cite{BORGES2023169221}.
In this case,
one can use the `adiabatic procedure' to extract the singular part of the quantum correction solely in terms of the coupling functions evaluated at that angle.
While not all quantum corrections can be computed this way,
we simplify those that are amenable to the adiabatic procedure.

First, let us consider the quantum correction that arises from the fermion self-energy in \eq{Z1_logmu_convolution}.
This quantum correction 
 exhibits an infrared divergence only from the point ${\bf L}=0$ and $\theta=\theta_1$ in the space of internal momentum,
and the infrared divergence is logarithmic at $d=5/2$\cite{DENNIS}. Therefore, one can extract the IR singular part through the adiabatic procedure\cite{BORGES2023169221} 
where $\theta$ in
$\left\{ \KFthetadim,
v_{F,\theta},
\edim_{\theta_1,\theta}
\right\}$
is replaced  with $\theta_1$\footnote{
One can expand the coupling function in powers of 
$(\theta-\theta_1)$,
where $\theta$ is the loop variable and $\theta_1$ is the angle of the external fermion.
Those with positive powers of 
$(\theta-\theta_1)$
are not infrared singular near $d=5/2$.
}.
Then, the dynamical critical exponent and the anomalous dimensions of the fields become (see Appendix
\ref{sec:Quantum_Corrections}
for details)
\bqa
z  =  1+u_1(d)g_{0}, ~~~
\eta_{\phi} =
    -\frac{d-1}{2}u_1(d)g_{0}, ~~~
\eta_{\psi,\theta} =
     \frac{u_1(d)}{2}
     \left(g_{\theta }-d  g_0\right).
     \label{eq:z_and_etas_in_termsofg}
\eqa
Here, $g_\theta$  
is the effective Yukawa coupling given by the ratio between the diagonal Yukawa coupling and the Fermi velocity at angle $\theta$,
\bqa
   g_{\theta } &=&
   \frac{1}{N}
   \frac{e^{4/3}_{\theta }\left|X_{\theta}\right||\chi_{\vartheta^{-1}(\theta)}|^{1/3}}{v^{1/3}_{F,\theta }K^{1/3}_{F,\theta }},
    \label{eq:gtheta}  
    \eqa
where $X_{\theta} = \sin\left(\vartheta^{-1}(\theta)-\theta\right)$.
$g_{\theta }$ controls the physical observables such as the anomalous dimensions.
Henceforth, we will use this effective coupling in place of the original Yukawa coupling.
$u_1(d) = \frac{1}{3\sqrt{3}} \frac{\Omega_{d-1}}{(2\pi)^{d-1}}\frac{\Gamma(\frac{d}{2})\Gamma(\frac{d-1}{3})\Gamma(\frac{d-1}{2})
\Gamma\left(\frac{11-2d}{6}\right)
}{\Gamma(\frac{1}{2})\Gamma(\frac{d-1}{6})\Gamma(\frac{5d-2}{6})} 
$ is $O(1)$ number.
Because the typical momentum 
 carried by the critical boson is much smaller than the Fermi momentum at low energies,
the quantum correction 
for electron at an angle 
is controlled by the diagonal Yukawa coupling and the Fermi velocity at that angle. 
To the leading order, 
the quantum corrections to 
the Fermi momentum,
Fermi velocity and the Yukawa coupling function are all governed by the small-angle scatterings,
and their beta functionals become
\bqa
 \frac{d \KFtheta}{dl}  &=& \KFtheta, \label{eq:betaKf} \\
\frac{ d v_{F,\theta}}{dl}  &=&
u_1(d)
\left(
g_{0}-g_{\theta}\right)
v_{F,\theta }, 
\label{eq:betavf} \\
\frac{\mathrm{d}g_{\theta_1 ,\theta_2}}{\mathrm{d}l} &=& g_{\theta_1,\theta_2}\left[\frac{5-2d}{3}+\frac{5-2d}{3}u_1(d)g_{0}-\frac{u_1(d)}{2}\left(g_{\theta_1}+g_{\theta_2}\right)\right],
\label{eq:betae}
\eqa
where
         $g_{\theta_1,\theta_2} = e_{\theta_1, \theta_2 }^2\sqrt{\frac{g_{\theta_1} g_{\theta_2}}{e_{\theta_1}^2e_{\theta_2}^2}}$
is the effective off-diagonal Yukawa coupling.
According to \eq{eq:betae},
the effective Yukawa coupling,
which controls the anomalous dimension of fermions,
is relevant (irrelevant) for $d<5/2$ ($d>5/2)$ at the non-interacting fixed point.
This implies that the upper critical dimension for the Yukawa coupling is $5/2$, 
which is consistent with the patch theory\cite{DENNIS}.
It is noted that the upper critical dimension is shifted from $d=3$ at which
the engineering scaling dimension of the Yukawa coupling vanishes 
(see \eq{eq:rescaledcouplings}).
This shift is caused by the Fermi momentum:
the screening of the Yukawa coupling by particle-hole excitations is enhanced by $\KFthetadim$
and quantum corrections from the Yukawa coupling are logarithmically divergent at $d=5/2$ not $d=3$.

%
\begin{figure}[th]
\centering
\begin{subfigure}{.45\textwidth}
  \centering
  \includegraphics[width=1.0\linewidth]{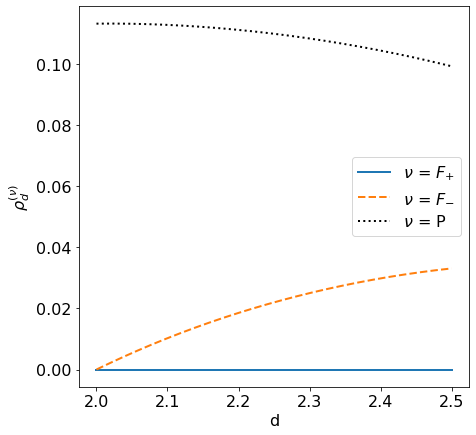}
  \caption{}
  \label{rho_d}
\end{subfigure}%
\begin{subfigure}{.45\textwidth}
  \centering\includegraphics[width=1.0\linewidth]{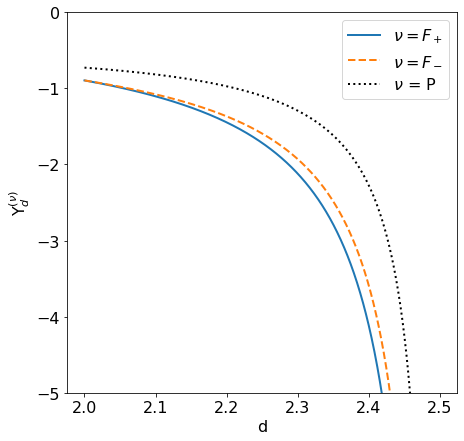}
  \caption{}
  \label{Upsilon_d}
\end{subfigure}
\caption{
The dimension dependence of the magnitude of
the crossover function in \eq{eq:Climits}.
}
\label{rho_upsilon}
\end{figure}

Near the upper critical dimension, the vertex correction for the four-fermion coupling generated by the critical boson can be also evaluated through the adiabatic procedure because the infrared singularity only arises from the contribution of the critical boson with small momenta.
Near $d=2$, one can not in priori apply the adiabatic procedure to the vertex correction 
because an infrared singularity can arise within a one-dimensional subspace of internal momentum in the pairing channel\cite{BORGES2023169221}.
This follows from the fact that virtual Cooper pairs can stay arbitrarily close to the Fermi surface and give rise to infrared singularities  irrespective of momentum on the Fermi surface
when the co-dimension of Fermi surface is close to one.
In this paper, we focus on the physics in $d$ close to the upper critical dimension 
in which the adiabatic procedure is justified for this vertex correction. 
This allows us to write 
the singular part of 
$       \left(h_{i;d}\cdot\frac{\mathscr{K}^{(\nu)}_{i;d}}{4}
\cdot 
\lambda^{(\nu,s)}
+\lambda^{(\nu,s)}\cdot\frac{\mathscr{K}^{(\nu)}_{i;d}}{4} \cdot h_{i;d}^{\dagger}\right)_{\theta_1,\theta_2}$
in  \eq{eq:betalambda_convolution}
solely in terms of
$\lambda^{(\nu,s)}_{\theta_1,\theta_2}$,
$g_{\theta_1}$
and
$g_{\theta_2}$.
Because the vertex correction arises from the critical boson 
that mediates small-angle scatterings of fermions, 
the effective Yukawa coupling at angles $\theta_1$ and $\theta_2$ governs the vertex correction for the four-fermion coupling functions at those angles.
For the vertex correction generated by the short-range four-fermion interaction,
$\lambda^{(\nu,s_1)}\cdot\mathscr{K}^{(\nu)}_{i;d}\cdot \lambda^{(\nu,s_2)}$
in  \eq{eq:betalambda_convolution},
the angular integration in it is kept as it is.
In the end, the beta functional for the four-fermion coupling function can be written as
\bqa
\frac{\mathrm{d} \lambda^{(\nu,s)}}{dl} 
    &=&
W_d \lambda^{(\nu,s)}
- J^{(\nu)}_d \cdot\lambda^{(\nu,s)} - \lambda^{(\nu,s)}\cdot
J^{(\nu)}_d
 -M^{(\nu,s)}_{s_1,s_2}\lambda^{\left(\nu,s_1\right)}\cdot\mathscr{K}^{(\nu)}_d\cdot\lambda^{\left(\nu,s_2\right)}+ S_d^{(\nu,s)},
\label{eq:betalambda}
\eqa
where
\begin{equation}
    \begin{aligned}
J^{(\nu)}_{d;\theta_1,\theta_2}\left(\vec{q}\right) = 
\frac{2 \pi v_{F,\theta_1}}{K_{F,\theta_1}}
\left[
\eta_{\psi,\theta_1}+\eta_{\psi,\Theta\left(\theta_1,\vec{q}\right)}
-     g_{\theta_1}\mathcal{C}^{(\nu)}_d\left(\mathscr{L}_{\mu,\theta_1}(\vec q)\right)
\right]   
        \delta\left(\theta_1-\theta_2\right)
    \end{aligned}
    \label{eq:J}
\end{equation}
represents the net anomalous dimension of the four-fermion coupling that arises from the anomalous dimension of the field
($\eta_{\psi,\theta}$)
and the vertex correction.
$\mathcal{C}^{(\nu)}_d\left(\mathscr{L}_{\mu,\theta}(\vec q)\right)$ 
is a function of
$\mathscr{L}_{\mu,\theta}(\vec q) = \frac{L_{\theta}(\vec q)}{\mu}$
that describes the crossover caused by the non-zero momentum $\vec q$ carried by the external fermion pairs.
The full expression of the crossover function can be found in 
\eq{appendix:crossover}.
What is important for us is the fact that
$\mathcal{C}^{(\nu)}_d\left(\mathscr{L}_{\mu,\theta}\right)$ approaches a non-zero constant
in the small  $\mathscr{L}_{\mu,\theta}$ limit
while it vanishes in the large
$\mathscr{L}_{\mu,\theta}$ limit.
The asymptotic limits of the crossover function is summarized as
(see Appendix \ref{lambda_1_order} for details)
\begin{equation}
    \mathcal{C}^{(\nu)}_d\left(\mathscr{L}_{\mu,\theta}(\vec q)\right)=
    \begin{cases}
      \rho_d^{(\nu)}, &\mathscr{L}_{\mu,\theta}(\vec q)\ll 1,~2\leq d \leq d_c\\ 
      \Upsilon_d^{(\nu)}\left(\mathscr{L}_{\mu,\theta}(\vec q)\right)^{-2}, &\mathscr{L}_{\mu,\theta}(\vec q)\gg 1, ~2\leq d <d_c \\
      \bar{\Upsilon}_{d_c}^{(\nu)}
      \log\left(\mathscr{L}_{\mu,\theta}(\vec q)\right)\left(\mathscr{L}_{\mu,\theta}(\vec q)\right)^{-2}, &\mathscr{L}_{\mu,\theta}(\vec q)\gg 1, ~d=d_c,
    \end{cases},
    \label{eq:Climits}
\end{equation}
where $\rho_d^{(\nu)}$ and $\Upsilon_d^{(\nu)}$ 
 are $d$-dependent constants plotted in Fig. \ref{rho_upsilon}\footnote{
It is noted that the low energy limit and $d\rightarrow d_c$ limit do not commute\cite{SCHLIEF2}.
%
}.
While the vertex correction enhances the four-fermion coupling both in the forward scattering and pairing channels,
it is non-singular in the forward scattering channel at $d=2$.
In particular, the exact forward scattering does not receive the vertex correction in two dimensions.
As a result,
$\rho_d^{(\nu)}$,
which is proportional to
$A_+^{(\nu)}$,
vanishes at $d=2$
for $\nu=F_\pm$.
$\mathscr{K}_{d}^{(\nu)}=
\mathscr{K}_{+;d}^{(\nu)}(\vec q;\mu) 
+
\mathscr{K}_{-;d}^{(\nu)}(\vec q;\mu)$ can be written as
\bqa
\mathscr{K}_{d;\theta_1,\theta_2}^{(\nu)}(\vec q;\mu) &=2\pi
\left[\frac{A_+^{(\nu)}(d) T_+(d)}{\left(\left(\mathscr{L}_{\mu,\theta_1}(\vec q)\right)^2+4\right)^{\frac{4-d}{2}}}+\frac{A_-^{(\nu)}(d) T_-(d)\left(\left(\mathscr{L}_{\mu,\theta_1}(\vec q)\right)^2+2(d-2)\right)}{\left(\left(\mathscr{L}_{\mu,\theta}(\vec q)\right)^2+4\right)^{\frac{6-d}{2}}}\right]
\frac{v_{F,\theta_1}}{K_{F,\theta_1}}
\delta\left(\theta_1-\theta_2\right).
\label{eq:Kdthetas}
  \eqa
%
%
%
%
%
Finally, the source for the four-fermion coupling generated from the critical boson becomes 
(see Appendix \ref{lambda_0_order} for details),
\begin{equation}
    \begin{aligned}
&S_{d;\theta_1,\theta_2}^{(\nu,s)}(\vec q;\mu) =   &
\delta_{s,s_{\nu}}\times
      \begin{cases}
      
       -
       \frac{1}{6 \pi N^{2}}
       \frac{\Omega_{d-1}}{(2\pi)^{d-1}\beta_d}
       \frac{e^{2}_{\theta_1}v_{F,\theta_1}\left|X_{\theta_1}\right|\left|\chi_{\vartheta^{-1}\left(\theta_1\right)}\right|}{K_{F,\theta_1}}
       \left(\frac{
       A^{(\nu)}_+(d)
       }{\left(\mathscr{L}_{\mu,\theta_1}(\vec q)\right)^2+16}+
       \frac{
       A_-^{(\nu)}(d)
       }{2(d-1)}\frac{\left(\left(\mathscr{L}_{\mu,\theta_1}(\vec q)\right)^2-4\right)}{\left(\left(\mathscr{L}_{\mu,\theta_1}(\vec q)\right)^2+4\right)^2}\right)
       \\
       ~&\hspace{-130pt} \text{for~~~}
                q\left(\theta_1,\theta_2\right) \ll
                \sqrt{\frac{Ng\KFdim \mu}{v_F}}, \\ 
       \\
\frac{\snu}{4}
\frac{e^{2}_{\theta_1,\theta_2}}{N}
\frac{
\left[
g_{\theta_1}\mathcal{C}^{(\nu)}_d\left(\mathscr{L}_{\mu,\theta_1}(\vec q)\right)+g_{\theta_2}\mathcal{C}^{(\nu)}_d\left(\mathscr{L}_{\mu,\theta_2}(\vec q)\right)\right]\mu^2
}
{q\left(\theta_1,\theta_2\right)^2 
      +
\beta_d\left(
\frac{Ng_{\mtheta}\mathbf{K}_{F,\mtheta}}{\left|X_{\mtheta}\right|\left|\chi_{\vartheta^{-1}\left(\mtheta\right)}\right|v_{F,\mtheta}}
\right)^{\frac{3}{2}}
        \frac{\mu^{\frac{3}{2}}}{{{
q\left(\theta_1,\theta_2\right)
 }}}}
        &
        \hspace{-130pt}
                 \text{for~~~} \sqrt{\frac{Ng\KFdim \mu}{v_F}} \ll q\left(\theta_1,\theta_2\right) \ll \KFdim
      \end{cases}
      .
    \end{aligned}
    \label{eq:Snu12}
\end{equation}
Here, $\mtheta = \frac{\theta_1+\theta_2}{2}$.
Since $L_{\theta}(\vec q)$
sets the IR cutoff for quantum corrections generated from fermion pairs with momentum $\vec q$,
both \eq{eq:Kdthetas} 
and \eq{eq:Snu12} 
vanishes
for 
$\mathscr{L}_{\mu,\theta}(\vec q) \gg 1$.
$\mathcal{C}^{(\nu)}_d\left(\mathscr{L}_{\mu,\theta}(\vec q)\right)$ 
in \eq{eq:Climits}
also controls the crossover for the source term in \eq{eq:Snu12} 
in the large-angle limit.
This is because the source is dominated by the process where one boson mediates a small-angle scattering and the other mediates a large-angle scattering in the ladder diagram shown in
\fig{ladder1}.
The small-angle scattering is 
governed by the exactly same virtual process that renormalizes the 
the short-ranged four-fermion interaction.

\begin{figure}[h]
 \centering
  \includegraphics[width=0.4\linewidth]{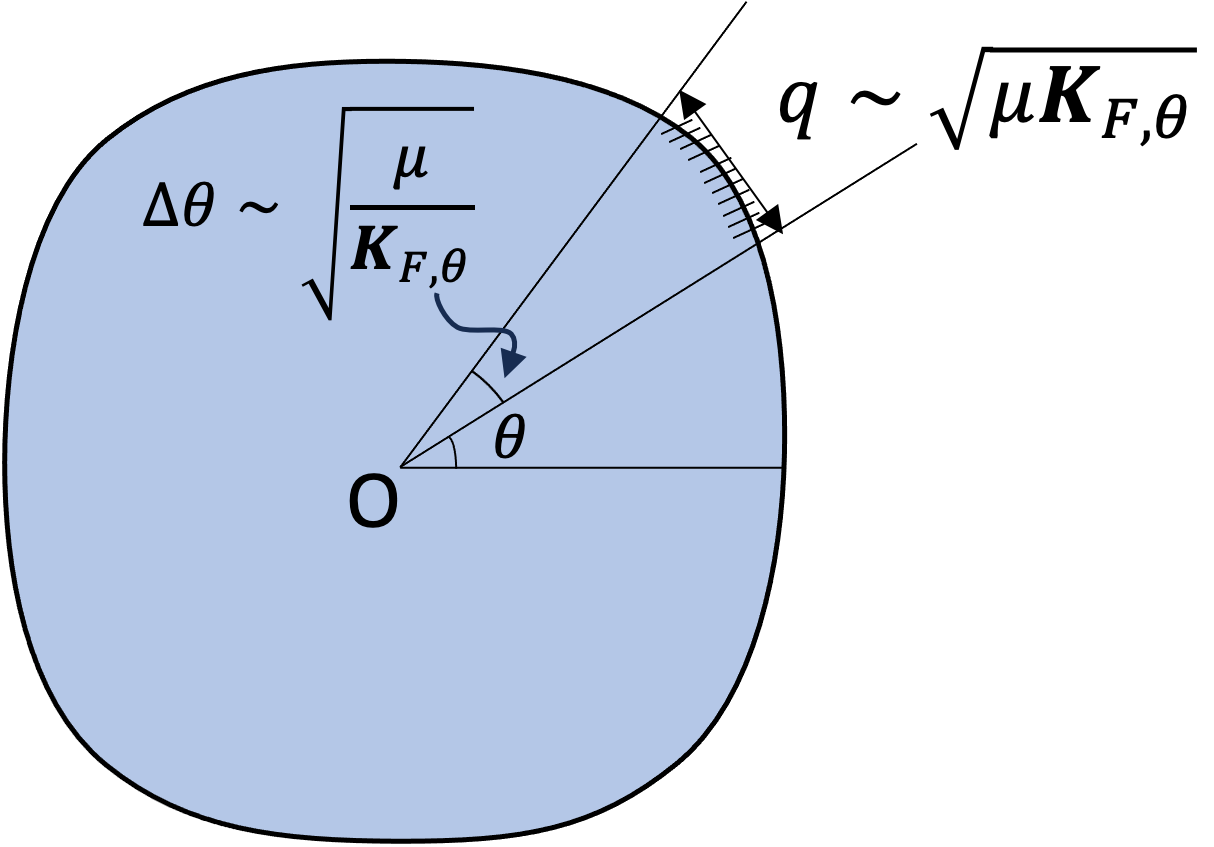}
\caption{
At energy scale $\mu$,
a boson can carry a momentum that is order of 
$q \sim \sqrt{\mathbf{K}_{F,\vartheta(\varphi)}\mu}$ in direction $\varphi$.
Around each angle, a region within the reach of that momentum defines a patch. 
Fermions can be readily scattered within each patch by absorbing or emitting a critical boson at that energy scale.
Fermions that lie with an angular separation greater than $\sqrt{\mu/\KFthetadim}$ are in distinct patches.
}
\label{fig:patch}
\end{figure}

The source in \eq{eq:Snu12} takes different forms depending on  
whether $q(\theta_1,\theta_2)$ is smaller or larger than
a crossover momentum given by
$q_\mu = \sqrt{\frac{Ng\KFdim \mu}{v_F}}$.
It corresponds to the typical momentum that a boson carries at energy scale $\mu$\footnote{
The momentum of boson can be much greater than $\mu$ because the particle-hole pairs  with which the boson mixes have energy $L_\theta(\vec q) \sim q^2/\KFdim$  near angle $\theta$ where $\vec q$ is tangential to the Fermi surface.  $q^2/\KFdim \sim \mu$ gives the momentum scale  $q \sim \sqrt{\mu \KFdim}$.
}.
This can be understood as a crossover in the angular dependence of the four-fermion coupling generated from the boson.
The angular width
$|\theta_1-\theta_2| \sim 
 q_\mu/\KFdim \sim K_{F}^{-1/2}$
defines the notion of patch as is shown in \fig{fig:patch}. 
Two points 
that are within angular separation $\KFtheta^{-1/2}$ 
near angle $\theta$ 
on the Fermi surface 
can be connected by the momentum carried by a boson with energy $\mu$.
Fermions can be scattered between those points by absorbing or emitting a boson readily available at that energy scale,
and they are regarded to be within one `patch'.
On the other hand, two points separated by a larger angle can not be connected through the interaction mediated by low-energy bosons.
Those points are in distinct patches.
It is noted that 
whether two points are within one patch or not depends on energy scale $\mu$
because the width of a patch decreases as $\KFtheta^{-1/2} \sim \sqrt{\mu/\KFthetadim}$ with decreasing energy.
Two points that are separated by a fixed angle are within one patch at high energies but move outside one patch at lower energies\cite{PhysRevD.59.094019,PhysRevB.91.115111}.
With decreasing energy scale,
the size of each patch decreases, and the number of distinct patches increases.
This translates into an increase in the number of patches with decreasing energy scale\cite{PhysRevB.91.115111}.

While the inter-patch coupling is weaker than the intra-patch one at each angle,
inter-patch scatterings are not necessarily negligible 
because the number of patches increases with decreasing energy. 
The abundant phase space for inter-patch scatterings can 
 overweight the weakness of its magnitude 
 if the inter-patch scattering amplitude decays slower than the rate at which the number of patches increases.
Therefore, the relevance of the inter-patch scattering is a dynamical issue,
which is to be determined from the angular profile of the coupling functions.
Inter-patch couplings arise both from the interaction mediated by the critical boson and the four-fermion coupling.
The former decays as $1/|\theta_1-\theta_2|^2$ at large angular separations to the leading order in the $\epsilon$ expansion.
Due to the fast decay of the boson propagator, this alone does not create a strong inter-patch coupling.
However, an anomalous dimension of the boson, 
which may arise at higher orders\cite{HOLDER}, can enhance the boson-mediated inter-patch scatterings.
The more significant source of inter-patch coupling at a small $\epsilon$ is the four-fermion coupling. 
In particular, the four-fermion coupling receives an anomalous dimension even at the one-loop order from the critical boson through the processes depicted in 
\fig{fig:small_large_angle}.
It modifies the exponent with which the universal four-fermion coupling decays at large angular separation.
As will be shown later, 
the leading-order calculation suggests that the inter-patch coupling becomes relevant
below a critical dimension that lies between $2$ and $5/2$.

\vspace{1cm}
\section{
Functional renormalization group flow
}
\label{sec:iv}

In this section, 
we analyze the functional RG flow of each coupling function.

\subsection{Fermi momentum, Yukawa coupling and Fermi velocity}

According to \eq{eq:betaKf},
the Fermi momentum grows under the RG flow as
\bqa
\KFtheta(l)=\KFtheta(0) e^l.
\label{eq:kFrunning}
\eqa
We can view
\eq{eq:kFrunning} as
one relevant parameter, 
$\kFAV = \frac{1}{2\pi} \int d\theta
\KFtheta
$
that flows to infinity in the low-energy limit,
and one marginal coupling function,
$\kappa_{F,\theta}= \KFtheta/\kFAV$.

While the beta functions for the Fermi velocity and the Yukawa coupling  are coupled, 
the beta function for the effective coupling $g_\theta$ only depends on itself.
In particular, the beta function for $g_0$ depends only on $g_0$ because the clock has been chosen with the reference to the Fermi velocity at $\theta=0$.
This leads us to first consider the beta function for $g_0$,
$     \frac{\mathrm{d} g_{0}}{\mathrm{d}l} = 
g_0\left(\frac{5-2d}{3}-
\frac{2d-2}{3}u_1(d)g_{0}\right)$
whose solution is 
\begin{equation}
    \begin{aligned}
        g_0(l) = \frac{ g_0^{UV} e^{\frac{5-2d}{3}l}}{\frac{2(d-1)}{5-2d}
        u_1(d)g_0^{UV}\left(e^{\frac{5-2d}{3}l}-1\right)+1},
    \end{aligned}
\end{equation}
where
$g_\theta^{UV} \equiv g_\theta(l=0)$ denotes the UV value of the effective coupling at $\theta$.
On the other hand, the beta function of $g_\theta$ for general $\theta$ depends on both $g_0$ and $g_\theta$ due to the contribution from the dynamical critical exponent determined from $g_0$,
$        \frac{\mathrm{d} g_{\theta}}{\mathrm{d}l} = 
       g_{\theta }\left(
       \frac{5-2d}{3}
       +\frac{5-2d}{3}u_1(d)g_{0}
       -u_1(d)g_{\theta}\right)$.
Using $g_0(l)$, we readily obtain its solution to be\footnote{
It is convenient to consider the ratio between the effective coupling at angle $\theta$  and $0$, 
 $g^r_\theta = g_{\theta}/g_0$.
 Its beta function  reads
$\frac{1}{g^r_\theta}\frac{\mathrm{d}g^r_{\theta}}{\mathrm{d}l} = g_0u_1(d)\left(1-g^r_{\theta}\right)$.
The straightforward integration 
        leads to the solution.
},
\begin{equation}
 \begin{aligned}
 g_\theta(l) = \frac{
        g^{UV}_\theta
        g^{UV}_0e^{\frac{5-2d}{3}l}\left(\frac{2(d-1)}{5-2d}
        u_1(d)g_0^{UV}\left(e^{\frac{5-2d}{3}l}-1\right)+1\right)^{\frac{5-2d}{2(d-1)}}}{
        g^{UV}_{\theta}\left(\frac{2(d-1)}{5-2d}
        u_1(d)g_0^{UV}\left(e^{\frac{5-2d}{3}l}-1\right)+1\right)^{\frac{3}{2(d-1)}}+
        \left (g^{UV}_0-g^{UV}_\theta\right)
        }.
\label{eq:gthetal}
\end{aligned}
\end{equation}

Now, we can solve the beta functional for the Fermi velocity.
The interaction between the critical boson and electron enhances the effective mass of electrons, 
which gives rise to a momentum-dependent Fermi velocity.
Since $v_{F,\theta}$ represents the Fermi velocity at angle $\theta$ measured in the unit of $v_{F,0}$,
\eq{eq:betavf} vanishes at $\theta=0$.
For $\theta \neq 0$, the Fermi velocity runs under the RG flow as far as $g_\theta \neq g_0$.
If  $g_\theta > g_0$ ($g_\theta < g_0$),
the stronger (weaker) Yukawa coupling relative to $\theta=0$ slows down 
 (accelerate) the electron at angle $\theta$, resulting in the decrease (increase) of $v_{F,\theta}$.
By rewriting the beta function for the Fermi velocity as
     $\frac{1}{v_{F,\theta}}
     \frac{\mathrm{d}v_{F,\theta}}{\mathrm{d}l} 
     =\frac{1}{g_\theta} \frac{\mathrm{d}g_{\theta}}{\mathrm{d}l}- \frac{1}{g_0} \frac{\mathrm{d}g_{0}}{\mathrm{d}l}$,
one readily obtains $v_{F,\theta}(l)$ in terms of the scale dependent effective coupling,
\begin{equation}
    \begin{aligned}
        v_{F,\theta}(l) =
        \frac{g_0^{UV}}{g_\theta^{UV}}\frac{g_{\theta}(l)}{g_{0}(l)}~
        v_{F,\theta}^{UV}.
    \end{aligned}
\end{equation}
From the solutions for the diagonal Yukawa coupling and the Fermi velocity,
we can write the beta functional for the off-diagonal Yukawa coupling as
$ \frac{1}{g_{\theta_1 ,\theta_2 }}\frac{\mathrm{d}g_{\theta_1 ,\theta_2}}{\mathrm{d}l} = g_{\theta_1,\theta_2}\left[\frac{5-2d}{3}+\frac{5-2d}{3}u_1(d)g_{0}-\frac{u_1(d)}{2}\left(g_{\theta_1}+g_{\theta_2}\right)\right]$.
By rewriting the beta function for the effective off-diagonal coupling  as
     $ \frac{1}{g_{\theta_1 ,\theta_2 }}\frac{\mathrm{d}g_{\theta_1 ,\theta_2}}{\mathrm{d}l} 
     =\frac{1}{2}\left[\frac{1}{g_{\theta_1}} \frac{\mathrm{d}g_{\theta_1}}{\mathrm{d}l}+\frac{1}{g_{\theta_2}} \frac{\mathrm{d}g_{\theta_2}}{\mathrm{d}l}\right]$,
$g_{\theta_1 ,\theta_2 }(l)$ can be expressed in terms of the scale dependent diagonal effective coupling,
$   g_{\theta_1,\theta_2}(l) = \frac{g_{\theta_1,\theta_2}^{UV}}{\sqrt{g_{\theta_1}^{UV}g_{\theta_2}^{UV}}}\sqrt{g_{\theta_1}(l)g_{\theta_2}(l)}$.

\begin{figure}[ht]
\centering
\begin{subfigure}{.45\textwidth}
\includegraphics[width=.95\linewidth]{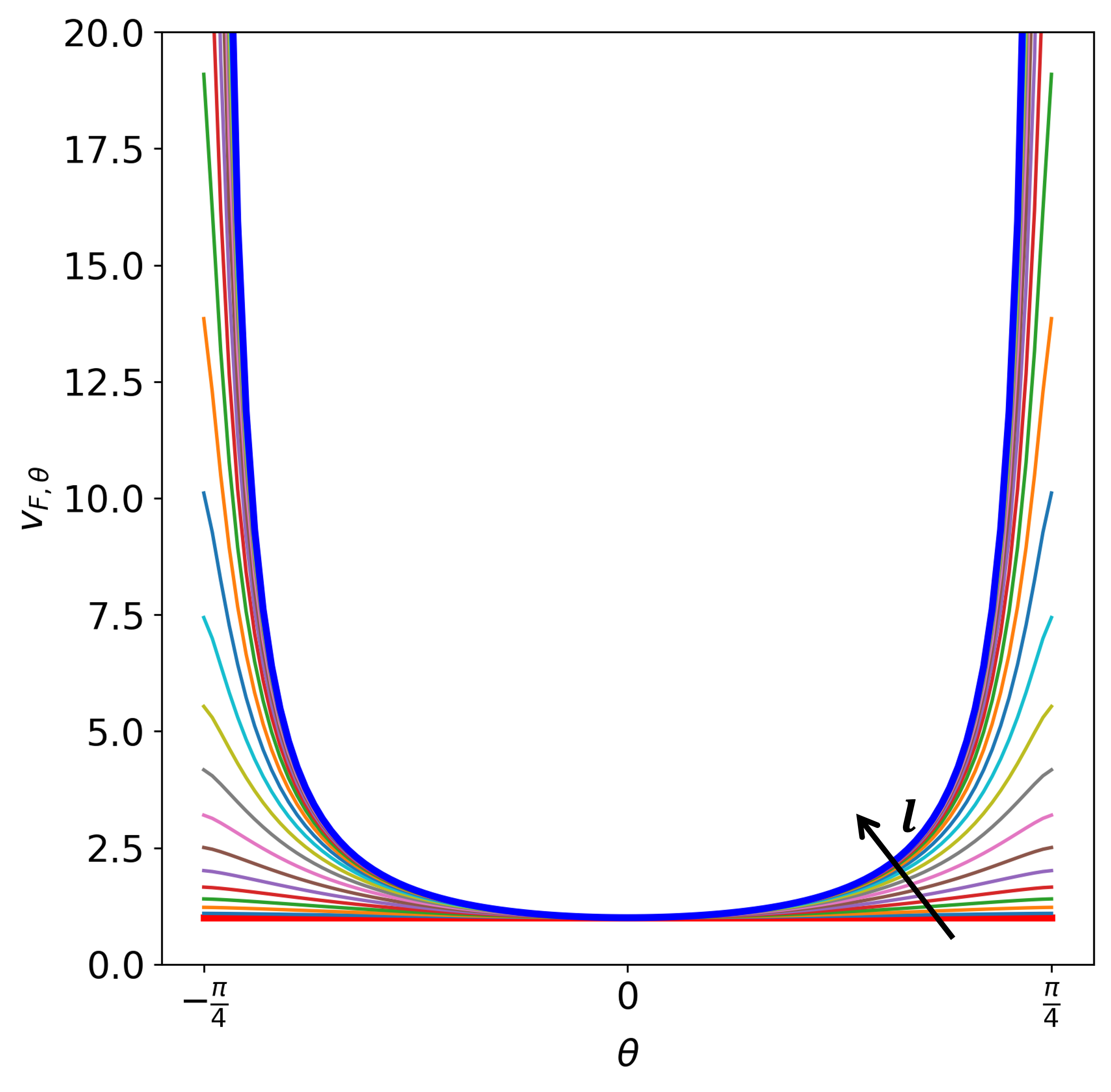}
  \caption{}
  \label{}
\end{subfigure}
\hfill
%
\begin{subfigure}{.45\textwidth}
\includegraphics[width=.95\linewidth]{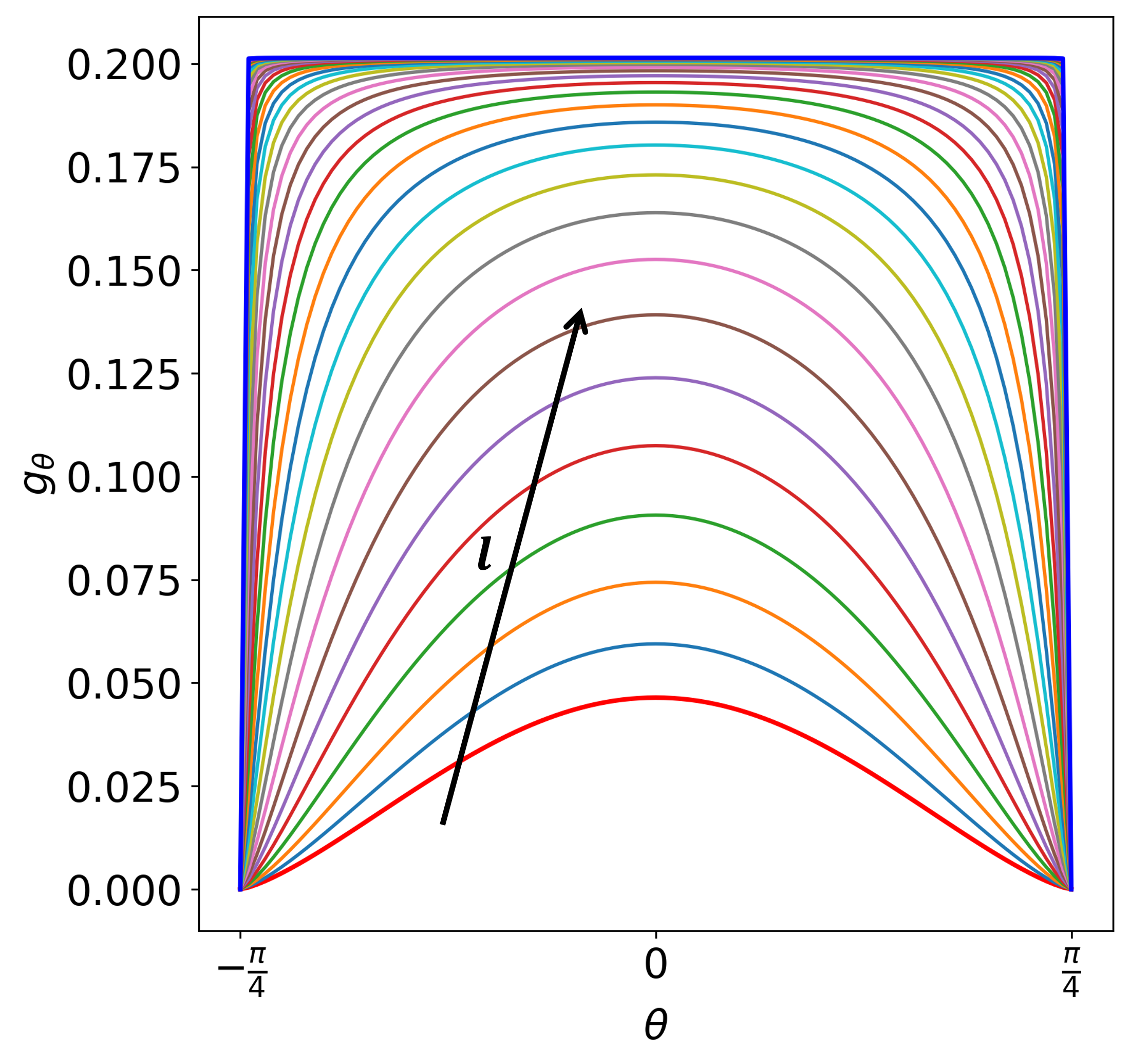}
  \caption{}
  \label{}
\end{subfigure}
\caption{
The functional RG evolution of 
(a) $v_{F,\theta}$,
(b) $g_\theta$
for $\epsilon=10^{-2}$.
At the UV scale ($l=0$), 
we assume that
the Fermi velocity is independent of angle 
and the Yukawa coupling has the d-wave form factor associated with the Ising-nematic order parameter :
$v_{F,\theta}(0)=1$ 
and
$e_\theta(0) =0.1\cos(2\theta)$,
which translates to
$g_\theta(0) =
(0.1\cos 2 \theta)^{\frac{4}{3}}$ 
for a circular Fermi surface with $\KFdim$ = 1 and $N$=1 
.
The arrows denote the direction of increasing length scale.
The Fermi velocity near the cold spots ($\pm \pi/4$) becomes much larger than the Fermi velocity at $\theta=0$ which is set to be $1$.
In the low-energy limit,
the effective coupling $g_\theta$, which is given by a ratio between the Yukawa coupling and the Fermi velocity, becomes independent of angle except at the cold spots.
The anomalous dimension of electron at angle $\theta$ is solely determined by $g_\theta$.
}
\label{fig:flowoffv}
\end{figure}

In \fig{fig:flowoffv}, 
we plot the RG evolution of 
$v_{F,\theta}(l)$
and
$g_{\theta}(l)$.
Since the coupling functions at general angles can be related to those within 
$-\frac{\pi}{4} \leq \theta \leq \frac{\pi}{4}$ through the $C_4$ symmetry,
we only show the profiles of the couplings within one quarter of the full range of angle.
At a UV cutoff scale ($l=0)$, we choose an angle-independent Fermi velocity and a small but non-zero Yukawa coupling with the d-wave form factor 
 related to the Ising-nematic order parameter.
Accordingly, the 
 Yukawa coupling 
 vanishes in the nodal direction
 ($\theta=\pm \pi/4$)
and is strongest in the 
anti-nodal direction ($\theta=0$).
As the energy scale is lowered, 
the Fermi velocity near the nodal direction becomes larger as electrons are slowed down less.
At the same time,
the Yukawa coupling, which is relevant at the non-interacting fixed point, grows.
It grows faster 
near the nodal direction
because electrons there are less incoherent.
Interestingly, 
the Yukawa coupling near the nodal direction 
eventually exceeds the coupling in the anti-nodal direction in the low-energy limit.
This is a consequence of the fact that 
the Yukawa coupling saturates at larger values for faster electrons due to the weaker screening effect.
The anomalous dimension is determined by the effective coupling which is the ratio between $e_\theta$ and $v_{F,\theta}$.
%
In the low-energy limit,
the effective coupling converges to an angle-independent value
\bqa
g^{*}_{\theta}
= 
\frac{5-2d}{2(d-1)}
\frac{1}{u_1(d)}
\label{eq:gstar}
\eqa
except at the cold spots.
It is noted that $\theta \rightarrow \pi/4$ limit 
and the low-energy limit do not commute.
Since our main interest lies in the majority of the `hot' Fermi surface, we will focus on the region away from the cold spots which form a set of measure zero.
As the effective coupling becomes angle-independent,
so does the anomalous dimensions of electrons.
In the low-energy limit,
the dynamical critical exponent 
and the anomalous dimensions become
\bqa
z^{*}= \frac{3}{2(d-1)}, ~~~~~
\eta^{*}_{\phi} = \frac{2d-5}{4}, ~~~~~
\eta^{*}_{\psi,\theta} = \frac{2d-5}{4}
\eqa
to the leading order in $\epsilon$.

It is noted that the  $\kappa_{F,\theta}$ and $v_{F,\theta}$ are marginal coupling functions.
This can be understood from the fact that the profiles of 
$\kappa_{F,\theta}$ and $v_{F,\theta}$ 
that emerge in the low-energy limit depend on  their profiles at the UV energy scale.
Namely, distinct low-energy fixed points have distinct 
$\kappa_{F,\theta}$ and
$v_{F,\theta}$.
On the contrary, 
the Yukawa coupling function is not marginal as its profile is tied to $\kappa_{F,\theta}$ and $v_{F,\theta}$
through the effective coupling that attains the universal value at low energies.

\subsection{Four-fermion coupling functions}

Let us turn to the beta-functionals for the four fermion coupling functions.
In the flavour space, 
the beta functionals take the diagonal form in the following basis,
  \bqa
\lambda^{(F_\pm),t}_{\theta_1,\theta_2}(\vec q) =  N\lambda^{(F_\pm,d)}_{\theta_1,\theta_2}(\vec q)+\lambda^{(F_\pm,e)}_{\theta_1,\theta_2}(\vec q), ~~~~~~
\lambda^{(F_\pm),a}_{\theta_1,\theta_2}(\vec q) = \lambda^{(F_\pm,e)}_{\theta_1,\theta_2}(\vec q), ~~~~~~
\lambda^{(P),\pm}_{\theta_1,\theta_2}(\vec q) = 
\lambda^{(P,d)}_{\theta_1,\theta_2}(\vec q)\pm \lambda^{(P,e)}_{\theta_1,\theta_2}(\vec q).
\label{eq:coupling_combination_betafunc}
\eqa
In two dimensions,
$\lambda^{(F_\pm),t}_{\theta_1,\theta_2}(\vec q)$
describes the forward scatterings 
for particle-hole pairs in the trivial (t) representation of the $SU(N)$ flavour group.
$F_\pm$ denotes the even ($+$) and odd ($-$) angular momentum channels, respectively. 
Similarly, 
$\lambda^{(F_\pm),a}_{\theta_1,\theta_2}(\vec q)$ represents forward scattering for particle-hole pairs in the adjoint (a) representation.
$\lambda^{(P),\pm}_{\theta_1,\theta_2}(\vec q)$ denotes the pairing interaction for particle-particle pairs in the symmetric ($+$) and anti-symmetric ($-$) representations, respectively.
Since the beta functional does not depend on the representation of the flavour group,
we will henceforth drop the superscript that specifies the representation  
in the coupling function.

Just as the effective 
 cubic coupling is given by the ratio between the Yukawa coupling and the Fermi velocity,
a ratio between the four-fermion coupling and the Fermi velocity plays the role of an effective quartic fermion coupling.
It is also convenient to absorb factors of $\KFtheta$ to simplify the angular measure.
For the effective coupling defined by
\bqa
\tilde{ \lambda}^{(\nu)}_{\theta_1,\theta_2}
= 
\sqrt{\frac{K_{F,\theta_1}K_{F,\theta_2}}{v_{F,\theta_1}v_{F,\theta_2}}}
\lambda^{(\nu)}_{\theta_1,\theta_2},
\label{eq:lambdatilde}
\eqa
the beta functional becomes
\begin{equation}
    \begin{aligned}
        \frac{\mathrm{d} \tilde{ \lambda}^{(\nu)}_{\theta_1,\theta_2}(\vec q)}{dl} 
         &=
        -\tilde{H}_{d; \theta_1,\theta_2}\left(\vec{q}\right)\tilde{ \lambda}^{(\nu)}_{\theta_1,\theta_2}(\vec q)
        +\left[g_{\theta_1}\mathcal{C}^{(\nu)}_d\left(\mathscr{L}_{\mu,\theta_1}(\vec q)\right)+g_{\theta_2}\mathcal{C}^{(\nu)}_d\left(\mathscr{L}_{\mu,\theta_2}(\vec q)\right)\right]\tilde{ \lambda}^{(\nu)}_{\theta_1,\theta_2}(\vec q)
        +\tilde{S}^{(\nu)}_{d;\theta_1,\theta_2}(\vec q)\\
        &-
        \bigintss_{-\frac{\pi}{2}}^{\frac{\pi}{2}}\frac{d\theta}{2\pi}\tilde{ \lambda}^{(\nu)}_{ \theta_1,\theta}(\vec q)\tilde{ \lambda}^{(\nu)}_{ \theta,\theta_2}(\vec q)\left[\frac{A_+^{(\nu)}(d) T_+(d)}{\left(\left(\mathscr{L}_{\mu,\theta}(\vec q)\right)^2+4\right)^{\frac{4-d}{2}}}+\frac{A_-^{(\nu)}(d) T_-(d)\left(\left(\mathscr{L}_{\mu,\theta}(\vec q)\right)^2+2(d-2)\right)}{\left(\left(\mathscr{L}_{\mu,\theta}(\vec q)\right)^2+4\right)^{\frac{6-d}{2}}}\right].
        \label{eq:betaforlambda}
    \end{aligned}
\end{equation}
Here, 
\begin{equation}\begin{aligned}
 \tilde{H}_{d; \theta_1,\theta_2}\left(\vec{q}\right)
= d-1 
+ (3d-3)(z-1)
+\eta_{\psi,\theta_1}+\eta_{\psi,\Theta\left(\theta_1,\vec{q}\right)}+\eta_{\psi,\theta_2}+\eta_{\psi,\Theta\left(\theta_2,\vec{q}\right)} -1+u_1(d)\left(g_0-\frac{g_{\theta_1}+g_{\theta_2}}{2}\right)
\end{aligned}
\label{eq:Hd}
\end{equation}
denotes the channel-independent part of the scaling dimension 
for the four-fermion coupling that includes 
the tree-level scaling dimension,
the dynamical critical exponent, the anomalous dimensions of the fields
and the extra contributions from the beta functions of $\KFtheta$ and $v_{F,\theta}$.
The second term on the right hand side is the contribution of the vertex correction generated by the critical boson.
The third term represents the source for the four-fermion coupling generated from the critical boson,
\begin{equation}
    \begin{aligned}
      &\tilde{S}_{d;\theta_1,\theta_2}^{(\nu)}(\vec q;\mu) = 
      \begin{cases}
      -
      \frac{1}{6 \pi N^{2}}
       \frac{\Omega_{d-1}}{(2\pi)^{d-1}\beta_d}
       e^{2}_{\theta_1}\left|X_{\theta_1}\right|\left|\chi_{\vartheta^{-1}\left(\theta_1\right)}\right|
       \left(\frac{
       A^{(\nu)}_+(d)
       }{\left(\mathscr{L}_{\mu,\theta_1}(\vec q)\right)^2+16}
       +\frac{
       A_-^{(\nu)}(d)
       }{2(d-1)}\frac{\left(\left(\mathscr{L}_{\mu,\theta_1}(\vec q)\right)^2-4\right)}{\left(\left(\mathscr{L}_{\mu,\theta_1}(\vec q)\right)^2+4\right)^2}\right)
       
       \\~&\hspace{-100pt} 
       \mbox{for~~~}
                       q\left(\theta_1,\theta_2\right) \ll
                \sqrt{\frac{Ng\KFdim \mu}{v_F}}, \\ 
       
      \frac{\snu}
      {4}\frac{e^{{}2}_{\theta_1,\theta_2}}{N}\sqrt{\frac{K_{F,\theta_1}K_{F,\theta_2}}{v_{F,\theta_1}v_{F,\theta_2}}}
      \frac{\left[g_{\theta_1}\mathcal{C}^{(\nu)}_d\left(\mathscr{L}_{\mu,\theta_1}(\vec q)\right)+g_{\theta_2}\mathcal{C}^{(\nu)}_d\left(\mathscr{L}_{\mu,\theta_2}(\vec q)\right)\right]\mu^2}{
       q\left(\theta_1,\theta_2\right)^2 
      +
\beta_d
\left(\frac{Ng_{\mtheta}\mathbf{K}_{F,\mtheta}}{\left|X_{\mtheta}\right|\left|\chi_{\vartheta^{-1}\left(\mtheta\right)}\right|v_{F,\mtheta}}\right)
^{\frac{3}{2}}
        \frac{\mu^{\frac{3}{2}}}{
          q\left(\theta_1,\theta_2\right) 
        }
        }
        \\~
        &
        \hspace{-100pt}
       \mbox{for~~~}
                \sqrt{\frac{Ng\KFdim \mu}{v_F}}
                \ll
q\left(\theta_1,\theta_2\right) 
    \ll
   \KFdim
      \end{cases}\label{eq:sourceforlambda}
      .
    \end{aligned}
\end{equation}
The channel-dependent sign $\snu$ 
is defined in Eq. (\ref{s_nu}).
The last term, which is quadratic in $\lambda$, represents the Fermi-liquid contribution.
The four-fermion coupling generated by the critical boson is 
$\tilde{ \lambda}^{(\nu)}(\vec q)
\sim
\tilde S^{(\nu)}_{d}(\vec q)$
whose eigenvalue as a matrix is order of $g^2$.
Then, 
$\int d\theta
 \tilde{ \lambda}^{(\nu)}_{ \theta_1,\theta}(\vec q)
 \tilde{ \lambda}^{(\nu)}_{ \theta,\theta_2}(\vec q)$
is negligible compared to the rest of the terms
in \eq{eq:betaforlambda}
near the upper critical dimension in which $g \ll 1$.
Therefore, the term that is quadratic in $\tilde \lambda$ can be ignored to the leading order in $\epsilon$ 
as far as the bare value of $\tilde \lambda$ is small and
$\tilde \lambda$ itself does not grow under the renormalization group flow.
Here, we consider the case where the bare four-fermion coupling is weak. 
Later, we will see that this approximation breaks down in dimensions close to $d=2$.

In solving the beta function of the four-fermion coupling, 
we suppose that the energy scale is sufficiently low that  $v_{F,\theta}$
and $g_\theta$ are already close to their fixed point profiles.
In this case, we can ignore the flow of 
$v_{F,\theta}$ and $g_\theta$  
because their beta functions
do not depend on $\KFtheta$ that keeps growing.
In the low-energy limit,
$g_\theta^*$ in \eq{eq:gstar} becomes independent of $\theta$ 
and will be denoted as $g^*$ henceforth.
Despite the angle-independent effective Yukawa coupling, Fermi velocity and Fermi momentum 
 generally have non-trivial angular dependence.
Since $v_{F,\theta}$ and $\kappa_{F,\theta}$ 
(defined in \eq{eq:KFkappaF})
are marginal coupling functions,
they can have general angular profiles in the low-energy limit as far as they respect the $C_4$ symmetry. 
The beta functional of the four-fermion coupling does not respect the continuous rotational symmetry because the source term in  \eq{eq:sourceforlambda} depends on $v_{F,\theta}$ and $\KFthetadim$.
Nonetheless, the angle-independent anomalous dimension of fermion points  toward a subtle emergent symmetry.
To take advantage of this symmetry, we note that
$\tilde{S}_{d;\theta_1,\theta_2}^{(\nu)}(\vec q;\mu)$ 
as a function of $\theta_1$
is sharply peaked around $\theta_2$ with
the width that goes to zero in the small $\mu$ limit as
$ \Delta \theta  \sim\left(\frac{N g^* \left|X_{\mtheta}\right|
\mu}
{\mathbf{K}_{F,\mtheta}
\left|\chi_{\vartheta^{-1}\left(\mtheta\right)}\right|v^{*}_{F,\mtheta}}\right)^{\frac{1}{2}}$. 
Away from the cold spots, the variations of $v_{F,\theta}$ and $\kappa_{F,\theta}$ are negligible within $\Delta \theta$ in the small $\mu$ limit. 
Namely, one can regard $v_{F,\theta}$ and $\kappa_{F,\theta}$ constants within the range of angles in which the source is non-negligible.
This allows one to remove the angular dependence in the source
by `stretching' the angular space in a $\theta$ dependent manner.
To implement this,
we introduce a new angular coordinate $\bar \theta$ related to the original one through a non-linear transformation,
\bqa
\theta = a(\bar{\theta})
\label{eq:nonlinear_theta_transformation}
\eqa
with 
$a^{\prime}(\bar{\theta}) \equiv\frac{d\theta}{d\bar{\theta}} =  \beta_d^{\frac{1}{3}}\left(\frac{N
g^{*}
\left|X_{\theta}\right|}{\KFthetadim|\chi_{\vartheta^{-1}(\theta)}|
v^{*}_{F,\theta}
}
\right)^{\frac{1}{2}}$
and $a(0)=0$.
Under this transformation,  $\bar \theta$ acquires scaling dimension $1/2$.
Physically, this captures the fact that the number of patches increases as $(\KFAV/\mu)^{1/2}$ with decreasing energy scale
as is shown in \fig{fig:patch}.
Ignoring 
$a^{\prime\prime}(\bar \theta)$ and higher derivatives,
which are suppressed by 
$\frac{\left( \partial_\theta v_{F,\theta} \right) 
\Delta \theta}{v_{F,\theta}}$
and
$\frac{\left( \partial_\theta \KFthetadim \right) 
\Delta \theta}{\KFthetadim}$,
we can write the boson propagator in the source term as
$
\left[
\frac{1}{\beta_d^{\frac{2}{3}}
\frac{N
g^{*}
\mathbf{K}_{F,\mtheta}
}{
\left|X_{\mtheta}\right|
\left|\chi_{\vartheta^{-1}\left(\mtheta\right)}\right|
v^{*}_{F,\theta_2}
}
}
\right] \times
$
$
\frac{1}
{
|\bar \theta_2-\bar \theta_1|^2
      +
        \frac{\mu^{\frac{3}{2}}}{
 |\bar \theta_2-\bar \theta_1|
        }}
$.
%
%
The remaining angle-dependent factor 
and $\mu^{1/2}$ can be absorbed to
the four-fermion coupling functions through
\begin{equation}
\begin{aligned}
\bar{\lambda}^{(\nu)}_{\bar{\theta}_1,\bar{\theta}_2}( \vec q )
    \equiv
\sqrt{
\mu
a^\prime(\bar{\theta}_1) a^\prime(\bar{\theta}_2)}
    \tilde{ \lambda}^{(\nu)}_{\theta_1,  \theta_2}(\vec q)
\end{aligned}
\label{eq:lambdabar}
\end{equation}
along with a trivial relabeling of other coupling functions,
\begin{equation}
\begin{aligned}
\bar{e}_{\bar{\theta}_1,\bar{\theta}_2} =   e_{
    \theta_1 ,  \theta_2}, ~~
    \bar{{\bf K}}_{F,\bar{\theta}_1} =  {\bf K}_{F,\theta_1},
    ~~ 
    \bar{v}_{F,\bar{\theta}_1} =  v_{F,   \theta_1}.
    \label{rescaled_couplings}
\end{aligned}
\end{equation}
To the leading order in 
$\partial_\theta v_{F,\theta}$
and
$\partial_\theta \KFtheta$,
the beta functional for $\bar \lambda$ in the the rescaled angular space becomes
\begin{equation}
    \begin{aligned}
 \frac{\mathrm{d} \bar{\lambda}^{(\nu)}_{\bar{\theta}_1,\bar{\theta}_2}(\vec q)}{dl} 
         &=
- \HD
\bar{\lambda}^{(\nu)}_{\bar{\theta}_1,\bar{\theta}_2}(\vec q)
        +g^{*}\left[\mathcal{C}^{(\nu)}_d\left(\bar{\mathscr{L}}_{\mu,\bar{\theta}_1}( \vec q )\right)+\mathcal{C}^{(\nu)}_d\left(\bar{\mathscr{L}}_{\mu,\bar{\theta}_2}( \vec q )\right)\right]\bar{\lambda}^{(\nu)}_{\bar{\theta}_1,\bar{\theta}_2}(\vec q)
        +\bar{S}_{d;\bar{\theta}_1,\bar{\theta}_2}^{(\nu)}\left(\vec q;\Lambda e^{-l}\right) \\
&-
\bigintss_{-\bar \theta_{max}}^{\bar \theta_{max}}
\frac{d\bar \theta}{2\pi\sqrt{\mu}}
\bar{ \lambda}^{(\nu)}_{ \bar \theta_1,\bar \theta}(\vec q)\bar{ \lambda}^{(\nu)}_{ \bar \theta,\bar \theta_2}(\vec q)\left[\frac{A_+^{(\nu)}(d) T_+(d)}{\left(\left(\bar{\mathscr{L}}_{\mu,\bar \theta}(\vec q)\right)^2+4\right)^{\frac{4-d}{2}}}+\frac{A_-^{(\nu)}(d) T_-(d)\left(\left(\bar{\mathscr{L}}_{\mu,\bar \theta}(\vec q)\right)^2+2(d-2)\right)}{\left(\left(\bar{\mathscr{L}}_{\mu,\bar \theta}(\vec q)\right)^2+4\right)^{\frac{6-d}{2}}}\right].
\label{beta_rescaled_lambda}
    \end{aligned}
\end{equation}
Here
$\HD=
\tilde{H}_{d; \bar \theta_1,\bar \theta_2}\left(\vec{q}\right)+\frac{1}{2} $
is independent of angle in the low-energy limit.
The new source term is written as
\begin{equation}
    \begin{aligned}
\bar{S}_{d;\bar{\theta}_1,\bar{\theta}_2}^{(\nu)}(\vec q;\mu) = 
         \begin{cases}
         -
         \frac{1}{6 \pi}
         \frac{g^{*2}}{\beta_d^{\frac{2}{3}}}\frac{\Omega_{d-1}}{(2\pi)^{d-1}}
          \left[
         \frac{A^{(\nu)}_+(d)}{\left(\bar{\mathscr{L}}_{\mu,\bar{\theta}_1}( \vec q )\right)^2+16}+\frac{A^{(\nu)}_-(d)}{2(d-1)}\frac{\left(\left(\bar{\mathscr{L}}_{\mu,\bar{\theta}_1}( \vec q )\right)^2-4\right)}{\left(\left(\bar{\mathscr{L}}_{\mu,\bar{\theta}_1}( \vec q )\right)^2+4\right)^2}
        \right] ~&~\mbox{for}~~~ |\bar{\theta}_1-\bar{\theta}_2|\ll \sqrt{\mu},\\
   
   \frac{\snu}
   {4}
   \frac{g^{*2}}{\beta_d^{\frac{1}{3}}}
        \frac{|\bar{\theta}_1 - \bar{\theta}_2 |\mu}{|\bar{\theta}_1 - \bar{\theta}_2 |^{3}+
        \mu^{\frac{3}{2}}}
        \left[
\mathcal{C}^{(\nu)}_d\left(\bar{\mathscr{L}}_{\mu,\bar{\theta}_1}( \vec q )\right)+
\mathcal{C}^{(\nu)}_d\left(\bar{\mathscr{L}}_{\mu,\bar{\theta}_2}( \vec q )\right)
\right]
        ~&~\mbox{for}~~~
        \sqrt{\mu} \ll
        |\bar{\theta}_1-\bar{\theta}_2| 
        \ll \sqrt{\KFdim}
         \end{cases},
         \label{Rescaled_Source}
    \end{aligned}
\end{equation}
where
$\bar{\mathscr{L}}_{\mu,\bar{\theta}}( \vec q ) =  \mathscr{L}_{\mu,\theta}(\vec q)$
(
$\bar{L}_{\bar{\theta}}( \vec q ) =  L_{\theta}(\vec q)$
),
$\bar{\mathscr{F}}_{\varphi,\bar{\theta}} = \mathscr{F}_{\varphi,\theta}$,
        $\bar{\mathscr{G}}_{\varphi,\bar{\theta}} = \mathscr{G}_{\varphi,\theta}$, 
$\bar \theta_{max}=
a^{-1}(\pi/2)
\sim \sqrt{\KFdim}$.
In this rescaled coordinate, the intra-patch to inter-patch crossover occurs at $\Delta \bar \theta \sim \sqrt{\mu}$.
The beta functional for $\bar{\lambda}^{(\nu)}_{\bar{\theta}_1,\bar{\theta}_2}(\vec{q} = 0)$  is manifestly invariant under 
$\bar{\theta} \rightarrow \bar{\theta} + \Delta\bar{\theta}$.
It is noted that the rigid translation in $\bar \theta$
corresponds to a non-linear diffeomorphism in $\theta$.
This diffeomorphism appears as a sliding transformation 
within the patch theory\cite{MAX0,DENNIS} because the non-linearity is negligible within a patch.

Near $d_c$, we drop the term that is quadratic in $\bar{\lambda}$ in Eq. (\ref{beta_rescaled_lambda}) and
its solution 
can be written as
\begin{equation}
    \begin{aligned}
\bar{\lambda}^{(\nu)}_{\bar{\theta}_1,\bar{\theta}_2}(\vec q;l) 
& =
\bar {\mathscr{V}}^{(\nu)}_{\bar{\theta}_1,\bar{\theta}_2}(\vec q,l) 
    \Bigg[
    \bar{\lambda}^{UV(\nu)}_{\bar{\theta}_1,\bar{\theta}_2}(\vec q)
+ \int_0^l dl' 
\frac{
\bar{S}_{d;\bar{\theta}_1,\bar{\theta}_2}^{(\nu)}(\vec q;\Lambda e^{-l'})  
}{
\bar{\mathscr{V}}^{(\nu)}_{\bar{\theta}_1,\bar{\theta}_2}(\vec q,l') 
}
    \Bigg],
    \end{aligned}
\label{lambda_general_solution}
\end{equation}
where
$\bar{\lambda}^{UV(\nu)}_{\bar{\theta}_1,\bar{\theta}_2}(\vec q)$ represents the UV four-fermion coupling function at $l=0$
and
\bqa
\bar{\mathscr{V}}^{(\nu)}_{\bar{\theta}_1,\bar{\theta}_2}(\vec q,l) =
     e^{ -\int_0^l dl^{\prime} \left\{ \HD
     -
     g^{*}
     \left[
     \mathcal{C}^{(\nu)}_d\left(e^{l^{\prime}-l^{*}_{\bar{\theta}_1,\vec{q}}}\right)
     +
     \mathcal{C}^{(\nu)}_d\left(e^{l^{\prime}-l^{*}_{\bar{\theta}_2,\vec{q}}}\right)\right]\right\}
     }
\eqa
describes the contribution of the anomalous dimension and the $\vec q$-dependent vertex correction.
The beta functional exhibit three crossovers
at logarithmic length scales 
$l^{*}_{\bar{\theta}_1,\bar{\theta}_2} $,
$l^{*}_{\bar{\theta}_1,\vec{q}}$
and
$l^{*}_{\bar{\theta}_2,\vec{q}}$,
where
\begin{equation}
    \begin{aligned}
l^{*}_{\bar{\theta}_1,\bar{\theta}_2} = 2\log\left| \frac{\Lambda^{\frac{1}{2}}}{\bar{\theta}_1-\bar{\theta}_2}\right|, 
~~~~
l^{*}_{\bar{\theta},\vec{q}} =  \mathrm{log}\left(\frac{\Lambda}{\bar{L}_{\bar{\theta}}( \vec q )}\right).
    \end{aligned}
\end{equation}
$l^{*}_{\bar{\theta}_1,\bar{\theta}_2}$ represents 
the logarithmic length scale above which two points at angle $\theta_1$ and $\theta_2$ move outside one patch.
$l^{*}_{\bar{\theta}_1,\vec{q}}$ and $l^{*}_{\bar{\theta}_2,\vec{q}}$
represent the length scales
above which the vertex corrections to the four-fermion coupling 
at angle $\theta_1$
and $\theta_2$
are turned off,
respectively.
In general,
$\mathcal{C}^{(\nu)}_d\left( e^{l-l^{*}_{\bar{\theta}_1,\vec{q}}}\right)$
and
$\mathcal{C}^{(\nu)}_d\left( e^{l-
l^{*}_{\bar{\theta}_2,\vec{q}}
}\right)$
turn off at different length scales:
if $\vec q$ is tangential to 
$\theta_1$ but not to $\theta_2$,
fermion pairs with momentum $\vec q$ has much lower energy 
near $\theta_1$ than
those near $\theta_2$,
which leads to
$l^{*}_{\bar{\theta}_1,\vec{q}} 
\gg
l^{*}_{\bar{\theta}_2,\vec{q}}$.
These scales create a rich crossover behaviours for the universal four-fermion coupling function.
In particular, the relative magnitudes among
$l^{*}_{\bar{\theta}_1,\bar{\theta}_2}$,
$l^{*}_{\bar{\theta}_1,\vec{q}}$
and
$l^{*}_{\bar{\theta}_2,\vec{q}}$
determine a series of crossovers the four-fermion coupling undergoes with decreasing energy.

\subsubsection{Intra-patch crossover}

If
$l^{*}_{\bar{\theta}_1,\bar{\theta}_2} 
\gg
l^{*}_{\bar{\theta}_1,\vec{q}} \sim
l^{*}_{\bar{\theta}_2,\vec{q}}$,
fermions at angles $\theta_1$ and $\theta_2$ are close to each other so that they stay within a patch 
until quantum corrections are cut off by non-zero $q$ at low energies.
At short distance scale with
$l \ll  l^{*}_{\bar{\theta}_1,\vec{q}}, l^{*}_{\bar{\theta}_2,\vec{q}}$, 
the four-fermion coupling becomes (see Appendix \ref{sec:case1})
\begin{equation}
    \begin{aligned}
       \bar{\lambda}^{(\nu)}_{\bar{\theta}_1,\bar{\theta}_1}(\vec{q}; l) =  e^{ -\DD l}
        \left[\bar{\lambda}^{UV(\nu)}_{\bar{\theta}_1,\bar{\theta}_1}+\frac{g^{*2}\varsigma_d^{(\nu)}}{\DD}\right] -\frac{g^{*2}\varsigma_d^{(\nu)}}{\DD}\label{eq:high_energy_lambda}
    \end{aligned}
\end{equation}
where
\bqa
\DD =  \HD
- \eta_d^{(\nu)},
~~~~~
\varsigma_d^{(\nu)} =
\frac{1}
{48\pi}
\frac{\Omega_{d-1}}{\beta_d^{\frac{2}{3}}(2\pi)^{d-1}}\left[\frac{A_+^{(\nu)}(d)}{2}-\frac{A^{(\nu)}_-(d)}{d-1}\right],
~~~~~
\eta_d^{(\nu)} = 
2
g^{*}\rho_d^{(\nu)}.
\label{eq:DD}
\eqa
At long distance scales with
$l \gg  l^{*}_{\bar{\theta}_1,\vec{q}}, l^{*}_{\bar{\theta}_2,\vec{q}}$, 
the interaction is suppressed as 
the energy of virtual fermions carrying momentum $\vec q$ in the loop becomes greater than the energy scale $\Lambda e^{-l}$ 
(see Appendix \ref{sec:case2})
\begin{equation}
    \begin{aligned}
\bar{\lambda}^{(\nu)}_{\bar{\theta}_1,\bar{\theta}_1}\left(\vec q;l\right)  = 
e^{-\DD l} 
        \left[\bar{\lambda}^{UV(\nu)}_{\bar{\theta}_1,\bar{\theta}_1}( \vec q )+\frac{g^{*2}\varsigma_d^{(\nu)}}{\DD}\right]\left(\frac{\mu}{\bar{L}_{\bar{\theta}_1}( \vec q )}\right)^{\eta_d^{(\nu)}} -\frac{g^{*2}\varsigma_d^{(\nu)}}{\DD}\left(\frac{\mu}{\bar{L}_{\bar{\theta}_1}( \vec q )}\right)^{
\DD+  \eta_d^{(\nu)}
} .
\label{eq:intra_low}
    \end{aligned}
\end{equation}
Interestingly, 
there is no single power 
 of $q$ with which the coupling decays as $q$ becomes larger relative to $\mu$ for a given $\KFthetadim$.
If $\vec q$ is tangential to the patch, the coupling function is proportional to 
$\left( \frac{\mu \KFdim}{q^2}
 \right)^{
\DD+  \eta_d^{(\nu)}
}$ 
and it takes a scale invariant form when $\frac{\sqrt{\mu}}{q}$ is fixed in the low-energy limit.
%
This patch scaling is valid within $|\varphi - \bar \vartheta^{-1}(\bar \theta_1)| < \sqrt{\mu/\KFdim}$.
On the other hand,
the coupling function scales as
$\left( \frac{\mu}{q}
 \right)^{
 \DD+  \eta_d^{(\nu)}
 }$ 
 for $\vec q$ that is not tangential to the patch at angle $\theta_1 \approx \theta_2$.
In this case, the coupling takes a scale invariant form when the low-energy limit is taken with
$\frac{\mu}{q}$ fixed.
%
This forces us to introduce different dynamical critical exponents for different observables.
We emphasize that this is different from the cases
in which the dynamical critical exponent runs under the renormalization group flow.
In the present case, 
different low-energy observables exhibit different dynamical critical exponents at {\it one} infrared fixed point.
This lack of a unique dynamical critical exponent is a manifestation of the extra relevant scale in the problem, the Fermi momentum.

\subsubsection{Inter-patch crossover}

If $l^{*}_{\bar{\theta}_1,\bar{\theta}_2}  \ll l^{*}_{\bar{\theta}_1,\vec{q}}, l^{*}_{\bar{\theta}_2,\vec{q}}$, two fermions that are engaged in the interactions are far from each other that
as the energy scale is lowered they first move outside the common patch before the vertex correction due to the Yukawa coupling is turned off at each angle. 
Below, we consider the case with 
$l^{*}_{\bar{\theta}_1,\vec{q}} \gg
l^{*}_{\bar{\theta}_2,\vec{q}}$.
At distance scales shorter than any of the crossover length scales,
the four-fermion coupling takes the same form as \eq{eq:high_energy_lambda}.
Within the first window of intermediate length scales 
with
$l^{*}_{\bar{\theta}_1,\vec{q}},~l^{*}_{\bar{\theta}_2,\vec{q}}
 \gg l \gg l^{*}_{\bar{\theta}_1,\bar{\theta}_2} $,
the source term from the ladder diagram is suppressed by a power of 
$1/(\bar \theta_1-\bar \theta_2)$
but the vertex correction is still on, giving rise to the following profile (see Appendix \ref{sec:case3})
\begin{equation}
    \begin{aligned}
        \bar{\lambda}^{(\nu)}_{\bar{\theta}_1,\bar{\theta}_2}(\vec{q}; l) \approx  Y_{d}^{(\nu)} \left|\frac{\sqrt{\mu}}{\bar{\theta}_1-\bar{\theta}_2}\right|^{ 2\DD 
        }
        +
e^{ -\DD l}
        \left[\bar{\lambda}^{UV(\nu)}_{\bar{\theta}_1,\bar{\theta}_2}\left(\vec q\right)+\frac{g^{*2}\varsigma_d^{(\nu)}}{\DD}\right],
    \label{eq:inter_mid1}
    \end{aligned}
\end{equation}
where
$ Y_{d}^{(\nu)} = g^{*}\left(\frac{\snu\eta_d^{(\nu)}} 
{
12
\beta_d^{\frac{1}{3}}}
\bigg\{\psi ^{(0)}\left(
\frac{2\DD   +1}{6}\right)-\psi ^{(0)}\left(\frac{\DD   +2}{3}\right)+2 \pi~
\mathrm{cosec} \left(
\frac{\pi }{3}  \left(2\DD+1 \right)\right)\bigg\}-\frac{ g^{*}\varsigma_d^{(\nu)}}{ \DD }\right)$\footnote{
Here, the expression for $Y_d^{(\nu)}$ has been obtained by taking the low-energy limit for 
$\DD\neq1$. 
For $\DD = 1$, which happens at $d=2$ 
for $\nu = F_-$
and 
at $d=5/2$ for all $\nu$,
the expression is 
 modified by a finite amount 
 due to the non-commutativity between the low-energy limit and 
the $d \rightarrow 2, 5/2$ limits.
At $d=2$, for example,
$Y_{d}^{(F_-)}$ becomes
$-\frac{g^{*2}\varsigma_d^{(\nu)}}{\DD }$.
}
with $\psi^{(m)}(z)$ being the polygamma function of order $m$.

The angle-dependent coupling functions become scale invariant when the low-energy limit is taken with 
$\bar \theta e^{l/2}$ fixed :
$\bar{\lambda}^{{}^*(\nu)}_{\bar{\theta}_1,\bar{\theta}_2} \approx  Y_{d}^{(\nu)} \left|
\frac{\sqrt{\Lambda}}{\hat{\theta}_1-\hat{\theta}_2}\right|^{ 
2\DD }
$,
where
$\hat \theta \equiv \bar \theta e^{l/2}$. 
For $q=0$,  \eq{eq:inter_mid1} is valid down to the zero-energy limit.
In particular, the universal exponent
$\DD $
determines how the Landau function ($\nu = F_{\pm}$)
and the pairing interaction ($\nu = P$) with zero center of mass momentum
decay as a function of 
 $\bar \theta_1 - \bar \theta_2$.
%

For $\vec q \neq 0$, there are additional crossovers.
In the second intermediate length scale with
$l^{*}_{\bar{\theta}_2,\vec{q}} \gg l \gg 
 l^{*}_{\bar{\theta}_1,\vec{q}} \gg l^{*}_{\bar{\theta}_1,\bar{\theta}_2}
 $,
 the vertex correction at angle $\theta_1$ is turned off 
 and the four-fermion coupling is further suppressed by $\mu/\bar L_{\bar \theta_1}(\vec q)$ (see Appendix \ref{sec:case4}),
\begin{equation}
    \begin{aligned}
    \bar{\lambda}^{(\nu)}_{\bar{\theta}_1,\bar{\theta}_2}\left(\vec q;l\right) \approx
    \left[
    Y_{d}^{(\nu)}
    \left|\frac{\sqrt{\mu}}{\bar{\theta}_1-\bar{\theta}_2}\right|^{
       2\DD }
               +
e^{ -\DD l}
        \left(\bar{\lambda}^{UV(\nu)}_{\bar{\theta}_1,\bar{\theta}_2}\left(\vec q\right)+\frac{g^{*2}\varsigma_d^{(\nu)}}{\DD}\right)
\right]
\left(\frac{\mu}{\bar{L}_{\bar{\theta}_1}( \vec q )}\right)^{\frac{\eta_d^{(\nu)}}{2}}.
    \label{eq:inter_mid2}
    \end{aligned}
\end{equation}
In the low-energy limit with
$l \gg 
l^{*}_{\bar{\theta}_2,\vec{q}} 
\gg 
l^{*}_{\bar{\theta}_1,\vec{q}} \gg l^{*}_{\bar{\theta}_1,\bar{\theta}_2}$,
there is an additional suppression due to the decoupling of the vertex correction at both angles (see Appendix \ref{sec:case5}),
\begin{equation}
    \begin{aligned}
    \bar{\lambda}^{{}^{*}(\nu)}_{\bar{\theta}_1,\bar{\theta}_2}\left(\vec{q};l\right) \approx
    \left[
    Y_{d}^{(\nu)}
    \left|\frac{\sqrt{\mu}}{\bar{\theta}_1-\bar{\theta}_2}\right|^{
       2\DD }
                 +
e^{ -\DD l}
        \left(\bar{\lambda}^{UV(\nu)}_{\bar{\theta}_1,\bar{\theta}_2}\left(\vec q\right)+\frac{g^{*2}\varsigma_d^{(\nu)}}{\DD}\right)
\right]
       \left(\frac{\mu^2}{\bar{L}_{\bar{\theta}_1}\left(\vec{q}\right)\bar{L}_{\bar{\theta}_2}\left(\vec{q}\right)}\right)^{\frac{\eta_d^{(\nu)}}{2}}.
    \label{eq:inter_low}
    \end{aligned}
\end{equation}
Like \eq{eq:intra_low},
there isn't a unique way of scaling $q$ that makes 
Eqs.
\eqref{eq:inter_mid1}
-\eqref{eq:inter_low}
invariant in the low energy limit.
The scaling dimension of $q$ depends on the relative angle between 
$\vec q$ and the Fermi surface at which the external fermions are located.

\begin{figure}[ht]
\centering
\begin{subfigure}{.4\textwidth}
  \centering
  \includegraphics[width=1\linewidth]{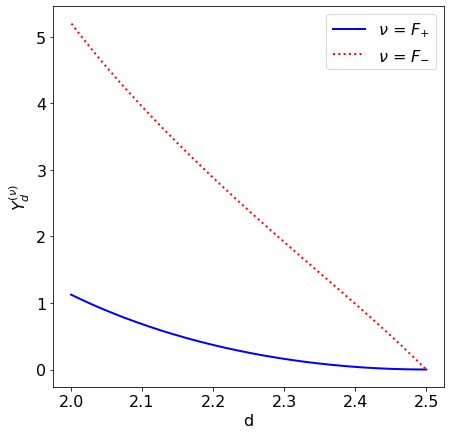}
  \caption{}
  \label{fig:M1d_F+-}
\end{subfigure}%
\begin{subfigure}{.4\textwidth}
  \centering\includegraphics[width=1\linewidth]{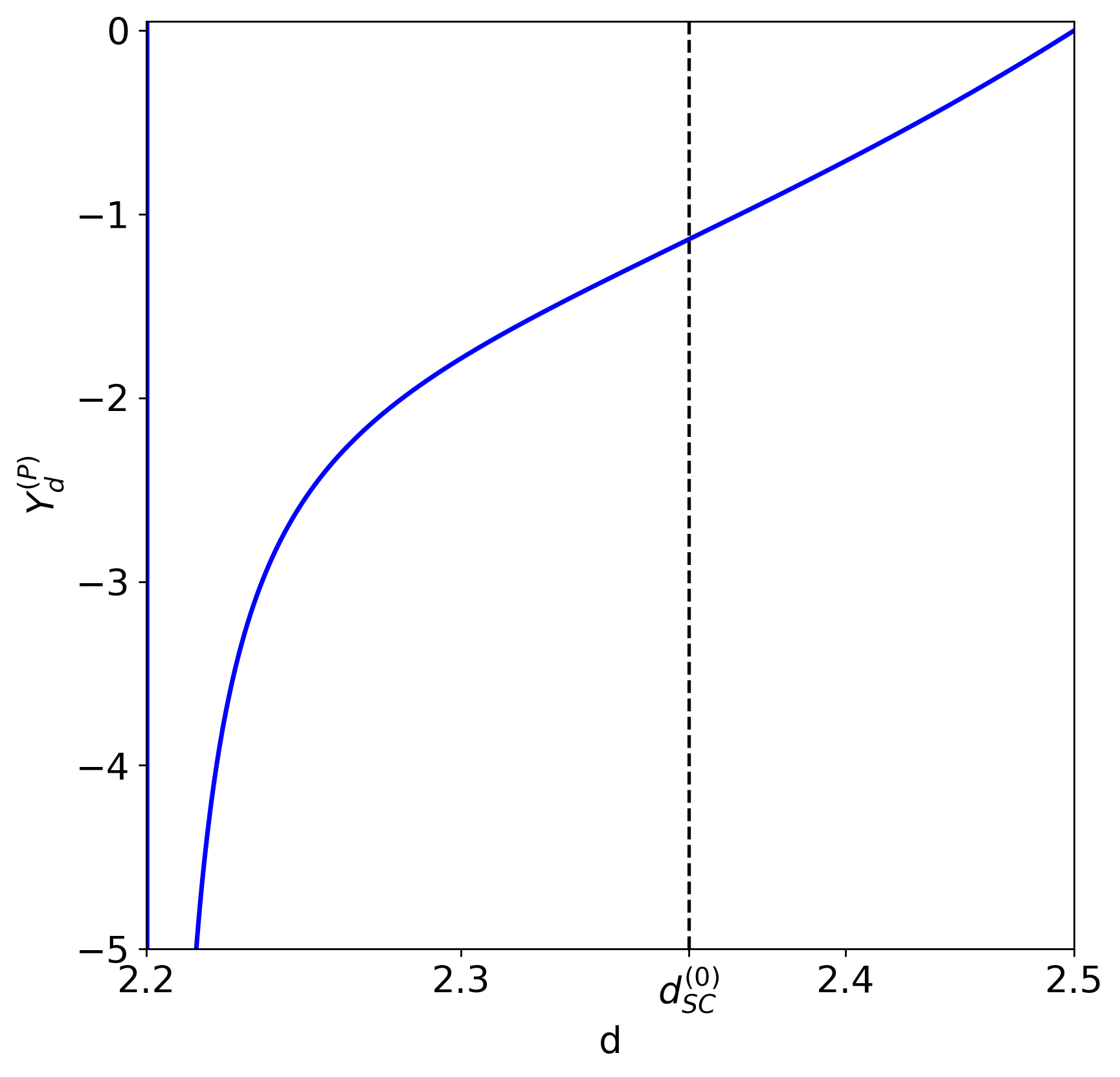}
  \caption{}
  \label{fig:M1d_FP}
\end{subfigure}
\caption{
The magnitude of the universal four-fermion interaction computed to the leading order in the $\epsilon$ expansion
in the large-angle limit. 
The interaction is repulsive in the forward scattering channels ($\nu= F_{\pm}$) while attractive in the paring channel ($\nu =P$). 
The amplitude of the pairing interaction diverges around $d=2.25$ at which its scaling dimension vanishes.
However, the pairing interaction becomes relevant well above that dimension around $\dsco=2.37$, 
at which its scaling dimension is $-1/2$.
}
\label{fig:M1d}
\end{figure}

In  Fig. \ref{fig:M1d}, we show the magnitude of the universal coupling 
($Y_{d}^{(\nu)}$)
generated from the critical boson 
as a function of space dimension.
In the forward scattering channels, 
the strength of interaction remains finite in all $d$.
On the other hand, the overall strength of the pairing interaction diverges around $d\approx 2.25$.
This is the dimension at which
the scaling dimension of the four-fermion coupling ($-\DD$) vanishes 
and becomes nominally marginal.
In $d<2.25$, the pairing interaction has a positive scaling dimension,
and the attractive interaction generated from the critical boson grows indefinitely under the renormalization group flow.
Clearly, the present computation that ignores the quadratic term in $\lambda$ is not valid in  $d<2.25$.
One may naively conclude that 
the four-fermion coupling is irrelevant 
and the current result is 
 qualitatively valid in $d>2.25$. 
However, this is not true because large-angle scatterings kick in and
make the four-fermion coupling relevant below 
 a larger critical dimension, 
 $d_{SC}^{(0)} = 2.37$.
In Sec. \ref{sec:NFLfp}, 
we will show that the four-fermion coupling actually turns relevant when its scaling dimension becomes greater than $-1/2$, which happens around $d_{SC}^{(0)} =2.37$ denoted as the vertical dashed line in \fig{fig:M1d_FP}.
This mismatch between the scaling dimension and the relevancy arises due to the projective nature of the metallic fixed point where the Fermi momentum never ceases to run.

\subsubsection{
Crossover from the intra-patch to the inter-patch crossovers
}


The crossovers discussed in the previous two subsections can be summarized as
\begin{equation}
    \begin{aligned}
\bar        \lambda^{(\nu)}_{\bar \theta_1,\bar \theta_2}(\vec q;l)=
         \left[
         F^{(\nu)}_{\bar \theta_1,\bar \theta_2}\left(\vec{q}\right) 
          +
e^{-\DD l } 
\left(\bar{\lambda}^{UV(\nu)}_{\bar{\theta}_1,\bar{\theta}_2}\left(\vec{q}\right)+\frac{g^{*2}\varsigma_d^{(\nu)}}{\DD}\right)
\right]
{\cal V}^{(\nu)}_{\mu;\bar \theta_1}(\vec q)
{\cal V}^{(\nu)}_{\mu;\bar \theta_2}(\vec q), 
\label{eq:full_lambda}
    \end{aligned}
\end{equation}
where
\begin{equation}
    \begin{aligned}
        F^{(\nu)}_{\bar \theta_1,\bar \theta_2}\left(\vec{q}\right) 
        =
        \begin{cases}
    -\frac{g^{*2}\varsigma_d^{(\nu)}}{\DD}~~&
  \frac{\mu^{1/2}}{|\bar \theta_1-\bar \theta_2|}
  \gg 1  ~\& ~
     \frac{\mu}{\bar{L}_{\btheta}(\vec q)} \gg 1 \\
           -\frac{g^{*2}\varsigma_d^{(\nu)}}{\DD}
           \left(\frac{\mu}{\bar{L}_{\btheta}(\vec q)}
           \right)^{\DD}
           ~~& 
  \frac{\mu^{1/2}}{|\bar \theta_1-\bar \theta_2|} \gg 1  ~\& ~
     \frac{\mu}{\bar{L}_{\btheta}(\vec q)} \ll 1 \\
           Y_{d}^{(\nu)}
          \left|
           \frac{\sqrt{\mu}}
           {
           \bar \theta_1-\bar \theta_2 }
           \right|^{2 \DD }
           ~~&
   \frac{\mu^{1/2}}{|\bar \theta_1-\bar \theta_2|} \ll
   1 
       \end{cases},
       ~~~~~~
        {\cal V}^{(\nu)}_{\mu;\bar \theta}(\vec q) = 
        \begin{cases}
           1~~&
     \frac{\mu}{\bar{L}_{\bar{\theta}}(\vec q)} \gg 1 \\
 \left(
\frac{\mu}{\bar{L}_{\bar{\theta}}(\vec q)}
\right)^{
 \frac{\eta_d^{(\nu)}}{2}
  }~~& 
     \frac{\mu}{\bar{L}_{\bar{\theta}}(\vec q)} \ll 1 \\
        \end{cases} 
\end{aligned}
    \label{eq:Vnuq}
\end{equation}
with
$\btheta  = \frac{\bar \theta_1+\bar \theta_2}{2}$.
The first term 
in \eq{eq:full_lambda}
is the universal part of the four-fermion coupling that is generated from the critical boson.
$F^{(\nu)}_{\bar \theta_1,\bar \theta_2}(\vec q)$
determines its profile as a function of
$|\bar \theta_1-\bar \theta_2|$.
The universal coupling takes distinct forms,
depending on whether the two angles are 
well within one patch 
($\frac{\mu^{1/2}}{|\bar \theta_1-\bar \theta_2|} \gg 1$)
or 
well outside one patch 
($\frac{\mu^{1/2}}{|\bar \theta_1-\bar \theta_2|} \ll 1$)
at energy scale $\mu = \Lambda e^{-l}$.
The additional suppression of
$\left(\frac{\mu}{\bar{L}_{\btheta}(\vec q)} \right)^{\DD}$
for 
$F^{(\nu)}_{\bar \theta_1,\bar \theta_2}\left(\vec{q}\right)$ in the second case of \eq{eq:Vnuq} is due to the fact that 
even if two incoming fermions are within one patch, the outgoing fermions move out of that patch if their momentum exchange $\vec q$ is large enough.
The suppression is controlled by the ratio between energy scale $\mu$ and 
$\bar L_{\bar \theta}(\vec q)$ 
that denotes the energy of the fermions that have scattered outside the patch with momentum $\vec q$.
${\cal V}^{(\nu)}_{\mu;\bar \theta}(\vec q)$,
which is also 
controlled by $\mu/\bar L_{\bar \theta}(\vec q)$,
is the crossover function that changes depending on whether the vertex correction from angle $\bar \theta$ is turned on  or off.
For $\bar L_{\bar \theta}(\vec q) \gg \mu$, the lack of the vertex correction suppresses the four-fermion coupling.
The way 
${\cal V}^{(\nu)}_{\mu;\bar \theta}(\vec q)$ scales with $q$
depends on whether $\vec q$ is tangential to the Fermi surface at angle $\theta$ or not.
For a more detailed discussion of the full crossover, see Appendix
\ref{sec:intra_to_inter}.
The term $\bar{\lambda}^{UV(\nu)}_{\bar{\theta}_1,\bar{\theta}_2}\left(\vec{q}\right)$ in \eq{eq:full_lambda} is the contribution of the UV coupling.

\section{
Projective fixed points
}
\label{sec:NFLfp}

Due to the nonstop growth of the Fermi momentum under the RG flow,
a metallic fixed point is defined only modulo a rescaling of the Fermi momentum.
In this section, 
we discuss consequences of the projective nature of the fixed points  
for physical observables in more details.

\subsection{ 
Mismatch between scaling dimension and relevancy of couplings
}

Although the tree-level scaling dimension of the four-fermion coupling is $1-d$, it does not mean that it is irrelevant in $d>1$.
For one thing, interactions introduce anomalous dimensions.
More importantly, the Fermi momentum that runs to infinity in the IR can modify the actual degree of relevance from the scaling dimension.
A foolproof way to determine the relevancy of the four-fermion coupling is to examine its contribution to physical observables such as 
the anomalous dimension of an operator.
Any operator would suit for this purpose, 
but it is convenient for us to use the four-fermion operator itself.
In the beta functional computed in \eq{beta_rescaled_lambda}, 
the term that is quadratic in $\lambda$ can be viewed as the
 contribution of the four-fermion coupling to the anomalous dimension of the four-fermion coupling itself.
Suppose that a UV four-fermion coupling  
flows to a particular IR coupling under the RG flow.
If a small deformation is added to the UV coupling, its RG flow is described by the linearized beta functional.
From \eq{beta_rescaled_lambda},
the beta functional for the small deformation 
$\delta \bar{ \lambda}^{(\nu)}(\vec q=0)$
is obtained to be
\begin{equation}
    \begin{aligned}
 \frac{\mathrm{d} 
 }{dl} 
 \delta \bar{\lambda}^{(\nu)}(0)
         &=
- \DD
\delta \bar{\lambda}^{(\nu)}(0)
-
R_d^{(\nu)}
\left(
\bar{ \lambda}^{(\nu)}(0)
\circ
\delta \bar{ \lambda}^{(\nu)}(0)
+ 
\delta \bar{ \lambda}^{(\nu)}(0)
\circ
 \bar{ \lambda}^{(\nu)}(0)
\right),
\label{eq:linearized_beta_rescaled_lambda}
    \end{aligned}
\end{equation}
where
$\delta \bar{ \lambda}^{(\nu)}(0)$
and
$\bar{ \lambda}^{(\nu)}(0)$
are viewed as matrices of 
$\bar \theta_1$
and
$\bar \theta_2$
with
the contraction of the angular matrix index given by 
$(A \circ B)_{\bar \theta_1, \bar \theta_2} =
\int_{-\bar \theta_{max}}^{\bar \theta_{max}}
\frac{d\bar \theta}{2\pi\sqrt{\mu}}
A_{\bar \theta_1, \bar \theta}
B_{\bar \theta, \bar \theta_2}
$
with $\bar \theta_{max} \sim \sqrt{\KFdim}$,
and $R_d^{(\nu)}= 2^{d-4}\left(A_+^{(\nu)}(d)T_+(d)+\frac{d-2}{2}A_-^{(\nu)}(d)T_-(d)\right)$. 
To extract the eigenmodes of the linearized beta functional, it is convenient to use the singular value decomposition to write the four-fermion coupling function as
\bqa
\bar \lambda^{(\nu)}_{\bar \theta_1,\bar \theta_2}(0)
= \sum_m
f^{(\nu)}_{m,\bar \theta_1}
\bar \lambda^{(\nu)}_m
f^{(\nu)*}_{m,\bar \theta_2},
\eqa
where
$\bar \lambda^{(\nu)}_m$ denotes the eigenvalue associated with the $m$-th  eigenvector 
$f^{(\nu)}_{m,\bar \theta}$
that satisfies
$\bar \lambda^{(\nu)}(0)
\circ
f^{(\nu)}_m
=
\bar \lambda^{(\nu)}_m 
f^{(\nu)}_m$
with the normalization
$f^{(\nu)\dagger}_n \circ 
f^{(\nu)}_m = \delta_{n,m}$.
In general, both eigenvectors and eigenvalues are $l$-dependent.
Eigenmodes of the linearized beta functional 
 at scale $l$ are then labeled by a pair of indices $(m,n)$,
\bqa
\delta \bar{ \lambda}^{(\nu);(m,n)}(0) =
f^{(\nu)}_m 
f^{(\nu)\dagger}_n
\eqa
with eigenvalue
\bqa
-\DD
- 
R_d^{(\nu)}
\left(\bar \lambda^{(\nu)}_m + \bar \lambda^{(\nu)}_n\right).
\label{eq:dimensionoflambda}
\eqa
This is the scaling dimension of the four-fermion coupling in the channel $(\nu,m,n)$.
$-\DD$ is the 
scaling dimension evaluated to the zeroth order in the four-fermion coupling,
and
$- 
R_d^{(\nu)}
\left(\bar \lambda^{(\nu)}_m + \bar \lambda^{(\nu)}_n\right)$
is the anomalous dimension that arises at the linear order of the four-fermion coupling.
Here, our main concern is 
 to understand how the latter depends on the former.
If the four-fermion coupling is irrelevant (relevant), 
the contribution of the UV four-fermion coupling to the scaling dimension would 
 be negligible (significant) in the low-energy limit.
One may naively expect that the four-fermion coupling is relevant (irrelevant) for 
 $-\DD > 0$ 
 ($-\DD < 0$). 
However, this is not the case because of the running Fermi momentum, as will be shown below.

A lower bound for the 
largest $\bar \lambda^{(\nu)}_m$
can be obtained from a trial function $f$ as
$\bar \lambda^{(\nu)}_{f}
= 
\frac{
f^{\dagger} \circ \bar \lambda^{(\nu)}(0) \circ f
}
{ f^{\dagger} \circ f }$.
With a choice of 
$f_{ \bar \theta}=
\left( \frac{\pi \sqrt{\mu}}{\bar \theta_{max}}\right)^{1/2}$,
the lower bound becomes
\bqa
\bar \lambda^{(\nu)}_{f}
= 
\int_{-\bar \theta_{max}}^{\bar \theta_{max}}
\frac{d\bar \theta_1}{2\pi\sqrt{\mu}}
\frac{d\bar \theta_2}{2\pi\sqrt{\mu}}~
 \frac{\pi \sqrt{\mu}}{\bar \theta_{max}}
 ~
\bar \lambda^{(\nu)}_{\bar \theta_1,\bar \theta_2}(0).
\label{eq:lower}
\eqa
Regarding \eq{eq:lower},
two main questions are
(1) whether the contribution of the UV coupling survives or not in the low-energy limit,
and
(2) whether the contribution of the large-angle scattering is divergent or convergent with increasing angular cutoff.
The first determines whether the four-fermion coupling is relevant or irrelevant at low energies,
and
the second determines the role of the large-angle scatterings.
For the purpose of answering these questions, we consider a simple angle-independent profile for the UV coupling function,
$\bar \lambda^{UV(\nu)}_{{\bar \theta}_1,{\bar \theta}_2}(0)
=\bar \lambda^{UV(\nu)}$\footnote{Generic angle dependence that arises from short-range interactions do not change the following conclusion.}.
From \eq{eq:full_lambda},
the lower bound is estimated to be
\bqa
\bar \lambda^{(\nu)}_{f}
& \sim &
~ e^{ (1/2-\DD) l} \left( \frac{\KFdim}{\Lambda} \right)^{1/2}  \bar \lambda^{UV(\nu)}
+
Y_{d}^{(\nu)}
\Lambda^{\frac{-1+2\DD}{2}}
\int_{\sqrt{\Lambda}}^{\sqrt{\KFdim} e^{l/2}}
d\hat \theta ~ 
\hat \theta^{-2 \DD}.
\label{eq:etamin}
\eqa
Here, a change of integration variable is made :
$\hat \theta = \bar \theta e^{l/2}$.
We also used the fact that 
$\bar \theta_{max} \sim \sqrt{\KFdim}$
and
$\mu = \Lambda e^{-l}$.
The additional factors of $e^{l/2}$ in the first term 
and the upper bound of the integral in the second term
arise from the measure of $\bar \theta$ integration.
It reflects the fact that the number of patches increases as $e^{l/2}$ as the length scale increases.
Interestingly, both terms in 
\eq{eq:etamin} exhibit qualitatively different behaviours depending on whether
$\DD > 1/2$,
$\DD = 1/2$ or
$\DD < 1/2$.

\begin{itemize}

\item $\DD > 1/2$ :

In this case,
the first term
in  \eq{eq:etamin},
which represents the contribution of the UV coupling,
vanishes in the low-energy limit.
Therefore, the scaling dimension
in \eq{eq:dimensionoflambda}
is independent of the UV 
 four-fermion coupling, and
the four-fermion coupling is irrelevant.
Furthermore, the universal four-fermion coupling function
falls off quickly in the angular separation,
and the contribution of the large $\hat \theta$ 
to the anomalous dimension in \eq{eq:etamin} is negligible. 
Due to the power-law suppression of the large-angle scatterings with a sufficiently large power,
the inter-patch coupling is negligible 
at low energies.
Therefore, 
one can understand the low-energy physics that is local in the momentum space 
(such as the $n$-point function of fermions clustered within two anti-podal patches) within a theory that only includes the relevant patches.
The low-energy theory exhibits the locality in the momentum space.

\item $\DD < 1/2$ : 

The UV contribution in \eq{eq:etamin} diverges in the low-energy limit.
This implies that the four-fermion coupling is relevant.
To properly understand the fate of the theory in the low-energy limit,
one needs to take into account the feedback of the four-fermion coupling to the flow of the coupling by including the $\lambda^2$-term in the beta functional.
At the same time, the 
  contribution of the universal four-fermion coupling generated from the critical boson becomes sensitive to the cutoff of $\hat \theta$ :
$\bar \lambda^{(\nu)}_{f}$ is proportional to $(\KFdim/\mu)^{\frac{1-2\DD}{2}}$,
which diverges in the low-energy limit.
Due to the strong inter-patch coupling,
the momentum-space locality and the patch description break down.

\item $\DD = 1/2$ : 

The contribution of the UV coupling barely survives in the IR and the four-fermion coupling is marginal.
Furthermore, the contribution of the  large-angle scatterings generated from the critical mode exhibits a logarithmic divergence, signifying a marginal breakdown of the momentum-space locality.
In this case,
one has to include 
sub-leading corrections in the analysis.

\end{itemize}

It is interesting to note that 
the four-fermion coupling turns relevant as its scaling dimension $-\DD$ rises above $-1/2$ not $0$.
This happens because 
the angular variable $\bar \theta$ acquires an anomalous dimension $1/2$ through  \eq{eq:nonlinear_theta_transformation} and 
so does the volume of the phase space associated with the extended Fermi surface.
Accordingly, the dimensionless phase space grows with decreasing energy scale. 
When the scaling dimension of the four-fermion coupling becomes greater than $-1/2$, 
the enhancement from the extended phase space becomes strong enough to overcome the negative scaling dimension.
{\it 
This mismatch between scaling dimensions of couplings and their relevancy
is a result of the projective nature of the metallic fixed point.}
In the next section, we discuss another consequence:  
a lack of universal scaling.

\subsection{
The absence of a unique
dynamical critical exponent
}

Near the upper critical dimension, 
$\DD \approx 1$ and the theory belongs to the first category.
In the low-energy limit,
the theory flows to one of the stable projective fixed points spanned by two marginal coupling functions,
$\kappa_{F,\theta}=\KFthetadim/\KFAV$ and $v_{F,\theta}$.
Nonetheless, the four-fermion coupling function  singularly depends on $\kFAV$ that runs to infinity.
This singular dependence on the Fermi momentum causes a lack of simple power-law scaling of physical observables.
To see this explicitly through the four-point function, 
we note that the dimensionless coupling function we computed by integrating the beta functional corresponds to the dimensionless crossover function in \eq{eq:fscaling},
and the four-point function of electron is given by
\begin{equation}
    \begin{aligned}
        \varGamma^{(4,0);(\nu)}\left( \omega, q, \theta_1, \theta_2, \varphi; [{\bf K}_F,v_F]; \Lambda \right) 
&\sim
\omega^{
\left(1-d-3(d-1)(z-1)-4\eta_\psi\right)/z
} 
f^{(4,0);(\nu)}\left(   \frac{q}{\omega^{1/z}}; \theta_1, \theta_2, \varphi;  \left[ \kappa_{F}, v_F \right]; \frac{\KFAV} {\omega^{1/z}} \right),
\label{eq:GammaLambda}
    \end{aligned}
\end{equation}
where 
$
f^{(4,0);(\nu)}\left(   \frac{q}{\omega^{1/z}}; \theta_1, \theta_2, \varphi;  \left[ \kappa_{F}, v_F \right];  \frac{\KFAV} {\omega^{1/z}} \right)
=
\left.
\bar{\lambda}^{(\nu)}_{\bar{\theta}_1,\bar{\theta}_2}( \vec q )
\right|_{
\mu = \omega^{1/z}  \Lambda^{1-1/z},
\bar \theta_i = a^{-1}(\theta_i)
}$ 
with 
$a^{-1}(\theta)$ being the inverse of \eq{eq:nonlinear_theta_transformation}\footnote{
This can be derived by relating
the four-fermion coupling function with the vertex function in \eq{eq:LQ}.
Through Eqs.
\eqref{eq:RG5},
\eqref{eq:lambdatilde}
and
\eqref{eq:lambdabar},
the four-fermion coupling function $\bar \lambda$ is related to the vertex function as
$
\bar{\lambda}^{(\nu)}_{\bar{\theta}_1,\bar{\theta}_2}( \vec q )
=
\mu^{d-1/2}
\sqrt{
\frac{
a^\prime\left(\bar{\theta}_1\right) a^\prime\left(\bar{\theta}_2\right)
K_{F,\theta_1}K_{F,\theta_2}}{v_{F,\theta_1}v_{F,\theta_2}}}
\varGamma^{(4,0);(\nu)}\left( \mu, q, \theta_1, \theta_2, \varphi; [k_F,v_F]; \mu \right)$ 
with $\mu = \Lambda e^{-l}$,
where the first argument in the vertex function represents frequencies of the external fermions
set by \eq{eq:4fmomenta},
and the last argument is the scale at which
the renormalized field and couplings are defined.
However,
$\varGamma^{(4,0);(\nu)}\left( \mu, q, \theta_1, \theta_2, \varphi; [k_F,v_F]; \mu \right)$ is not what can be readily measured experimentally.
The quantity that is directly relevant in experiments is the vertex function evaluated with respect to the bare field as a function of the bare frequency.
\eq{eq:GBG}
relates the vertex functions defined at scales $\Lambda$ and $\mu$ through 
$
\varGamma^{(4,0);(\nu)}\left( \omega, q, \theta_1, \theta_2, \varphi; [k_F,v_F]; \Lambda \right) =
  Z_{\tau;\mu}^{-3(d-1)}
Z^{-2}_{\psi;\mu}  
\varGamma^{(4,0);(\nu)}\left( \mu, q, \theta_1, \theta_2, \varphi; [k_F,v_F]; \mu \right)
$, where
the bare frequency $\omega$ and the renormalized frequency $\mu$ are related to each other through
$\omega = \mu^z/\Lambda^{z-1}$,
and
the multiplicative renormalization factors are given by
$ Z_{\tau;\mu} = 
\left( \frac{\mu}{\Lambda} \right)^{z-1} $,
$Z_{\psi;\mu} = \left( \frac{\mu}{\Lambda} \right)^{ 
2 \eta_{\psi}}$
through  \eq{dynam_crit_anomalous_dim}.
Here, it is assumed that the theory is already close to a projective fixed point at energy scale $\Lambda$ 
so that the anomalous dimension of the field is angle-independent.
Combining these, we obtain
\eq{eq:GammaLambda}.
}.
The vertex function in the low frequency limit is fully determined by 
two marginal coupling functions
$\kappa_{F,\theta}$ 
and
$v_{F,\theta}$
and one relevant scale
$\frac{\KFAV} {\omega^{1/z}}$.
The former determines the shape of the Fermi surface and angle-dependent Fermi velocity,
and the latter sets  the overall size of Fermi surface in the unit of the external frequency.
The space of projective fixed points is spanned by the two marginal coupling functions.
Interestingly,
the critical exponents exhibit a {\it super-universality} in that
they do not depend on  the marginal parameters.
%
%

In \eq{eq:GammaLambda},
the crossover as a function of $q$ is controlled by
$\left( \frac{\omega^{1/z}}{L_{\theta_1}(\vec q)} \right)^{ \frac{\eta_d^{(\nu)}}{2}}$
and
$\left( \frac{\omega^{1/z}}{L_{\theta_2}(\vec q)} \right)^{ \frac{\eta_d^{(\nu)}}{2}}$
%
through  
${\cal V}^{(\nu)}_{\mu;\bar \theta}(\vec q)$ defined in \eq{eq:Vnuq}.
%
In the low frequency limit,
the crossover function becomes
$\left(
\frac{
\omega^{1/z}}{
\cos(\varphi-\theta)
q} 
\right)^{ \frac{\eta_d^{(\nu)}}{2}}$
for 
$\cos(\varphi-\theta) \gg  q/\KFthetadim$
and
$\left(
\frac{
\omega^{1/z}
\KFthetadim
}{q^2}
\right)^{ \frac{\eta_d^{(\nu)}}{2}}$
for
$\cos(\varphi-\theta)  \ll q/\KFthetadim$.
%
%
In the angular region with
$\cos(\varphi-\theta) \gg  q/\KFthetadim$,
$\vec q$ is not sufficiently tangential to the Fermi surface that 
the curvature of the Fermi surface does not enter,
and the crossover function is independent of Fermi momentum.
In this kinematic region, the vertex function is scale invariant in the limit in which $\omega$ and $q$ are sent to zero with  $\omega^{1/z}/q$ fixed.
In the narrow region of $\cos(\varphi-\theta)  \ll q/\KFthetadim$, on the other hand, $\vec q$ is sufficiently tangential to the Fermi surface so that the curvature becomes important.
The vertex function in this region becomes scale invariant 
in the low-energy limit where $\omega^{1/z}/q^2$ is fixed.
Therefore, {\it there is no single dynamical critical exponent that sets the relative scaling between $\omega$ and $q$ for all low-energy observables.}
With decreasing energy scale, the range of the second region becomes narrower,
and
the crossover function becomes non-analytic in $\theta_i$ and $\varphi$ in the low-energy limit.

\section{
The rise of momentum-space non-locality
}
\label{sec:SC_locality}

The question of ultimate physical interest is the fate of the theory at $d=2$.
Since we don't have a non-perturbative access to the theory in $d=2$,
the next best thing we can do is to 
examine the trend
the perturbation theory points to 
as $d$ is lowered 
toward the physical dimension.
The scaling dimension of the four-fermion coupling ($-\DD$) depends on $d$ through two competing factors.
On the one hand, 
electrons become more incoherent in lower dimensions and the infrared singularity is weakened by it.
On the other hand, 
the tree-level scaling dimension ($1-d$) 
and the anomalous dimension 
 ($\eta_d^{(\nu)}$) 
 generated from the vertex correction tend to enhance the infrared singularity 
with decreasing $d$ in the `attractive' channel with $\eta_d^{(\nu)}>0$. 
The outcomes of the competition is summarized in the exponent $\DD$ in \eq{eq:DD}.
As is discussed in the previous section, the critical value of $\DD$ for the four-fermion coupling to become relevant is $1/2$.
With decreasing $d$,
the forward scattering and the pairing interaction behave differently.

\begin{figure}[th]
\centering
\begin{subfigure}{.4\textwidth}
  \centering
  \includegraphics[width=1.0\linewidth]{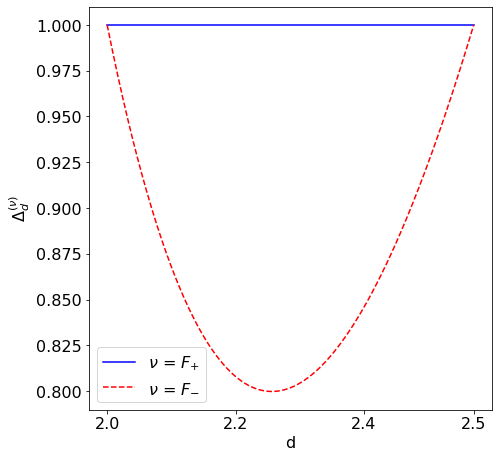}
  \caption{}
  \label{fig:15a}
  \end{subfigure}%
\begin{subfigure}{.4\textwidth}
  \centering
  \includegraphics[width=1.0\linewidth]{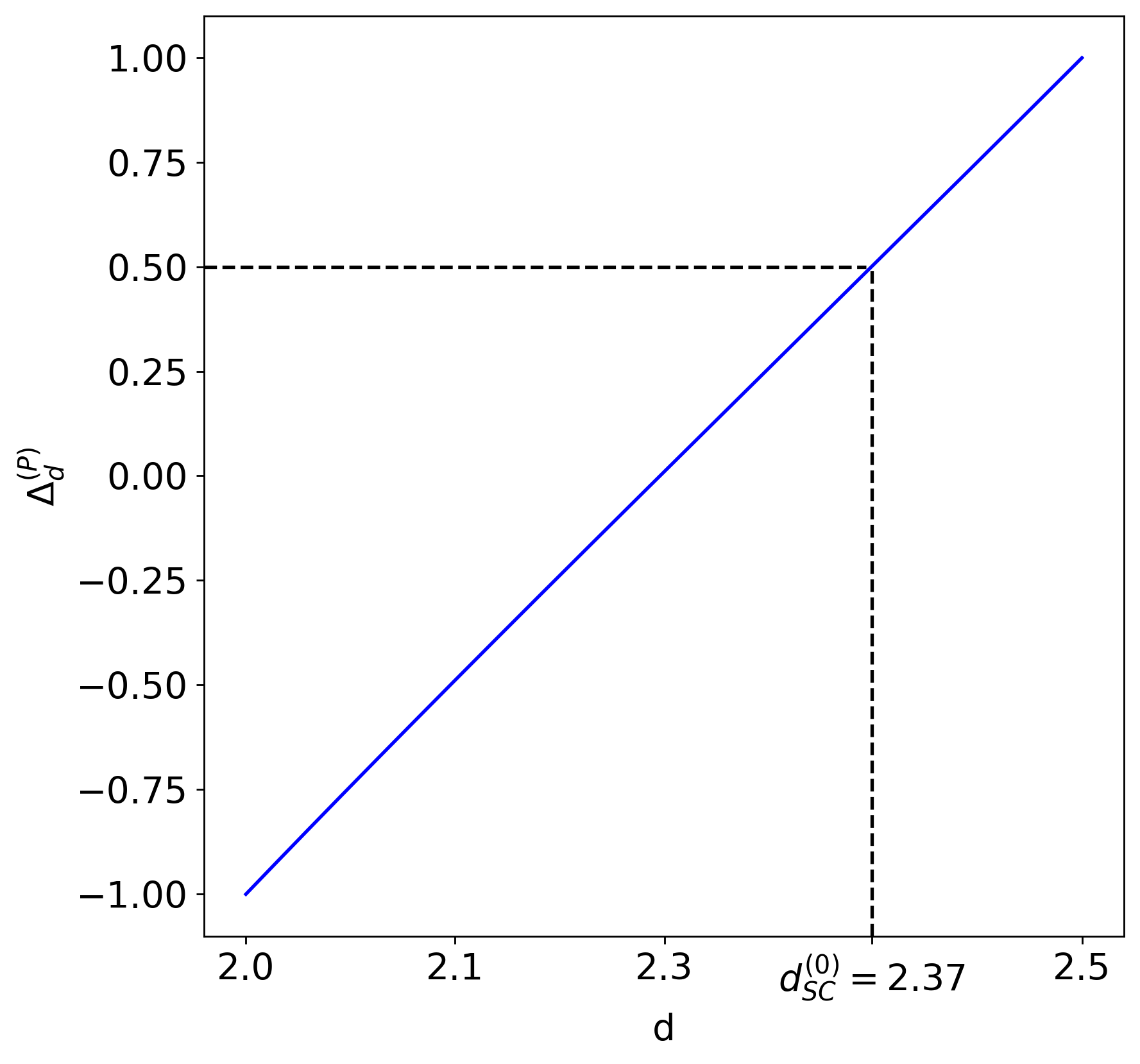}
  \caption{}
  \label{fig:15b}
\end{subfigure}
\caption{
The minus of the leading scaling dimension of the four-fermion coupling in the forward scattering and pairing channel.
$\DD$ 
is greater than the critical value $1/2$ in all $d$ 
for $\nu = F_{\pm}$,
which implies that the Landau function is irrelevant.
$\DD$ cross the critical value $1/2$ at
$
d_{SC}^{(0)}
\approx 2.37$.
The pairing interaction becomes relevant below this critical dimension.
}
\label{fig:Heta12}
\end{figure}

\subsection{Landau function}

\fig{fig:15a} shows  
$\Delta_d^{(F_\pm)}$
computed to the leading order in the $\epsilon$-expansion for the forward scattering channel.
Interestingly, $\Delta_d^{(F_\pm)}  > 1/2$   in all dimensions, which suggests that the forward scattering remains irrelevant even at $d=2$.
This is because the vertex correction is highly suppressed in the forward scattering channels.
In diagrams that do not involve interaction in the pairing channel,
quantum corrections to the forward scattering vertex at general angles $\theta'$ and $\theta$ arise from small-angle scatterings within each patch around those two angles.
Each of those vertex corrections is then controlled by the single-patch theory. 
However, the single-patch theory does not have any singular vertex correction
due to the kinematic constraint associated with the chiral nature\cite{SHOUVIK2}.

\subsection{
Superconducting fluctuations
and UV/IR mixing 
}

In \fig{fig:15b}, 
$\Delta_d^{(P)}$
is shown in the pairing channel.
As expected,
$\Delta_d^{(P)}  > 1/2$  
and
the pairing interaction is irrelevant 
near the upper critical dimension.
However, 
$\Delta_d^{(P)}$
 monotonically decreases with decreasing $d$ due to the vertex correction 
 that becomes stronger.
The leading order computation suggests that 
$\Delta_d^{(P)}$
goes below the critical value $1/2$ around $\dsco =2.37$.
Below this critical dimension, the pairing interaction becomes relevant.
To understand the fate of the theory in the low-energy limit, one can not ignore higher order corrections in the four-fermion coupling.
The inclusion of the quadratic term in the four-fermion coupling in the beta functional is expected to make the pairing interaction  more relevant due to the BCS scatterings\footnote{
This can be seen from 
\eq{eq:dimensionoflambda} where the four-fermion coupling enhances the dimension of the four-fermion coupling in the 
 channel with $\bar \lambda_m^{(\nu)}<0$.
}, which likely implies superconducting instabilities\cite{LEDERER}.
We defer the analysis of the full beta functional and the ensuing superconducting instability to a separate paper.
Here, let us focus on the  
 physics within the window of intermediate energy scales where the quadratic term in the four-fermion coupling is still negligible.
%
The size of this window is parametrically large near $d_{SC}^{(0)}$.
This is because the energy scale below which the non-Fermi liquid physics sets in is determined by the interaction which is strictly relevant at the Gaussian fixed point while the superconducting transition temperature approaches zero as the critical dimension is approached 
(see \fig{fig:1}).
Within this window of energy scales, 
the physics is controlled by the non-Fermi liquid state that exhibits an approximate sense of scale invariance\cite{BORGES2023169221}.

In the non-Fermi liquid state realized in the intermediate energy scales, 
there exist intrinsic superconducting fluctuations which are captured by the four-fermion coupling generated from the quantum critical mode 
as is shown in  \eq{eq:full_lambda}.
With the slow decay 
($\Delta_d^{(P)} \leq 1/2$)
of the four-fermion coupling in $\theta_2-\theta_1$,
the mixing among low-energy modes with widely different momenta becomes important.
In this case, the low-energy physics becomes sensitive to the short-distance data associated with the four-fermion coupling function with large momentum transfers.
Because Cooper pairs at one angle can be scattered to anywhere on the Fermi surface, the number of fermions fails to be conserved in each patch.
Consequently, the low-energy symmetry is broken from LU(1) to \OLU.
Two remarks are in order regarding the symmetry.
First, this is not an artifact of the dimensional regularization scheme. 
With decreasing $d$, the LU(1) symmetry breaking terms that arise from the dimensional regularization scheme become weaker while the symmetry breaking effect from the large-angle Cooper pair scatterings becomes stronger.
This suggests that the LU(1) symmetry is likely to remain broken to \OLU at $d=2$. 
Second, the larger-angle scatterings that break the LU(1) symmetry is not a `leading sub-leading effect' 
- an effect that dominates physical observables but is suppressed by powers of energy scales at low energies.
This can be seen from the fact that 
the contribution of the large-angle scatterings to the anomalous dimension  in \eq{eq:etamin} is not suppressed at low energies.

The strong UV/IR mixing has experimental consequences.
First, the relevant large-angle scatterings  generate a fast relaxation of neutral collective excitations.
In particular, there will be a universal component in the decay rate of deformations of Fermi surface in the even angular momentum channels at low temperatures.
This is in contrast to 
Fermi liquids in which
 the relaxation of excitations in even angular momentum channels is mainly created by the head-on collisions which is marginally irrelevant 
 for repulsive interactions
\cite{
PhysRevB.45.1259,
LEDWITH2019167913}.
Second,
the scaling dimension of superconducting fluctuations can be measured through the pair susceptibility.
Due to the dependence of the scaling dimension on the Fermi momentum as discussed in Sec. \ref{sec:NFLfp}, 
the scaling dimension `crawls' as the energy is lowered.
This will result in
a strong enhancement of superconducting fluctuations and a deviation from a pure power-law scaling.

\section{Conclusion}
\label{sec:conclusion}

Low-energy effective theories of metals are characterized by a set of `coupling functions' 
that determine 
the size and shape of the Fermi surface, 
the angle-dependent Fermi velocity
and interactions.
The functional renormalization group flow defined in the space of these functions does not exhibit a fixed point in the strict sense 
because the size of the Fermi surface continues to grow under the renormalization group flow.
Therefore, a notion of fixed point can be defined only modulo rescaling of the Fermi momentum.
Due to this projective nature of fixed points, 
there is no unique way of scaling frequency and momentum such that the entire low-energy observables exhibit a scale invariance.
Furthermore, 
even couplings with negative scaling dimensions can give rise to infrared singularity through non-perturbative enhancement from the running Fermi momentum.


In this paper, we illustrate these points by charting out the space of projective fixed points and characterizing their universal low-energy physics
for the Ising-nematic quantum critical metal.
Near the upper critical dimension,
the theory flows to stable projective fixed points whose space is spanned by two exactly marginal coupling functions.
The marginal parameters are the angle-dependent functions that specify the shape of the Fermi surface and the Fermi velocity.
Compared to Fermi liquids, the non-Fermi liquid fixed points exhibit two salient 
 universal features.
First, the Landau function is no longer tunable at 
 low energies as it is completely fixed by the 
 shape of the Fermi surface and the Fermi velocity.
Second, there exist   universal superconducting fluctuations intrinsic to the non-Fermi liquid state.
As the space dimension is lowered, the vertex corrections enhance the critical superconducting fluctuations.
Below a critical space dimension, which is estimated to be around $2.37$,
the patch theory breaks down due to 
strong inter-patch coupling mediated by large-angle scatterings of Cooper pairs. 
Consequently,
the number of fermions in each patch is no longer conserved,
and the emergent symmetry of the normal state becomes a proper subgroup of that of Fermi liquids.
In the low-energy limit,
the theory is expected to become unstable toward superconducting states near two dimensions.
The details on the superconducting instability will be discussed in a separate paper.
We conclude with some remarks and open questions.

\begin{itemize}

\item Quasi-fixed point

In $d$ close to $2$, the large-angle scatterings generated in the pairing channel become important.
At the same time, the non-Fermi liquid state becomes unstable against superconducting states.
Nonetheless, there exists a finite window of energy scales in which the renormalization group flow is stalled due to the proximity to the non-Hermitian fixed points\cite{BORGES2023169221}.
Physical observables within the window of energy scale is controlled by a quasi(pseudo)-fixed point that arises at the bottleneck of the renormalization group flow.
It is of great interest to  understand the nature of the non-Fermi liquid state that arises
above the superconducting transition temperature in $d=2$.

\item Feedback of superconducting fluctuations 

To the leading order in the $\epsilon$ expansion, the feedback of the universal four-fermion coupling to the electron self-energy and the Yukawa coupling can be ignored.
However, the feedback effect is expected to become important near $d=2$.
It will be of great interest to understand the effect of the critical superconducting fluctuations to 
the electron spectral function,
the collective modes
and transport phenomena
in view of the possibility that
the large-angle scatterings
 can give rise to a pseudogap behaviour 
 of normal electrons
 and open up new channels for dissipating collective modes of Fermi surface and the electric current combined with the Umklapp scatterings\cite{
PhysRevLett.69.2001,
PhysRevB.108.235125,
PhysRevB.108.235125,
2022arXiv220407585D,
2023arXiv231103458G,
PhysRevB.109.195110,
PhysRevB.108.045107,
2024arXiv240401534K}.

\item Alternative routes to superconductivity 

In the Ising-nematic quantum critical metal, 
the superconducting instability arises below the critical dimension in which the UV/IR mixing sets in.
This is because the superconducting instability is driven by the large-angle scatterings.
How general is this as the mechanism to drive non-Fermi liquids into superconductivity
\cite{
PhysRevD.59.094019,
PhysRevB.91.115111,
LEDERER,
BORGES2023169221,
PhysRevB.102.024524,
PhysRevB.102.024525,
PhysRevB.102.094516,
PhysRevB.103.024522,
PhysRevB.103.184508,
PhysRevB.104.144509,
PhysRevB.95.165137,
LEDERERREV,
PhysRevB.95.174520}?
In principle, it is possible that normal states become unstable against pairing through singular small-angle scatterings,
which can be captured within a patch theory.

\item{
Non-Fermi liquids with 
 LU(1) symmetry 
}

A system that is closely related to the Ising-nematic quantum critical metal is the critical Fermi surface coupled with an emergent U(1) gauge field.
It, for instances, arises as the low-energy description of the half-filled Landau level\cite{HALPERINHALF,YBKIM} 
and a gapless spin liquid that supports Fermi surface of fractionalized degrees of freedom\cite{
PhysRevB.72.045105,
PhysRevLett.95.036403,
SSLEE}.
In the Coulomb gauge, the gauge theory takes the same form as \eq{eq:action_2d} with the following changes.
First, the boson field $\phi$ is replaced with the transverse part of the U(1) gauge field.
Second, the boson is now coupled with the current and the Yukawa coupling satisfies
$e_\theta = -e_{\theta+\pi}$.
This makes the vertex correction repulsive in the pairing channel,
while it is attractive 
for the Ising-nematic case.
Consequently, $\DD$ is expected to stay greater than $1/2$ at $d=2$ at least to the leading-order.
This suggests that the critical Fermi surface coupled with the U(1) gauge field may exhibit 
the momentum space locality 
and the LU(1) symmetry\cite{PhysRevX.11.021005} 
unlike the Ising-nematic counterpart where the large-angle scatterings lowers the symmetry to OLU(1).

\item Disorder

In the presence of disorder that breaks the translational invariance, large-angle scatterings become more important. 
Even at the upper critical dimension $d=5/2$, the disorder is expected to generate non-perturbative effects beyond logarithmic corrections\cite{2024arXiv240310148K}.
Recent studies suggest that strong disorder can create universal behaviours that are largely insensitive to the clean parent non-Fermi liquids within an intermediate energy scale\cite{ doi:10.1126/science.abq6011, 2023arXiv231206751P, 2023arXiv231007768B}.
It will be interesting to understand the low-energy limit of the disordered non-Fermi liquids.

\end{itemize}

%% file: appendix.tex
\appendix


\newpage

\section{ Derivation of Eq. \ref{eq:ThetaDelta} }
\label{app:Momentum_conservation}

In the Fermi-polar coordinate system,
momentum of a fermion is represented by $(\delta, \theta)$,
where $\theta$ is the usual polar angle and $\delta$ is the radial momentum measured with respect to Fermi momentum at that angle.
In this appendix, we derive the expression for the final momentum 
$(\deltaq, \thetaq$) 
when a fermion at 
$(\delta, \theta)$ absorbs a boson with momentum 
$\vec q = q \cos\varphi \hat x + q \sin \varphi \hat y$.
In the Cartesian coordinate, 
the momentum conservation gives
\begin{equation}
\begin{aligned}
\big(\KFthetadim  +\delta \big)\cos(\theta  )+q \cos(\varphi )&=\big(\mathbf{K}_{F, \thetaq}+ \deltaq\big)\cos( \thetaq),\\
\big(\KFthetadim  +\delta \big)\sin(\theta  )+q \sin(\varphi )&=\big(\mathbf{K}_{F, \thetaq}+ \deltaq\big)\sin( \thetaq).
\label{appendix:momentum_conservation}
\end{aligned}
\end{equation}
The exact expression for $\thetaq$ and $\deltaq$ are given by
\bqa
\tan( \thetaq) &=&  \frac{\left(\KFthetadim +\delta\right)\sin(\theta  )+q \sin(\varphi )}{\left(\KFthetadim +\delta\right)\cos(\theta  )+q \cos(\varphi )}, 
\label{appendix:Tan(thetaq)} \\
\deltaq &=&
\left[\left(\KFthetadim +\delta\right)^2+q^2+2q\left(\KFthetadim +\delta\right)\cos\left(\varphi-\theta\right)\right]^{\frac{1}{2}}-\mathbf{K}_{F,\thetaq}.
\label{appendix:deltaq}
\eqa
At low energies, 
$\delta \ll \KFthetadim $. 
If the momentum transfer $\vec q$ is also smaller than the Fermi momentum, we can expand the above expressions in powers of 
$\delta/\KFthetadim $
and
$q/\KFthetadim $ as
\bqa
\thetaq & \approx  &
\theta+\mathscr{A}_{\varphi,\theta}\frac{q}{\KFthetadim }-\mathscr{A}_{\varphi,\theta}\frac{q\delta}{\KFthetadim^2}+\mathscr{B}_{\varphi,\theta}\left(\frac{q}{\KFthetadim }\right)^2+\mathrm{O}\left(\frac{q^2\delta}{\KFthetadim^3}\right),
        \label{appendix:thetaq_initial} \\
 \deltaq &\approx& \delta+q
   \mathscr{F}_{\varphi,\theta}
   +
    \frac{q^2}{\KFthetadim }\mathscr{G}_{\varphi,\theta}
    +\frac{q\delta}{\KFthetadim }\mathscr{I}_{\varphi,\theta}+\mathrm{O}\left(\frac{q^2\delta}{\KFthetadim^2}\right),
\label{appendix:deltaq_smallq_initial}
\eqa
where
\begin{equation}
\begin{aligned}
\mathscr{A}_{\varphi,\theta} = 
       \sin(\varphi-\theta), ~~~~
\mathscr{B}_{\varphi,\theta} =
        -\frac{\sin\left(2\left(\varphi-\theta\right)\right)}{2}, ~~~~
\mathscr{F}_{\varphi,\theta} = \cos(\varphi -\theta )-\sin(\varphi -\theta )\frac{\KFthetadim^{\prime}}{\KFthetadim }, \\
\mathscr{G}_{\varphi,\theta} = \frac{1}{2}\left( 
        \sin^2(\varphi -\theta )\left[1-\frac{\KFthetadim^{\prime\prime}}{\KFthetadim }\right]+\sin\left(2\left(\varphi-\theta\right)\right)\frac{\KFthetadim^{\prime}}{\KFthetadim }\right), ~~~~
~\mathscr{I}_{\varphi,\theta} = \frac{\KFthetadim ^\prime}{\KFthetadim }\sin\left(\varphi-\theta\right). 
\label{appendix:curlyF_G}
    \end{aligned}
\end{equation}
Here, $\KFthetadim^{\prime}$ and $\KFthetadim^{\prime\prime}$ represent the first and second-derivative of Fermi momentum with respect to angle. 
When $\vec q$ represents momentum of the critical boson that is almost tangential to the Fermi surface at angle $\theta$, 
$q$ can be much bigger than the energy of fermion $\delta$ as the energy of the scattered fermion increases as $q^2/\KFtheta$.
At a low energy scale $\mu$,
$\delta \sim \mu$ while
$q\sim \sqrt{\KFthetadim \mu}$.
For $\delta \ll q \ll \KFthetadim $, 
one can further simplify the expressions as
\begin{equation}\begin{aligned}
    \thetasq \approx  \theta+\mathscr{A}_{\varphi,\theta}\frac{q}{\KFthetadim },
    ~~&~~
    \deltaq \approx \delta+q
   \mathscr{F}_{\varphi,\theta}
   +
    \frac{q^2}{\KFthetadim }\mathscr{G}_{\varphi,\theta}.
    \label{appendix:thetaq_deltaq_smallq}
\end{aligned}
\end{equation}

When we compute quantum corrections, 
it is convenient to consider 
\eq{appendix:thetaq_deltaq_smallq} in the limit in which $\vec q$ is nearly tangential to a point on the Fermi surface because the largest contribution arises from that kinematic region.
Let $\vartheta(\varphi)$ be the 
 function 
 that maps the angle of boson $\varphi$
to the angle of fermion 
$\theta \in [-\pi/2, \pi/2)$ 
at which $\vec q = q (\cos \varphi, \sin \varphi)$ is tangential to the Fermi surface,
\begin{equation}\begin{aligned}
      \vartheta(\varphi ) =  \varphi -\arctan\left(\frac{\mathbf{K}_{F,\vartheta(\varphi ) }}{\mathbf{K}^{\prime}_{F,\vartheta(\varphi)}}\right).
       \label{appendix:varthea}
\end{aligned}\end{equation}
We assume that the Fermi surface is globally convex,
in which case  \eq{appendix:varthea} has a unique value in $[-\pi/2, \pi/2)$. 
For a circular Fermi surface, 
this reduces to
$
\vartheta(\varphi ) = \varphi  - \frac{\pi}{2}(2n+1)$ for some integer $n$ as expected.
If $\theta$ deviates from $\vartheta(\varphi)$ by $\Delta \theta = \theta  - \vartheta(\varphi )$,
\eq{appendix:thetaq_deltaq_smallq}
becomes \begin{equation}\begin{aligned}
\deltaq
=
\delta+q\Delta\theta  \chi_{\varphi}+\frac{q^2}{\mathbf{K}_{F,\vartheta(\varphi ) }} \tilde{B}_{\varphi}
\label{appendix:dispersion_external_boson}
\end{aligned}\end{equation}
to the quadratic order in $q$ and $\Delta \theta$,
where 
\begin{equation}\begin{aligned}
    \chi_{\varphi} &=
    \sin( \varphi -\vartheta(\varphi ))\left[1-\frac{\mathbf{K}^{\prime\prime}_{F,\vartheta(\varphi ) }}{\mathbf{K}_{F,\vartheta(\varphi ) }}+\left(\frac{\mathbf{K}^{\prime}_{F,\vartheta(\varphi )}}{\mathbf{K}_{F,\vartheta(\varphi ) }}\right)^2\right]+\cos( \varphi -\vartheta(\varphi ))\frac{\mathbf{K}^{\prime}_{F,\vartheta(\varphi )}}{\mathbf{K}_{F,\vartheta(\varphi ) }},
    \\
    \tilde{B}_{\varphi} &= 
    \frac{1}{2
    }\left(\sin^2(\varphi -\vartheta(\varphi ) )\left[1-\frac{\mathbf{K}^{\prime\prime}_{F,\vartheta(\varphi ) }}{\mathbf{K}_{F,\vartheta(\varphi ) }}\right]+
    \sin\left(2\left(\varphi-\vartheta(\varphi )\right)\right)\frac{\mathbf{K}^{\prime}_{F,\vartheta(\varphi )}}{\mathbf{K}_{F,\vartheta(\varphi ) }}\right).
    \label{appendix:chi_tildeB}
\end{aligned}\end{equation}

Alternatively, one can also expand  
\eq{appendix:thetaq_deltaq_smallq}
in powers of $ \Delta\varphi = \varphi - \vartheta^{-1}(\theta)$
when the angle of boson $\varphi$ deviates away from $\vartheta^{-1}(\theta)$ for a fixed $\theta$.
To the leading order in $ \Delta\varphi$,
Eq. (\ref{appendix:thetaq_deltaq_smallq}) becomes
 \begin{equation}\begin{aligned}
      \deltaq
      =
      \delta-q\Delta\varphi F_{\theta}+\frac{q^2}{\KFthetadim } G_{\theta},
\label{appendix:dispersion,external_fermion}
\end{aligned}\end{equation}
where
\begin{equation}\begin{aligned}
    F_{\theta} = \sin( \vartheta^{-1}(\theta)-\theta )+\cos( \vartheta^{-1}(\theta)-\theta )\frac{\KFthetadim^{\prime}}{\KFthetadim }, ~&~
   G_{\theta} = \frac{1}{2
   }\left(\sin^2(\vartheta^{-1}(\theta) -\theta )\left[1-\frac{\KFthetadim^{\prime\prime}}{\KFthetadim }\right]+
   \sin\left(2\left(\vartheta^{-1}(\theta)-\theta\right)\right)
   \frac{\KFthetadim^{\prime}}{\KFthetadim }\right). \\
\label{appendix:F_G_external_fermion}
\end{aligned}\end{equation}

\section{Fierz Transformation}
\label{sec:Fierz_Transformation}

The four-fermion operator in 
Eq. (\ref{action_generald}) takes the form of 
$\mathbf{O}^{(\nu)} = \bar{\Psi}_{1}
        I_m^{(\nu)}
       \Psi_{2}~\bar{\Psi}_{3}
     I_m^{(\nu)}
     \Psi_{4}$,
where
$m$ is summed over 
$1$, $(d-1)$
and $(4-d)$ components
for each $\nu=F_+,F_-,P$,  respectively,
\begin{equation}\begin{aligned}
   I^{(F_+)} = i\gamma_{d-1}, ~~~
   I^{(F_-)} = 
   (\gamma_0,..,\gamma_{d-2}), 
    ~~~
    I^{(P)} 
    &= \left(\mathbbm{1},i\gamma_d,...,i\gamma_2\right).
    \label{appendix:4f_channels}
\end{aligned}\end{equation}
Since 
$\{ \mathbf{O}^{(\nu)} \}$
forms a complete basis of the four-fermion operators that
respect the $SO(d-1) \times SO(4-d)$ symmetry
(see 
Sec. 
\ref{sec:symmetry} for discussion on symmetry),
one can express operators of the form
$\mathbf{O}^{{}^\prime(\nu)} = \bar{\Psi}_{1}
        I^{(\nu)}_m
       \Psi_{4}~\bar{\Psi}_{3}
        I^{(\nu)}_m
     \Psi_{2}$
as 
\begin{equation}
    \begin{aligned}
        \mathbf{O}^{{}^\prime(\nu)} = -C_{\nu\nu^{\prime}}\mathbf{O}^{\left(\nu^{\prime}\right)},
        \label{appendix:change_basis_4f_operator}
    \end{aligned}
\end{equation}
where the negative sign is due to the anti-commuting nature of the fermionic fields. 
We can calculate matrix $C$ by considering rank-4 orthogonal basis tensors, 
\begin{equation}
    \begin{aligned}
        \mathscr{T}^{(\nu)}_{abcd} = \left(I^{(\nu)}_m\right)_{ab}\left(I^{(\nu)}_m\right)_{cd}, 
        \end{aligned}
\end{equation}
where $a,b,c$ and $d$ are the spinor indices. Here, m is summed over (but not $\nu$).
The inner product of the basis tensors is defined as
$      \Big\langle\mathscr{T}^{(\nu)}
       , \mathscr{T}^{\left(\nu^{\prime}\right)}
       \Big\rangle
        = \sum_{m,m'} \left(\Tr\bigg\{\left(I_m^{(\nu)}\right)^{\dagger}I_{m^{\prime}}^{\left(\nu^{\prime}\right)}\bigg\}\right)^2
       = 
\mathscr{N}_\nu^2
\delta_{\nu\nu^{\prime}}
       $
with
$\mathscr{N}_\nu = \left(
2, 2\sqrt{d-1}
,
2\sqrt{4-d} \right)$.
From
$\mathscr{T}^{(\nu)}_{adcb} = C_{\nu\nu^{\prime}}\mathscr{T}^{\left(\nu^{\prime}\right)}_{abcd}$,
we readily obtain
\bqa
 C_{\nu\nu^{\prime}} =  \mathscr{N}_{\nu^{\prime}}^{-2}\Tr\Bigg\{I^{(\nu)}_m\left(I^{\left(\nu^{\prime}\right)}_{m^{\prime}}\right)^{\dagger}I^{(\nu)}_m\left(I^{\left(\nu^{\prime}\right)}_{m^{\prime}}\right)^{\dagger}\Bigg\}
 =
 \begin{pmatrix}
     \frac{1}{2} & \frac{1}{2} & -\frac{1}{2} 
     \\
     \vspace{4pt}
    \frac{d-1}{2} & \frac{3-d}{2} & \frac{d-1}{2}
    \\
    \vspace{4pt}
   -\frac{4-d}{2} & \frac{4-d}{2} & \frac{d-2}{2} 
  \end{pmatrix}_{ij}.
  \eqa

\section{Critical exponents and beta functionals}
\label{app:deriv_beta}

In this section, we derive the formal expressions for the critical exponents and the beta functionals in terms of the counter terms.
These expressions combined with the counter terms derived in the next appendix give rise to
Eqs.      (\ref{eq:z_and_etas_in_convolution}),
(\ref{eq:betaKf_1}), (\ref{eq:betavf_general}), (\ref{eq:betae_general}) and (\ref{eq:betalambda_convolution}), respectively.

From   \eq{dynam_crit_anomalous_dim},
the dynamical critical exponent and the anomalous dimensions of the fields are written as
$z = 1+
\frac{\mathrm{d}~\mathrm{ln}(Z_1(0))}{\mathrm{d}~\mathrm{ln}~\mu}
-\frac{\mathrm{d}~\mathrm{ln}(Z_2(0))}{\mathrm{d}~\mathrm{ln}~\mu}$, 
$\eta_{\phi} = 
\frac{1}{2}
\left[
\frac{\mathrm{d}~\mathrm{ln}(Z_3)}{\mathrm{d}~\mathrm{ln}~\mu}+(d-1)\left(\frac{\mathrm{d}~\mathrm{ln}(Z_2(0))}{\mathrm{d}~\mathrm{ln}~\mu}-\frac{\mathrm{d}~\mathrm{ln}(Z_1(0))}{\mathrm{d}~\mathrm{ln}~\mu}\right)
\right]$,
$\eta_{\psi,\theta}
=
  \frac{1}{2}
  \left[
  \frac{\mathrm{d}~\mathrm{ln}(Z_1(\theta ))}{\mathrm{d}~\mathrm{ln}~\mu}+d\left(\frac{\mathrm{d}~\mathrm{ln}(Z_2(0))}{\mathrm{d}~\mathrm{ln}~\mu}-\frac{\mathrm{d}~\mathrm{ln}(Z_1(0))}{\mathrm{d}~\mathrm{ln}~\mu}\right)
  \right]$.

We can proceed to calculate the beta functional of the Fermi momentum starting from the multiplicative relationship as
$    \mathbf{K}_{FB,\theta_B} = \mu
     \left[
     \KFtheta-
     \frac{c(\theta)}{Z_2(\theta)}
     \right]$.
At the one-loop order, $c(\theta) = 0$, 
and the beta functional for $\KFtheta$ is given in Eq. 
(\ref{eq:betaKf_1}).

The multiplicative renormalization for the bare Fermi velocity can be explicitly written from Eqs. (\ref{eq:baretorenorm}) and (\ref{eq:Z_multiplicative}) as $v_{FB,\theta_B} = 
\frac{Z_2(\theta )}{Z_1(\theta )}\frac{Z_1(0)}{Z_2(0)}v_{F,\theta }$.
Taking the derivative on the logarithm of 
this expression
with respect to $\ln$ $\mu$ for fixed bare Fermi-velocity we obtain
$  \frac{\beta_{v_F}(\theta )}{v_{F,\theta }}+\frac{d~\mathrm{ln}(Z_2(\theta ))}{d~\mathrm{ln}~\mu}- \frac{d~\mathrm{ln}(Z_1(\theta ))}{d~\mathrm{ln}~\mu}+\frac{d~\mathrm{ln}(Z_1(0))}{d~\mathrm{ln}~\mu}- \frac{d~\mathrm{ln}(Z_2(0))}{d~\mathrm{ln}~\mu} = 0$.
At the one-loop order, 
$Z_1$ is dominant and
the beta functional becomes 
$   \frac{1}{v_{F,\theta }}\beta_{v_F}(\theta ) = \frac{1}{Z_1(\theta )}\frac{\mathrm{d}(Z_1(\theta ))}{d~\mathrm{ln}~\mu}-\frac{1}{Z_1(0)}\frac{\mathrm{d}(Z_1(0))}{d~\mathrm{ln}~\mu}$.
To the leading order in $\epsilon$,
the IR beta functional $\frac{ d v_{F,\theta}}{dl}$ with $l = \log\left(\frac{\Lambda}{\mu}\right)$ is 
given in Eq. (\ref{eq:betavf_general}).

The RG flow of the Yukawa coupling function can be computed starting from the multiplicative relationship, 
$    \edim_{B,\theta_{1B},\theta_{2B}} = Z_{e}(\theta_1 ,\theta_2  ) \mu^{\frac{3-d}{2}}e_{\theta_1 ,\theta_2 }\equiv
 \frac{Z_4(\theta_1,\theta_2)}{\sqrt{Z_1(\theta_1 )Z_1(\theta _2  )Z_3}}\left(\frac{Z_1(0)}{Z_2(0)}\right)^{\frac{3-d}{2}}\mu^{\frac{3-d}{2}}e_{\theta_1,\theta_2 }$.
At one loop order, $Z_3$ = 1 and $Z_{4}(\theta_1,\theta_2) = 1$ for $\forall$ $\theta_1 ,\theta_2$. 
From this, we obtain
$     \frac{1}{e_{\theta_1 ,\theta_2 }}\beta_{e}(\theta_1 ,\theta_2 ) =  \frac{d-3}{2}+\frac{d-3}{2}\frac{1}{Z_1(0)}\frac{\mathrm{d}(Z_1(0))}{d~\mathrm{ln}~\mu}+\frac{1}{2}\left[\frac{1}{Z_1(\theta_1 )}\frac{\mathrm{d}(Z_1(\theta_1 ))}{d~\mathrm{ln}~\mu}+ \frac{1}{Z_1(\theta _2)}\frac{\mathrm{d}Z_1(\theta_2)}{d~\mathrm{ln}~\mu}\right]$.
To the leading order, the beta-functional of the Yukawa coupling is 
given in Eq. (\ref{eq:betae_general}).
The beta functional for $g_\theta$ 
defined in Eq. (\ref{eq:gtheta}),
can be readily obtained from
       $ \frac{1}{g_\theta}\frac{d~g_\theta}{d~\ln~\mu} = \frac{4}{3}\frac{1}{e_\theta}\frac{d~e_\theta}{d~\ln~\mu}-\frac{1}{3}\left[\frac{1}{v_{F,\theta }}\frac{d~v_{F,\theta }}{d~\ln~\mu}+\frac{1}{\KFtheta}\frac{d~\KFtheta}{d~\ln~\mu}\right]$.
Here, we have used the fact that the $\ln$ $\mu$ derivative of  $X_\theta$ and $\chi_{\vartheta^{-1}(\theta)}$ is $0$ as 
these functions depend only on ratio of $\KFthetadim $ and its derivatives. 
From Eqs. (\ref{eq:betaKf_1}), (\ref{eq:betavf_general})  and (\ref{eq:betae_general}), 
we obtain $ \frac{\mathrm{d} g_{\theta}}{\mathrm{d}l} = 
       g_{\theta }\left(
       \frac{5-2d}{3}
       +\frac{5-2d}{3}u_1(d)g_{0}
       -u_1(d)g_{\theta}\right)$.

In general, one can also define the effective coupling 
at general angles as
$g_{\theta_1,\theta_2} = e_{\theta_1, \theta_2 }^2\sqrt{\frac{g_{\theta_1} g_{\theta_2}}{e_{\theta_1}^2e_{\theta_2}^2}}$.
From
        $\frac{1}{g_{\theta_1,\theta_2}}\frac{d~g_{\theta_1,\theta_2}}{d~\ln~\mu} = 2\frac{1}{e_{\theta_1,\theta_2}}\frac{d~e_{\theta_1,\theta_2}}{d~\ln~\mu}+\frac{1}{2}\left[\frac{1}{g_{\theta_1}}\frac{d~g_{\theta_1}}{d~\ln~\mu}+\frac{1}{g_{\theta_2}}\frac{d~g_{\theta_2}}{d~\ln~\mu}\right]-\left[\frac{1}{e_{\theta_1}}\frac{d~e_{\theta_1}}{d~\ln~\mu}+\frac{1}{e_{\theta_2}}\frac{d~e_{\theta_2}}{d~\ln~\mu}\right]$,
the beta functional of $ g_{\theta_1,\theta_2}$ is 
given in Eq. (\ref{eq:betae}).

Now, we turn to the beta functional for the four-fermion coupling function.
The bare four-fermion coupling is related to the renormalized one through
$    \lambdadim^{(\nu,s)}_{B, \theta_{1B},\theta_{2B}}(q_B,\varphi_B) = \left(Z_{\tau}\right)^{3-3d}\prod_{i=1}^{4}Z_{\psi}(\theta_i )^{-\frac{1}{2}}Z^{(\nu,s)}_{5}\left(\theta_1,\theta_2,\vec{q}\right)\mu^{1-d}\lambda^{(\nu,s)}_{\theta_1,\theta_2}\left(\vec{q}\right)$.
To the leading order,
the beta functional can be calculated by 
taking the derivative on the logarithm of the expression for the bare coupling
with respect to $\ln$ $\mu$
for a fixed $\boldsymbol{\lambda}^{(\nu,s)}_{\theta_1,\theta_2}\left(\vec{q}\right)$,
\begin{equation}\begin{aligned}
\beta^{(\nu,s)}_{\lambda}(\theta_1,\theta_2, \vec q)
&= \left[(d-1)+ (3d-3)(z-1)+\left(\eta_{\psi,\theta_1}+\eta_{\psi,\Theta\left(\theta_1,\vec{q}\right)}+\eta_{\psi,\theta_2}+\eta_{\psi,\Theta\left(\theta_2,\vec{q}\right)}  \right)\right]\lambda^{(\nu,s)}_{\theta_1,\theta_2}\left(\vec{q}\right)
\\&-\mu^{d-1}
    \frac{d\left(\mu^{1-d}A^{(\nu,s)}_{5}\left(\theta_1,\theta_2,\vec{q}\right)\lambda^{(\nu,s)}_{\theta_1,\theta_2}\left(\vec{q}\right)\right)}{d~\ln~\mu},
\end{aligned}\end{equation}
where we have used the definitions of dynamical critical exponent $z$ and anomalous dimension of fermionic field $\eta_{\psi}\left(\theta\right)$ given in Eq. (\ref{dynam_crit_anomalous_dim})
and  
\begin{equation}
    \begin{aligned}
    \mu^{d-1}\frac{d\left(\mu^{1-d}A^{(\nu,s)}_{5}\left(\theta_1,\theta_2,\vec{q}\right)\lambda^{(\nu,s)}_{\theta_1,\theta_2}\left(\vec{q}\right)\right)}{d~\ln~\mu} &= \mu^{d-1}\left\{\frac{d~\Gamma^{CT;(2);(\nu,s)}_{\mu,\theta_1,\theta_2}(\vec q)}{d~\ln~\mu}+\frac{d~\Gamma^{CT;(1);(\nu,s)}_{\mu,\theta_1,\theta_2}(\vec q)}{d~\ln~\mu}
        +\frac{d~\Gamma^{CT;\left(1'\right);(\nu,s)}_{\mu,\theta_1,\theta_2}(\vec q)}{d~\ln~\mu}
        \right.\\&\left.
        +\frac{d~\Gamma^{CT;(0);(\nu,s)}_{\mu,\theta_1,\theta_2}(\vec q)}{d~\ln~\mu}
        \right\}.
        \label{appendix:4f_generic_beta_functional_contribution}
    \end{aligned}
\end{equation}
Plugging 
Eqs. (\ref{(2)_betafunctional_contribution}), 
(\ref{appenidx:(1)_general_beta_functional})
, 
(\ref{appenidx:(1')_general_beta_functional})
and (\ref{appendix:(0)_general_beta_functional})
to Eq. (\ref{appendix:4f_generic_beta_functional_contribution}) 
leads to the beta functional  given by Eq. (\ref{eq:betalambda_convolution}).


\section{Quantum Corrections}
\label{sec:Quantum_Corrections}

In this appendix, we compute the quantum corrections 
that determine the beta functionals to the leading order in the $\epsilon$ expansion.
We first compile a list of formulae used later in this appendix.
\bqa
&& \frac{1}{A^mB^n} = \frac{\Gamma\left(m+n\right)}{\Gamma\left(m\right)\Gamma\left(n\right)}\int_0^1~dx\frac{x^{m-1}(1-x)^{n-1}}{\left(xA+(1-x)B\right)^{m+n}}, 
\\ &&
 \int_0^1dx~x^m(1-x)^n = \frac{\Gamma\left(1+m\right)\Gamma\left(1+n\right)}{\Gamma\left(2+m+n\right)},
\\ &&
 \int_{0}^{\infty} dx \frac{x^m}{(x^2+A^2)^n} = \frac{\Gamma\left(\frac{m+1}{2}\right)\Gamma\left(n-\frac{m+1}{2}\right)}{2(A)^{2n-m-1}\Gamma\left(n\right)},
\\ &&
 \int_{-\infty}^{\infty}\frac{dx}{2\pi}\frac{(x+a)^2+b}{\left((x+c)^2+d\right)^2} = \frac{(a-c)^2+b+d}{4d^{\frac{3}{2}}},
\\ &&
 \int_{-\infty}^{\infty}\frac{dx}{2\pi}\frac{|x|}{|x|^3+a} = \frac{2}{3\sqrt{3}}a^{-\frac{1}{3}},
\\ &&
 \int_{-\infty}^{\infty}dx~\frac{x^2-a^2}{\left(x^2+a^2\right)^2} = 0,
\\ &&
 \int_{-\infty}^{\infty}\frac{dx}{2\pi}\frac{|x|^2}{(|x|^3+a)(|x|^3+b)} = \frac{1}{3\pi}\frac{\ln\left(\frac{a}{b}\right)}{a-b},
\\ &&
 \int_{-\infty}^{\infty}\frac{dx}{2\pi}\frac{1}{((x+a)^2+A^2)(((x+b)^2+B^2)} = \frac{(|A|+|B|)}{2|A||B|\left((a-b)^2+(|A|+|B|)^2\right)},
\\ &&
 \int_{-\infty}^{\infty}\frac{dx}{2\pi}\frac{(x+a)(x+b)}{((x+a)^2+A^2)(((x+b)^2+B^2)} = \frac{(|A|+|B|)}{2\left((a-b)^2+(|A|+|B|)^2\right)}.
\eqa

\subsection{Boson Self-Energy}
\label{sec:boson_self_energy}

One loop boson energy shown in Fig. \ref{Fig:BSE} is given by
\begin{equation}\begin{aligned}
\Pi_1(\mathbf{q}) = -\left[-\mu^{3-d}\int\frac{d\mathbf{L}d\delta \KFthetadim d\theta}{(2\pi)^{d+1}}\left(-
i
e_{\thetasq,\theta}
\right)\left(-
i
e_{\theta,\thetasq}
\right)
\mathrm{\mathrm{Tr}}\left\{\gamma_{d-1}G_{0}(\mathbf{l}+\mathbf{q})\gamma_{d-1}G_{0}(\mathbf{l})\right\}\right],
\label{Boson_SE_formal}
\end{aligned}\end{equation}
where
$   \mathbf{l} = (\mathbf{L},\delta,\theta)$,
$\mathbf{q} =  (\mathbf{Q},\vec{q})$,
$\mathbf{l+q} = (\mathbf{L}+\mathbf{Q},
\deltaq, \thetasq )$.
Here, $\deltaq$ and $\thetasq$ are given in Eq. (\ref{appendix:thetaq_deltaq_smallq}).
We write the trace inside 
Eq. (\ref{Boson_SE_formal}) as 
$\mathrm{\mathrm{Tr}}\{\gamma_{d-1}G_0(\mathbf{l}+\mathbf{q})\gamma_{d-1}G_0(\mathbf{l})\} = \frac{\mathcal{N}}{\mathcal{D}}$
with
\begin{equation}\begin{aligned}
    \mathcal{N} &= \mathrm{\mathrm{Tr}}\{-\mathbf{L}\cdot(\mathbf{L}+\mathbf{Q})-\mathbf{\Gamma}\cdot(\mathbf{L}+\mathbf{Q})v_{F,\theta }\delta\gamma_{d-1}-\mathbf{\Gamma}\cdot\mathbf{L}v_{F,\thetasq }\deltaq\gamma_{d-1}+v_{F,\thetasq}v_{F,\theta }\deltaq\delta\},\\
    \mathcal{D} &= i^2\left(|\mathbf{L}+\mathbf{Q}|^2+v^2_{F,\thetasq}\deltaq^{2}\right)\left(|\mathbf{L}|^2+v^2_{F,\theta }\delta^{2}\right).
\end{aligned}\end{equation}
With $\Tr\left\{\gamma_i,\gamma_j\right\}$ = 2$\delta_{ij}$,
the trace  becomes
\begin{equation}\begin{aligned}
    \mathrm{\mathrm{Tr}}\{\gamma_{d-1}G_0(\mathbf{l}+\mathbf{q})\gamma_{d-1}G_0(\mathbf{l})\} = -2\frac{-\mathbf{L}\cdot(\mathbf{L}+\mathbf{Q})+v_{F,
    \thetasq}v_{F,\theta }\deltaq\delta}{\left(|\mathbf{L}+\mathbf{Q}|^2+v^2_{F,\thetasq}\deltaq^{2}\right)\left(|\mathbf{L}|^2+v^2_{F,\theta }\delta^{2}\right)}.
\end{aligned}\end{equation}
For $q \ll \KFthetadim $,
we can use
$e_{\thetasq,\theta}\approx  e_{\theta,\theta}$
and
$v_{F,\thetasq}\approx v_{F,\theta }$
to rewrite
Eq. (\ref{Boson_SE_formal}) 
as 
\begin{equation}\begin{aligned}
    \Pi_1(\mathbf{q})=2\frac{\Gamma(2)}{\Gamma(1)\Gamma(1)}\int \frac{d\mathbf{L}d\delta \KFthetadim  d\theta}{(2\pi)^{d+1}}  \boldsymbol{e}_\theta^2\int_{0}^{1}dx \frac{-\mathbf{L}\cdot(\mathbf{L}+\mathbf{Q})+v^2_{F,\theta }\delta\deltaq}{(x|\mathbf{L}+\mathbf{Q}|^{2}+(1-x)|\mathbf{L}|^{2}+v_{F,\theta }^2\left(x\deltaq^2+(1-x)\delta^2\right))^{2}}.
\end{aligned}
\end{equation}
Here, 
we use the Feynmann parametrization,
and
$\edim_{\theta}\equiv \edim_{\theta,\theta}$ denotes the dimensionful diagonal Yukawa coupling.
Shifting $\mathbf{L}\rightarrow\mathbf{L}-x\mathbf{Q}$ and dropping terms 
that are odd in $\mathbf{L}$,
we write the self-energy as
\begin{equation}\begin{aligned}
    \Pi_1(\mathbf{q})=2\int \frac{d\mathbf{L}d\delta \KFthetadim  d\theta}{(2\pi)^{d+1}}  \boldsymbol{e}_\theta^2\int_{0}^{1}dx \frac{-\left(|\mathbf{L}|^{2}-x(1-x)|\mathbf{Q}|^{2}\right)+v^2_{F,\theta }\delta\deltaq}{(|\mathbf{L}|^{2}+x(1-x)|\mathbf{Q}|^{2}+v_{F,\theta }^2\left(x\deltaq^2+(1-x)\delta^2\right))^{2}}.
\end{aligned}\end{equation}
Completing the square in numerator and denominator 
to perform integration over $\delta$, we obtain
\begin{equation}
    \begin{aligned}
        \Pi_1(\mathbf{q})=\int_0^1 dx~x(1-x)\int \frac{d\mathbf{L} d\theta}{(2\pi)^{d}}\frac{\KFthetadim }{v_{F,\theta }} \boldsymbol{e}_\theta^2\frac{|\mathbf{Q}|^{2}}{\left(|\mathbf{L}|^{2}+x(1-x)\left( \left(L_{\theta}\left(\vec{q}\right)\right)^2+|\mathbf{Q}|^{2}\right)\right)^{3/2}},
    \end{aligned}
\end{equation}
where
$        L_{\theta}\left(\vec{q}\right)= v_{F,\theta}\left(\mathscr{F}_{\varphi,\theta}q+\mathscr{G}_{\varphi,\theta}\frac{q^2}{\KFthetadim }\right)$.
The integrations of $\mathbf{L}$ and $x$ give
\begin{equation}
    \begin{aligned}
       \Pi_1(\mathbf{q}) =v_d  \int_{-\frac{\pi}{2}}^{\frac{\pi}{2}}\frac{d\theta}{2\pi}\frac{\KFthetadim }{v_{F,\theta }} \boldsymbol{e}_\theta^2\frac{Q^2}{\left(Q^2+ \left(L_{\theta}\left(\vec{q}\right)\right)^2\right)^{\frac{4-d}{2}}},
       \label{appendix:BSE_General}
    \end{aligned}
\end{equation}
where
$        v_d = \frac{1}{2}\frac{\Omega_{d-1}\Gamma\left(\frac{d-1}{2}\right)\Gamma\left(\frac{4-d}{2}\right)\Gamma^2\left(\frac{d}{2}\right)}{\Gamma\left(\frac{3}{2}\right)(2\pi)^{d-1}\Gamma(d)}$
and
$\Omega_d = \frac{2\pi^{\frac{d}{2}}}{\Gamma\left(\frac{d}{2}\right)}$ is the solid angle in $d$ dimensions.

The result of the $\theta$ integration depends on the relative magnitude between $Q$ and $q$.
Let us first consider the limit in which $q \gg Q$.
This is the limit that is relevant for the processes where the boson scatters fermions almost tangentially 
with momentum transfer $q$ much larger than the energy transfer.
In this case, the largest contribution arises from angle around $\vartheta(\varphi)$ where particle-hole excitations with momentum $\vec q$ is almost tangential to the Fermi surface.
Let us perform gradient expansion of $\frac{\KFthetadim }{v_{F,\theta }}\edim^2_\theta$ around $\vartheta(\varphi)$,
\begin{equation}
    \begin{aligned}
        \frac{\KFthetadim }{v_{F,\theta }}\edim^2_\theta =  \frac{\mathbf{K}_{F,\vartheta(\varphi)}}{v_{F,\vartheta(\varphi)}}\edim^2_\vartheta(\varphi)
        \sum\limits_{n=0}F^{(n)}_{\vartheta(\varphi)}\left(\theta-\vartheta(\varphi)\right)^n,
        \label{appendix:gradient_expansion_BSE}
    \end{aligned}
\end{equation}
where $F^{(n)}_{\vartheta(\varphi)}$ is the ratio of $n^{th}$ derivative and zeroth order term in the gradient expansion.
For instance,
$        F^{(0)}_{\vartheta(\varphi)} = 1$,
$F^{(1)}_{\vartheta(\varphi)}= \left(2\frac{\edim^{\prime}_{\vartheta(\varphi)}}{\edim_{\vartheta(\varphi)}}+\frac{\mathbf{K}^{\prime}_{F,\vartheta(\varphi)}}{\mathbf{K}_{F,\vartheta(\varphi)}}-\frac{v^\prime_{F,\vartheta(\varphi)}}{v_{F,\vartheta(\varphi)}}\right)$.
Similarly, $ L_{\theta}\left(\vec{q}\right)$ can be approximated in the tangential direction as 
\begin{equation}
    \begin{aligned}
        L_{\theta}\left(\vec{q}\right) \approx v_{F,\vartheta(\varphi ) }\left(\chi_{\varphi}q\left(\theta-\vartheta(\varphi)\right)+\tilde{B}_{\varphi}\frac{q^{2}}{\mathbf{K}_{F,\vartheta(\varphi ) }}\right),
        \label{appendix_l_theta_tangential}
    \end{aligned}
\end{equation}
where $\chi_{\varphi}$ and $\tilde{B}_{\varphi}$ are given in Eq. (\ref{appendix:chi_tildeB}), respectively.
Using Eqs. (\ref{appendix:BSE_General}), (\ref{appendix:gradient_expansion_BSE}) and (\ref{appendix_l_theta_tangential}), and shifting $\theta \rightarrow \theta +\vartheta(\varphi)$ followed by a variable change $\tilde{\theta} = \mathbf{K}_{F,\vartheta(\varphi)}\theta$, 
we write the boson self-energy as
\begin{equation}
    \begin{aligned}
       \Pi_1(\mathbf{q}) =v_d\frac{ \boldsymbol{e}^2_{\vartheta\left(\varphi\right)}}{v_{F,\vartheta\left(\varphi\right)}}Q^2\sum\limits_{n=0}\frac{F^{(n)}_{\vartheta(\varphi)}}{\left(\mathbf{K}_{F,\vartheta(\varphi)}\right)^n}
       \int_{-\tilde \theta_{max}}^{\tilde \theta_{max}}
       \frac{d\tilde{\theta}}{2\pi}\frac{\tilde{\theta}^n}{\left\{Q^2+ \left[v_{F,\vartheta(\varphi ) }\left(\chi_{\varphi}\tilde{\theta}\frac{q}{\mathbf{K}_{F,\vartheta(\varphi ) }}+\tilde{B}_{\varphi}\frac{q^{2}}{\mathbf{K}_{F,\vartheta(\varphi ) }}\right)\right]^2\right\}^{\frac{4-d}{2}}},
    \end{aligned}
\end{equation}
where  
$\tilde \theta_{max} \sim \mathbf{K}_{F,\vartheta(\varphi ) }$.
Shifting $\tilde{\theta} \rightarrow \tilde{\theta}-\frac{
\tilde{B}_{\varphi}}{\chi_{\varphi}}q$, we obtain
\begin{equation}
    \begin{aligned}
       \Pi_1(\mathbf{q}) =v_d\frac{ \boldsymbol{e}^2_{\vartheta\left(\varphi\right)}}{v_{F,\vartheta\left(\varphi\right)}}Q^2\sum\limits_{n=0} 
       \frac{F^{(n)}_{\vartheta(\varphi)}}
       {\left(\mathbf{K}_{F,\vartheta(\varphi)}\right)^n}
              \int_{-\tilde \theta_{max}}^{\tilde \theta_{max}}
\frac{d\tilde{\theta}}{2\pi}\frac{\left(\tilde{\theta}-\frac{
       \tilde{B}_{\varphi}}{\chi_{\varphi}}q\right)^n}{\left(Q^2+ \left(q\frac{v_{F,\vartheta(\varphi ) }\chi_{\varphi}}{\mathbf{K}_{F,\vartheta(\varphi ) }}\right)^2 \tilde{\theta}^2\right)^{\frac{4-d}{2}}}.
       \label{eq:bse_taylor}
    \end{aligned}
\end{equation}
The term which are most singular in Q and $q$, can be identified from Eq. (\ref{eq:bse_taylor})\footnote{
The integral in Eq. (\ref{eq:bse_taylor}) can be performed using the binomial expansion of the numerator $\sim q^m\tilde{\theta}^{m^\prime}$, with $m+m^\prime = n$. 
The integral is non-vanishing only for even $m'$. 
In $2\leq d<3$, the integral is UV convergent (divergent) for $m^{\prime} = 0$ ($m^{\prime}\geq2$). 
For $m^{\prime} = 0$, 
the contribution to  $\Pi_1({\bf q})$ goes as
$
\frac{\mathbf{K}_{F,\vartheta(\varphi)}}{q}\left(\frac{q}{\mathbf{K}_{F,\vartheta(\varphi)}}\right)^n Q^{d-1}$. 
The most singular correction arises at $n=0$ which is given by first term 
in Eq. (\ref{eq:bse_singular}). 
The terms from $n>0$ are suppressed by 
$q/\KFthetadim $.
The contribution to  $\Pi_1({\bf q})$ from the terms with 
$m^{\prime} \geq 2$ is order of 
$
Q^2
q^{d-3}\left(\frac{q}{\mathbf{K}_{F,\vartheta(\varphi)}}\right)^{m-1}$, all of which are suppressed by either positive powers of
$q/\KFthetadim $
or
$Q/q$.
},
\begin{equation}
    \begin{aligned}
          \Pi_1(\mathbf{q})= \beta_d\frac{ \boldsymbol{e}^2_{\vartheta\left(\varphi\right)}\mathbf{K}_{F,\vartheta(\varphi)}}{\left|\chi_\varphi\right| v^2_{F,\vartheta\left(\varphi\right)}}\frac{Q^{d-1}}{q}+\delta\Pi_1(\mathbf{q}),\label{eq:bse_singular}
    \end{aligned}
\end{equation}
where
$ \beta_d = 
\frac{\Gamma^{2}(\frac{d}{2})}{2^{d-1}\pi^{\frac{d-1}{2}}\Gamma(d)\Gamma\left(\frac{d-1}{2}\right)|cos(\frac{\pi d}{2})|}$
and
$        \delta\Pi_1(\mathbf{q}) \sim \frac{ \boldsymbol{e}^2_{\vartheta\left(\varphi\right)}}{\left|\chi_\varphi\right|^{4-d} v^{5-d}_{F,\vartheta\left(\varphi\right)}}Q^2 q^{d-3}\left(\frac{\mathbf{K}_{F,\vartheta(\varphi)}}{q}-\frac{\tilde{B}_\varphi
          }{\chi_\varphi}\right)$.
Every term in $\delta\Pi_1(\mathbf{q})$ is
subleading compared to Eq. (\ref{eq:bse_singular}) in the limit $Q\ll q \ll \KFthetadim $. 
Therefore, the most singular correction is given by
\begin{equation}\begin{aligned}
    \Pi_1(\mathbf{q})  = \boldsymbol{f}_{d,\varphi }\frac{|\mathbf{Q}|^{d-1}}{q},
    \label{Boson_SE_Final_Expression}
\end{aligned}\end{equation}
where
$\boldsymbol{f}_{d,\varphi } = \beta_d\frac{\mathbf{K}_{F,\vartheta(\varphi ) } \boldsymbol{e}^2_{\vartheta(\varphi )}}{|\chi_{\varphi}|v^2_{F,\vartheta(\varphi ) }}$.
In the $Q\gg q$ limit, Eq. (\ref{appendix:BSE_General})  becomes
\begin{equation}
    \begin{aligned}
         \Pi_1(\mathbf{q}) = v_d Q^{d-2}\int_{-\frac{\pi}{2}}^{\frac{\pi}{2}}\frac{d\theta}{2\pi}\frac{\KFthetadim }{v_{F,\theta }}\boldsymbol{e}_\theta^2.
    \end{aligned}
\end{equation}

\subsection{Fermion Self-energy and the Cubic vertex correction}
\label{sec:fermion_self_energy}

\begin{figure}[h]
 \centering
  \includegraphics[width=0.4\linewidth]{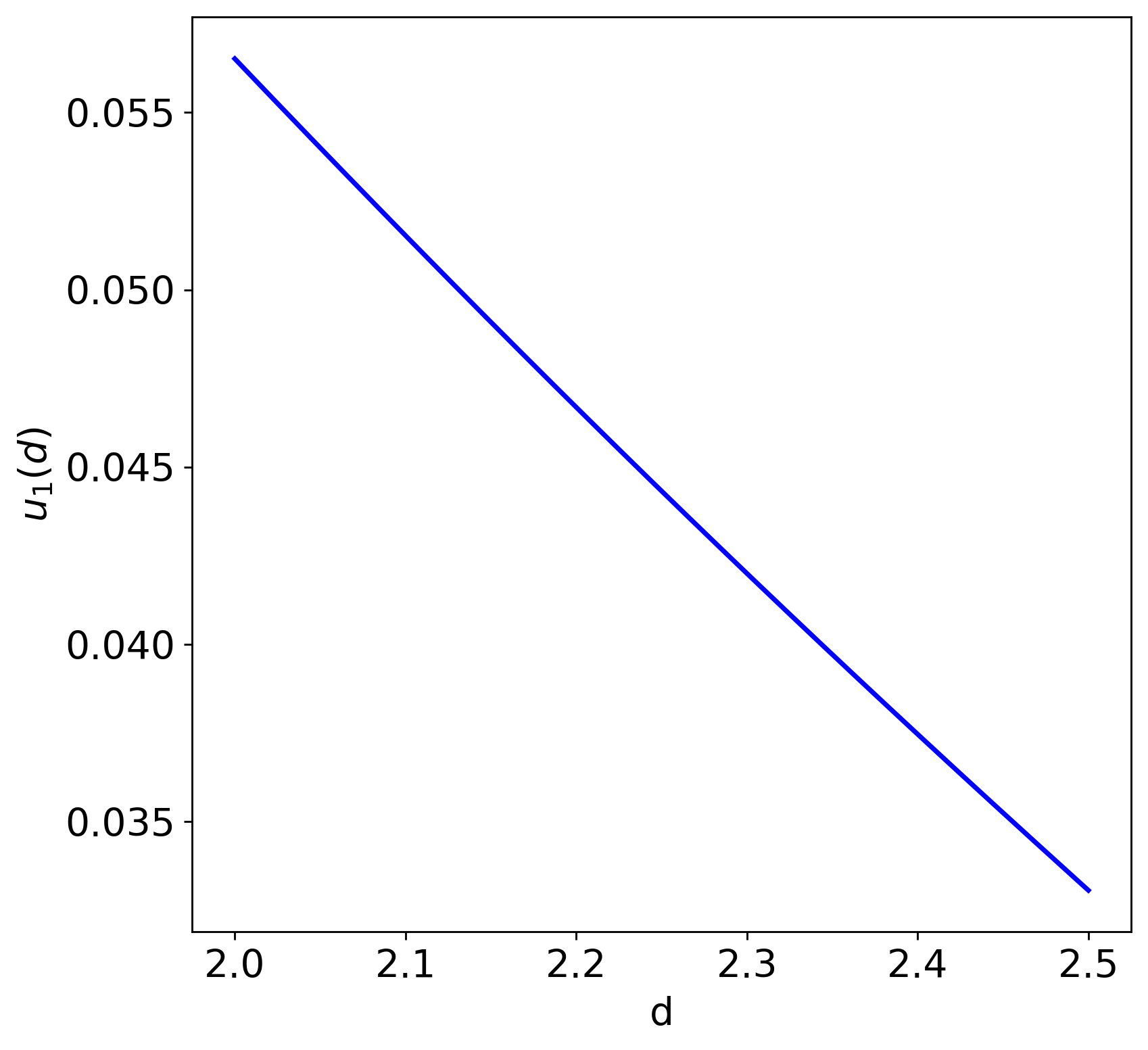}
\caption{
$u_1(d)$ in \eq{appendix:u1_d} plotted as a function of dimension $d$.
}
\label{fig:u1_d}
\end{figure}

The one-loop fermion self-energy (Fig. \ref{Fig:FSE}) reads 
\begin{equation}\begin{aligned}
\Sigma_{1}(\mathbf{k}) = -
\frac{2}{2!}
\frac{\mu^{3-d}}{N}\int\frac{d\mathbf{L}d\delta \KFthetadim d\theta}{(2\pi)^{d+1}}(-ie_{\theta_1 ,\theta})(-ie_{\theta,\theta_1 })(\gamma_{d-1}G_{0}(\mathbf{l})\gamma_{d-1})D_1(\mathbf{l}-\mathbf{k}),
\label{fermion_SE_formal}
\end{aligned}\end{equation}
where
$\mathbf{k} = (\mathbf{K},\delta_1,\theta_1)$ is the external momentum,
$\mathbf{l} = (\mathbf{L},\delta,\theta)$ is the loop momentum
and
\begin{equation}
    \begin{aligned}
   D_{1}(\mathbf{q}) =\frac{q}{q^{3}+ \boldsymbol{f}_{d,\varphi}
        |\mathbf{Q}|
        ^{d-1}}
   \label{boson_propagator_Q}
    \end{aligned}
\end{equation}
is the one-loop dressed boson propagator.
In Eq. (\ref{boson_propagator_Q}),
we use the boson self-energy that is valid in the $q \gg |{\bf Q}|$ limit because
the typical loop momentum is much larger than the frequency.
Eq. (\ref{fermion_SE_formal}) now reads
\begin{equation}\begin{aligned}
 \Sigma_{1}(\mathbf{k}) = -\frac{i}{N}\int\frac{d\mathbf{L}d\delta \KFthetadim d\theta}{(2\pi)^{d+1}}
 D_{1}\left(\mathbf{l}-\mathbf{k}\right)
 \frac{-\mathbf{\Gamma} \cdot
 \mathbf{L}
 +v_{F,\theta } \delta \gamma_{d-1}}
{
\left|\mathbf{L}\right|
^2 +v^2_{F,\theta }\delta^2} \boldsymbol{e}_{\theta_1 ,\theta} \boldsymbol{e}_{\theta,\theta_1 }.
\end{aligned}\end{equation}
The counter term $A_1\left(\theta_1\right)$ that is analytic in momentum space for $\mu \neq 0$ and 
cancels the IR divergence of the quantum correction can be expressed as
\begin{equation}
    \begin{aligned}
        A_1\left(\theta_1\right) =
         \frac{1}{2N(d-1)}
        \tr\left\{\mathbf{\Gamma}\cdot\nabla_{\mathbf{K}}
        \int\frac{d\mathbf{L}d\delta \KFthetadim d\theta}{(2\pi)^{d+1}}
 D_{1;\mu}\left(\mathbf{L}-\mathbf{K},\theta,\theta_1\right)
 \frac{-\mathbf{\Gamma} \cdot
 \mathbf{L}
 +v_{F,\theta } \delta \gamma_{d-1}}
{
\left|\mathbf{L}\right|
^2 +v^2_{F,\theta }\delta^2} \boldsymbol{e}_{\theta_1 ,\theta} \boldsymbol{e}_{\theta,\theta_1 }\right\}\bigg|_{\mathbf{K} = \boldsymbol{\mu}
        }.
        \label{appendix:Z1}
    \end{aligned}
\end{equation}
Here,
$D_{1;\mu}\left(\mathbf{L},
\theta_1,\theta_2\right) = \frac{1}{q(\theta_1,\theta_2)^{2}+ \boldsymbol{f}_{d,\vartheta^{-1}\left(
\mtheta
\right)}
        \frac{|\mathbf{L}
        |
        ^{d-1}}{\sqrt{q(\theta_1,\theta_2)^{2}+\mu^2}}}$ 
corresponds to the regularized boson propagator
with
$q\left(\theta_1,\theta_2\right) = \sqrt{\mathbf{K}^2_{F,\theta_1}+\mathbf{K}^2_{F,\theta_2}-2\mathbf{K}_{F,\theta_1}\mathbf{K}_{F,\theta_2}\cos\left(\theta_1-\theta_2\right)}$
denoting the momentum that connects the Fermi surface at angles 
$\theta_1$ and $\theta_2$. 
To make sure that the counter term is analytic in momentum, the boson propagator has been 
regularized with energy $\mu$ for small momentum
$q(\theta_1,\theta_2)$.
This infrared cutoff arises naturally from the full boson propagator. 
We also have used the fact  that the boson propagator is strongly suppressed as $(\Delta\theta)^{-2}$ at large $\Delta\theta$, so $\boldsymbol{f}_{d,\varphi}$
is evaluated at the angle which is tangential to the average of the angles,
$\mtheta = \frac{\theta_1+\theta_2}{2}$. 
This does not affect the singular part of the self-energy.
The log $\mu$ derivative of Eq. (\ref{appendix:Z1}) can be formally written as
\begin{equation}
    \begin{aligned}
        \bar{\omega}_{\theta_1}\equiv \frac{\partial Z_1\left(\theta_1\right)}{\partial\log\mu} = \int_{-\frac{\pi}{2}}^{\frac{\pi}{2}}\frac{d\theta}{2\pi\mu}\frac{\KFthetadim }{v_{F,\theta}} \omega_{d;\theta,\theta_1}(\mu),
    \end{aligned}
\end{equation}
where
\begin{equation}
    \begin{aligned}
    \omega_{d;\theta,\theta_1}(\mu) = -\frac{\edim^2_{\theta_1,\theta}\mu}{4N(d-1)}\partial_{\log\mu}\left(\tr\left\{\mathbf{\Gamma}\cdot\nabla_{\mathbf{K}}\int\frac{d\mathbf{L}}{\left(2\pi\right)^{d-1}}
 D_{1;\mu}\left(\mathbf{L}-\mathbf{K},\theta,\theta_1\right)
 \frac{\mathbf{\Gamma}\cdot\mathbf{L}}{|\mathbf{L}|}\right\}\bigg|_{\mathbf{K} = \boldsymbol{\mu}}\right).
 \label{appendix:self_energy_weight}
    \end{aligned}
\end{equation}
Since the boson propagator is sharply peaked at $\theta_1$ (as it is suppressed as $\frac{1}{(\Delta\theta)^2}$ at large $\Delta\theta$), 
we can extract the singular part of the self-energy using the adiabatic approximation\cite{BORGES2023169221}. 
We use $\edim_{\theta,\theta_1}\rightarrow \edim_{\theta_1,\theta_1} $, $\mathbf{K}_{F,\theta}\rightarrow \mathbf{K}_{F,\theta_1}$, $v_{F,\theta}\rightarrow v_{F,\theta_1}$, 
$\boldsymbol{f}_{d,\vartheta^{-1}\left(\frac{\theta_1+\theta}{2}\right)}\rightarrow\boldsymbol{f}_{d,\vartheta^{-1}\left(\theta_1\right)}$, and
$q(\theta,\theta_1)\approx \pmb{\mathcal{K}}_{\theta_1}|\theta-\theta_1|$
with $\pmb{\mathcal{K}}_{\theta} = \frac{\KFthetadim }{\left|X_{\theta}\right|}$ and $X_{\theta} = \mathrm{sin}\left(\vartheta^{-1}\left(\theta\right)-\theta\right)$ to perform integration over $\theta$\footnote{
In performing the $\theta$ integration,
we use the fact the $q\left(\theta,\theta_1\right)\gg\mu$ as the dominant contribution arises from virtual bosons whose momenta are almost tangential to the Fermi surface at angle $\theta_1$. 
} to calculate 
Eq. (\ref{appendix:self_energy_weight})
as
\begin{equation}
    \begin{aligned}
        \bar{\omega}_{\theta_1} = -\frac{\mathbf{g}_{\theta_1}}{6\sqrt{3}(d-1)\beta_d^{\frac{1}{3}}}
        \partial_{\log\mu}\left(\tr\left\{\mathbf{\Gamma}\cdot\nabla_{\mathbf{K}}\int\frac{d\mathbf{L}}{\left(2\pi\right)^{d-1}}\frac{\mathbf{\Gamma}\cdot\mathbf{L}}{ |\mathbf{L}-\mathbf{K}|
 ^{\frac{d-1}{3}}|\mathbf{L}|}\right\}\bigg|_{\mathbf{K} = \boldsymbol{\mu}}\right)
    \end{aligned}
\end{equation}
where
$        \boldsymbol{g}_{\theta} = \frac{1}{N} \frac{\KFthetadim \boldsymbol{e}^2_{\theta} \beta_d^{1/3}}{\pmb{\mathcal{K}}_{\theta}v_{F,\theta }\left(\boldsymbol{f}_{d,\vartheta^{-1}(\theta)}\right)^{\frac{1}{3}}}
        \equiv
        \frac{1}{N} \frac{\edim^{4/3}_{\theta }\left|X_{\theta}\right||\chi_{\vartheta^{-1}(\theta)}|^{1/3}}{v^{1/3}_{F,\theta }\KFthetadim^{1/3}}$.
Finally, integration over $\mathbf{L}$ through Feynman integration gives
\begin{equation}\begin{aligned}
   \bar{\omega}_{\theta_1}  =g_{\theta_1}u_1(d),
   \label{appendix:omega_bar_adiabatic}
\end{aligned}\end{equation}
to the leading order in $\epsilon = \frac{5}{2}-d$. Here,
\begin{equation}\begin{aligned}
    u_1(d) = \frac{1}{3\sqrt{3}} \frac{1}{\beta_d^{\frac{1}{3}}}\frac{\Omega_{d-1}}{(2\pi)^{d-1}}\frac{\Gamma(\frac{d}{2})\Gamma\left(\frac{d-1}{3}\right)\Gamma\left(\frac{d-1}{2}\right)\Gamma\left(\frac{11-2d}{6}\right)}{\Gamma\left(\frac{1}{2}\right)\Gamma\left(\frac{d-1}{6}\right)\Gamma\left(\frac{5d-2}{6}\right)}
    \label{appendix:u1_d} \end{aligned}\end{equation}
is an $O(1)$ dimension dependent number which is shown in Fig. 
\ref{fig:u1_d}
and
        $g_{\theta} = \boldsymbol{g}_{\theta}\mu^{\frac{2d-5}{3}}
        \equiv
        \frac{1}{N}\frac{e^{4/3}_{\theta }\left|X_{\theta}\right||\chi_{\vartheta^{-1}(\theta)}|^{1/3}}{v^{1/3}_{F,\theta }\KFtheta^{1/3}}$
is the dimensionless effective Yukawa coupling with
$\KFtheta = \KFthetadim /\mu$.

At one loop order,
the fermion self-energy is independent of spatial momentum :
$A_2(\theta) =
c(\theta) = 0$. Due to Ward's identity, $ \Gamma(\mathbf{k},\mathbf{0})  =   \frac{ \partial G^{-1}(\mathbf{k}) }{\partial \delta  }$, 
the vertex correction in Fig.  \ref{Fig:Yukawa_Vertex} vanishes and $A_4(\thetasq,\theta) = 0$, at the one-loop order.

\subsection{Four-fermion coupling 
in group 1 and group 2}
\label{app:lambda012}

The quantum corrections to the four-fermion vertex can be written as
\begin{equation}
    \begin{aligned}
        \delta S_4 &= 
         \int^{'}
        \prod_{i=1}^{4}d_{f}^{d+1} {\bf k}_i ~
       \Gamma_{\{j_i\};abcd }\left(\mathbf{k}_1, \mathbf{k}_2, \mathbf{k}_3, \mathbf{k}_4\right)
       ~~
    \bar{\Psi}_{j_1;a}
    (\mathbf{k}_1)
    \Psi_{j_4;b}
    \left(\mathbf{k}_4  \right)
    \bar{\Psi}_{j_2;c}\left(\mathbf{k}_2
\right)
    \Psi_{j_3;d}(\mathbf{k}_3)
    \delta(\mathbf{k}_1+\mathbf{k}_2-\mathbf{k}_3-\mathbf{k}_4).
    \end{aligned}
\end{equation}
The counter term is to be added to cancel the quantum correction evaluated at a set of external momenta as is shown in \eq{eq:RG5}.
To make sure that the counter term is Hermitian,
it is convenient to write the counter term to cancel the average of the quantum corrections evaluated at
two sets of external momenta with different frequencies but with the same two-dimensional spatial momentum,
$\left\{ {\bf k}^{A}_i \right\}$
and
$\left\{ {\bf k}^{B}_i \right\}$,
\begin{equation}
    \begin{aligned}
        \mathbf{k}^{A}_1 =\left(3\boldsymbol{\mu},
        0,\theta_1 \right),~~ 
        \mathbf{k}^{A}_2=\left(-\boldsymbol{\mu},\Delta(0,\theta_2,\vec q),\Theta(\theta_2,\vec q)\right),  ~~
        \mathbf{k}^{A}_3= \left( \boldsymbol{\mu}, 0,\theta_2 \right),   ~~
\mathbf{k}^{A}_4=\left(\boldsymbol{\mu},\Delta(0,\theta_1,\vec q),\Theta(\theta_1,\vec q)\right),~~
        \\
    \mathbf{k}^{B}_1 =
    \left(-\boldsymbol{\mu},
    0,\theta_1 \right),~~
\mathbf{k}^{B}_2=\left(3\boldsymbol{\mu},\Delta(0,\theta_2,\vec q),\Theta(\theta_2,\vec q)\right),  ~~
    \mathbf{k}^{B}_3=
\left(
\boldsymbol{\mu},
0,\theta_2 \right), ~~
    \mathbf{k}^{B}_4=\left( \boldsymbol{\mu}, \Delta(0,\theta_1,\vec q),\Theta(\theta_1,\vec q) \right), ~~
\label{appendix_mu2}
    \end{aligned}
\end{equation}
where 
$\boldsymbol{\mu}$
is a $(d-1)$-dimensional vector with magnitude $\mu$.
The external frequencies are chosen so that the frequency running in the s, t and u channels are all non-zero, which guarantees that all IR divergence is cut off by $\mu \neq 0$.
The resulting angle-dependent vertex function is written as 
\bqa
\Gamma_{\{j_i\};abcd}
\left(\mathbf{k}_1^{A;B}, \mathbf{k}_2^{A;B}, \mathbf{k}_3^{A;B}, \mathbf{k}_4^{A;B}\right)
=
\sum_{\nu,s}
\Gamma^{(\nu,s)}_{\mu^{A;B}}\left(\theta_1,\theta_2,\vec{q};\mu\right)
\left(I_{m}^{(\nu)}\right)_{ab} 
\left(I_{m}^{(\nu)}\right)_{cd} 
T^{(\nu,s)}_{\left(\begin{smallmatrix}     j_1  & j_2      \\ j_4      & j_3       \end{smallmatrix}\right)}
+
\mbox{regular terms}.
\label{eq:generalQC4}
\eqa
$I^{(\nu)}_m$ and $T^{(\nu,s)}_{\left(\begin{smallmatrix}     j_1  & j_2      \\ j_4      & j_3       \end{smallmatrix}\right)}$   are defined in  Eqs.  (\ref{4f_channels}) and (\ref{Flavour_Tensor}), respectively.
The first term in \eq{eq:generalQC4} is the most general form of the four-fermion vertex function allowed by the
$SO(d-1)\times SO(4-d)$ symmetry.
The `regular terms',
which are not singular in the small $\mu$ limit,
can take a form that is not symmetric due to the asymmetric choice of 
external frequency.
The counter term in \eq{eq:SCT} is chosen as
\begin{equation}\begin{aligned}
\Gamma^{CT;(\nu,s)}_{\mu,\theta_1,\theta_2}\left(\vec{q}\right)= -\frac{1}{2}\left(\Gamma^{(\nu,s)}_{\mu^{A}}\left(\theta_1,\theta_2,\vec{q};\mu\right)+\Gamma^{(\nu,s)}_{\mu^{B}}\left(\theta_1,\theta_2,\vec{q};\mu\right)\right).
\label{appendix:4f_counter_term}
\end{aligned}\end{equation}
With this choice, 
the renormalization condition in \eq{eq:RG5} is satisfied.

\subsubsection{\texorpdfstring{$\lambda^2$}{Lg} order}
\label{lambda_2_order}

In this section, we derive 
the counter term for the quantum correction that arises from Fig. \ref{fish1}.
The quantum correction evaluated at external momenta $\mu^{A}$ 
can be written as 
\begin{equation}
    \begin{aligned}
        \Gamma^{(2)}_{\{j_i\};abcd}
       \left(\mathbf{k}_1^{A}, \mathbf{k}_2^{A}, \mathbf{k}_3^{A}, \mathbf{k}_4^{A}\right) &= 
        -
        \frac{4}{2!}T^{(\nu_1,s_1)}_{\left(\begin{smallmatrix}     j_1  & j_2^{\prime}      \\ j_4      & j_1^{\prime}       \end{smallmatrix}\right)}T^{(\nu_2,s_2)}_{\left(\begin{smallmatrix}     j_1^{\prime}  & j_2      \\ j_2^{\prime}      & j_3       \end{smallmatrix}\right)}\left(I^{(\nu_1)}_m\right)_{ab}\left(I^{(\nu_2)}_{m^{\prime}}\right)_{cd} \times 
        \\ &
        \left\{- \int^{'}d^{d+1}_f\mathbf{l} 
        \left(-\boldsymbol{\lambda}^{(\nu_1,s_1)}_{\theta_1,\theta}\left(\vec{q}\right)\right)\left(-\boldsymbol{\lambda}^{(\nu_2,s_2)}_{\theta,\theta_2}\left(\vec{q}\right)\right)
       \mathrm{Tr}\left\{I^{(\nu_1)}_m G_0(\mathbf{l})I^{(\nu_2)}_{m^{\prime}} G_0(\mathbf{l+q})\right\}
       \right\},
        \label{appendix:S4_(2)}
    \end{aligned}
\end{equation}
where
$\mathbf{l} = \left(\mathbf{L},\delta,\theta\right)$ is the loop momentum 
and $\mathbf{l+q} = (\mathbf{L}-2\boldsymbol{\mu},\deltaq,\thetasq)$. 
For $\nu_1\neq\nu_2$,
Eq. (\ref{appendix:S4_(2)}) is non-singular in the small $\mu$ limit.
To show this, we write
\begin{equation}\begin{aligned}
   \left(I^{(\nu_1)}_m\right)_{ab}
   \mathrm{Tr}\left\{I^{(\nu_1)}_mG_0(\mathbf{l})I^{(\nu_2)}_{m^{\prime}}G_0(\mathbf{l+q})\right\}
   \left(I^{(\nu_2)}_{m^{\prime}}\right)_{cd}= \frac{\mathcal{N}^{(2);\left(\nu_1,\nu_2\right)}_{abcd}}{\mathcal{D}^{(2)}},
\label{Quartic_kernel}
 \end{aligned}\end{equation}
where the denominator is 
$     \mathcal{D}^{(2)} = i^2{\left(|\mathbf{L}-2\boldsymbol{\mu}|^{2}+v_{F,\thetasq}^2\deltaq^2\right)\left(|\mathbf{L}|^{2}+v_{F,\theta }^2\delta^2\right)}$.
For the numerator, we consider the off-diagonal and diagonal elements in $\nu_1$
and $\nu_2$ separately.
For $\nu_1\neq\nu_2$,
the numerator becomes
\begin{equation}
     \begin{aligned}
        \mathcal{N}^{(2);\left(\nu_1,\nu_2\right)}_{abcd} &=  \left(I^{(\nu_1)}_m\right)_{ab}
        \mathrm{Tr}\left\{I^{(\nu_1)}_m \left(\Gamma_{j} L_{j}+v_{F,\theta }\delta\gamma_{d-1}\right)I^{(\nu_2)}_{m^\prime}\left(\Gamma_{k}(L-2\mu)_{k}+v_{F,\thetasq}\deltaq\gamma_{d-1}\right)\right\}
        \left(I^{(\nu_2)}_{m^\prime}\right)_{cd}\\
     &= \left(I^{(\nu_1)}_m\right)_{ab}\mathrm{Tr}\left\{I^{(\nu_1)}_m \Gamma_{j}I^{(\nu_2)}_{m^\prime}\Gamma_{k}\right\} L_{j}(L-2\mu)_{k}\left(I^{(\nu_2)}_{m^\prime}\right)_{cd}\\
     &+\left(I^{(\nu_1)}_m\right)_{ab} \mathrm{Tr}\left\{I^{(\nu_1)}_m \gamma_{d-1}I^{(\nu_2)}_{m^\prime}\Gamma_{k}\right\} v_{F,\theta }\delta (L-2\mu)_{k}\left(I^{(\nu_2)}_{m^\prime}\right)_{cd}\\
     &+\left(I^{(\nu_1)}_m\right)_{ab}\mathrm{Tr}\left\{I^{(\nu_1)}_m \Gamma_{j}I^{(\nu_2)}_{m^\prime}\gamma_{d-1}\right\}v_{F,\thetasq}\deltaq L_{j}\left(I^{(\nu_2)}_{m^\prime}\right)_{cd}\\
     &+\left(I^{(\nu_1)}_m\right)_{ab}\mathrm{Tr}\left\{I^{(\nu_1)}_m \gamma_{d-1}I^{(\nu_2)}_{m^\prime}\gamma_{d-1}\right\}v_{F,\theta }v_{F,\thetasq}\delta\deltaq\left(I^{(\nu_2)}_{m^\prime}\right)_{cd}.
     \label{appendix:N_nu_nuprime}
     \end{aligned}
 \end{equation}
The second and third terms are odd in $\mathbf{L}$ in the $\mu \rightarrow 0$ limit, and don't generate a singular contribution.
For the first and the fourth terms,
we can make the following replacement,
\begin{equation}\begin{aligned}
\left(\Gamma_{j}\right)_{ab}
\left(\Gamma_{k}\right)_{cd}
    L_{j}
    (L-2\mu)_{k} 
    \rightarrow 
    \frac{\delta_{jk}}{d-1}
   ({\bf \Gamma}_{ab} 
   \cdot
   {\bf \Gamma}_{cd} )
    \left[
    \mathbf{L}\cdot(\mathbf{L}-2\boldsymbol{\mu})
    \right].
    \label{appendix:non_vanishing_freqprod}
\end{aligned}\end{equation}
This shows that they vanish for $\nu_1 \neq \nu_2$. 

Using Eq. (\ref{Flavour_Tensor}),
we can write Eq. (\ref{appendix:S4_(2)}) as
\begin{equation}
    \begin{aligned}
          \Gamma^{(2)}_{\{j_i\};abcd}
       \left(\mathbf{k}_1^{A}, \mathbf{k}_2^{A}, \mathbf{k}_3^{A}, \mathbf{k}_4^{A}\right) &= 2T^{(\nu,s)}_{\left(\begin{smallmatrix}     j_1  & j_2      \\ j_4     & j_3       \end{smallmatrix}\right)}M^{(\nu,s)}_{s_1,s_2} \int^{'}d_{f}^{d+1}\mathbf{l} ~\boldsymbol{\lambda}^{(\nu,s_1)}_{
   \theta_1,\theta}\left(\vec{q}\right)\boldsymbol{\lambda}^{(\nu,s_2)}_{\theta,\theta_2}\left(\vec{q}\right)\left(I^{(\nu)}_m\right)_{ab}
   \mathrm{Tr}\{I^{(\nu)}_m G_0(\mathbf{l})I^{(\nu)}_{m^{\prime}} G_0(\mathbf{l+q})\}\left(I^{(\nu)}_{m^{\prime}}\right)_{cd}\\
   &+\text{regular terms},
    \end{aligned}
\end{equation}
where the matrix that determines the mixing between different flavour channels is 
 obtained to be
\begin{equation}
    \begin{aligned}
       M^{(F_{\pm},d)}_{s_1,s_2} = 
        \begin{cases}
            N,&~
            (s_1, s_2) = (d,d)\\
            1,&~
            ~(s_1,s_2)=(e,d) ~~\text{or}~~(d,e)\\
            0,&~\text{otherwise}
        \end{cases}, & ~~~~~~
       M^{(F_{\pm},e)}_{s_1,s_2} = 
        \begin{cases}
            1,&~
            (s_1,s_2) = (e,e)\\
            0,&~\text{otherwise}
        \end{cases},\\
        M^{(P,d)}_{s_1,s_2} = 
        \begin{cases}
            1,&~
            (s_1,s_2)=(d,d) ~\text{or}~~(e,e)\\
            0,&~\text{otherwise}
        \end{cases}, & ~~~~~~
        M^{(P,e)}_{s_1,s_2} = 
        \begin{cases}
            1,&~
            (s_1,s_2)=(e,d) ~~\text{or}~~(d,e) \\
            0&~\text{otherwise}
        \end{cases}.
    \end{aligned}
\end{equation}
Let us now evaluate the the diagonal elements one by one.
\begin{itemize}
\item $\nu_1=\nu_2=F_+$

In the forward scattering channel with even angular momentum,
the numerator becomes
\begin{equation}\begin{aligned}
     \mathcal{N}^{(2);(F_+,F_+)}_{abcd} &=  \left(i\gamma_{d-1}\right)_{ab}
     \mathrm{Tr}\bigl\{\left(i\gamma_{d-1}\right)\left(\Gamma_{i} L_{i}+v_{F,\theta }\delta\gamma_{d-1}\right)\left(i\gamma_{d-1}\right)\left(\Gamma_{j}(L-2\mu)_{j}+v_{F,\thetasq}\deltaq\gamma_{d-1}\right)\bigl\}
     \left(i\gamma_{d-1}\right)_{cd}\\
     &= -\left(i\gamma_{d-1}\right)_{ab}\left(i\gamma_{d-1}\right)_{cd}
     \left[\mathrm{Tr}\left\{\gamma_{d-1}\Gamma_{i}\gamma_{d-1}\Gamma_{j}\right\}
     L_{i}(L-2\mu)_{j}
     +
     \mathrm{Tr}\left\{\gamma_{d-1}\gamma_{d-1}\right\}v_{F,\theta }v_{F,\thetasq}\delta\deltaq\right],
 \end{aligned}\end{equation}
and the singular contribution of
Eq. (\ref{Quartic_kernel}) becomes
 \begin{equation}\begin{aligned}
    \frac{\mathcal{N}^{(2);(F_+,F_+)}_{abcd}}{\mathcal{D}^{(2)}}
     &=
     -2\frac{\mathbf{L}\cdot(\mathbf{L}-2\boldsymbol{\mu})-v_{F,\theta }v_{F,\thetasq}\delta\deltaq}{(|\mathbf{L}-2\boldsymbol{\mu}|^2+v_{F,\thetasq}^2\deltaq^2)(|\mathbf{L}|^{2}+v_{F,\theta }^2\delta^2)}\left(i\gamma_{d-1}\right)_{ab}\left(i\gamma_{d-1}\right)_{cd}.
     \label{appendix:quartic_kernel_nu1}
 \end{aligned}\end{equation}
 
\item $\nu_1=\nu_2=F_-$

In the forward scattering channel with odd angular momentum,
we obtain
\begin{equation}\begin{aligned}
&\mathcal{N}^{(2);(F_-,F_-)}_{abcd} = \left(\Gamma_{i}\right)_{ab}
\mathrm{Tr}\left\{\Gamma_{i}\left(\Gamma_{j} L_{j}+v_{F,\theta }\delta\gamma_{d-1}\right)\Gamma_{k}\left(\Gamma_{l}(L-2\mu)_{l}+v_{F,\thetasq}\deltaq\gamma_{d-1}\right)\right\}
\left(\Gamma_{k}\right)_{cd}\\
&=\left(\Gamma_{i}\right)_{ab}\mathrm{Tr}\bigl\{\Gamma_{i}\Gamma_{j}\Gamma_{k}\Gamma_{l}\bigl\}
L_{j}(L-2\mu)_{l}
\left(\Gamma_{k}\right)_{cd}
+\left(\Gamma_{i}\right)_{ab}\mathrm{Tr}\bigl\{\Gamma_{i}\gamma_{d-1}\Gamma_{k}\gamma_{d-1}\bigl\}v_{F,\theta }v_{F,\thetasq}\delta\deltaq\left(\Gamma_{k}\right)_{cd}.\\
&=2\left[ 
L_{j}(L-2\mu)_{l}
\left[\left(\Gamma_{j}\right)_{ab}\left(\Gamma_{l}\right)_{cd}+\left(\Gamma_{l}\right)_{ab}\left(\Gamma_{j}\right)_{cd}\right]-\left(\boldsymbol{\Gamma}\right)_{ab}\cdot\left(\boldsymbol{\Gamma}\right)_{cd}\mathbf{L}\cdot(\mathbf{L}-2\boldsymbol{\mu})-v_{F,\theta }v_{F,\thetasq}\delta\deltaq\left(\boldsymbol{\Gamma}\right)_{ab}\cdot\left(\boldsymbol{\Gamma}\right)_{cd}\right]
\label{trace_group1_nu=2}
\end{aligned}\end{equation}
and
\begin{equation}\begin{aligned}
    \frac{\mathcal{N}^{(2);(F_-,F_-)}_{abcd}}{\mathcal{D}^{(2)}}
     &= -2\frac{\frac{3-d}{d-1}\mathbf{L}\cdot(\mathbf{L}-2\boldsymbol{\mu})-v_{F,\theta }v_{F,\thetasq}\delta\deltaq}{(|\mathbf{L}-2\boldsymbol{\mu}|^2+v_{F,\thetasq}^2\deltaq^2)(|\mathbf{L}|^{2}+v_{F,\theta }^2\delta^2)}\left(\boldsymbol{\Gamma}\right)_{ab}\cdot\left(\boldsymbol{\Gamma}\right)_{cd}.
     \label{appendix:quartic_kernel_nu2}
 \end{aligned}\end{equation}
 
\item $\nu_1=\nu_2=P$

In the pairing channel,
we have
\begin{equation}
    \begin{aligned}
        \frac{ 
        \mathcal{N}^{(2);(P,P)}_{abcd}
        }{\mathcal{D}^{(2)}} = \left(\mathbbm{1}\right)_{ab}
        \mathrm{Tr}\left\{\mathbbm{1}G_0(\mathbf{l})\mathbbm{1}G_0(\mathbf{l+q})\right\}
        \left(\mathbbm{1}\right)_{cd}
        +
        \left(i\Gamma^{\prime}_m\right)_{ab}
        \mathrm{Tr}\left\{\left(i\Gamma^{\prime}_m\right)G_0(\mathbf{l})\left(i\Gamma^{\prime}_{m^{\prime}}\right)G_0(\mathbf{l+q})\right\}
        \left(i\Gamma^{\prime}_{m^{\prime}}\right)_{cd},
        \label{Appendix:Quartic_Kernel_Pairing_Interaction}
    \end{aligned}
\end{equation}
where 
$\boldsymbol{\Gamma}^{\prime} \equiv \left(\gamma_d, ...,\gamma_2\right)$ 
and non-singular terms are dropped.
The numerator of the first term on the right hand side of Eq. (\ref{Appendix:Quartic_Kernel_Pairing_Interaction}) can be explicitly written as
\begin{equation}\begin{aligned}
     &\left(\mathbbm{1}\right)_{ab}
     \mathrm{Tr}\left\{\left(\Gamma_{i} L_{i}+v_{F,\theta }\delta\gamma_{d-1}\right)\left(\Gamma_{j}(L-2\mu)_{j}+v_{F,\thetasq}\deltaq\gamma_{d-1}\right)\right\}
     \left(\mathbbm{1}\right)_{cd}\\
     &= \left(\mathbbm{1}\right)_{ab}\left(\mathbbm{1}\right)_{cd}
     \left\{\mathrm{Tr}\left\{\Gamma_{i}\Gamma_{j}\right\} 
     L_{i}(L-2\mu)_{j}
     +
     \mathrm{Tr}\{\mathbbm{1}\}v_{F,\theta }v_{F,\thetasq}\delta\deltaq
     \right\}.
     \label{appendix:Identity_channel_quartic}
 \end{aligned}\end{equation}
 Eq. (\ref{appendix:Identity_channel_quartic}) then gives
 \begin{equation}\begin{aligned}
    \left(\mathbbm{1}\right)_{ab}
    \mathrm{Tr}\{\mathbbm{1}G_0(\mathbf{l})\mathbbm{1}G_0(\mathbf{l}+\mathbf{q})\}\left(\mathbbm{1}\right)_{cd}
     &= -2\frac{\mathbf{L}\cdot(\mathbf{L}-2\boldsymbol{\mu})+v_{F,\theta }v_{F,\thetasq}\delta\deltaq}{(|\mathbf{L}-2\boldsymbol{\mu}|^2+v_{F,\thetasq}^2\deltaq^2)(|\mathbf{L}|^{2}+v_{F,\theta }^2\delta^2)}\left(\mathbbm{1}\right)_{ab}\left(\mathbbm{1}\right)_{cd}.
     \label{appendix:quartic_kernel_nu3}
 \end{aligned}\end{equation}
The numerator of the second term on the right hand side of Eq. (\ref{Appendix:Quartic_Kernel_Pairing_Interaction}) can be expanded as
\begin{equation}\begin{aligned}
  &\left(i\Gamma^{\prime}_i\right)_{ab}
  \mathrm{Tr}\left\{\left(i\Gamma^{\prime}_i\right)\left(\Gamma_{j} L_{j}+v_{F,\theta }\delta\gamma_{d-1}\right)\left(i\Gamma^{\prime}_k\right)\left(\Gamma_{l}(L-2\mu)_{l}+v_{F,\thetasq}\deltaq\gamma_{d-1}\right)\right\}
   \left(i\Gamma^{\prime}_k\right)_{cd}\\
  &=\left(i\Gamma^{\prime}_i\right)_{ab}
  \left(i\Gamma^{\prime}_k\right)_{cd}
  \left\{-\mathrm{Tr}\left\{\Gamma^{\prime}_i\Gamma_{j}\Gamma^{\prime}_k\Gamma_{l}\right\} 
  L_{j}(L-2\mu)_{l} 
  +
  \mathrm{Tr}\bigl\{\Gamma^{\prime}_i\Gamma^{\prime}_k\bigl\}v_{F,\theta }v_{F,\thetasq}\delta\deltaq
  \right\}
  .
\label{trace_group2_nu=4}
\end{aligned}\end{equation}
It results in
\begin{equation}\begin{aligned}
    \left(i\Gamma^{\prime}_m\right)_{ab}
    \mathrm{Tr}\left\{\left(i\Gamma^{\prime}_m\right)G_0(\mathbf{l})\left(i\Gamma^{\prime}_{m^{\prime}}\right)G_0(\mathbf{l+q})\right\}
    \left(i\Gamma^{\prime}_{m^{\prime}}\right)_{cd}
     &= -2\frac{\mathbf{L}\cdot(\mathbf{L}-2\boldsymbol{\mu})+v_{F,\theta}v_{F,\thetasq}\delta\deltaq}{(|\mathbf{L}-2\boldsymbol{\mu}|^2+v_{F,\thetasq }^2\deltaq^2)(|\mathbf{L}|^{2}+v_{F,\theta }^2\delta^2)}\left(i\boldsymbol{\Gamma}^{\prime}\right)_{ab}\cdot\left(i\boldsymbol{\Gamma}^{\prime}\right)_{cd}.\label{appendix:quartic_kernel_nu4}
 \end{aligned}\end{equation}
From Eqs. (\ref{appendix:quartic_kernel_nu3}) and (\ref{appendix:quartic_kernel_nu4}), Eq. (\ref{Appendix:Quartic_Kernel_Pairing_Interaction}) becomes
\begin{equation}
    \begin{aligned}
        \frac{ \mathcal{N}^{(2);(P,P)}_{abcd}}{\mathcal{D}^{(2)}} = -2\frac{\mathbf{L}\cdot(\mathbf{L}-2\boldsymbol{\mu})+v_{F,\theta}v_{F,\thetasq}\delta\deltaq}{(|\mathbf{L}-2\boldsymbol{\mu}|^2+v_{F,\thetasq }^2\deltaq^2)(|\mathbf{L}|^{2}+v_{F,\theta }^2\delta^2)}
        \left(I^{(P)}_m\right)_{ab}\left(I^{(P)}_m\right)_{cd}.
        \label{appendix:quartic_kernel_pairing_final}
    \end{aligned}
\end{equation}
\end{itemize}
Using Eqs. (\ref{appendix:quartic_kernel_nu1})-(\ref{appendix:quartic_kernel_pairing_final}),
we write the quantum correction for the forward scattering ($\nu = F_{\pm}$) and pairing channel ($\nu = P$) as
 \begin{equation}\begin{aligned}
   \Gamma^{(2);(\nu,s)}_{\mu^{A}}\left(\theta_1,\theta_2,\vec{q};\mu\right) &= -4
    M^{(\nu,s)}_{s_1,s_2}\int \frac{d\mathbf{L}d\delta \KFthetadim  d\theta}{(2\pi)^{d+1}} \boldsymbol{\lambda}^{\left(\nu,s_1\right)}_{
   \theta_1,\theta}\left(\vec{q}\right)\boldsymbol{\lambda}^{\left(\nu,s_2\right)}_{
   \theta,\theta_2}\left(\vec{q}\right)
   K^{(\nu)}_d\left(\mathbf{L},\boldsymbol{\mu},\delta,\theta,\vec{q}\right),
   \label{appendix:(2)_mu1_1}
\end{aligned}\end{equation}
where
\begin{equation}
    \begin{aligned}
K^{(\nu)}_d\left(\mathbf{L},\boldsymbol{\mu},\delta,\theta,\vec{q}\right) = \frac{A^{(\nu)}_1(d)\mathbf{L}\cdot(\mathbf{L}-2\boldsymbol{\mu})+A^{(\nu)}_2(d)v_{F,\theta }v_{F,\thetasq}\delta\deltaq}{(|\mathbf{L}-2\boldsymbol{\mu}|^{2}+v_{F,\thetasq}^2\deltaq^2)(|\mathbf{L}|^{2}+v_{F,\theta }^2\delta^2)}
\label{appendix:general_4fkernel} \end{aligned} \end{equation}
and
\begin{equation}
    \begin{aligned}
        A_1^{(F_+)}(d) = 1,~~ A_2^{(F_+)}(d)  = -1,~~
        A_1^{(F_-)}(d) = \frac{3-d}{d-1},~~ A_2^{(F_-)}(d)  = -1,~~
         A_1^{(P)}(d) = 1,~~ A_2^{(P)}(d)  = 1
        .
        \label{A1A2}
    \end{aligned}
\end{equation}
Here, $\lambdadim^{(\nu,s)}_{
\theta_1,\theta_2}(\vec{q})$ is the dimensionful four-fermion coupling. 
For $q \ll \KFthetadim $,
we can use $v_{F,\thetasq} \approx v_{F,\theta }$
to simplify the expression as
\begin{equation}\begin{aligned}   
 \Gamma^{(2);(\nu,s)}_{\mu^{A}}\left(\theta_1,\theta_2,\vec{q};\mu\right) =-
 4
  M^{(\nu,s)}_{s_1,s_2}\int \frac{d\mathbf{L}d\delta \KFthetadim  d\theta}{(2\pi)^{d+1}} \boldsymbol{\lambda}^{\left(\nu,s_1\right)}_{
   \theta_1,\theta}\left(\vec{q}\right)\boldsymbol{\lambda}^{\left(\nu,s_2\right)}_{
   \theta,\theta_2}\left(\vec{q}\right)
 \frac{A^{(\nu)}_1(d)\mathbf{L}\cdot(\mathbf{L}-2\boldsymbol{\mu})+A^{(\nu)}_2(d)v^2_{F,\theta }\delta\deltaq}{(|\mathbf{L}-2\boldsymbol{\mu}|^{2}+v_{F,\theta }^2\deltaq^2)(|\mathbf{L}|^{2}+v_{F,\theta }^2\delta^2)}.\\
\end{aligned}\end{equation}
We then use the Feynman parametrization to transform the expression into
\begin{equation}\begin{aligned}
    \Gamma^{(2);(\nu,s)}_{\mu^{A}}\left(\theta_1,\theta_2,\vec{q};\mu\right)
    &=
    -
    4
     M^{(\nu,s)}_{s_1,s_2}\int \frac{d\mathbf{L}d\delta \KFthetadim  d\theta}{(2\pi)^{d+1}} \boldsymbol{\lambda}^{\left(\nu,s_1\right)}_{
   \theta_1,\theta}\left(\vec{q}\right)\boldsymbol{\lambda}^{\left(\nu,s_2\right)}_{
   \theta,\theta_2}\left(\vec{q}\right)
    \\
   &\times
   \int_{0}^{1}dx
   \frac{A_1^{(\nu)}(d)\mathbf{L}\cdot(\mathbf{L}-2\boldsymbol{\mu})+A^{(\nu)}_2(d)v^2_{F,\theta }\delta\deltaq}{(x|\mathbf{L}|^{2}+(1-x)|\mathbf{L}-2\boldsymbol{\mu}|^{2}+v_{F,\theta }^2\left(x\delta^2+(1-x)\deltaq^2\right))^{2}}
    \label{appendix:feyman_param_(2)}
\end{aligned}
\end{equation}
and a shift of
$\mathbf{L}\rightarrow\mathbf{L}+2(1-x)\boldsymbol{\mu}$ gives
\begin{equation}\begin{aligned}
    \Gamma^{(2);(\nu,s)}_{\mu^{A}}\left(\theta_1,\theta_2,\vec{q};\mu\right)
    &=
    -
    4
     M^{(\nu,s)}_{s_1,s_2}\int \frac{d\mathbf{L}d\delta \KFthetadim  d\theta}{(2\pi)^{d+1}} \boldsymbol{\lambda}^{\left(\nu,s_1\right)}_{
   \theta_1,\theta}\left(\vec{q}\right)\boldsymbol{\lambda}^{\left(\nu,s_2\right)}_{
   \theta,\theta_2}\left(\vec{q}\right)
   \\
   &\times
    \int_{0}^{1}dx \frac{(A^{(\nu)}_1(d)\left(|\mathbf{L}|^{2}-x(1-x)|2\boldsymbol{\mu}|^{2}\right)+A^{(\nu)}_2(d)v^2_{F,\theta }\delta\deltaq)}{(|\mathbf{L}|^{2}+x(1-x)|2\boldsymbol{\mu}|^{2}+v_{F,\theta }^2\left(x\delta^2+(1-x)\deltaq^2\right))^{2}}.
    \label{appendix:shift_freq_(2)}
\end{aligned}\end{equation}
Completing the square in numerator and denominator 
to perform integration over $\delta$, we obtain
\begin{equation}
    \begin{aligned}
       \Gamma^{(2);(\nu,s)}_{\mu^{A}}\left(\theta_1,\theta_2,\vec{q};\mu\right) =-
        M^{(\nu,s)}_{s_1,s_2}\int_0^1 dx\int \frac{d\mathbf{L} d\theta}{(2\pi)^{d}}\frac{\KFthetadim }{v_{F,\theta }}
       \boldsymbol{\lambda}^{\left(\nu,s_1\right)}_{\theta_1,\theta}\left(\vec{q}\right)\boldsymbol{\lambda}^{\left(\nu,s_2\right)}_{\theta,\theta_2}\left(\vec{q}\right)
       \frac{A^{(\nu)}_+(d)|\mathbf{L}|^{2}-4A^{(\nu)}_-(d)x(1-x)|\boldsymbol{\mu}|^{2}}{\left(|\mathbf{L}|^{2}+x(1-x)\left( \left(L_{\theta}\left(\vec{q}\right)\right)^2+4|\boldsymbol{\mu}|^{2}\right)\right)^{3/2}},
    \end{aligned}
\end{equation}
where $L_{\theta}\left(\vec{q}\right)$ is given by 
\eq{eq:Lqvarphi}
and  
$A^{(\nu)}_\pm(d) = A_1^{(\nu)}(d) \pm A_2^{(\nu)}(d)$.
Finally, the integrations of $\mathbf{L}$ and x gives
\begin{equation}
    \begin{aligned}
       \Gamma^{(2);(\nu,s)}_{\mu^{A}}\left(\theta_1,\theta_2,\vec{q};\mu\right) = 
        M^{(\nu,s)}_{s_1,s_2}
        \bigintss_{-\frac{\pi}{2}}^{\frac{\pi}{2}}\frac{d\theta}{2\pi}\frac{\KFthetadim }{v_{F,\theta }}
       \boldsymbol{\lambda}^{\left(\nu,s_1\right)}_{\theta_1,\theta}\left(\vec{q}\right)\boldsymbol{\lambda}^{\left(\nu,s_2\right)}_{\theta,\theta_2}\left(\vec{q}\right)
       \left[-\frac{1}
       {4(2-d)}
       \frac{A_+^{(\nu)}(d) T_+(d)}{\left( \left(L_{\theta}\left(\vec{q}\right)\right)^2+4\mu^{2}\right)^{\frac{2-d}{2}}}
       +\frac{A_-^{(\nu)}(d) T_-(d)\mu^2}
       {2\left( \left(L_{\theta}\left(\vec{q}\right)\right)^2+4\mu^{2}\right)^{\frac{4-d}{2}}}
       \right], \label{group1_fish_quantum_correction}
    \end{aligned}
\end{equation}
where
\begin{equation}
    \begin{aligned}
        T_+(d)  =
        \frac{
        4
        \Omega_{d-1}\Gamma\left(\frac{d+1}{2}\right)\Gamma\left(\frac{4-d}{2}\right)\Gamma^2\left(\frac{d}{2}\right)}{\Gamma\left(\frac{3}{2}\right)(2\pi)^{d-1}\Gamma(d)},
        ~~&~~ 
        T_-(d)  =  \frac{
        4
        \Omega_{d-1}\Gamma\left(\frac{d-1}{2}\right)\Gamma\left(\frac{4-d}{2}\right)\Gamma^2\left(\frac{d}{2}\right)}{\Gamma\left(\frac{3}{2}\right)(2\pi)^{d-1}\Gamma(d)}.\label{T_hat_fish}
    \end{aligned}
\end{equation}
\begin{figure}[h]
\begin{subfigure}{.4\textwidth}
  \centering
\includegraphics[width=1\linewidth]{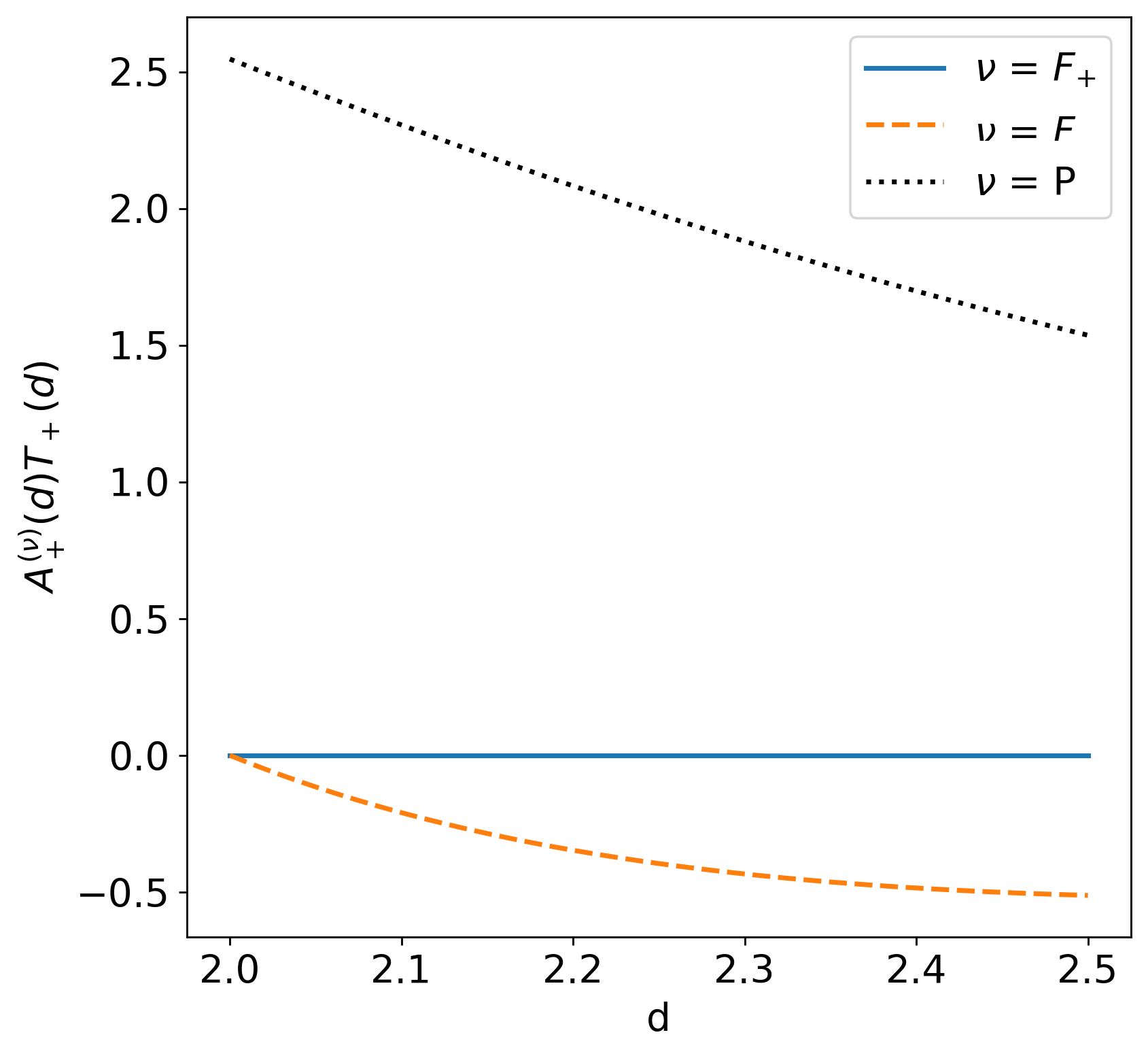}
  \caption{}
  \label{T1hat}
\end{subfigure}%
\begin{subfigure}{.4\textwidth}
\centering
\includegraphics[width=1\linewidth]{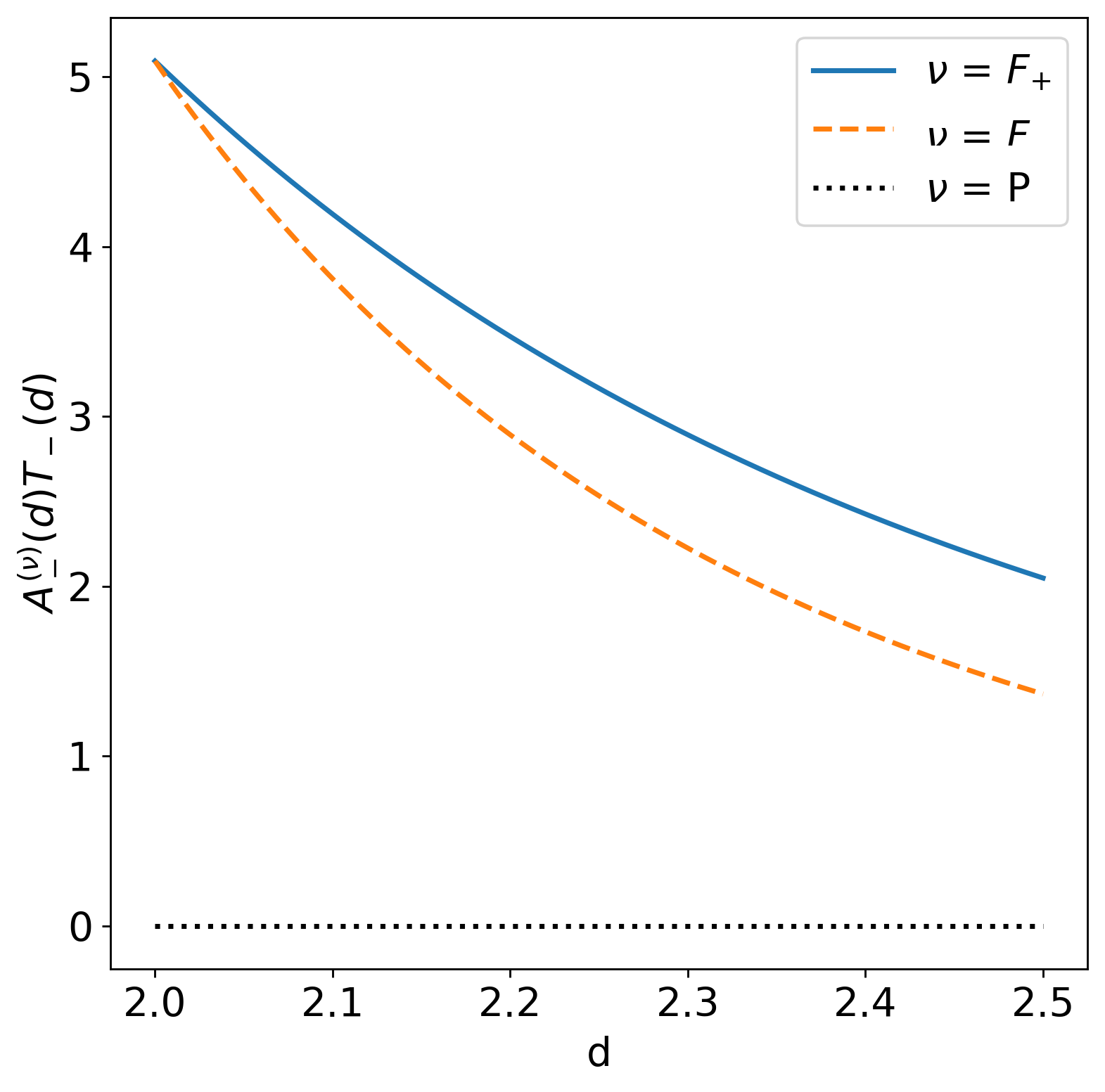}
  \caption{}
  \label{T2hat}
\end{subfigure}
\caption{
$A_+^{(\nu)}(d) T_+(d)$ is the residue of the $1/(d-2)$ pole that arises from the BCS singularity in the vertex correction for the four-fermion coupling.
$A_-^{(\nu)}(d) T_-(d)$ represents a non-singular Fermi liquid correction. 
}
\label{THat}
\end{figure}$A_+^{(\nu)}(d)T_+(d)$ and $A_-^{(\nu)}(d)T_-(d)$ are shown in Figs. 
\ref{T1hat}
and
\ref{T2hat}, respectively.
Similarly, the quantum correction can be evaluated for $\mu^{B}$ 
as
\begin{equation}\begin{aligned}
   \Gamma^{(2);(\nu,s)}_{\mu^{B}}\left(\theta_1,\theta_2,\vec{q};\mu\right) &= -
   4
   M^{(\nu,s)}_{s_1,s_2}\int \frac{d\mathbf{L}d\delta \KFthetadim  d\theta}{(2\pi)^{d+1}} \boldsymbol{\lambda}^{\left(\nu,s_1\right)}_{
   \theta_1,\theta}\left(\vec{q}\right)\boldsymbol{\lambda}^{\left(\nu,s_2\right)}_{
   \theta,\theta_2}\left(\vec{q}\right)
   K^{(\nu)}_d\left(\mathbf{L},-\boldsymbol{\mu},\delta,\theta,\vec{q}\right)
\end{aligned}\end{equation}
and we find that\footnote{This can be realized by replacing $\boldsymbol{\mu} \rightarrow-\boldsymbol{\mu}$ in Eq. (\ref{appendix:feyman_param_(2)}) and shifting  $\mathbf{L}\rightarrow\mathbf{L}-2(1-x)\boldsymbol{\mu}$ to get an equation same as Eq. (\ref{appendix:shift_freq_(2)}).}
 $        \Gamma^{(2);(\nu,s)}_{\mu^{A}}\left(\theta_1,\theta_2,\vec{q};\mu\right) = \Gamma^{(2);(\nu,s)}_{\mu^{B}}\left(\theta_1,\theta_2,\vec{q};\mu\right)$.
From Eqs. (\ref{group1_fish_quantum_correction})
and  (\ref{appendix:4f_counter_term}), the counter term reads
\begin{equation}
    \begin{aligned}
        \Gamma^{CT;(2);(\nu,s)}_{\mu,\theta_1,\theta_2}\left(\vec{q}\right) = - M^{(\nu,s)}_{s_1,s_2}
        \bigintss_{-\frac{\pi}{2}}^{\frac{\pi}{2}}\frac{d\theta}{2\pi}\frac{\KFthetadim }{v_{F,\theta }}
       \boldsymbol{\lambda}^{\left(\nu,s_1\right)}_{\theta_1,\theta}\left(\vec{q}\right)\boldsymbol{\lambda}^{\left(\nu,s_2\right)}_{\theta,\theta_2}\left(\vec{q}\right)
       \left[-\frac{1}{4(2-d)}\frac{A_+^{(\nu)}(d) T_+(d)}{\left( \left(L_{\theta}\left(\vec{q}\right)\right)^2+4\mu^{2}\right)^{\frac{2-d}{2}}}
       +\frac{A_-^{(\nu)}(d) T_-(d)\mu^2}{2\left( \left(L_{\theta}\left(\vec{q}\right)\right)^2+4\mu^{2}\right)^{\frac{4-d}{2}}}\right].
    \end{aligned}
\end{equation}
The contribution of the counter term to the beta function is given by the derivative of the counter term with respect to $\log \mu$ with the bare coupling fixed.
To the leading order,
the bare coupling is 
$\boldsymbol{\lambda}^{(\nu,s)}_{ \theta_1,\theta_2}$,
and the derivative is given by
\begin{equation}
    \begin{aligned}
        \mu^{d-1}\frac{d~\Gamma^{CT;(2);(\nu,s)}_{\mu,\theta_1,\theta_2}\left(\vec{q}\right)}{d~\mathrm{ln}~\mu} = - M^{(\nu,s)}_{s_1,s_2}
        \bigintss_{-\frac{\pi}{2}}^{\frac{\pi}{2}}\frac{d\theta}{2\pi\mu}\frac{\KFthetadim }{v_{F,\theta }}
        \lambda^{\left(\nu,s_1\right)}_{ \theta_1,\theta}\left(\vec{q}\right)\lambda^{\left(\nu,s_2\right)}_{ \theta,\theta_2}\left(\vec{q}\right)\left[\frac{A_+^{(\nu)}(d) T_+(d)}{\left(\left(\mathscr{L}_{\mu,\theta}\left(\vec{q}\right)\right)^2+4\right)^{\frac{4-d}{2}}}
        \right.\\\left.
        +\frac{A_-^{(\nu)}(d) T_-(d)\left(\left(\mathscr{L}_{\mu,\theta}\left(\vec{q}\right)\right)^2+2(d-2)\right)}{\left(\left(\mathscr{L}_{\mu,\theta}\left(\vec{q}\right)\right)^2+4\right)^{\frac{6-d}{2}}}\right],
        \label{(2)_betafunctional_contribution}
    \end{aligned}
\end{equation}
where
$        \mathscr{L}_{\mu,\theta}\left(\vec{q}\right) = \frac{L_{\theta}\left(\vec{q}\right)}{\mu}$.
\subsubsection{\texorpdfstring{$\lambda^1$}{Lg} order}
\label{lambda_1_order}

Now, we derive the counter term to the four-fermion coupling that is linear in the four-fermion coupling.
The quantum correction from Fig. \ref{vertex3_1} evaluated at $\mu^{A}$ can be written as
\begin{equation}\begin{aligned}
  \Gamma^{(1)}_{\{j_i\};abcd}
       \left(\mathbf{k}_1^{A}, \mathbf{k}_2^{A}, \mathbf{k}_3^{A}, \mathbf{k}_4^{A}\right)
 &=
  -\frac{3\times2}{3!}
  \frac{1}{N}
  T^{(\nu,s)}_{\left(\begin{smallmatrix}     j_1  & j_2      \\ j_4     & j_3       \end{smallmatrix}\right)}
  \int^{\prime} d_{f}^{d+1}\mathbf{l}~ 
  \left\{\left(-\lambdadim^{(\nu,s)}_{\theta_1,\theta}\left(\vec{q}\right) \right)\left(-i\edim_{\Theta\left(\theta_2,\vec{q}\right),\thetasq}\right)\left(-i\edim_{\theta,\theta_2}\right)
  D_1\left(\mathbf{l}-\mathbf{k}_3^{A}\right)\right.\\&\left.
  \times\left(I^{(\nu)}_m\right)_{ab}\left(\gamma_{d-1}G_0(\mathbf{l+q})I^{(\nu)}_mG_0(\mathbf{l})\gamma_{d-1}\right)_{cd}
  \right\}.
\label{quantum_correction_(1)_expression}
\end{aligned}\end{equation}
Here, momenta $\mathbf{k}_i^{A}$'s  are defined in Eq. (\ref{appendix_mu2}),  $\mathbf{l} = \left(\mathbf{L},\delta,\theta\right)$ and $\mathbf{l+q} = (\mathbf{L}-2\boldsymbol{\mu},\deltaq,\thetasq)$. 
We single out the integrand that comes from the fermion propagators as
\begin{equation}
    \begin{aligned}
        \left(I^{(\nu)}_m\right)_{ab}\left(\gamma_{d-1}G_0(\mathbf{l+q})I^{(\nu)}_m
        G_0(\mathbf{l})
        \gamma_{d-1}\right)_{cd} = \frac{\mathcal{N}^{(1);(\nu)}_{abcd}}{\mathcal{D}^{(1)}}, \label{vertex_(1)}
    \end{aligned}
\end{equation}
where
\bqa
 D^{(1)} &=& i^2\left(|\mathbf{L}-2\boldsymbol{\mu}|^2+v_{F,\thetasq}^2\deltaq^2\right)\left(|\mathbf{L}|^2+v_{F,\theta }^2\delta^2\right),
  \label{D(1)} \nn
 \mathcal{N}^{(1);(F_+)}_{abcd} 
        &=&
        \frac{1}{d-1}
        \left(i\gamma_{d-1}\right)_{ab}
        (\gamma_{d-1}\Gamma_{i}
        \left(i\gamma_{d-1}\right)\Gamma_{i}\gamma_{d-1})_{cd}
        \mathbf{L}\cdot(\mathbf{L}-2\boldsymbol{\mu})+\left(i\gamma_{d-1}\right)_{ab}\left(i\gamma_{d-1}\right)_{cd}v_{F,\theta }v_{F,\thetasq}\delta\deltaq, \nn
 \mathcal{N}^{(1);(F_-)}_{abcd} &=& \frac{1}{d-1}
     \left( \Gamma_{j} \right)_{ab}
     \left(
     \gamma_{d-1}\Gamma_{i}\Gamma_{j}\Gamma_{i}\gamma_{d-1}\right)_{cd}
     \mathbf{L}\cdot(\mathbf{L}-2\boldsymbol{\mu})+\left(\mathbf{\Gamma}\right)_{ab}\cdot\left(\mathbf{\Gamma}\right)_{cd}v_{F,\theta }v_{F,\thetasq}\delta\deltaq, \nn
 \mathcal{N}^{(1);(P)}_{abcd} 
 &=& \frac{1}{d-1}
        \left(\mathbbm{1}\right)_{ab}
        (\gamma_{d-1}\Gamma_{i}\Gamma_{i}\gamma_{d-1})_{cd}
        \mathbf{L}\cdot(\mathbf{L}-2\boldsymbol{\mu})+\left(\mathbbm{1}\right)_{ab}\left(\mathbbm{1}\right)_{cd}v_{F,\theta}v_{F,\thetasq}\delta\deltaq \nn
        &&
        +\frac{1}{d-1}
\left(i\Gamma_j^{\prime}\right)_{ab}
     \left(\gamma_{d-1}\Gamma_{i}i\Gamma_j^{\prime}\Gamma_{i}\gamma_{d-1}\right)_{cd}
     \mathbf{L}\cdot(\mathbf{L}-2\boldsymbol{\mu})+\left(i\mathbf{\Gamma}^{\prime}\right)_{ab}\cdot\left(i\mathbf{\Gamma}^{\prime}\right)_{cd}v_{F,\theta}v_{F,\thetasq}\delta\deltaq.
\eqa
We can isolate the singular part of the quantum correction using Eq. (\ref{appendix:non_vanishing_freqprod}).
The singular part of Eq. (\ref{vertex_(1)}) becomes
\bqa
 \frac{\mathcal{N}^{(1);(F_+)}_{abcd}}{\mathcal{D}^{(1)}}  &=& \frac{\mathbf{L}\cdot(\mathbf{L}-2\boldsymbol{\mu})-v_{F,\theta }v_{F,\thetasq}\delta\deltaq}{(|\mathbf{L}-2\boldsymbol{\mu}|^{2}+v_{F,\thetasq}^2\deltaq^2)(|\mathbf{L}|^{2}+v_{F,\theta }^2\delta^2)}\left(i\gamma_{d-1}\right)_{ab}\left(i\gamma_{d-1}\right)_{cd},  
          \label{appendix:(2)_kernel_nu1} \\
 \frac{\mathcal{N}^{(1);(F_-)}_{abcd}}{\mathcal{D}^{(1)}}  &=& \frac{\frac{3-d}{d-1}\mathbf{L}\cdot(\mathbf{L}-2\boldsymbol{\mu})-v_{F,\theta }v_{F,\thetasq}\delta\deltaq}{(|\mathbf{L}-2\boldsymbol{\mu}|^{2}+v_{F,\thetasq}^2\deltaq^2)(|\mathbf{L}|^{2}+v_{F,\theta }^2\delta^2)}\left(\mathbf{\Gamma}\right)_{ab}\cdot\left(\mathbf{\Gamma}\right)_{cd},
          \label{appendix:(2)_kernel_nu3} \\
 \frac{\mathcal{N}^{(1);(P)}_{abcd}}{\mathcal{D}^{(1)}}  &=& -\frac{\mathbf{L}\cdot(\mathbf{L}-2\boldsymbol{\mu})+v_{F,\theta}v_{F,\thetasq}\delta\deltaq}{(|\mathbf{L}-2\boldsymbol{\mu}|^{2}+v_{F,\thetasq }^2\deltaq^2)(|\mathbf{L}|^{2}+v_{F,\theta }^2\delta^2)}
\left(I^{(P)}_m\right)_{ab}\left(I^{(P)}_m\right)_{cd}.
\label{appendix:(2)_kernel_nu4}
\eqa
From Eqs. (\ref{appendix:(2)_kernel_nu1})-(\ref{appendix:(2)_kernel_nu4}), the quantum correction is written as
\begin{equation}
    \begin{aligned}
        &\Gamma^{(1);(\nu,s)}_{\mu^{A}}\left(\theta_1,\theta_2,\vec{q};\mu\right) = 
        & -\frac{\snu}{N}\int \frac{d\mathbf{L}d\delta \KFthetadim  d\theta}{(2\pi)^{d+1}}\boldsymbol{\lambda}^{(\nu,s)}_{
        \theta_1,\theta}\left(\vec{q}\right)
        \edim_{\Theta\left(\theta_2,\vec{q}\right),\thetasq}~\edim_{\theta,\theta_2}
         D_1\left(\mathbf{l}-\mathbf{k}_3^{A}\right)K^{(\nu)}_d\left(\mathbf{L},\boldsymbol{\mu},\delta,\theta,\vec{q}\right),
        \label{quantum_correction_(1)_explicit}
    \end{aligned}
\end{equation}
where $K^{(\nu)}_d\left(\mathbf{L},\boldsymbol{\mu},\delta,\theta,\vec{q}\right)$ is defined in Eq. (\ref{appendix:general_4fkernel}) and 
\begin{equation}
    \begin{aligned}
        \snu = 
        \begin{cases}
            1~~&\nu = F_{\pm}\\
            -1~~&\nu = P
        \end{cases}.
        \label{appendix:s_nu}
    \end{aligned}
\end{equation}
Like the fermion self-energy, an IR singularity arises below $d_c$.

In the low energy limit, 
the momentum transfer $q$ is at most $\sqrt{\KFthetadim \mu}$.
If the couplings vary slowly within an angular width of $\sqrt{\frac{\mu}{\KFthetadim }}$, 
which will turn out to be the case,
we can use the adiabatic approximation\cite{BORGES2023169221} to make replacements,
\begin{equation}
    \begin{aligned}
        e_{\Theta\left(\theta_2,\vec{q}\right),\thetasq}\approx e_{\theta_2,\theta},~~ 
        v_{F,\thetasq} \approx v_{F,\theta}
        \label{appendix:slow_varying_approx}
    \end{aligned}
\end{equation}
within the loop.
Similarly the quantum correction at $\mu^{B}$ can be obtained by replacing $\boldsymbol{\mu}$ with $-\boldsymbol{\mu}$ in $K_d^{(\nu)}$ in Eq. (\ref{quantum_correction_(1)_explicit}).

In the following, 
we first obtain the formal expression for the counter term written as an angular integration,
which enters in the beta functional in \eq{eq:betalambda_convolution}.
Later, we explicitly compute the quantum correction.

\begin{center}
    \MakeUppercase{\romannumeral 1}. 
    Linear mixing matrix for the vertex correction
\end{center}

Using Eq. (\ref{appendix:4f_counter_term}) and Eq. (\ref{quantum_correction_(1)_explicit}), 
we write down a counter term that is analytic in momentum space for $\mu \neq 0$ and 
cancels the IR divergence of the quantum correction, 
\begin{equation}
    \begin{aligned}
        \Gamma^{CT;(1);(\nu,s)}_{\mu,\theta_1,\theta_2}\left(\vec{q}\right) = \frac{\snu}{2N}\int \frac{d\mathbf{L}d\delta \KFthetadim  d\theta}{(2\pi)^{d+1}}\boldsymbol{\lambda}^{(\nu,s)}_{
        \theta_1,\theta}\left(\vec{q}\right)
        \edim^2_{\theta,\theta_2}
        D_{1;\mu}\left(\mathbf{L}-\boldsymbol{\mu},\theta,\theta_2 \right)\left(K^{(\nu)}_d\left(\mathbf{L},\boldsymbol{\mu},\delta,\theta,\vec{q}\right)+K^{(\nu)}_d\left(\mathbf{L},-\boldsymbol{\mu},\delta,\theta,\vec{q}\right)\right),
        \label{eq:C76}
    \end{aligned}
\end{equation}
The derivative of this counter term with respect to $\log \mu$,
which determines the beta functional, becomes
\begin{equation}
    \begin{aligned}
        \mu^{d-1}\frac{d~\Gamma^{CT;(1);(\nu,s)}_{\mu,\theta_1,\theta_2}\left(\vec{q}\right)}{d~\log\mu} &= -\frac{\snu}{4}\int \frac{d\theta}{2\pi\mu}\frac{\KFthetadim }{v_{F,\theta }}\lambda^{(\nu,s)}_{
        \theta_1,\theta}\left(\vec{q}\right)\left[\frac{A_+^{(\nu)}(d) T_+(d)}{\left(\left(\mathscr{L}_{\mu,\theta}\left(\vec{q}\right)\right)^2+4\right)^{\frac{4-d}{2}}}h^{\dagger}_{+;d;\theta,\theta_2}\left(\vec{q};\mu\right)
        \right.\\&\left.+\frac{A_-^{(\nu)}(d) T_-(d)\left(\left(\mathscr{L}_{\mu,\theta}\left(\vec{q}\right)\right)^2+2(d-2)\right)}{\left(\left(\mathscr{L}_{\mu,\theta}\left(\vec{q}\right)\right)^2+4\right)^{\frac{6-d}{2}}}h^{\dagger}_{-;d;\theta,\theta_2}\left(\vec{q};\mu\right)\right].
        \label{appenidx:(1)_general_beta_functional}
    \end{aligned}
\end{equation}
Here, a power of $\mu$ is multiplied to make expression dimensionless.
Similarly, the contribution of Fig. \ref{vertex3_2} can be written as
\begin{equation}
    \begin{aligned}
        \mu^{d-1}\frac{d~\Gamma^{CT;\left(1^{\prime}\right);(\nu,s)}_{\mu,\theta_1,\theta_2}\left(\vec{q}\right)}{d~\log\mu} &= -\frac{\snu}{4}\int \frac{d\theta}{2\pi\mu}\frac{\KFthetadim }{v_{F,\theta }}\left[h_{+;d;\theta_1,\theta}\left(\vec{q};\mu\right)\frac{A_+^{(\nu)}(d) T_+(d)}{\left(\left(\mathscr{L}_{\mu,\theta}\left(\vec{q}\right)\right)^2+4\right)^{\frac{4-d}{2}}}
        \right.\\&\left.+h_{-;d;\theta_1,\theta}\left(\vec{q};\mu\right)\frac{A_-^{(\nu)}(d) T_-(d)\left(\left(\mathscr{L}_{\mu,\theta}\left(\vec{q}\right)\right)^2+2(d-2)\right)}{\left(\left(\mathscr{L}_{\mu,\theta}\left(\vec{q}\right)\right)^2+4\right)^{\frac{6-d}{2}}}\right]\lambda^{(\nu,s)}_{
        \theta,\theta_2}\left(\vec{q}\right).
        \label{appenidx:(1')_general_beta_functional}
    \end{aligned}
\end{equation}
The expressions for $h_{+;d;\theta_1,\theta_2}\left(\vec{q};\mu\right)$ and $h_{-;d;\theta_1,\theta_2}\left(\vec{q};\mu\right)$ are shown in Eqs. (\ref{main:h1_def}) and (\ref{main:h2_def}), respectively,
and
$L_{\theta }\left(\vec{q}\right)$ is defined in Eq. (\ref{eq:Lqvarphi}).

\begin{center}
\MakeUppercase{\romannumeral 2}. 
The explicit computation of the vertex correction
\end{center}

In this subsection, we employ the adiabatic procedure to evaluate the vertex correction explicitly.
The singular part of 
\eq{eq:C76}
can be evaluated through 
$\boldsymbol{e}_{\theta_2,\theta} \rightarrow   \boldsymbol{e}_{\theta_2,\theta_2},~\boldsymbol{\lambda}^{(\nu,s)}_{\theta_1,\theta} \rightarrow \boldsymbol{\lambda}^{(\nu,s)}_{\theta_1,\theta_2},~v_{F,\theta } \rightarrow v_{F,\theta_2},~\KFthetadim  \rightarrow\mathbf{K}_{F,\theta_2}$, $\boldsymbol{f}_{d,\vartheta^{-1}\left( 
\frac{\theta+\theta_2}{2}
\right)}\rightarrow \boldsymbol{f}_{d,\vartheta^{-1}\left( 
\theta_2
\right)}$
and $\mathscr{F}_{\varphi,\theta}\rightarrow\mathscr{F}_{\varphi,\theta_2},~~\mathscr{G}_{\varphi,\theta}\rightarrow\mathscr{G}_{\varphi,\theta_2}$
for the loop variable $\theta$ because the integrand exhibits an IR singularity 
only at  $\theta = \theta_2$.
This allows us write the singular part of the counter term as
\begin{equation}
    \begin{aligned}
       \Gamma^{CT;(1);(\nu,s)}_{\mu,\theta_1,\theta_2}\left(\vec{q}\right) =  \frac{\snu}{3\sqrt{3}}\frac{1}{\beta_d^{\frac{1}{3}}}\boldsymbol{\lambda}^{(\nu,s)}_{\theta_1,\theta_2}(\vec{q})\boldsymbol{g}_{\theta_2}
       I^{(\nu)}_{\mu,\theta_2}\left(\vec{q};d\right) 
       ,
       \label{appendix:(1)_adiabatic_counter_term_theta_integrated}
    \end{aligned}
\end{equation}
where
\begin{equation}
    \begin{aligned}
         I^{(\nu)}_{\mu,\theta_2}\left(\vec{q};d\right) &=\int \frac{d\mathbf{L}dE }{(2\pi)^{d}}\frac{1}{|\mathbf{L}-\boldsymbol{\mu}|^{\frac{d-1}{3}}}
        \left\{
        \frac{A^{(\nu)}_1(d)\mathbf{L}\cdot(\mathbf{L}-2\boldsymbol{\mu})+A^{(\nu)}_2(d)E\left(E+L_{\theta_2}(\vec{q})\right)}{\left(|\mathbf{L}-2\boldsymbol{\mu}|^{2}+\left(E+L_{\theta_2}(\vec{q})\right)^2\right)\left(|\mathbf{L}|^{2}+E^2\right)} \right. \\ & \left.
        \hspace{3.5cm}
        + \frac{A^{(\nu)}_1(d)\mathbf{L}\cdot(\mathbf{L}+2\boldsymbol{\mu})+A^{(\nu)}_2(d)E\left(E+L_{\theta_2}(\vec{q})\right)}{\left(|\mathbf{L}+2\boldsymbol{\mu}|^{2}+\left(E+L_{\theta_2}(\vec{q})\right)^2\right)\left(|\mathbf{L}|^{2}+E^2\right)}
        \right\}.
        \label{eq:C80}
   \end{aligned}
\end{equation}
Here, the integration over $\theta$ has been performed and
$ E = v_{F,\theta_2}\delta$.
Using the Feynman parametrization,
we rewrite \eq{eq:C80} as
\begin{equation}
    \begin{aligned}
         I^{(\nu)}_{\mu,\theta_2}\left(\vec{q};d\right)  &=
         \int_0^1 dx\int \frac{d\mathbf{L}dE}{(2\pi)^{d}}\frac{1}{|\mathbf{L}-\boldsymbol{\mu}|^{\frac{d-1}{3}}}
        \left\{
        \frac{A^{(\nu)}_1(d)\mathbf{L}\cdot(\mathbf{L}-2\boldsymbol{\mu})+A^{(\nu)}_2(d)E\left(E+L_{\theta_2}(\vec{q})\right)}{\left[(1-x)\left(|\mathbf{L}-2\boldsymbol{\mu}|^{2}+\left(E+L_{\theta_2}(\vec{q})\right)^2\right)+x\left(|\mathbf{L}|^{2}+E^2\right)\right]^2}
        \right.\\&\left.+
        \frac{A^{(\nu)}_1(d)\mathbf{L}\cdot(\mathbf{L}+2\boldsymbol{\mu})+A^{(\nu)}_2(d)E\left(E+L_{\theta_2}(\vec{q})\right)}{\left[x\left(|\mathbf{L}+2\boldsymbol{\mu}|^{2}+\left(E+L_{\theta_2}(\vec{q})\right)^2\right)+(1-x)\left(|\mathbf{L}|^{2}+E^2\right)\right]^2}\right\}.
    \end{aligned}
\end{equation}
The integral over $E$ leads to
\begin{equation}
    \begin{aligned}
          I^{(\nu)}_{\mu,\theta_2}\left(\vec{q};d\right) &= \frac{1}{4}\int_0^1 dx \int \frac{d\mathbf{L}}{(2\pi)^{d-1}}\frac{1}{|\mathbf{L}-\boldsymbol{\mu}|^{\frac{d-1}{3}}}
          \left\{\frac{A^{(\nu)}_1(d)\mathbf{L}\cdot(\mathbf{L}-2\boldsymbol{\mu})+A^{(\nu)}_2(d)\left(x|\mathbf{L}|^2+(1-x)|\mathbf{L}-2\boldsymbol{\mu}|^2\right)}{\left[x|\mathbf{L}|^2+(1-x)|\mathbf{L}-2\boldsymbol{\mu}|^2+x(1-x) \left(L_{\theta_2}\left(\vec{q}\right)\right)^2\right]^{\frac{3}{2}}}\right.\\&\left.+\frac{A^{(\nu)}_1(d)\mathbf{L}\cdot(\mathbf{L}+2\boldsymbol{\mu})+A^{(\nu)}_2(d)\left(x|\mathbf{L}+2\boldsymbol{\mu}|^2+(1-x)|\mathbf{L}|^2\right)}{\left[x|\mathbf{L}+2\boldsymbol{\mu}|^2+(1-x)|\mathbf{L}|^2+x(1-x) \left(L_{\theta_2}\left(\vec{q}\right)\right)^2\right]^{\frac{3}{2}}}\right\}. \label{I_mu_Q}
    \end{aligned}
\end{equation}
With another Feynman parameterization,
Eq. (\ref{I_mu_Q}) can be written as
\begin{equation}
    \begin{aligned}
        I^{(\nu)}_{\mu,\theta_2}\left(\vec{q};d\right)  =  \frac{1}{4}\frac{\Gamma(\frac{d+8}{6})}{\Gamma(\frac{d-1}{6})\Gamma(\frac{3}{2})}{}\int_{x,y}\mathcal{I}^{(\nu)}_{\mu,\theta_2}\left(x,y;\vec{q};d\right),
        \label{appendix:(1)_intermediate1}
    \end{aligned}
\end{equation}
where $\int_{x,y} \equiv \int_0^1\int_0^1 dx~dy~ y^{\frac{d-7}{6}}(1-y)^{\frac{1}{2}}$ and
\begin{equation}
    \begin{aligned}
        \mathcal{I}^{(\nu)}_{\mu,\theta_2}\left(x,y;\vec{q};d\right)&=\int\frac{d\mathbf{L}}{(2\pi)^{d-1}}
         \left\{\frac{A^{(\nu)}_1(d)\mathbf{L}\cdot(\mathbf{L}-2\boldsymbol{\mu})+A^{(\nu)}_2(d)\left(x|\mathbf{L}|^2+(1-x)|\mathbf{L}-2\boldsymbol{\mu}|^2\right)}{\left[y|\mathbf{L}-\boldsymbol{\mu}|^2+(1-y)\left(x(|\mathbf{L}|^2+(1-x)|\mathbf{L}-2\boldsymbol{\mu}|^2+x(1-x) \left(L_{\theta_2}\left(\vec{q}\right)\right)^2\right)\right]^{\frac{d+8}{6}}}
         \right.\\&\left.+
         \frac{A^{(\nu)}_1(d)\mathbf{L}\cdot(\mathbf{L}+2\boldsymbol{\mu})+A^{(\nu)}_2(d)\left(x|\mathbf{L}+2\boldsymbol{\mu}|^2+(1-x)|\mathbf{L}|^2\right)}{\left[y|\mathbf{L}-\boldsymbol{\mu}|^2+(1-y)\left(x(|\mathbf{L}+2\boldsymbol{\mu}|^2+(1-x)|\mathbf{L}|^2+x(1-x) \left(L_{\theta_2}\left(\vec{q}\right)\right)^2\right)\right]^{\frac{d+8}{6}}}
         \right\}.
         \label{appendix:(1)_mu1_res_doublefeynman}
    \end{aligned}
\end{equation}
%
Shifting
$\mathbf{L}\rightarrow \mathbf{L}-(y+2x(1-y)-2)\boldsymbol{\mu}$ and  $\mathbf{L}\rightarrow \mathbf{L}+(y-2x(1-y))\boldsymbol{\mu}$ for the first and second terms in brackets, respectively, 
followed by the integration over $\mathbf{L}$ gives
\begin{equation}
    \begin{aligned}
        \mathcal{I}^{(\nu)}_{\mu,\theta_2}\left(x,y;\vec{q};d\right) = \frac{\Omega_{d-1}}{2(2\pi)^{d-1}\Gamma\left(\frac{d+8}{6}\right)}\left(\mathscr{H}_{\mu,\theta_2,\vec{q}}^{(\nu)}(x,y,d)+\mathscr{J}_{\mu,\theta_2,\vec{q}}^{(\nu)}(x,y,d)+\tilde{\mathscr{H}}_{\mu,\theta_2,\vec{q}}^{(\nu)}(x,y,d)+\tilde{\mathscr{J}}_{\mu,\theta_2,\vec{q}}^{(\nu)}(x,y,d)\right),
         \label{appendix:(1)_intermediate2}
    \end{aligned}
\end{equation}
where
\begin{equation}
    \begin{aligned}
   \mathscr{H}_{\mu,\theta,\vec{q}}^{(\nu)}(x,y,d)&=
   \frac{3A^{(\nu)}_+(d)\Gamma\left(\frac{d+1}{2}\right)\Gamma\left(\frac{11-2d}{6}\right)}{\left(a_2(x,y)\mu^2+x(1-x)(1-y)\left(L_{\theta}\left(\vec{q}\right)\right)^2\right)^{\frac{5-2d}{6}}}\frac{1}{\epsilon},\\
   \mathscr{J}_{\mu,\theta,\vec{q}}^{(\nu)}(x,y,d) &= \frac{a_1^{(\nu)}(x,y,d)\Gamma\left(\frac{d-1}{2}\right)\Gamma\left(\frac{11-2d}{6}\right)\mu^2}{\left(a_2(x,y)\mu^2+x(1-x)(1-y)\left(L_{\theta}\left(\vec{q}\right)\right)^2\right)^{\frac{11-2d}{6}}},\\
   \tilde{\mathscr{H}}_{\mu,\theta_2,\vec{q}}^{(\nu)}(x,y,d) &= \mathscr{H}_{\mu,\theta_2,\vec{q}}^{(\nu)}(x,y,d)\left(a_2(x,y)\rightarrow a_4(x,y)\right),\\
   \tilde{\mathscr{J}}_{\mu,\theta_2,\vec{q}}^{(\nu)}(x,y,d) &= \mathscr{J}_{\mu,\theta_2,\vec{q}}^{(\nu)}(x,y,d)\left(a_2(x,y)\rightarrow a_4(x,y),~ a_1^{(\nu)}(x,y,d)\rightarrow a_3^{(\nu)}(x,y,d)\right).
        \label{H_J_Vertex}
    \end{aligned}
\end{equation}
and
\begin{equation}\begin{aligned}
    a^{(\nu)}_{1}(x,y,d) &= A^{(\nu)}_1(d)(y+2x(1-y))((y+2x(1-y))-2)\\&~~+A^{(\nu)}_2(d)\left(x(2-(y+2x(1-y)))^2+(1-x)(y+2x(1-y))^2\right),\\
    a_{2}(x,y) &= (2x-1)^{2}(-y^{2}+2y-1)-y+1,\\
     a^{(\nu)}_{3}(x,y,d) &= A^{(\nu)}_1(d)(-y+2x(1-y))((-y+2x(1-y))-2)\\&~~+A^{(\nu)}_2(d)\left(x(2-(-y+2x(1-y)))^2+(1-x)(-y+2x(1-y))^2\right),\\
    a_{4}(x,y) &= -(2xy-2x+y+1)^2+3y+1.
\end{aligned}\end{equation}
From Eqs. (\ref{appendix:(1)_adiabatic_counter_term_theta_integrated}), (\ref{appendix:(1)_intermediate1}) and (\ref{appendix:(1)_intermediate2}), 
the counter term is obtained to be
\begin{equation}
    \begin{aligned}
        \Gamma^{CT;(1);(\nu,s)}_{\mu,\theta_1,\theta_2}\left(\vec{q}\right) =
        \boldsymbol{g}_{\theta_2}B^{d;(\nu)}_{\mu,\theta_2}\left(\vec{q}\right)\boldsymbol{\lambda}^{(\nu,s)}_{\theta_1,\theta_2}(\vec{q}), 
    \end{aligned}
\end{equation}
where
\begin{equation}
    \begin{aligned}
        B^{ d;(\nu)}_{\mu,\theta}\left(\vec{q}\right) = \frac{\snu}{24\sqrt{3}}\frac{\Omega_{d-1}}{(2\pi)^{d-1}\beta_d^{\frac{1}{3}}\Gamma\left(\frac{d-1}{6}\right)\Gamma\left(\frac{3}{2}\right)}\int_{x,y}\left(\mathscr{H}_{\mu,\theta,\vec{q}}^{(\nu)}(x,y,d)+\mathscr{J}_{\mu,\theta,\vec{q}}^{(\nu)}(x,y,d)+\tilde{\mathscr{H}}_{\mu,\theta,\vec{q}}^{(\nu)}(x,y,d)+\tilde{\mathscr{J}}_{\mu,\theta,\vec{q}}^{(\nu)}(x,y,d)\right).
        \label{appendix:B_d_theta}
    \end{aligned}
\end{equation}
The derivative of the counter term with respect to $\log \mu$ becomes 
\begin{equation}
    \begin{aligned}
         \frac{d~\Gamma^{CT;(1);(\nu,s)}_{\mu,\theta_1,\theta_2}\left(\vec{q}\right)}{d~\mathrm{ln}~\mu} =
         \boldsymbol{\lambda}^{(\nu,s)}_{\theta_1,\theta_2}\left(\vec{q}\right)
         \boldsymbol{g}_{\theta_2}\left( \mathscr{Y}^{d;\left(\nu\right)}_{\mu,\theta_2}\left(\vec{q}\right)+ \mathscr{W}^{d;\left(\nu\right)}_{\mu,\theta_2}\left(\vec{q}\right)\right),
         \label{(1)_counterterm_logmu}
    \end{aligned}
\end{equation}
where
\begin{equation}
    \begin{aligned}
        \mathscr{Y}^{d;\left(\nu\right)}_{\mu,\theta}\left(\vec{q}\right) =-\snu\Xi_d A^{(\nu)}_+(d)\Gamma\left(\frac{d+1}{2}\right)
        \bigints_{x,y}
        \left[
        \frac{a_2(x,y)\mu^2}{\left(a_2(x,y)\mu^2+x(1-x)(1-y)\left(L_{\theta}\left(\vec{q}\right)\right)^2\right)^{\frac{11-2d}{6}}}
        +\left(a_2(x,y)\rightarrow a_4(x,y)\right)
        \right],
        \label{Y_2b}
    \end{aligned}
\end{equation}
\begin{equation}
    \begin{aligned}
       \mathscr{W}^{d;\left(\nu\right)}_{\mu,\theta}\left(\vec{q}\right) &=
       \snu\Xi_d\Gamma\left(\frac{d-1}{2}\right)
       \bigints_{x,y}
       \Bigg[
       \frac{a_1^{(\nu)}(x,y,d)\left(\frac{2d-5}{6}a_2(x,y)\mu^2+x(1-x)(1-y)\left(L_{\theta}\left(\vec{q}\right)\right)^2\right)\mu^2}{\left(a_2(x,y)\mu^2+x(1-x)(1-y)\left(L_{\theta}\left(\vec{q}\right)\right)^2\right)^{\frac{17-2d}{6}}}\\&+\left(a_2(x,y) \leftrightarrow a_4(x,y), a_1^{(\nu)}(x,y,d)\leftrightarrow
       a_3^{(\nu)}(x,y,d)
       \right)
       \Bigg]
       \label{W_2b}
    \end{aligned}
\end{equation}
and 
        $\Xi_d =\frac{1}{12\sqrt{3}}\frac{\Omega_{d-1}\Gamma\left(\frac{11-2d}{6}\right)}{(2\pi)^{d-1}\beta_d^{\frac{1}{3}}\Gamma\left(\frac{d-1}{6}\right)\Gamma\left(\frac{3}{2}\right)}$.
Finally, Eq. (\ref{(1)_counterterm_logmu}) can be written as
\begin{equation}
    \begin{aligned}
         \frac{d~\Gamma^{CT;(1);(\nu,s)}_{\mu,\theta_1,\theta_2}\left(\vec{q}\right)}{d~\mathrm{ln}~\mu} = \lambdadim^{(\nu,s)}_{\theta_1,\theta_2}\left(\vec{q}\right)g_{\theta_2}\mathcal{C}^{(\nu)}_d\left(\mathscr{L}_{\mu,\theta_2}\left(\vec{q}\right)\right).
         \label{(1)_logmu}
    \end{aligned}
\end{equation}
Here, $\mathcal{C}^{(\nu)}_d\left(\mathscr{L}_{\mu,\theta}\left(\vec{q}\right)\right)$ is the crossover function given by
\begin{equation}
    \begin{aligned}
        \mathcal{C}^{(\nu)}_d\left(\mathscr{L}_{\mu,\theta}\left(\vec{q}\right)\right) 
        &= 
        -\snu\Xi_d
        \bigints_{x,y}\left[A^{(\nu)}_+(d)\Gamma\left(\frac{d+1}{2}\right)\times\left(\frac{a_2(x,y)}{\left(a_2(x,y)+x(1-x)(1-y)\left(\mathscr{L}_{\mu,\theta}\left(\vec{q}\right)\right)^2\right)^{\frac{11-2d}{6}}}+\left(a_2(x,y)\rightarrow a_4(x,y)\right)\right)
        \right.\\&\left.
        \hspace{-65pt}-
        \Gamma\left(\frac{d-1}{2}\right)\left(\frac{a_1^{(\nu)}(x,y,d)\left(
        -\frac{\epsilon}{3}
        a_2(x,y)+x(1-x)(1-y)\left(\mathscr{L}_{\mu,\theta}\left(\vec{q}\right)\right)^2\right)}{\left(a_2(x,y)+x(1-x)(1-y)\left(\mathscr{L}_{\mu,\theta}\left(\vec{q}\right)\right)^2\right)^{\frac{17-2d}{6}}}
        +\left(a_2(x,y)\rightarrow a_4(x,y), a_1^{(\nu)}(x,y,d)\rightarrow a_3^{(\nu)}(x,y,d)\right)\right)\right].
        \label{appendix:crossover_initial_vertex}
    \end{aligned}
\end{equation}
Dropping terms that are sub-leading in $\epsilon$, we obain
\begin{equation}
    \begin{aligned}
        &
        \mathcal{C}^{(\nu)}_d\left(\mathscr{L}_{\mu,\theta}\left(\vec{q}\right)\right) \approx 
        -\snu\Xi_d
        \bigints_{x,y}\left[A^{(\nu)}_+(d)\Gamma\left(\frac{d+1}{2}\right)\times\left(\frac{a_2(x,y)}{\left(a_2(x,y)+x(1-x)(1-y)\left(\mathscr{L}_{\mu,\theta}\left(\vec{q}\right)\right)^2\right)^{\frac{11-2d}{6}}}+\left(a_2(x,y)\rightarrow a_4(x,y)\right)\right)\right.\\
        & \left.
        \hspace{-6pt}-
        \Gamma\left(\frac{d-1}{2}\right)\left(\frac{a_1^{(\nu)}(x,y,d)\left(x(1-x)(1-y)\left(\mathscr{L}_{\mu,\theta}\left(\vec{q}\right)\right)^2\right)}{\left(a_2(x,y)+x(1-x)(1-y)\left(\mathscr{L}_{\mu,\theta}\left(\vec{q}\right)\right)^2\right)^{\frac{17-2d}{6}}}
        +\left(a_2(x,y)\rightarrow a_4(x,y), a_1^{(\nu)}(x,y,d)\rightarrow a_3^{(\nu)}(x,y,d)\right)\right)\right]
        \label{appendix:crossover}
    \end{aligned}
\end{equation}
to the leading order in $\epsilon$.
In the small
$\mathscr{L}_{\mu,\theta}$ limit,
$\mathcal{C}^{(\nu)}_d$
approaches a non-zero constant. 
In the large $\mathscr{L}_{\mu,\theta}$ limit,
$\mathcal{C}^{(\nu)}_d$
goes to zero as
$\left(\mathscr{L}_{\mu,\theta}\left(\vec{q}\right)\right)^{-2}$ in $d<d_c$
and
$\left(\mathscr{L}_{\mu,\theta}\left(\vec{q}\right)\right)^{-2}
  \log \mathscr{L}_{\mu,\theta}\left(\vec{q}\right)$ at $d=d_c$\footnote{
  In the large $\mathscr{L}_{\mu,\theta}\left(\vec{q}\right)$ limit, 
it is tempting to drop $a_2(x,y)$ or $a_4(x,y)$ in the denominator.
However, it gives rise to divergences at $x=0$ and $x=1$. 
The leading contribution can be extracted from those two regions by expanding $a_2(x,y)$ and $a_4(x,y)$ in the denominator near $x=0$ and $x=1$ to the linear order 
and performing the integration over $\delta x$ and $y$.}. 
This shows that the low energy limit and $d\rightarrow d_c$ limit do not commute with each other. The asymptotic limits of $\mathcal{C}^{(\nu)}_d\left(\mathscr{L}_{\mu,\theta}\left(\vec{q}\right)\right)$ are summarized as
\begin{equation}
    \mathcal{C}^{(\nu)}_d\left(\mathscr{L}_{\mu,\theta}\left(\vec{q}\right)\right)=
    \begin{cases}
      \rho_d^{(\nu)}, &\mathscr{L}_{\mu,\theta}\left(\vec{q}\right)\ll 1,~2\leq d \leq d_c \\
      \Upsilon_d^{(\nu)}\left(\mathscr{L}_{\mu,\theta}\left(\vec{q}\right)\right)^{-2}, &\mathscr{L}_{\mu,\theta}\left(\vec{q}\right)\gg 1, ~2\leq d <d_c \\
      \bar{\Upsilon}_{d_c}^{(\nu)}
      \log\left(\mathscr{L}_{\mu,\theta}\left(\vec{q}\right)\right)\left(\mathscr{L}_{\mu,\theta}\left(\vec{q}\right)\right)^{-2}, &\mathscr{L}_{\mu,\theta}\left(\vec{q}\right)\gg 1, ~d=d_c,
    \end{cases},
    \label{appendix:crossover_asymptotics}
\end{equation}
where
\begin{equation}
    \begin{aligned}
        \rho_d^{(\nu)} = -\snu A^{(\nu)}_+(d)\zeta(d)
        \label{appendix:rho_d_nu}
    \end{aligned}
\end{equation}
with 
$         \zeta(d) =  \Xi_d\Gamma\left(\frac{d+1}{2}\right)
         \int_{x,y}
         \left(a_2(x,y)^{\frac{2d-5}{6}}+a_4(x,y)^{\frac{2d-5}{6}}\right)$
         and
\begin{equation}
    \begin{aligned}
        \Upsilon_d^{(\nu)} &=
        -\snu\Xi_d\int_0^{\infty}d\tilde{x}\int_0^1 dy y^{\frac{d-7}{6}}(1-y)^{\frac{1}{2}}\left[A^{(\nu)}_+(d)\Gamma\left(\frac{d+1}{2}\right)\left(\frac{a_2(0,y)}{\left(a_2(0,y)+(1-y)\tilde{x}\right)^{\frac{11-2d}{6}}}+\frac{a_2(1,y)}{\left(a_2(1,y)+(1-y)\tilde{x}\right)^{\frac{11-2d}{6}}}
        \right.\right.\\&\left.\left.
        +
        \left(a_2\rightarrow a_4\right)\right)
        -
        \Gamma\left(\frac{d-1}{2}\right)\left(\frac{a_1^{(\nu)}(0,y,d)\left(\tilde{x}(1-y)\right)}{\left(a_2(0,y)+\tilde{x}(1-y)\right)^{\frac{17-2d}{6}}}
        +\frac{a_1^{(\nu)}(1,y,d)\left(\tilde{x}(1-y)\right)}{\left(a_2(1,y)+\tilde{x}(1-y)\right)^{\frac{17-2d}{6}}}
        +\left(a_2\rightarrow a_4, a_1^{(\nu)}\rightarrow a_3^{(\nu)}\right)\right)\right].
        \label{appendix:upsilon_d_nu}
    \end{aligned}
\end{equation}
Finally, the contribution to the beta functional reads
\begin{equation}
    \begin{aligned}
         \mu^{d-1}\frac{d~\Gamma^{CT;(1);(\nu,s)}_{\mu,\theta_1,\theta_2}\left(\vec{q}\right)}{d~\mathrm{ln}~\mu} = \lambda^{(\nu,s)}_{\theta_1,\theta_2}\left(\vec{q}\right)g_{\theta_2}\mathcal{C}^{(\nu)}_d\left(\mathscr{L}_{\mu,\theta_2}\left(\vec{q}\right)\right).
         \label{(1)_quantum_correction_aiabatic}
    \end{aligned}
\end{equation}
The quantum correction that arises from Fig. \ref{vertex3_2} 
 reads
\begin{equation}
\begin{aligned}
  \Gamma^{\left(1^\prime\right)}_{\{j_i\};abcd}
       \left(\mathbf{k}_1^{A}, \mathbf{k}_2^{A}, \mathbf{k}_3^{A}, \mathbf{k}_4^{A}\right)
 &=
  -\frac{3\times2}{3!}
  \frac{1}{N}
  T^{(\nu,s)}_{\left(\begin{smallmatrix}     j_1  & j_2      \\ j_4     & j_3       \end{smallmatrix}\right)}
  \int^{\prime} d_{f}^{d+1}\mathbf{l}
  \left(-i\edim_{\theta_1,\theta}\right)\left(-i\edim_{\thetasq,\Theta\left(\theta_1,\vec{q}\right)}\right) 
  \left(-\lambdadim^{(\nu,s)}_{
  \theta,\theta_2}\right)D_1\left(\mathbf{l}-\mathbf{k}_1^{A}\right)\\&\hspace{100pt}\times
  \left(\gamma_{d-1}G_0(\mathbf{l})I^{(\nu)}_mG_0(\mathbf{l+q})\gamma_{d-1}\right)_{ab} 
  \left(I^{(\nu)}_m\right)_{cd},
\label{quantum_correction_(1')_expression}
\end{aligned}
\end{equation}
where $\mathbf{k}_i^{A}$ is defined in Eq. (\ref{appendix_mu2}).
Proceeding in the same way,
we obtain
\begin{equation}
    \begin{aligned}
        \Gamma^{\left(1'\right);(\nu,s)}_{\mu^{A}}\left(\theta_1,\theta_2,\vec{q};\mu\right) &= -\frac{\snu}{N}\int \frac{d\mathbf{L}d\delta \KFthetadim  d\theta}{(2\pi)^{d+1}}\boldsymbol{e}^2_{\theta_1,\theta}\boldsymbol{\lambda}^{(\nu,s)}_{
        \theta,\theta_2}\left(\vec{q}\right)
        D_1\left(\mathbf{l}-\mathbf{k}_1^{A}\right)K^{(\nu)}_d\left(\mathbf{L},\boldsymbol{\mu},\delta,\theta,\vec{q}\right),\\
        \Gamma^{\left(1'\right);(\nu,s)}_{\mu^{B}}\left(\theta_1,\theta_2,\vec{q};\mu\right) &= -\frac{\snu}{N}\int \frac{d\mathbf{L}d\delta \KFthetadim  d\theta}{(2\pi)^{d+1}}\boldsymbol{e}^2_{\theta_1,\theta}\boldsymbol{\lambda}^{(\nu,s)}_{
        \theta,\theta_2}\left(\vec{q}\right)
        D_1\left( \mathbf{l}-\mathbf{k}_1^{B}\right)K^{(\nu)}_d\left(\mathbf{L},-\boldsymbol{\mu},\delta,\theta,\vec{q}\right).\label{quantum_correction_(1')_mu1_mu2_explicit}
    \end{aligned}
\end{equation}
where $\snu$ is defined in Eq. (\ref{appendix:s_nu}).
The counter term reads
\begin{equation}
    \begin{aligned}
        \Gamma^{CT;\left(1^\prime\right);(\nu,s)}_{\mu,\theta_1,\theta_2}\left(\vec{q}\right) &= \frac{\snu}{2N}\int \frac{d\mathbf{L}d\delta \KFthetadim  d\theta}{(2\pi)^{d+1}}\boldsymbol{\lambda}^{(\nu,s)}_{
        \theta_1,\theta}\left(\vec{q}\right)
        \edim^2_{\theta,\theta_2}
        \left\{D_{1;\mu}\left(\mathbf{L}-3\boldsymbol{\mu},\theta_1,\theta \right)K^{(\nu)}_d\left(\mathbf{L},\boldsymbol{\mu},\delta,\theta,\vec{q}\right)
        \right.\\&\left.+D_{1;\mu}\left(\mathbf{L}+\boldsymbol{\mu},\theta_1,\theta \right)K^{(\nu)}_d\left(\mathbf{L},-\boldsymbol{\mu},\delta,\theta,\vec{q}\right)\right\}.
    \end{aligned}
\end{equation}
Using the adiabatic approximation for the coupling functions,
we obtain an expression that is identical to
Eq. (\ref{appendix:(1)_mu1_res_doublefeynman})
with $|\mathbf{L}-\boldsymbol{\mu}|$ ($|\mathbf{L}-\boldsymbol{\mu}|$) replaced with $|\mathbf{L}-3\boldsymbol{\mu}|$ ($|\mathbf{L}+\boldsymbol{\mu}|$) in the first (second) term of the integrand.
Shifting $\mathbf{L}\rightarrow \mathbf{L}-(-y+2x(1-y)-2)\boldsymbol{\mu}$ ($\mathbf{L}\rightarrow \mathbf{L}-(y+2x(1-y))\boldsymbol{\mu}$) and integrating over $\mathbf{L}$,
we reach\footnote{
Notice that there will be an interchange of  $\mathscr{H}\leftrightarrow\tilde{\mathscr{H}}$ and  $\mathscr{J}\leftrightarrow\tilde{\mathscr{J}}$ for $\mu^{A}$ and $\mu^{B}$ contributions of the counter term as compared to counter term for quantum correction shown in Fig. \ref{vertex3_1}. 
}

\begin{equation}
    \begin{aligned}
        \Gamma^{CT;\left(1^{\prime}\right);(\nu,s)}_{\mu,\theta_1,\theta_2}\left(\vec{q}\right) =\boldsymbol{g}_{\theta_1}B^{d;(\nu)}_{\mu,\theta_1}\left(\vec{q}\right)\boldsymbol{\lambda}^{(\nu,s)}_{\theta_1,\theta_2}(\vec{q}),
    \end{aligned}
\end{equation}
where $B^{d;(\nu)}_{\mu,\theta}$ is defined in Eq. (\ref{appendix:B_d_theta}),
and the contribution to the beta functional becomes
\begin{equation}
    \begin{aligned}
        \mu^{d-1}\frac{d~\Gamma^{CT;\left(1'\right);(\nu,s)}_{\mu,\theta_1,\theta_2}\left(\vec{q}\right)}{d~\mathrm{ln}~\mu} = g_{\theta_1}\lambda^{(\nu,s)}_{\theta_1,\theta_2}\left(\vec{q}\right)\mathcal{C}^{(\nu)}_d\left(\mathscr{L}_{\mu,\theta_1}\left(\vec{q}\right)\right).
        \label{(1')_adiabatic}
    \end{aligned}
\end{equation}
It is noted that the weight from the convolution integration in Eqs. (\ref{appenidx:(1)_general_beta_functional}) and (\ref{appenidx:(1')_general_beta_functional}) 
corresponds to Eqs. (\ref{(1)_quantum_correction_aiabatic}) and (\ref{(1')_adiabatic}), respectively. 
In the $\mathscr{L}_{\mu,\theta}\left(\vec{q}\right)\ll1$ limit, the weight of Eq. (\ref{main:h1_def}) gives leading contribution in $\epsilon$ compared to that of Eq. (\ref{main:h2_def}). 
This can be understood from the fact that the former exhibits logarithmic singularity at $d=d_c$ while the latter is finite.

\subsubsection{\texorpdfstring{$\lambda^0$}{Lg} order}
\label{lambda_0_order}

Here, we derive 
the counter term that is independent of the four-fermion coupling 
(Fig. \ref{ladder1}).
The quantum correction 
reads
\begin{equation}\begin{aligned}
   &\delta S^{(0)}_4 
    = -\frac{6\times2}{4!}
    \frac{1}{N^2}
\int^{'}
d_{f}^{d+1} {\bf p}_1
d_{f}^{d+1} {\bf p}_2
\int
d_{b}^{d+1} {\bf q}
\int^{'}
d^{d+1}_{f}\mathbf{l}~
    (-i\edim_{\theta_1,\theta})(-i\edim_{\theta,\theta_2})(-i\edim_{\Theta\left(\theta_2,\vec{q}\right),\thetasq})(-i\edim_{\thetasq,\Theta\left(\theta_1,\vec{q}\right)}) D_1\left(\mathbf{l}-\mathbf{p}_2\right) \\&\hspace{10pt}\times D_1\left(\mathbf{l}-\mathbf{p}_1\right)\left(\gamma_{d-1}G_0(\mathbf{l})\gamma_{d-1}\right)_{ad}
    \left(\gamma_{d-1}G_0(\mathbf{l+q})\gamma_{d-1}\right)_{cb}
    \bar{\Psi}_{j_1;a}
    (\mathbf{p}_1)
     \Psi_{j_1;d}(\mathbf{p}_2)
     \bar{\Psi}_{j_2;c}\left(\mathbf{p}_2
     + {\bf q}\right)
    \Psi_{j_2;b}
    \left(\mathbf{p}_1+{\bf q}  \right),
    \label{(0)_mu1_formal}
\end{aligned}\end{equation}
where 
$\mathbf{l} = \left(\mathbf{L},\delta,\theta\right)$ is the loop momentum and $\mathbf{l+q} = 
(\mathbf{L}+\mathbf{Q},\deltaq,\thetasq)
$.
We single out the part of the integrand that depends on the fermion propagator as
\begin{equation}
    \begin{aligned}
        \left(\gamma_{d-1}G_0(\mathbf{l})\gamma_{d-1}\right)_{ad}\left(\gamma_{d-1}G_0(\mathbf{l+q})\gamma_{d-1}\right)_{cb} = \frac{\mathcal{N}^{(0)}_{adcb}}{\mathcal{D}^{(0)}},
    \end{aligned}
\end{equation}
where 
\begin{equation}\begin{aligned}
   \mathcal{D}^{(0)} &= i^2\left(|\mathbf{L}+\mathbf{Q}|^2+v_{F,\thetasq}^2\deltaq^2\right)\left(|\mathbf{L}|^2+v_{F,\theta }^2\delta^2\right),\\
   \mathcal{N}^{(0)}_{adcb} &= \left(-\mathbf{\Gamma}\cdot\mathbf{L}+v_{F,\theta }\delta\gamma_{d-1}\right)_{ad}\left(-\mathbf{\Gamma}\cdot(\mathbf{L}+\mathbf{Q})+v_{F,\thetasq}\deltaq\gamma_{d-1}\right)_{cb}.
   \label{appendix:numerator_(0)}
\end{aligned}\end{equation}
From Eq. (\ref{appendix:non_vanishing_freqprod}),
the singular contribution in Eq. (\ref{appendix:numerator_(0)}) arises from
\begin{equation}\begin{aligned}
    \mathcal{N}^{(0)}_{adcb} = \frac{1}{d-1}\mathbf{L} \cdot \left(\mathbf{L}+\mathbf{Q}\right)\mathbf{\Gamma}_{ad}\cdot\mathbf{\Gamma}_{cb} + v_{F,\theta }v_{F,\thetasq}\delta\deltaq \left(\gamma_{d-1}\right)_{ad}\left(\gamma_{d-1}\right)_{cb}.
    \label{(0)_tensors}
\end{aligned}\end{equation}
Using 
the Fierz transformation 
(see Appendix \ref{sec:Fierz_Transformation}), 
we turn Eq. (\ref{(0)_mu1_formal}) into
\begin{equation}\begin{aligned}
   &\delta S^{(0)}_4 
    = -
    \frac{1}{4N^2}\sum_{\nu}
    \int^{'}
    d_{f}^{d+1} {\bf p}_1
    d_{f}^{d+1} {\bf p}_2
    \int
    d_{b}^{d+1} {\bf q}
    \int^{'}
    d_{f}^{d+1}
    \mathbf{l}~
    \left\{
    \edim_{\theta_1,\theta}~\edim_{\theta,\theta_2}~\edim_{\Theta\left(\theta_2,\vec{q}\right),\thetasq}~\edim_{\thetasq,\Theta\left(\theta_1,\vec{q}\right)}K^{(\nu)}_d\left(\mathbf{L},-\mathbf{Q}/2,\delta,\theta,\vec{q}\right) \right.\\&\left.\hspace{10pt}\times
    D_1\left(\mathbf{l}-\mathbf{p}_2\right) D_1\left(\mathbf{l}-\mathbf{p}_1\right)\left(I_m^{\left(\nu\right)}\right)_{ab}\left(I_m^{\left(\nu\right)}\right)_{cd}+\text{regular terms}\right\}
    \bar{\Psi}_{j_1;a}\left(\mathbf{p}_1\right)
    \Psi_{j_2;b}
    \left(\mathbf{p}_1+{\bf q}  \right)
    \bar{\Psi}_{j_2;c}\left(\mathbf{p}_2+{\bf q}\right)
    \Psi_{j_1;d}\left(\mathbf{p}_2\right),
    \label{(0)_mu1_formal_postFierz}
\end{aligned}\end{equation}
where 
$K^{(\nu)}_d$
is defined in Eq. (\ref{appendix:general_4fkernel}). It is noted from Eq. (\ref{(0)_mu1_formal_postFierz}) that the quantum correction generates four-fermion interaction in the $(P,d)$ and $(F_{\pm},e)$ channels only. 
Using Eq. (\ref{appendix:slow_varying_approx}), 
we rewrite the quantum correction evaluated at
$\mathbf{p}_1 = \mathbf{k}_1^A$,
$\mathbf{p}_1+{\bf q} = \mathbf{k}_4^A$,
$\mathbf{p}_2 = \mathbf{k}_3^A$,
$\mathbf{p}_2+{\bf q} = \mathbf{k}_2^A$ as
\begin{equation}\begin{aligned}
  \Gamma^{(0);(\nu,s)}_{\mu^{A}}\left(\theta_1,\theta_2,\vec{q};\mu\right) = -
  \frac{\delta_{s,s_{\nu}}}{4}\int \frac{d\mathbf{L}d\delta \KFthetadim  d\theta}{(2\pi)^{d+1}}\frac{\boldsymbol{e}^2_{\theta_1,\theta}}{N}\frac{\boldsymbol{e}^2_{\theta,\theta_2}}{N}
    K^{(\nu)}_d\left(\mathbf{L},\boldsymbol{\mu},\delta,\theta,\vec{q}\right)D_1\left(\mathbf{l}-\mathbf{k}_3^{A}\right) D_1\left(\mathbf{l}-\mathbf{k}_1^{A}\right), 
    \label{(0)_mu1_formal_concise} 
\end{aligned}\end{equation}
where 
$ s_{F_\pm}=e $,
$ s_{P}=d $.
Before we evaluate Eq. (\ref{(0)_mu1_formal_concise}) explicitly,
let us first write down the formal expression that enters in the full beta functional. 
The counter term that cancels the average of the quantum corrections evaluated at 
${\bf k}^{A}_i$
and
${\bf k}^{B}_i$
is written as
\begin{equation}
    \begin{aligned}
        \Gamma^{CT;(0);(\nu,s)}_{\mu,\theta_1,\theta_2}\left(\vec{q}\right) &= 
        \frac{\delta_{s,s_{\nu}}}{8}\int \frac{d\mathbf{L}d\delta \KFthetadim  d\theta}{(2\pi)^{d+1}}\frac{\boldsymbol{e}^2_{\theta_1,\theta}}{N}\frac{\boldsymbol{e}^2_{\theta,\theta_2}}{N}
        \left\{
        D_{1;\mu}\left(\mathbf{L}-3\boldsymbol{\mu},\theta_1,\theta\right)K^{(\nu)}_d\left(\mathbf{L},\boldsymbol{\mu},\delta,\theta,\vec{q}\right)
        D_{1;\mu}\left(\mathbf{L}-\boldsymbol{\mu},\theta,\theta_2\right)
        \right.\\&\left.+
        D_{1;\mu}\left(\mathbf{L}+\boldsymbol{\mu},\theta_1,\theta\right)K^{(\nu)}_d\left(\mathbf{L},-\boldsymbol{\mu},\delta,\theta,\vec{q}\right)
        D_{1;\mu}\left(\mathbf{L}-\boldsymbol{\mu},\theta,\theta_2\right)
        \right\}.
    \end{aligned}
\end{equation}
The contribution to the beta function then reads 
\begin{equation}
    \begin{aligned}
&\mu^{d-1}\frac{d~\Gamma^{CT;\left(0\right);(\nu,s)}_{\mu,\theta_1,\theta_2}\left(\vec{q}\right)}{d~\log\mu}=
        \frac{\delta_{s,s_{\nu}}}{16}
\mu^{6-d}
       \sum_{i=\pm}  A_i^{(\nu)}(d)
\int \frac{ d\theta}{2\pi\mu}\frac{\KFthetadim }{v_{F,\theta}}
   \frac{e^2_{\theta_1,\theta}
   e^2_{\theta,\theta_2}
   }{N^2}
        \int\frac{d\mathbf{L}dE }{(2\pi)^{d}}
        \partial_{\log\mu} \\&\times
        \left\{ 
D_{1;\mu}\left(\mathbf{L}-3\boldsymbol{\mu},\theta_1,\theta\right)
\tilde{K}_i\left(\mathbf{L},\boldsymbol{\mu},E,\theta,\vec{q}\right) 
D_{1;\mu}\left(\mathbf{L}-\boldsymbol{\mu},\theta,\theta_2\right)
 +
D_{1;\mu}\left(\mathbf{L}+\boldsymbol{\mu},\theta_1,\theta\right)
\tilde{K}_i\left(\mathbf{L},-\boldsymbol{\mu},E,\theta,\vec{q}\right) 
D_{1;\mu}\left(\mathbf{L}-\boldsymbol{\mu},\theta,\theta_2\right)
        \right\},
        \label{appendix:(0)_general_beta_functional}
    \end{aligned}
\end{equation}
where $\tilde{K}_\pm\left(\mathbf{L},\boldsymbol{\mu},E,\theta,\vec{q}\right)
=\frac{\mathbf{L}\cdot(\mathbf{L}-2\boldsymbol{\mu})\pm E\left(E+L_{\theta}(\vec{q})\right)}{\left(|\mathbf{L}-2\boldsymbol{\mu}|^{2}+\left(E+L_{\theta}(\vec{q})\right)^2\right)\left(|\mathbf{L}|^{2}+E^2\right)}$.

Now, we compute the quantum correction explicitly.
In general, it is hard to evaluate Eq. (\ref{(0)_mu1_formal_concise}) in a closed form.
%
Here, we compute the quantum correction in two limits,
one in the intra-patch limit with 
small $|\theta_1-\theta_2|$
and the other in the inter-patch limit with large angular separation.
Two points on the Fermi surface at angles $\theta_1$ and $\theta_2$ move outside one patch at energy scale $\mu$
if the typical momentum that the critical boson carries at that energy scale becomes comparable to $q(\theta_1,\theta_2)$.
This leads to the condition,
    $\pmb{\mathcal{K}}_{\theta}|\theta_1-\theta_2|\sim \left(\frac{Ng_{\theta}\KFthetadim }{\left|X_{\theta}\right||\chi_{\vartheta^{-1}(\theta)}|v_{F,\theta }}\right)^{\frac{1}{2}}\mu^{\frac{1}{2}}$.
At energy scale $\mu$,
two points are within (outside) one patch if their angular separation is smaller (larger) than
\begin{equation}
    \begin{aligned}
        |\theta_1-\theta_2|\sim \left(\frac{Ng_{\theta}\left|X_{\theta}\right|}{|\chi_{\vartheta^{-1}(\theta)}|v_{F,\theta }}\right)^{\frac{1}{2}}
        \KFtheta^{-\frac{1}{2}},
        \label{eq:C110}
    \end{aligned}
\end{equation}
where $\KFtheta
= \frac{\KFthetadim }{\mu}$. 
We emphasize that the crossover angular separation in \eq{eq:C110} is a function of energy scale $\mu$ through  $\KFtheta$.
With decreasing $\mu$, the width of a patch decreases as the momentum of the critical boson also decreases.

\begin{center}
    \MakeUppercase{\romannumeral 1}. $|\theta_1-\theta_2|\ll \KFtheta^{-\frac{1}{2}}$
\end{center}

In this region, fermions at angle $\theta_1$ and $\theta_2$ lie within a single patch. 
In the small angle difference limit, we use $\Gamma^{(0);(\nu,s)}_{\mu^{A}}\left(\theta_1,\theta_2,\vec{q};\mu\right) \approx \Gamma^{(0);(\nu,s)}_{\mu^{A}}\left(\theta_1,\theta_1,\vec{q};\mu\right) $ to express the quantum correction evaluated at $\mu^{A}$ as
\begin{equation}
    \begin{aligned}
       \Gamma^{(0);(\nu,s)}_{\mu^{A}}\left(\theta_1,\theta_1,\vec{q};\mu\right)
        =  -
        \frac{\delta_{s,s_{\nu}}}{4}
        \int \frac{d\mathbf{L}\KFthetadim d\delta d\theta}{(2\pi)^{d+1}}\frac{\boldsymbol{e}_{\theta_1,\theta}^2}{N}\frac{\boldsymbol{e}_{\theta,\theta_1}^2}{N} K^{(\nu)}_d\left(\mathbf{L},\boldsymbol{\mu},\delta,\theta,\vec{q}\right) 
        D_1\left(\mathbf{L}-\boldsymbol{\mu},\vec{l}-\vec{k}_1\right)
        D_1\left(\mathbf{L}-3\boldsymbol{\mu},\vec{l}-\vec{k}_1\right).
        \label{(0)_mu1_formal_concise_small_angle}
    \end{aligned}
\end{equation}
From the adiabatic approximation for the couplings functions, 
Eq. (\ref{(0)_mu1_formal_concise_small_angle}) becomes
\begin{equation}
    \begin{aligned}
     & \Gamma^{(0);(\nu,s)}_{\mu^{A}}\left(\theta_1,\theta_1,\vec{q};\mu\right) \approx \nn 
     & -\delta_{s,s_{\nu}}\frac{\boldsymbol{e}^4_{\theta_1}\mathbf{K}_{F,\theta_1}}
      {4N^2}
      \int \frac{d\mathbf{L}d\delta d\theta}{(2\pi)^{d+1}}\frac{\left(\pmb{\mathcal{K}}_{\theta_1}|\theta -\theta_1|\right)^2K^{(\nu)}_d\left(\mathbf{L},\boldsymbol{\mu},\delta,\theta_1,\vec{q}\right)}{\left[\left(\pmb{\mathcal{K}}_{\theta_1}|\theta -\theta_1|\right)^{3}+\boldsymbol{f}_{d,\vartheta^{-1}(\theta_1)}|\mathbf{L}-\boldsymbol{\mu}|^{d-1}\right]\left[\left(\pmb{\mathcal{K}}_{\theta_1}|\theta -\theta_1|\right)^{3}+\boldsymbol{f}_{d,\vartheta^{-1}(\theta_1)}|\mathbf{L}-3\boldsymbol{\mu}|^{d-1}\right]},
    \end{aligned}
\end{equation}
where both the boson propagators are approximated to the leading order in the expansion of $q(\theta,\theta_1)$ 
in $\theta-\theta_1$
as the singular contribution arises from small-angle scatterings.
Shifting $\theta\rightarrow \theta+\theta_1$, rescaling $\theta\rightarrow\theta/\pmb{\mathcal{K}}_{\theta_1}$ and integrating over $\theta$, 
we obtain the quantum correction to be
\begin{equation}
    \begin{aligned}
       \Gamma^{(0);(\nu,s)}_{\mu^{A}}\left(\theta_1,\theta_1,\vec{q};\mu\right) = 
         -\frac{\delta_{s,s_{\nu}}}
         {12\pi N^{2}}
         \frac{1}{\beta_d}\frac{\boldsymbol{e}^{2}_{\theta_1}v^2_{F,\theta_1}\left|X_{\theta_1}\right||\chi_{\vartheta^{-1}(\theta_1)}|}{\mathbf{K}_{F,\theta_1}}
         \int \frac{d\mathbf{L}d\delta}{(2\pi)^{d}}\frac{\log\left(\frac{|\mathbf{L}-3\boldsymbol{\mu}|^{d-1}}{|\mathbf{L}-\boldsymbol{\mu}|^{d-1}}\right)}{|\mathbf{L}-3\boldsymbol{\mu}|^{d-1}-|\mathbf{L}-\boldsymbol{\mu}|^{d-1}}
         K^{(\nu)}_d\left(\mathbf{L},\boldsymbol{\mu},\delta,\theta_1,\vec{q}\right).
    \end{aligned}
\end{equation}
Rescaling $\delta\rightarrow\delta/v_{F,\theta_1}$, integrating over $\delta$ and shifting $\mathbf{L}\rightarrow\mathbf{L}+\boldsymbol{\mu}$, we obtain
\begin{equation}
    \begin{aligned}
       \Gamma^{(0);(\nu,s)}_{\mu^{A}}\left(\theta_1,\theta_1,\vec{q};\mu\right) = 
         -\frac{\delta_{s,s_{\nu}}}
         {24\pi N^{2}}
         \frac{1}{\beta_d}
         \frac{\boldsymbol{e}^{2}_{\theta_1}v_{F,\theta_1}\left|X_{\theta_1}\right||\chi_{\vartheta^{-1}(\theta_1)}|}{\mathbf{K}_{F,\theta_1}}
        \int \frac{d\mathbf{L}}{(2\pi)^{d-1}}I^{(\nu)}_d(\mathbf{L},\boldsymbol{\mu},\vec{q}),
    \end{aligned}
\end{equation}
where
\begin{equation}
\begin{aligned}
I^{(\nu)}_d(\mathbf{L},\boldsymbol{\mu},\vec{q})=\frac{\log\left(\frac{|\mathbf{L}-2\boldsymbol{\mu}|^{d-1}}{|\mathbf{L}|^{d-1}}\right)}{|\mathbf{L}-2\boldsymbol{\mu}|^{d-1}-|\mathbf{L}|^{d-1}}
\frac{|\mathbf{L}+\boldsymbol{\mu}|+|\mathbf{L}-\boldsymbol{\mu}|}{\left(L_{\theta_1}\left(\vec{q}\right)\right)^2+\left(|\mathbf{L}+\boldsymbol{\mu}|+|\mathbf{L}-\boldsymbol{\mu}|\right)^2}
\left[A^{(\nu)}_1(d)\frac{(\mathbf{L}+\boldsymbol{\mu})\cdot(\mathbf{L}-\boldsymbol{\mu})}{|\mathbf{L}+\boldsymbol{\mu}||\mathbf{L}-\boldsymbol{\mu}|}+A^{(\nu)}_2(d)\right].\label{small_theta_frequency_kernel}
\end{aligned}\end{equation}
For $|\mathbf{L}|>>2|\boldsymbol{\mu}|$, Eq. (\ref{small_theta_frequency_kernel}) becomes
$I^{(\nu)}_d(\mathbf{L},\boldsymbol{\mu},\vec{q}) \approx 
\frac{2A^{(\nu)}_+(d)}{|\mathbf{L}|^{d-2}\left[\left(L_{\theta_1}\left(\vec{q}\right)\right)^2+4|\mathbf{L}|^2\right]}$
and its contribution becomes
\begin{equation}\begin{aligned}
\int \frac{d\mathbf{L}}{(2\pi)^{d-1}}I^{(\nu)}_d(\mathbf{L},\boldsymbol{\mu},\vec{q}) \approx 
\frac{A^{(\nu)}_+(d)\Omega_{d-1}}{(2\pi)^{d-1}}\frac{1}{L_{\theta_1}\left(\vec{q}\right)}\left[\frac{\pi}{2}-\arctan\left(\frac{4\mu}{L_{\theta_1}\left(\vec{q}\right)}\right)\right].
\end{aligned}\end{equation}
For $|\mathbf{L}|<<2|\boldsymbol{\mu}|$, 
Eq. (\ref{small_theta_frequency_kernel}) becomes
$I^{(\nu)}_d(\mathbf{L},\boldsymbol{\mu},\vec{q}) \approx 
-\frac{(d-1)A^{(\nu)}_-(d)}{2^{d-2}|\boldsymbol{\mu}|^{d-2}\left[\left(L_{\theta_1}\left(\vec{q}\right)\right)^2+4|\boldsymbol{\mu}|^2\right]}\log\left(\frac{2|\boldsymbol{\mu}|}{|\mathbf{L}|}\right)$
and
its contribution becomes
\begin{equation}\begin{aligned}
\int \frac{d\mathbf{L}}{(2\pi)^{d-1}}I^{(\nu)}_d(\mathbf{L},\boldsymbol{\mu},\vec{q}) &\approx 
-\frac{2A^{(\nu)}_-(d)\Omega_{d-1}}{(2\pi)^{d-1}(d-1)}\frac{\mu}{\left(L_{\theta_1}\left(\vec{q}\right)\right)^2+4|\boldsymbol{\mu}|^2}.
\end{aligned}\end{equation}
Finally, the quantum correction in Eq. (\ref{(0)_mu1_formal_concise_small_angle}) can be written as
\begin{equation}\begin{aligned}
&\Gamma^{(0);(\nu,s)}_{\mu^{A}}\left(\theta_1,\theta_1,\vec{q};\mu\right) = \nn& 
         -\frac{\delta_{s,s_{\nu}}}
          {12\pi N^{2}}
          \frac{\boldsymbol{e}^{2}_{\theta_1}v_{F,\theta_1}\left|X_{\theta_1}\right||\chi_{\vartheta^{-1}(\theta_1)}|}{\mathbf{K}_{F,\theta_1}}
          \frac{\Omega_{d-1}}{\beta_d(2\pi)^{d-1}}\left[\frac{A^{(\nu)}_+(d)}{2L_{\theta_1}\left(\vec{q}\right)}\left(\frac{\pi}{2}-\arctan\left(\frac{4\mu}{L_{\theta_1}\left(\vec{q}\right)}\right)\right)
         -\frac{A^{(\nu)}_-(d)}{(d-1)}\frac{\mu}{\left(L_{\theta_1}\left(\vec{q}\right)\right)^2+4\mu^2}\right].
         \label{appendix:(0)_mu1_qc}
\end{aligned}\end{equation}
Similarly, the quantum correction evaluated at $\mu^{B}$ reads
\begin{equation}
    \begin{aligned}
       \Gamma^{(0);(\nu,s)}_{\mu^{B}}\left(\theta_1,\theta_1,\vec{q};\mu\right) =  -
        \frac{\delta_{s,s_{\nu}}}{4}
       \int \frac{d\mathbf{L}\KFthetadim d\delta d\theta}{(2\pi)^{d+1}}
       \frac{\boldsymbol{e}_{\theta_1,\theta}^2}{N}\frac{\boldsymbol{e}_{\theta,\theta_1}^2}{N} K^{(\nu)}_d\left(\mathbf{L},-\boldsymbol{\mu},\delta,\theta,\vec{q}\right) D_1\left(\mathbf{L}-\boldsymbol{\mu},\vec{l}-\vec{k}_1\right)D_1\left(\mathbf{L}+\boldsymbol{\mu},\vec{l}-\vec{k}_1\right).\label{(0)_mu2_formal_concise_small_angle}
    \end{aligned}
\end{equation}
Proceeding in the exact same way as in the computation for $\mu^{A}$, we find out that\footnote{After integration over $\theta$ and $\delta$, we can shift $\mathbf{L}\rightarrow\mathbf{L}-\boldsymbol{\mu}$ to get same equation as Eq. (\ref{small_theta_frequency_kernel}).},
\begin{equation}
    \begin{aligned}
        \Gamma^{(0);(\nu,s)}_{\mu^{A}}\left(\theta_1,\theta_1,\vec{q};\mu\right) =\Gamma^{(0);(\nu,s)}_{\mu^{B}}\left(\theta_1,\theta_1,\vec{q};\mu\right).
        \label{appendix:(0)_mu2_qc}
    \end{aligned}
\end{equation} 

Collecting Eqs. (\ref{appendix:(0)_mu1_qc}) and (\ref{appendix:(0)_mu2_qc}), and using Eq. (\ref{appendix:4f_counter_term}), 
we obtain the counter term,
\begin{equation}
    \begin{aligned}
         \Gamma^{CT;(0);(\nu,s)}_{\mu,\theta_1,\theta_1}\left(\vec{q}\right) = \frac{\delta_{s,s_{\nu}}}
          {12\pi N^{2}}
          \frac{\boldsymbol{e}^{2}_{\theta_1}v_{F,\theta_1}\left|X_{\theta_1}\right||\chi_{\vartheta^{-1}(\theta_1)}|}{\mathbf{K}_{F,\theta_1}}
          \frac{\Omega_{d-1}}{\beta_d(2\pi)^{d-1}}\left[\frac{A^{(\nu)}_+(d)}{2L_{\theta_1}\left(\vec{q}\right)}\left(\frac{\pi}{2}-\arctan\left(\frac{4\mu}{L_{\theta_1}\left(\vec{q}\right)}\right)\right)
         -\frac{A^{(\nu)}_-(d)}{(d-1)}\frac{\mu}{\left(L_{\theta_1}\left(\vec{q}\right)\right)^2+4\mu^2}\right],
    \end{aligned}
\end{equation}
and its derivative with respect to $\log \mu$,
\begin{equation}
    \begin{aligned}
       \frac{d~\Gamma^{CT;(0);(\nu,s)}_{\mu,\theta_1,\theta_1}\left(\vec{q}\right)}{d~\log~\mu} &= -
       \frac{\delta_{s,s_{\nu}}}{6 \pi N^{2}}
       \frac{\Omega_{d-1}}{\beta_d(2\pi)^{d-1}}
       \frac{\boldsymbol{e}^{2}_{\theta_1}v_{F,\theta_1}\left|X_{\theta_1}\right||\chi_{\vartheta^{-1}(\theta_1)}|}{\mathbf{K}_{F,\theta_1}}\left(\frac{A^{(\nu)}_+(d)}{\left(\mathscr{L}_{\mu,\theta_1}\left(\vec{q}\right)\right)^2+16}
       +\frac{A^{(\nu)}_-(d)}{2(d-1)}\frac{\left(\left(\mathscr{L}_{\mu,\theta_1}\left(\vec{q}\right)\right)^2-4\right)}{\left(\left(\mathscr{L}_{\mu,\theta_1}\left(\vec{q}\right)\right)^2+4\right)^2}\right)
       \frac{1}{\mu} .
    \end{aligned}
\end{equation}
Finally, 
the contribution to the beta  functional reads
\begin{equation}
    \begin{aligned}
       \mu^{d-1}\frac{d~\Gamma^{CT;(0);(\nu,s)}_{\mu,\theta_1,\theta_1}\left(\vec{q}\right)}{d~\log~\mu} &= -
        \frac{\delta_{s,s_{\nu}}}{6 \pi N^{2}}
       \frac{\Omega_{d-1}}{\beta_d(2\pi)^{d-1}}
       \frac{e^{2}_{\theta_1}v_{F,\theta_1}\left|X_{\theta_1}\right||\chi_{\vartheta^{-1}(\theta_1)}|}{K_{F,\theta_1}}
       \left(\frac{A^{(\nu)}_+(d)}{\left(\mathscr{L}_{\mu,\theta_1}\left(\vec{q}\right)\right)^2+16}
       +\frac{A^{(\nu)}_-(d)}{2(d-1)}\frac{\left(\left(\mathscr{L}_{\mu,\theta_1}\left(\vec{q}\right)\right)^2-4\right)}{\left(\left(\mathscr{L}_{\mu,\theta_1}\left(\vec{q}\right)\right)^2+4\right)^2}\right).
       \label{(0)_patch_beta_functional}
    \end{aligned}
\end{equation}

\begin{center}
    \MakeUppercase{\romannumeral 2}. $|\theta_1-\theta_2|\gg \KFtheta^{-\frac{1}{2}}$
\end{center}

In this region, 
fermions at angle $\theta_1$ and $\theta_2$ lie outside one patch.
In Eq. (\ref{(0)_mu1_formal_concise}), we note that the integrand is peaked at
$\theta \sim \theta_1$ or $\theta_2$. 
For 
$\theta \sim \theta_1$ 
($\theta \sim \theta_2$),
$D_1\left(\mathbf{l}-\mathbf{k}_3^{A}\right)\approx 
\frac{1}{q(\theta_1,\theta_2)^2}$ 
$\left( D_1\left(\mathbf{l}-\mathbf{k}_1^{A}\right)\approx  \frac{1}{q(\theta_1,\theta_2)^2} \right)$
and the rest of the loop integration can be done by following the steps 
involved in the computation of 
$\Gamma^{\left(1^{\prime}\right);(\nu,s)}_{\mu^{A}}\left(\theta_1,\theta_2,\vec{q};\mu\right)$
$\left(\Gamma^{\left(1\right);(\nu,s)}_{\mu^{A}}\left(\theta_1,\theta_2,\vec{q};\mu\right)\right)$.
Summing up both the contributions, we can extract the infrared divergent contribution of the quantum correction by using the adiabatic approximation as
\begin{equation}
    \begin{aligned}
  \Gamma^{(0);(\nu,s)}_{\mu^{A}}\left(\theta_1,\theta_2,\vec{q};\mu\right)   &= -
  \frac{\delta_{s,s_{\nu}}}{48\sqrt{3}}\frac{\boldsymbol{e}^2_{\theta_1,\theta_2}}{N}\frac{\Omega_{d-1}}{(2\pi)^{d-1}\beta_d^{\frac{1}{3}}\Gamma\left(\frac{d-1}{6}\right)\Gamma\left(\frac{3}{2}\right)}
    \int_{x,y}\left[\frac{\boldsymbol{g}_{\theta_1}}
    {q(\theta_1,\theta_2)^2}
    \left(\tilde{\mathscr{H}}_{\mu,\theta_1,\vec{q}}(x,y,d)+\tilde{\mathscr{J}}_{\mu,\theta_1,\vec{q}}^{(\nu)}(x,y,d)\right)
    \right.\\&\left.
    +\frac{\boldsymbol{g}_{\theta_2}}
    {q(\theta_1,\theta_2)^2}\left(\mathscr{H}_{\mu,\theta_2,\vec{q}}(x,y,d)+\mathscr{J}_{\mu,\theta_2,\vec{q}}^{(\nu)}(x,y,d)\right)\right].
    \end{aligned}\label{quantumcorrec_(0)_mu1}
\end{equation}
Here $\mathscr{H}_{\mu,\theta,\vec{q}}^{(\nu)}(x,y,d)$, $\mathscr{J}_{\mu,\theta,\vec{q}}^{(\nu)}(x,y,d)$,  $\tilde{\mathscr{H}}_{\mu,\theta,\vec{q}}^{(\nu)}(x,y,d)$ and $\tilde{\mathscr{J}}_{\mu,\theta,\vec{q}}^{(\nu)}(x,y,d)$ are defined in Eqs. (\ref{H_J_Vertex}).
Similarly, the quantum correction evaluated at $\mu^{B}$ becomes
\begin{equation}
    \begin{aligned}
        \Gamma^{(0);(\nu,s)}_{\mu^{B}}\left(\theta_1,\theta_2,\vec{q};\mu\right) = \left.\Gamma^{(0);(\nu,s)}_{\mu^{A}}\left(\theta_1,\theta_2,\vec{q};\mu\right)\right\vert_{\mathscr{H}\leftrightarrow \tilde{\mathscr{H}},\mathscr{J}\leftrightarrow \tilde{\mathscr{J}}}.
        \label{quantumcorrec_(0)_mu2}
    \end{aligned}
\end{equation}

The counter term that satisfies the RG conditions can be written as
\begin{equation}
    \begin{aligned}
        \Gamma^{CT;(0);(\nu,s)}_{\mu,\theta_1,\theta_2}\left(\vec{q}\right) = 
        \frac{\snu\delta_{s,s_{\nu}}}{4}
        \frac{\boldsymbol{e}^2_{\theta_1,\theta_2}}{N}
        \left[\frac{B^{d;(\nu)}_{\mu,\theta_1}\left(\vec{q}\right)\boldsymbol{g}_{\theta_1}+B^{d;(\nu)}_{\mu,\theta_2}\left(\vec{q}\right)\boldsymbol{g}_{\theta_2}}{q(\theta_1,\theta_2)^2+
        \boldsymbol{f}_{d,\vartheta^{-1}(\mtheta)}
        \alpha_d\frac{\mu^{d-1}}{{{\left[\sqrt{q(\theta_1,\theta_2)^{2}+\mu^{2}}\right]}}}}
        \right],
        \label{(0)_counter_term}
    \end{aligned}
\end{equation}
where $\snu$ and $B^{ d;(\nu)}_{\mu,\theta}\left(\vec{q}\right)$  are defined in Eqs. (\ref{appendix:s_nu}) and  (\ref{appendix:B_d_theta}), respectively, and $\mtheta = \frac{\theta_1+\theta_2}{2}$.
$\alpha_d$ is a positive definite constant that is 
is scheme-dependent.
It is noted that the smoothening by $\mu$ to ensure analyticity in $q(\theta_1,\theta_2)$
does not play an important role in the inter-patch limit because
$\sqrt{\KFdim\mu} \gg \mu$.
Taking the derivative of 
Eq. (\ref{(0)_counter_term})
with respect to $\log \mu$,
we obtain
\begin{equation}
    \begin{aligned}
        \frac{d~\Gamma^{CT;(0);(\nu,s)}_{\mu,\theta_1,\theta_2}\left(\vec{q}\right)}{d~\mathrm{ln}~\mu} &=
        \frac{\snu\delta_{s,s_{\nu}}}{4}\frac{\boldsymbol{e}^2_{\theta_1,\theta_2}}{N}
        \left[\frac{\boldsymbol{g}_{\theta_1}\partial_{\mathrm{ln}~\mu}\left(B^{d;(\nu)}_{\mu,\theta_1}\left(\vec{q}\right)\right)+\boldsymbol{g}_{\theta_2}\partial_{\mathrm{ln}~\mu}\left(B^{d;(\nu)}_{\mu,\theta_2}\left(\vec{q}\right)\right)}{q(\theta_1,\theta_2)^2+\boldsymbol{f}_{d,\vartheta^{-1}(\mtheta)}\alpha_d\frac{\mu^{d-1}}{{{\left[\sqrt{q(\theta_1,\theta_2)^{2}+\mu^{2}}\right]}}}}\right.\\&\left.\hspace{-100pt}-\frac{\boldsymbol{f}_{d,\vartheta^{-1}(\mtheta)}\alpha_d\mu^{d-1}\left((d-1)
        q(\theta_1,\theta_2)^2
        +(d-2)\mu^2\right)\left\{\boldsymbol{g}_{\theta_1}B^{d;(\nu)}_{\mu,\theta_1}\left(\vec{q}\right)+\boldsymbol{g}_{\theta_2}B^{d;(\nu)}_{\mu,\theta_2}\left(\vec{q}\right)\right\}}{\left[\sqrt{q(\theta_1,\theta_2)^{2}+\mu^{2}}\right]\left[q(\theta_1,\theta_2)^2\left[\sqrt{q(\theta_1,\theta_2)^{2}+\mu^{2}}\right]+\boldsymbol{f}_{d,\vartheta^{-1}(\mtheta)}\alpha_d \mu^{d-1}\right]^2}
        \right].
\label{(0)_log_counter_term_1}
    \end{aligned}
\end{equation}
It can be readily seen from Eq. (\ref{(0)_log_counter_term_1}) and Eq. (\ref{appendix:B_d_theta}) that 
the leading contribution arises from the first term in
\eq{(0)_log_counter_term_1},
\begin{equation}
    \begin{aligned}
        \frac{d~\Gamma^{CT;(0);(\nu,s)}_{\mu,\theta_1,\theta_2}\left(\vec{q}\right)}{d~\mathrm{ln}~\mu} = \frac{\snu\delta_{s,s_{\nu}}}{4}\frac{\boldsymbol{e}^2_{\theta_1,\theta_2}}{N}
        \left[\frac{\boldsymbol{g}_{\theta_1}\left( \mathscr{Y}^{d;\left(\nu\right)}_{\mu,\theta_1}\left(\vec{q}\right)+ \mathscr{W}^{d;\left(\nu\right)}_{\mu,\theta_1}\left(\vec{q}\right)\right)+\boldsymbol{g}_{\theta_2}\left( \mathscr{Y}^{d;\left(\nu\right)}_{\mu,\theta_2}\left(\vec{q}\right)+ \mathscr{W}^{d;\left(\nu\right)}_{\mu,\theta_2}\left(\vec{q}\right)\right)}{q(\theta_1,\theta_2)^2+
        \alpha_d\beta_d\left(\frac{Ng_{\mtheta}\mathbf{K}_{F,\mtheta}}{\left|X_{\mtheta}\right|\left|\chi_{\vartheta^{-1}\left(\mtheta\right)}\right|v_{F,\mtheta}}\right)^{\frac{3}{2}}
        \frac{\mu^{\frac{3}{2}}}{{{\left[\sqrt{q(\theta_1,\theta_2)^{2}+\mu^{2}}\right]}}}}
        \right].
    \end{aligned}
\end{equation}
Here, $\mathscr{Y}^{d;(\nu)}_{\mu,\theta}\left(\vec{q}\right)$ and $\mathscr{W}^{d;(\nu)}_{\mu,\theta_1}\left(\vec{q}\right)$ are defined in Eqs.     (\ref{Y_2b}) and (\ref{W_2b}), respectively.
The contribution to the beta functional reads
\begin{equation}
    \begin{aligned}
        \mu^{d-1}\frac{d~\Gamma^{CT;(0);(\nu,s)}_{\mu,\theta_1,\theta_2}\left(\vec{q}\right)}{d~\mathrm{ln}~\mu} = 
        \frac{\snu\delta_{s,s_{\nu}}}{4}
        \frac{e^2_{\theta_1,\theta_2}}{N}\left(\frac{\left\{g_{\theta_1}\mathcal{C}^{(\nu)}_d\left(\mathscr{L}_{\mu,\theta_1}\left(\vec{q}\right)\right)+g_{\theta_2}\mathcal{C}^{(\nu)}_d\left(\mathscr{L}_{\mu,\theta_2}\left(\vec{q}\right)\right)\right\}\mu^2}{q(\theta_1,\theta_2)^2+
        \beta_d\left(\frac{Ng_{\mtheta}\mathbf{K}_{F,\mtheta}}{\left|X_{\mtheta}\right|\left|\chi_{\vartheta^{-1}\left(\mtheta\right)}\right|v_{F,\mtheta}}\right)^{\frac{3}{2}}
        \frac{\mu^{\frac{3}{2}}}{{{\left[\sqrt{q(\theta_1,\theta_2)^{2}+\mu^{2}}\right]}}}}
        \right).
        \label{(0)_adiabatic}
    \end{aligned}
\end{equation}
It can also be shown\footnote{
At large  $q(\theta_1,\theta_2)$,
the difference is strongly suppressed as
\begin{align*}
\mu^{d-1} \left(\left.\frac{d~\Gamma^{CT;(0);(\nu,s)}_{\mu,\theta_1,\theta_2}\left(\vec{q}\right)}{d~\mathrm{ln}~\mu}\right\vert_{\alpha_d}-\left.\frac{d~\Gamma^{CT;(0);(\nu,s)}_{\mu,\theta_1,\theta_2}\left(\vec{q}\right)}{d~\mathrm{ln}~\mu}\right\vert_{\tilde{\alpha}_d}\right) 
\sim
\frac{\beta_d\left(\frac{Ng_{\mtheta}\mathbf{K}_{F,\mtheta}}{\left|X_{\mtheta}\right||\chi_{\vartheta^{-1}\left(\mtheta\right)}|v_{F,\mtheta}}\right)^{\frac{3}{2}}
g_{\theta_1}
        \mathcal{C}^{(\nu)}_d\left(\mathscr{L}_{\mu,\theta_1}\left(\vec{q}\right)\right)\left(\tilde{\alpha}_d-\alpha_d\right)\frac{\mu^{\frac{7}{2}}}{\sqrt{q(\theta_1,\theta_2)^2+\mu^2}}}{\left[q(\theta_1,\theta_2)^2+\alpha_d\beta_d\left(\frac{Ng_{\mtheta}\mathbf{K}_{F,\mtheta}}{\left|X_{\mtheta}\right||\chi_{\vartheta^{-1}\left(\mtheta\right)}|v_{F,\mtheta}}\right)^{\frac{3}{2}}
        \frac{\mu^{\frac{3}{2}}}{\sqrt{q(\theta_1,\theta_2)^2+\mu^2}}\right]\left[\alpha_d\rightarrow \tilde{\alpha}_d\right]}.
\end{align*}
The finite difference that arises for small
$q(\theta_1,\theta_2)$
does not affect the low-energy universal physics
because the four-fermion coupling itself has a negative scaling dimension\cite{BORGES2023169221}.
} 
that different choices of 
$\alpha_d$ only alter terms that are suppressed by  
$\left(\sim 
\KFtheta^{-\frac{7}{2}}\right)$. 
Therefore, we set $\alpha_d$ to be $1$.

\section{Derivation of the fixed point profile of the four-fermion coupling}

In this appendix, we derive the fixed point profile of four-fermion coupling in different kinematic regions controlled by three characteristic length scales $l^{*}_{\bar{\theta}_1,\bar{\theta}_2}$, $l^{*}_{\bar{\theta}_1,\vec{q}}$ and $l^{*}_{\bar{\theta}_2,\vec{q}}$ case by case.
 
\subsection{
\texorpdfstring{$l^{*}_{\bar{\theta}_1,\bar{\theta}_2}$,  $l^{*}_{\bar{\theta}_1,\vec{q}},~l^{*}_{\bar{\theta}_2,\vec{q}}
 \gg l$}{Lg} :
  Eq. (\ref{eq:high_energy_lambda})
}
\label{sec:case1}
If the RG scale is smaller than all three characteristic length scales, 
the full source in Eq. (\ref{Rescaled_Source}) in regime $\left|\bar{\theta}_1-\bar{\theta}_2\right|\ll\sqrt{\mu}$ contributes to the four-fermion coupling.
In this case,
Eq. (\ref{lambda_general_solution}) can be explicitly solved to 
\begin{equation}
    \begin{aligned}
        \bar{\lambda}^{(\nu)}_{\bar{\theta}_1,\bar{\theta}_1}(l) 
        &= 
        e^{-\DD l}
        \left[\bar{\lambda}^{UV(\nu)}_{\bar{\theta}_1,\bar{\theta}_1}+\frac{g^{*2}\varsigma_d^{(\nu)}}{\DD }\right] -\frac{g^{*2}\varsigma_d^{(\nu)}}{\DD }.
        \label{appendix:4fcase1_solution}
    \end{aligned}
\end{equation}
where
        $\DD = \HD -
        \eta_d^{(\nu)}
        $
with $\eta_d^{(\nu)} = 2g^{*}\rho_d^{(\nu)}$
being
the channel-dependent anomalous dimension that arises from the vertex correction.
$g^{*} = \frac{5-2d}{2(d-1)}
\frac{1}{u_1(d)}$
is the fixed point value of the effective Yukawa coupling 
with $u_1(d)$ defined in Eq. (\ref{appendix:u1_d}). 
$\varsigma_d^{(\nu)}$ in Eq. (\ref{appendix:4fcase1_solution}) is an $O(1)$ constant that depends on dimension as
\begin{equation}
    \begin{aligned}
        \varsigma_d^{(\nu)} =
        \frac{1}
        {48\pi}
        \frac{\Omega_{d-1}}{\beta_d^{\frac{2}{3}}(2\pi)^{d-1}}\left[\frac{A_+^{(\nu)}(d)}{2}-\frac{A^{(\nu)}_-(d)}{d-1}\right].
    \end{aligned}
\end{equation}
Near the upper critical dimension with 
$\DD  > 1/2$,
the four-fermion coupling function takes the universal form in the low-energy limit,
        $\left(\bar{\lambda}^{(\nu)}_{\theta_1,\theta_1}\right)^{*} = -
        \frac{g^{*2}\varsigma_d^{(\nu)}}{\DD }$.

\subsection{
\texorpdfstring{$l^{*}_{\bar{\theta}_1,\bar{\theta}_2}
 \gg l\gg l^{*}_{\bar{\theta}_1,\vec{q}}$}{Lg} :
 Eq. (\ref{eq:intra_low})
}
\label{sec:case2}

When the RG scale is smaller than $l^{*}_{\bar{\theta}_1,\bar{\theta}_2}$ but greater than $l^{*}_{\bar{\theta}_1,\vec{q}}\sim l^{*}_{\bar{\theta}_2,\vec{q}}$, 
the source is suppressed 
at sufficiently low energies
due to the cross-over defined in Eq. (\ref{appendix:crossover}). 
This leads to
\begin{equation}
    \begin{aligned}
        \bar{\lambda}^{(\nu)}_{\bar{\theta}_1,\bar{\theta}_1}\left(\vec{q};l\right) 
        &= 
        e^{-\left(\HD l- \eta_d^{(\nu)}l^{*}_{\bar{\theta}_1,\vec{q}}\right)} \left[\bar{\lambda}^{UV(\nu)}_{\bar{\theta}_1,\bar{\theta}_1}\left(\vec{q}\right)
        -g^{*2}\varsigma_d^{(\nu)}\int_0^{l^{*}_{\bar{\theta}_1,\vec{q}}}
        dl^\prime e^{\DD l^\prime} 
         \right] \\
         &= 
       e^{-\left(\HD l- \eta_d^{(\nu)}l^{*}_{\bar{\theta}_1,\vec{q}}\right)} 
        \left[\bar{\lambda}^{UV(\nu)}_{\bar{\theta}_1,\bar{\theta}_1}\left(\vec{q}\right)+\frac{g^{*2}\varsigma_d^{(\nu)}}{\DD }\right] -\frac{g^{*2}\varsigma_d^{(\nu)}}{\DD }e^{\HD\left(l^{*}_{\bar{\theta}_1,\vec{q}}-l\right)} .
        \label{appendix:lambda_bar_case2}
    \end{aligned}
\end{equation}
We don't expect Eq. (\ref{appendix:lambda_bar_case2})
to be scale invariant 
because it depends on $l$ explicitly.
In order to see a scale invariance, one has to 
take the large $l$ limit while fixing the rescaled fermionic angle $\hat{\theta}$, momentum transfer (or center of mass momentum of cooper pair) $\hat{q} $ and Fermi momentum $\hat{\mathbf{K}}_{F,\hat{\theta}}$, where
\bqa
\hat{\theta} = \bar{\theta} e^{\frac{l}{2}}, ~\hat{q} = q e^{l},~\hat{\mathbf{K}}_{F,\hat{\theta}} = \bar{\mathbf{K}}_{F,\hat{\theta}e^{-\frac{l}{2}}}e^{l}.
\eqa
Accordingly, we define the coupling function in the rescaled space as
\begin{equation}
    \begin{aligned}
    \hat{\lambda}^{(\nu)}_{\hat{\theta}_1,\hat{\theta}_2}\left(\hat{q},\varphi\right) = \bar{\lambda}^{(\nu)}_{\hat{\theta}_1e^{-\frac{l}{2}},\hat{\theta}_2e^{-\frac{l}{2}}}\left(\hat{q}e^{-l},\varphi\right).
    \end{aligned}
\end{equation}
In the low-energy limit, the range of $\hat \theta$ is stretched out\cite{BORGES2023169221}.
In the rescaled space,
Eq. (\ref{appendix:lambda_bar_case2})  takes the scale-invariant form,
\begin{equation}
    \begin{aligned}
        \hat{\lambda}^{(\nu)}_{\hat{\theta}_1,\hat{\theta}_1}\left(\hat{q},\varphi;l\right) 
        =
       e^{-\DD l} 
        \left[\hat{\lambda}^{UV(\nu)}_{\hat{\theta}_1,\hat{\theta}_1}\left(\hat{q},\varphi\right)+\frac{g^{*2}\varsigma_d^{(\nu)}}{\DD }\right]\left(\frac{\Lambda}{\hat{L}_{\hat{\theta}_1}\left(\hat{q},\varphi\right)}\right)^{ \eta_d^{(\nu)}} -\frac{g^{*2}\varsigma_d^{(\nu)}}{\DD }\left(\frac{\Lambda}{\hat{L}_{\hat{\theta}_1}\left(\hat{q},\varphi\right)}\right)^{
        \DD+ \eta_d^{(\nu)}
        } ,
    \end{aligned}
\end{equation}
where
$\hat{L}_{\hat{\theta}}\left(\hat{q},\varphi\right) = \hat{v}_{F,\hat{\theta}}\left(\hat{\mathscr{F}}_{\varphi,\hat{\theta}}\hat{q}+
\hat{\mathscr{G}}_{\varphi,\hat{\theta}}
\frac{\hat{q}^2}{\hat K_{F,\hat \theta}}\right)$,
$\hat{v}_{F,\hat{\theta}} = \bar{v}_{F,\hat{\theta}e^{-\frac{l}{2}}}$,
$\hat{\mathscr{F}}_{\varphi,\hat{\theta}}= \bar{\mathscr{F}}_{\varphi,\hat{\theta}e^{-\frac{l}{2}}}$ and $\hat{\mathscr{G}}_{\varphi,\hat{\theta}}= \bar{\mathscr{G}}_{\varphi,\hat{\theta}e^{-\frac{l}{2}}}$.
Near $d_c$, the coupling function takes the universal form as
        $\hat{\lambda}^{{}^{*}(\nu)}_{\hat{\theta}_1,\hat{\theta}_1}\left(\hat{q},\varphi\right) \approx -\frac{g^{*2}\varsigma_d^{(\nu)}}{\DD }\left(\frac{\Lambda}{\hat{L}_{\hat{\theta}_1}\left(\hat{q},\varphi\right)}\right)^{
        \DD+ \eta_d^{(\nu)}}$.

\subsection{
\texorpdfstring{$l^{*}_{\bar{\theta}_1,\vec{q}},~l^{*}_{\bar{\theta}_2,\vec{q}}
 \gg l \gg l^{*}_{\bar{\theta}_1,\bar{\theta}_2} $}{Lg}
 :  Eq. (\ref{eq:inter_mid1})
}
\label{sec:case3}

In this region, 
Eq. (\ref{lambda_general_solution}) becomes
\begin{equation}
    \begin{aligned}
        \bar{\lambda}^{(\nu)}_{\bar{\theta}_1,\bar{\theta}_2}(l) 
        &= 
        e^{-\DD l} \left[\bar{\lambda}^{UV(\nu)}_{\bar{\theta}_1,\bar{\theta}_2}
        -g^{*2}\varsigma_d^{(\nu)}\int_0^{l^{*}_{\bar{\theta}_1,\bar{\theta}_2}}
        dl^\prime e^{\DD l^\prime} 
        +
        \frac{\snu g^{*2}\rho_d^{(\nu)}}
        {2\beta_d^{\frac{1}{3}}}\int_{l^{*}_{\bar{\theta}_1,\bar{\theta}_2}}^l
        dl'\frac{\Lambda|\bar{\theta}_1-\bar{\theta}_2|e^{-\left(1-\DD \right) l'}}{|\bar{\theta}_1-\bar{\theta}_2|^3+\Lambda^{\frac{3}{2}}e^{-\frac{3}{2}l'}} \right] \\
&= e^{-\DD l} \left[\bar{\lambda}^{UV(\nu)}_{\bar{\theta}_1,\bar{\theta}_2}
+\frac{g^{*2}\varsigma_d^{(\nu)}}{\DD }\left[1-e^{\DD l^{*}_{\bar{\theta}_1,\bar{\theta}_2}} \right]
+
\frac{\snu g^{*2}\rho_d^{(\nu)}}
{\left(1+2\DD\right) \beta_d^{\frac{1}{3}}}
        \frac{\left|\bar{\theta}_1-\bar{\theta}_2\right|}{\sqrt{\Lambda } }  
        \right.
        \\
        & \times
        \Bigg\{
        e^{
        l \left(
        \frac{1}{2}+\DD 
        \right)
        } 
        2F_1
        \left(
        1,\frac{1+2\DD }{3};\frac{4+2\DD }{3};
        -\frac{
        e^{3 l/2} 
        \left| \bar{\theta}_1- \bar{\theta}_2 \right|^3}{ \Lambda^{3/2} } \right)  \\
        &\left.
        -e^{l^{*}_{\bar{\theta}_1,\bar{\theta}_2} \left(
        \frac{1}{2}+\DD 
        \right)
        } \,_2F_1\left(1,\frac{1+2\DD }{3};\frac{4+2\DD }{3};-\frac{e^{\frac{3l^{*}_{\bar{\theta}_1,\bar{\theta}_2}}{2}}\left|\bar{\theta}_1-\bar{\theta}_2\right|^3}{\Lambda ^{3/2} }\right)
        \Bigg\}
        \right].
\end{aligned}
\label{appendix:lambda_bar_case3}
\end{equation}
In the rescaled space,
the coupling function takes the scale-invariant form as
\begin{equation}
    \begin{aligned}
        \hat{\lambda}^{(\nu)}_{\hat{\theta}_1,\hat{\theta}_2}(l) &= 
 e^{-\DD l}
\left[\hat{\lambda}^{UV(\nu)}_{\hat{\theta}_1,\hat{\theta}_2}+\frac{g^{*2}\varsigma_d^{(\nu)}}{\DD }\right] -\frac{g^{*2}\varsigma_d^{(\nu)}}{\DD }e^{\DD \hat{l}^{*}_{\hat{\theta}_1,\hat{\theta}_2} }
       \\ &
       +
       \frac{\snu g^{*2}\rho_d^{(\nu)}}
       {\left(1+2\DD\right) \beta_d^{\frac{1}{3}}}
        \frac{\left|\hat{\theta} _1-\hat{\theta} _2\right|}{\sqrt{\Lambda } }  
        \left[\,
        _2F_1\left(1,\frac{1+2\DD }{3};\frac{4+2\DD }{3};-\frac{\left|\hat{\theta} _1-\hat{\theta}
        _2\right|^3}{\Lambda ^{3/2} }\right) 
        \right. 
        \\
        &
        \left.
        ~~~ 
        -e^{\left(\DD+\frac{1}{2}\right) \hat{l}^{*}_{\hat{\theta}_1,\hat{\theta}_2} }
        \,_2F_1\left(1,\frac{1+2\DD }{3};\frac{4+2\DD }{3};-1\right)\right].
\label{appendix:4f_case3_rescaled_explicit}
    \end{aligned}
\end{equation}
In the $\left|\hat{\theta}_1-\hat{\theta}_2\right|\gg \sqrt{\Lambda}$ limit, 
Eq. (\ref{appendix:4f_case3_rescaled_explicit}) can be concisely written as
\begin{equation}
    \begin{aligned}
    \hat{\lambda}^{(\nu)}_{\hat{\theta}_1,\hat{\theta}_2}(l) = e^{ -\DD l}\left[\hat{\lambda}^{UV(\nu)}_{\hat{\theta}_1,\hat{\theta}_2}\left(\hat{q},\varphi\right)+\frac{g^{*2}\varsigma_d^{(\nu)}}{\DD}\right]
        +Y_{d}^{(\nu)} \left|\frac{\sqrt{\Lambda}}{\hat{\theta}_1-\hat{\theta}_2}\right|^{ 2\DD 
        }.
    \end{aligned}
\end{equation}
Near $d_c$, 
the  coupling function takes a universal form,
\begin{equation}
    \begin{aligned}
        \hat{\lambda}^{{}^*(\nu)}_{\hat{\theta}_1,\hat{\theta}_2} \approx 
       Y_{d}^{(\nu)}
      \left|\frac{\sqrt{\Lambda}}{\hat{\theta}_1-\hat{\theta}_2}\right|^{
       2\DD }.
        \label{eq:lambda_case_1}
    \end{aligned}
\end{equation}
where
\begin{equation}
    \begin{aligned}
         Y_{d}^{(\nu)} &= g^{*}\left(\frac{\snu\eta_d^{(\nu)}} 
         {
         12
         \beta_d^{\frac{1}{3}}}
         \bigg\{\psi ^{(0)}\left(
   \frac{2\DD   +1}{6}\right)-\psi ^{(0)}\left(
   \frac{\DD   +2}{3}
   \right)+2 \pi~
\mathrm{cosec} \left(
\frac{\pi }{3}  \left(2\DD+1 \right)
\right)
\bigg\}-\frac{ g^{*}\varsigma_d^{(\nu)}}{ \DD }
   \right),
   \label{appendix:M1d}
    \end{aligned}
\end{equation}
Here, $\psi^{(m)}(z)$ is the polygamma function of order m. 

\subsection{\texorpdfstring{$l^{*}_{\bar{\theta}_2,\vec{q}} \gg l \gg 
 l^{*}_{\bar{\theta}_1,\vec{q}} \gg l^{*}_{\bar{\theta}_1,\bar{\theta}_2}
 $}{Lg} 
:  Eq. (\ref{eq:inter_mid2})
 }\label{sec:case4}
When the RG length scale is much smaller than the crossover scale at $\theta_2$ and much bigger than the crossover length scale at $\theta_1$, 
the solution to Eq. (\ref{lambda_general_solution}) becomes
\begin{equation}
    \begin{aligned}
        \bar{\lambda}^{(\nu)}_{\bar{\theta}_1,\bar{\theta}_2}(\vec{q};l) &= e^{-\bar{\Delta}_d^{(\nu)}l+ \frac{\eta_d^{(\nu)}}{2}l^{*}_{\bar{\theta}_1,\vec{q}}}\left[\bar{\lambda}^{UV(\nu)}_{\bar{\theta}_1,\bar{\theta}_2}(\vec{q})
        +\frac{g^{*2}\varsigma_d^{(\nu)}}{\DD }\left(1-e^{\DD l^{*}_{\bar{\theta}_1,\bar{\theta}_2}} \right)\right]
        \\&
        \hspace{-60pt}
        +\frac{\snu g^{*2}\rho_d^{(\nu)}}
         {2\beta_d^{\frac{1}{3}}}\frac{\left|\bar{\theta}_1 - \bar{\theta}_2 \right|}{\sqrt{\Lambda } }
        \left\{
         \frac{2}
        {
       1+2\DD }\left(\frac{\,_2F_1\left(1,\frac{1+2\DD }{3};\frac{4+2\DD }{3};-\frac{e^{\frac{3 l^{*}_{\bar{\theta}_1,\vec{q}}}{2}} \left|\bar{\theta}_1 - \bar{\theta}_2 \right|^3}{\Lambda ^{3/2} }\right)}{e^{\bar{\Delta}_d^{(\nu)}l-\left(\bar{\Delta}_d^{(\nu)}+\frac{1}{2}\right)l^{*}_{\bar{\theta}_1,\vec{q}}}}
        -
        \frac{\,_2F_1\left(1,\frac{1+2\DD }{3};\frac{4+2\DD }{3};-\frac{e^{\frac{3l^{*}_{\bar{\theta}_1,\bar{\theta}_2}}{2}}\left|\bar{\theta}_1 - \bar{\theta}_2 \right|^3}{\Lambda ^{3/2} }\right)}{e^{\bar{\Delta}_d^{(\nu)}l-\left(\frac{1}{2}+\DD\right)l^{*}_{\bar{\theta}_1,\bar{\theta}_2}- \frac{\eta_d^{(\nu)}}{2}l^{*}_{\bar{\theta}_1,\vec{q}}}}\right)
        \right.\\&\left.
        +
        \frac{1}
        {
        1+2\bar{\Delta}_d^{(\nu)}}
        \left(e^{\frac{l}{2}}\,
        _2F_1\left(1,\frac{1+2\bar{\Delta}_d^{(\nu)}}{3};\frac{4+2\bar{\Delta}_d^{(\nu)}}{3};-\frac{e^{\frac{3 l}{2}} \left|\bar{\theta}_1 - \bar{\theta}_2 \right|^3}{\Lambda ^{3/2} }\right)
        -
        \frac{\,_2F_1\left(1,\frac{1+2\bar{\Delta}_d^{(\nu)}}{3};\frac{4+2\bar{\Delta}_d^{(\nu)}}{3};-\frac{e^{\frac{3l^{*}_{\bar{\theta}_1\vec{q}}}{2}}\left|\bar{\theta}_1 - \bar{\theta}_2 \right|^3}{\Lambda ^{3/2} }\right)}{e^{\bar{\Delta}_d^{(\nu)}l-\left(\bar{\Delta}_d^{(\nu)}+\frac{1}{2}\right)l^{*}_{\bar{\theta}_1,\vec{q}}}}\right) 
        \right\}, \label{solution_case2}
    \end{aligned}
\end{equation}
where
        $\bar{\Delta}_d^{(\nu)} = \DD + \frac{\eta_d^{(\nu)}}{2}$.
In the rescaled coordinate,
Eq. (\ref{solution_case2}) gives
\begin{equation}
    \begin{aligned}
        \hat{\lambda}^{(\nu)}_{\hat{\theta}_1,\hat{\theta}_2}\left(\hat{q},\varphi;l\right) &= e^{-\DD l}\left[\hat{\lambda}^{UV(\nu)}_{\hat{\theta}_1,\hat{\theta}_2}\left(\hat{q},\varphi \right)+\frac{g^{*2}\varsigma_d^{(\nu)}}{\DD }\right]e^{ \frac{\eta_d^{(\nu)}}{2}\hat{l}^{*}_{\hat{\theta}_1,\hat{q},\varphi}}-\frac{g^{*2}\varsigma_d^{(\nu)}}{\DD }e^{\DD \hat{l}^{*}_{\hat{\theta}_1,\hat{\theta}_2}+ \frac{\eta_d^{(\nu)}}{2}\hat{l}^{*}_{\hat{\theta}_1,\hat{q},\varphi}}\\&
        \hspace{-20pt}
        +
        \frac{\snu g^{*2}\rho_d^{(\nu)}}
        {2\beta_d^{\frac{1}{3}}}\frac{\left|\hat{\theta}_1-\hat{\theta}_2\right|}{\sqrt{\Lambda } }
        \left\{
         \frac{2}
        {
       1+2\DD }
       \left(
       \frac{\,
        _2F_1\left(1,\frac{1+2\DD }{3};\frac{4+2\DD }{3};-\frac{e^{\frac{3 \hat{l}^{*}_{\hat{\theta}_1,\hat{q},\varphi}}{2}} \left|\hat{\theta}_1-\hat{\theta}_2\right|^3}{\Lambda ^{3/2} }\right)}
        {e^{-\left(\bar{\Delta}_d^{(\nu)}+\frac{1}{2}\right)\hat{l}^{*}_{\hat{\theta}_1,\hat{q},\varphi}}}
        -
        \frac{\,_2F_1\left(1,\frac{1+2\DD }{3};\frac{4+2\DD }{3};-1\right)}{e^{-\left(\DD+\frac{1}{2}\right)\hat{l}^{*}_{\hat{\theta}_1,\hat{\theta}_2}- \frac{\eta_d^{(\nu)}}{2}\hat{l}^{*}_{\hat{\theta}_1,\hat{q},\varphi}}}
        \right)
        \right.\\&\left.+
          \frac{1}
        {
        1+2\bar{\Delta}_d^{(\nu)}}
        \left(\,
        _2F_1\left(1,\frac{1+2\bar{\Delta}_d^{(\nu)}}{3};\frac{4+2\bar{\Delta}_d^{(\nu)}}{3};-\frac{ \left|\hat{\theta}_1-\hat{\theta}_2\right|^3}{\Lambda ^{3/2} }\right)
        -
        \frac{\,_2F_1\left(1,\frac{1+2\bar{\Delta}_d^{(\nu)}}{3};\frac{4+2\bar{\Delta}_d^{(\nu)}}{3};-\frac{e^{\frac{3 \hat{l}^{*}_{\hat{\theta}_1,\hat{q},\varphi}}{2}} \left|\hat{\theta}_1-\hat{\theta}_2\right|^3}{\Lambda ^{3/2} }\right)}{e^{-\left(\bar{\Delta}_d^{(\nu)}+\frac{1}{2}\right)\hat{l}^{*}_{\hat{\theta}_1,\hat{q},\varphi}}}\right) 
        \right\}. 
        \label{appendix:4f_case4_rescaled_solution}
    \end{aligned}
\end{equation}
In the $\left|\hat{\theta}_1-\hat{\theta}_2\right|\gg \sqrt{\hat{L}_{\hat{\theta}_1}\left(\hat{q},\varphi\right)} \gg
\sqrt{\Lambda}$ limit,
Eq. (\ref{appendix:4f_case4_rescaled_solution}) can be written in a compact form as
\begin{equation}
    \begin{aligned}
    \hat{\lambda}^{(\nu)}_{\hat{\theta}_1,\hat{\theta}_2}\left(\hat{q},\varphi;l\right) =  
        \left[
    e^{ -\DD l}\left(\hat{\lambda}^{UV(\nu)}_{\hat{\theta}_1,\hat{\theta}_2}\left(\hat{q},\varphi\right)+\frac{g^{*2}\varsigma_d^{(\nu)}}{\DD}\right)+
    Y_{d}^{(\nu)}
    \left|\frac{\sqrt{\Lambda}}{\hat{\theta}_1-\hat{\theta}_2}\right|^{
       2\DD }       \right]
    \left(\frac{\Lambda}{\hat{L}_{\hat{\theta}_1}\left(\hat{q},\varphi\right)}\right)^{ \frac{\eta_d^{(\nu)}}{2}},
    \end{aligned}
\end{equation}
where $Y_{d}^{(\nu)}$ is given in Eq. (\ref{appendix:M1d}).
The coupling takes the following universal form near $d_c$,
\begin{equation}
    \begin{aligned}
    \hat{\lambda}^{{}^{*}(\nu)}_{\hat{\theta}_1,\hat{\theta}_2}\left(\hat{q},\varphi\right) \approx
    Y_{d}^{(\nu)}
    \left|\frac{\sqrt{\Lambda}}{\hat{\theta}_1-\hat{\theta}_2}\right|^{
       2\DD }\left(\frac{\Lambda}{\hat{L}_{\hat{\theta}_1}\left(\hat{q},\varphi\right)}\right)^{ \frac{\eta_d^{(\nu)}}{2}}.
    \end{aligned}
\end{equation}

\subsection{
\texorpdfstring{$l \gg 
l^{*}_{\bar{\theta}_2,\vec{q}} 
\gg 
l^{*}_{\bar{\theta}_1,\vec{q}} \gg l^{*}_{\bar{\theta}_1,\bar{\theta}_2}$}{Lg}
:  Eq. (\ref{eq:inter_low})
}
\label{sec:case5}

When the RG scale is much bigger than the crossover scale at $\theta_1$ and $\theta_2$, and the characteristic length scale that controls inter-patch scatterings, the solution to Eq. (\ref{lambda_general_solution})
becomes
\begin{equation}
    \begin{aligned}
        \bar{\lambda}^{(\nu)}_{\bar{\theta}_1,\bar{\theta}_2}(\vec{q};l) &= e^{-\HD l+ \frac{\eta_d^{(\nu)}}{2}\left(l^{*}_{\bar{\theta}_1,\vec{q}}+l^{*}_{\bar{\theta}_2,\vec{q}}\right)}\left[\bar{\lambda}^{UV(\nu)}_{\bar{\theta}_1,\bar{\theta}_2}(\vec{q})
        +\frac{g^{*2}\varsigma_d^{(\nu)}}{\DD }\left(1-e^{\DD l^{*}_{\bar{\theta}_1,\bar{\theta}_2}} \right)\right]\\&
        \hspace{-60pt}
        +
        \frac{\snu g^{*2}\rho_d^{(\nu)}}
        {2\beta_d^{\frac{1}{3}}}\frac{\left|\bar{\theta}_1 - \bar{\theta}_2 \right|}{\sqrt{\Lambda } } 
        \left\{
        \frac{2}
        {
       1+2\DD }
       \left(
         \frac{\,
        _2F_1\left(1,\frac{1+2\DD }{3};\frac{4+2\DD }{3};-\frac{e^{3 l^{*}_{\bar{\theta}_1,\vec{q}}/2} \left|\bar{\theta}_1 - \bar{\theta}_2 \right|^3}{\Lambda ^{3/2} }\right)}
        {e^{\HD l-\left(\bar{\Delta}_d^{(\nu)}+\frac{1}{2}\right)l^{*}_{\bar{\theta}_1,\vec{q}}- \frac{\eta_d^{(\nu)}}{2}l^{*}_{\bar{\theta}_2,\vec{q}}}}
        -
        \frac{\,_2F_1\left(1,\frac{1+2\DD }{3};\frac{4+2\DD }{3};-\frac{e^{\frac{3l^{*}_{\bar{\theta}_1,\bar{\theta}_2}}{2}}\left|\bar{\theta}_1 - \bar{\theta}_2 \right|^3}{\Lambda ^{3/2} }\right)}{ e^{\HD l-\left(\DD+\frac{1}{2}\right)l^{*}_{\bar{\theta}_1,\bar{\theta}_2}- \frac{\eta_d^{(\nu)}}{2}\left(l^{*}_{\bar{\theta}_1,\vec{q}}+l^{*}_{\bar{\theta}_2,\vec{q}}\right)}}\right)\right.\\&\left.+
         \frac{1}
        {
        1+2\bar{\Delta}_d^{(\nu)}}
        \left(
        \frac{\,_2F_1\left(1,\frac{1+2\bar{\Delta}_d^{(\nu)}}{3};\frac{4+2\bar{\Delta}_d^{(\nu)}}{3};-\frac{e^{\frac{3 l^{*}_{\bar{\theta}_2,\vec{q}}}{2}} \left|\bar{\theta}_1 - \bar{\theta}_2 \right|^3}{\Lambda ^{3/2} }\right)}{ e^{\HD \left(l-l^{*}_{\bar{\theta}_2,\vec{q}}\right)-\frac{l^{*}_{\bar{\theta}_2,\vec{q}}}{2}}}
        -
        \frac{\,_2F_1\left(1,\frac{1+2\bar{\Delta}_d^{(\nu)}}{3};\frac{4+2\bar{\Delta}_d^{(\nu)}}{3};-\frac{e^{3 l^{*}_{\bar{\theta}_1,\vec{q}}/2} \left|\bar{\theta}_1 - \bar{\theta}_2 \right|^3}{\Lambda ^{3/2} }\right)}{e^{\HD l-\left(\bar{\Delta}_d^{(\nu)}+\frac{1}{2}\right)l^{*}_{\bar{\theta}_1,\vec{q}}- \frac{\eta_d^{(\nu)}}{2}l^{*}_{\bar{\theta}_2,\vec{q}}}}\right) 
        \right\}. \label{solution_case3}
    \end{aligned}
\end{equation}
The coupling function in the rescaled coordinate becomes
\begin{equation}
    \begin{aligned}
        \hat{\lambda}^{(\nu)}_{\hat{\theta}_1,\hat{\theta}_2}\left(\hat{q},\varphi;l\right) &=
        e^{-\DD l}\left(\hat{\lambda}^{UV(\nu)}_{\hat{\theta}_1,\hat{\theta}_2}\left(\hat{q},\varphi \right)+\frac{g^{*2}\varsigma_d^{(\nu)}}{\DD }\right)e^{ \frac{\eta_d^{(\nu)}}{2}\left(\hat{l}^{*}_{\hat{\theta}_1,\hat{q},\varphi}+\hat{l}^{*}_{\hat{\theta}_2,\hat{q},\varphi}\right)}
        -\frac{g^{*2}\varsigma_d^{(\nu)}}{\DD }e^{\DD \hat{l}^{*}_{\hat{\theta}_1,\hat{\theta}_2}+ \frac{\eta_d^{(\nu)}}{2}\left(\hat{l}^{*}_{\hat{\theta}_1,\hat{q},\varphi}+\hat{l}^{*}_{\hat{\theta}_2,\hat{q},\varphi}\right)}
        \\&
        \hspace{-40pt}
        +
        \frac{\snu g^{*2}\rho_d^{(\nu)}}
        {2\beta_d^{\frac{1}{3}}}\frac{\left|\hat{\theta}_1-\hat{\theta}_2\right|}{\sqrt{\Lambda } }
        \left\{
        \frac{2}
        {
       1+2\DD }
       \left(
       \frac{\,
        _2F_1\left(1,\frac{1+2\DD }{3};\frac{4+2\DD }{3};-\frac{e^{\frac{3 \hat{l}^{*}_{\hat{\theta}_1,\hat{q},\varphi}}{2}} \left|\hat{\theta}_1-\hat{\theta}_2\right|^3}{\Lambda ^{3/2} }\right)}{e^{-\left(\bar{\Delta}_d^{(\nu)}+\frac{1}{2}\right)\hat{l}^{*}_{\hat{\theta}_1,\hat{q},\varphi}- \frac{\eta_d^{(\nu)}}{2}\hat{l}^{*}_{\hat{\theta}_2,\hat{q},\varphi}}}
        -
        \frac{\,_2F_1\left(1,\frac{1+2\DD }{3};\frac{4+2\DD }{3};-1\right)}{ e^{-\left(\DD +\frac{1}{2}\right)\hat{l}^{*}_{\hat{\theta}_1,\hat{\theta}_2}- \frac{\eta_d^{(\nu)}}{2}\left(\hat{l}^{*}_{\hat{\theta}_1,\hat{q},\varphi}+\hat{l}^{*}_{\hat{\theta}_2,\hat{q},\varphi}\right)}}
        \right)\right.\\&\left.+
        \frac{1}
        {
        1+2\bar{\Delta}_d^{(\nu)}}
        \left(
        \frac{\,_2F_1\left(1,\frac{1+2\bar{\Delta}_d^{(\nu)}}{3};\frac{4+2\bar{\Delta}_d^{(\nu)}}{3};-\frac{e^{\frac{3 \hat{l}^{*}_{\hat{\theta}_2,\hat{q},\varphi}}{2}} \left|\hat{\theta}_1-\hat{\theta}_2\right|^3}{\Lambda ^{3/2} }\right)}{  e^{-\left(\HD+\frac{1}{2}\right) \hat{l}^{*}_{\hat{\theta}_2,\hat{q},\varphi}}}
        -
        \frac{\,_2F_1\left(1,\frac{1+2\bar{\Delta}_d^{(\nu)}}{3};\frac{4+2\bar{\Delta}_d^{(\nu)}}{3};-\frac{e^{\frac{3 \hat{l}^{*}_{\hat{\theta}_1,\hat{q},\varphi}}{2}} \left|\hat{\theta}_1-\hat{\theta}_2\right|^3}{\Lambda ^{3/2} }\right)}{e^{-\left(\bar{\Delta}_d^{(\nu)}+\frac{1}{2}\right)\hat{l}^{*}_{\hat{\theta}_1,\hat{q},\varphi}- \frac{\eta_d^{(\nu)}}{2}\hat{l}^{*}_{\hat{\theta}_2,\hat{q},\varphi}}}
        \right) 
        \right\}. 
        \label{appendix:4f_case5_rescaled_solution}
    \end{aligned}
\end{equation}
In the 
$\left|\hat{\theta}_1-\hat{\theta}_2\right|\gg \mathrm{Max}\left\{\sqrt{\hat{L}_{\hat{\theta}_1}\left(\hat{q},\varphi\right)}, \sqrt{\hat{L}_{\hat{\theta}_2}\left(\hat{q},\varphi\right)}\right\}\gg \sqrt{\Lambda}$ limit,
Eq. (\ref{appendix:4f_case5_rescaled_solution}) can be written as
\begin{equation}
    \begin{aligned}
    \hat{\lambda}^{(\nu)}_{\hat{\theta}_1,\hat{\theta}_2}\left(\hat{q},\varphi;l\right) =  
        \left[
    e^{ -\DD l}\left(\hat{\lambda}^{UV(\nu)}_{\hat{\theta}_1,\hat{\theta}_2}\left(\hat{q},\varphi\right)+\frac{g^{*2}\varsigma_d^{(\nu)}}{\DD}\right)+
    Y_{d}^{(\nu)}
    \left|\frac{\sqrt{\Lambda}}{\hat{\theta}_1-\hat{\theta}_2}\right|^{
       2\DD }       \right]
   \left(\frac{\Lambda^2}{\hat{L}_{\hat{\theta}_1}\left(\hat{q},\varphi\right)\hat{L}_{\hat{\theta}_2}\left(\hat{q},\varphi\right)}\right)^{ \frac{\eta_d^{(\nu)}}{2}},
    \end{aligned}
\end{equation}
where $Y_{d}^{(\nu)}$ is given in Eq. (\ref{appendix:M1d}).
Near $d_c$, the fixed point profile of the coupling reads
\begin{equation}
    \begin{aligned}
    \hat{\lambda}^{{}^{*}(\nu)}_{\hat{\theta}_1,\hat{\theta}_2}\left(\hat{q},\varphi\right) \approx
     Y_{d}^{(\nu)}
    \left|\frac{\sqrt{\Lambda}}{\hat{\theta}_1-\hat{\theta}_2}\right|^{
       2\DD }\left(\frac{\Lambda^2}{\hat{L}_{\hat{\theta}_1}\left(\hat{q},\varphi\right)\hat{L}_{\hat{\theta}_2}\left(\hat{q},\varphi\right)}\right)^{ \frac{\eta_d^{(\nu)}}{2}}.
    \end{aligned}
\end{equation}



\subsection{Full crossover of the four-fermion coupling function
}
\label{sec:intra_to_inter}

\begin{figure}
  \centering
  \includegraphics[width=0.65\linewidth]{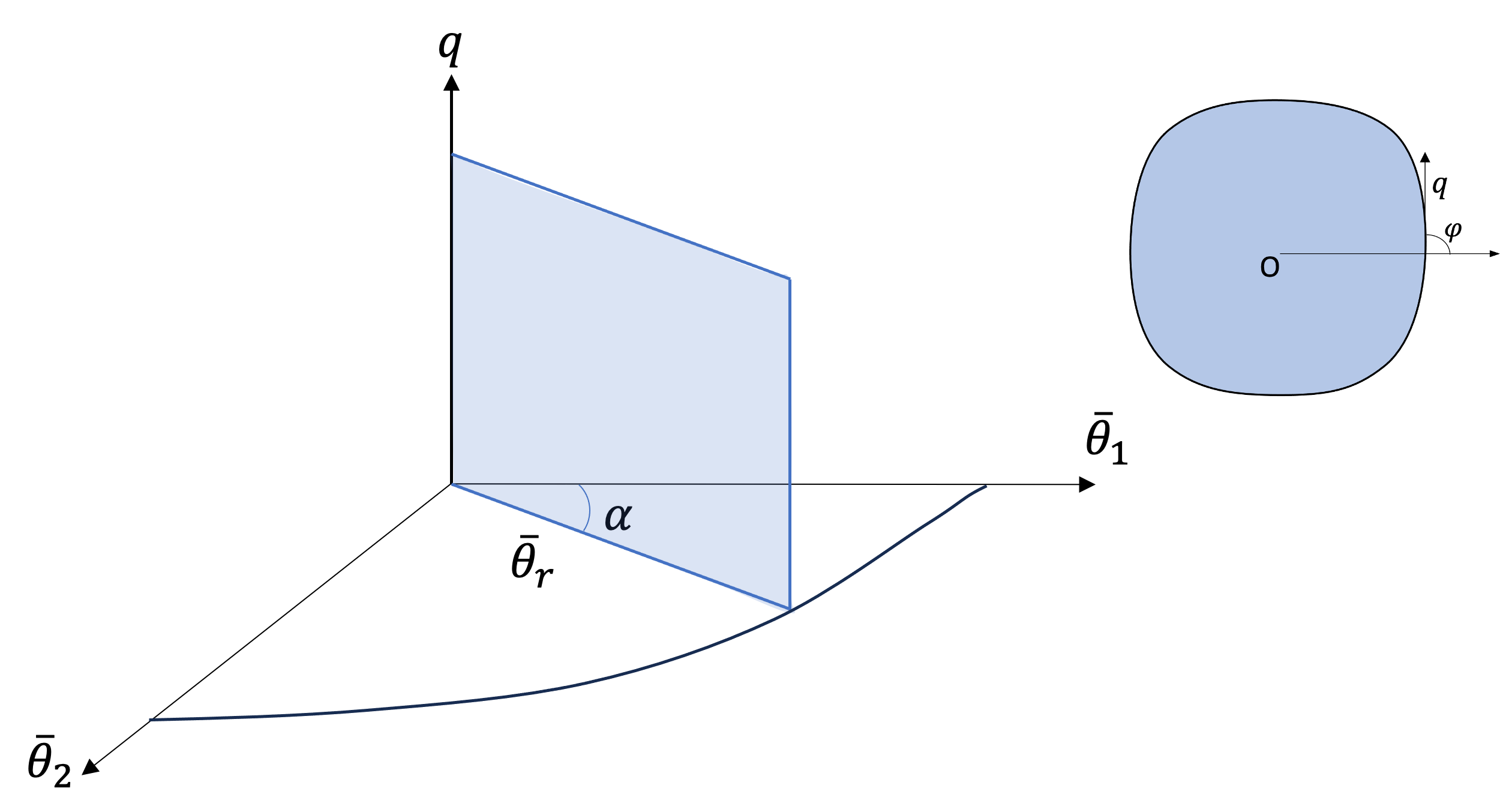}
  \caption{
The crossovers of the four-fermion coupling function are controlled 
 by $q$, $\bar \theta_1$
  and $\bar \theta_2$.
The space of these parameters can be divided into slices of two-dimensional planes composed of $\bar \theta_r = \sqrt{ \bar \theta_1^2 + \bar \theta_2^2}$ and $q$ for a fixed $\alpha = \tan^{-1} \bar \theta_2/\bar \theta_1$.
  }
  \label{Fig:4f_q_schematic}
\end{figure}

In this appendix, we summarize the full crossover behaviour that emerges as the crossover scales are independently varied as functions of 
$\bar \theta_1$, $\bar \theta_2$, $\vec q$.
Let $\vec q$ be tangential to the Fermi surface at angle $\bar \theta=0$\footnote{
The following discussion applies to $\vec q$ 
at sufficiently low energies 
unless $\vec q$ is tangential to the cold spots.
}.
Then we can focus on the three-dimensional space of  $\bar \theta_1$, $\bar \theta_2$ and $q$ as is depicted in 
  \fig{Fig:4f_q_schematic}.
To help visualize the global crossover structure, we parameterize the two angles as
$(\bar \theta_1,\bar \theta_2)= \bar \theta_r ( \cos \alpha, \sin \alpha)$ 
and focus on a two-dimensional plane spanned by $q$ and $\bar \theta_r$ for a fixed $\alpha$.

\begin{figure}
\centering
\begin{subfigure}{.5\textwidth}
  \centering
  \includegraphics[width=1\linewidth]{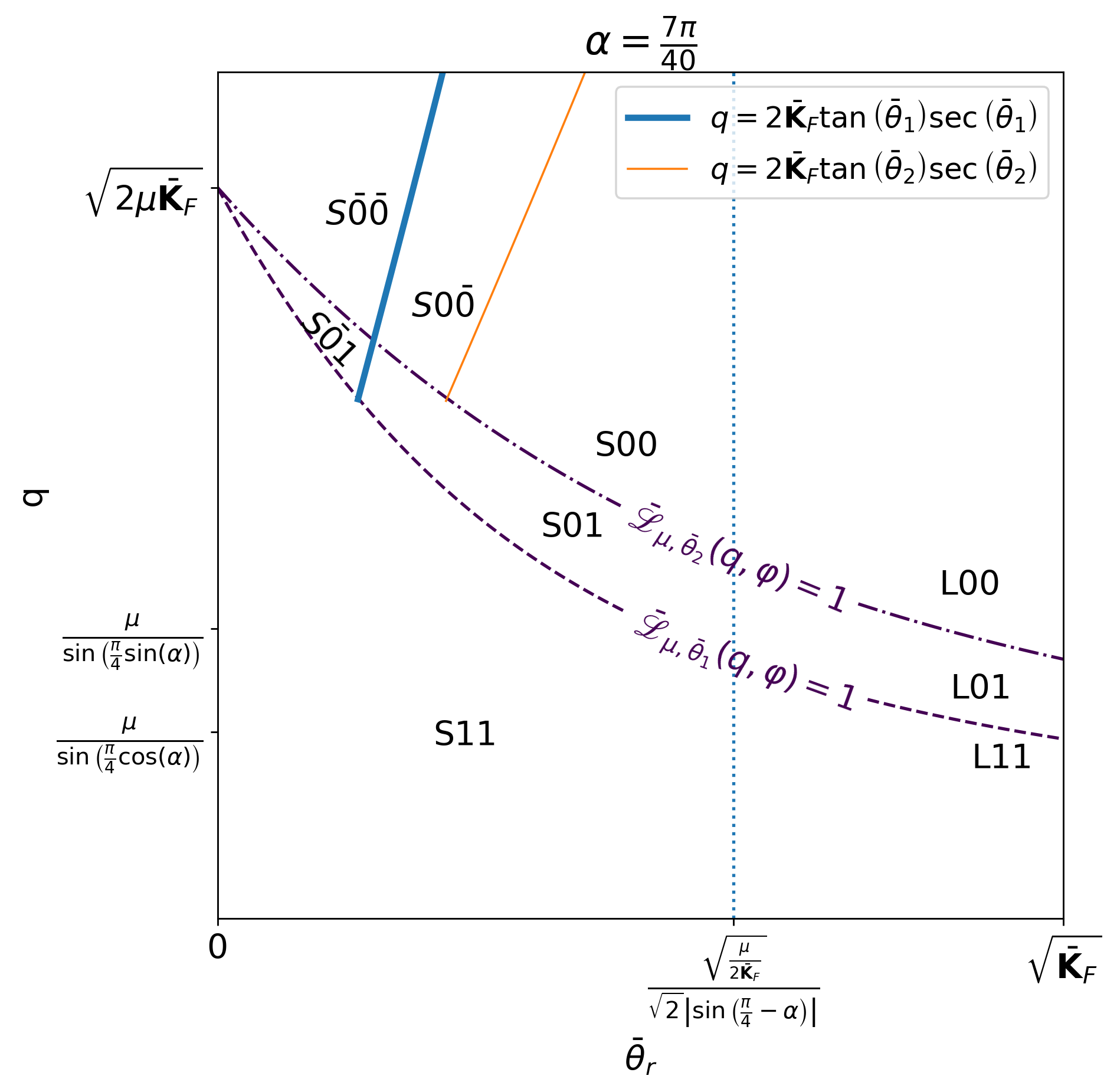}
  \caption{}
  \label{}
\end{subfigure}%
\begin{subfigure}{.5\textwidth}
  \centering\includegraphics[width=1\linewidth]{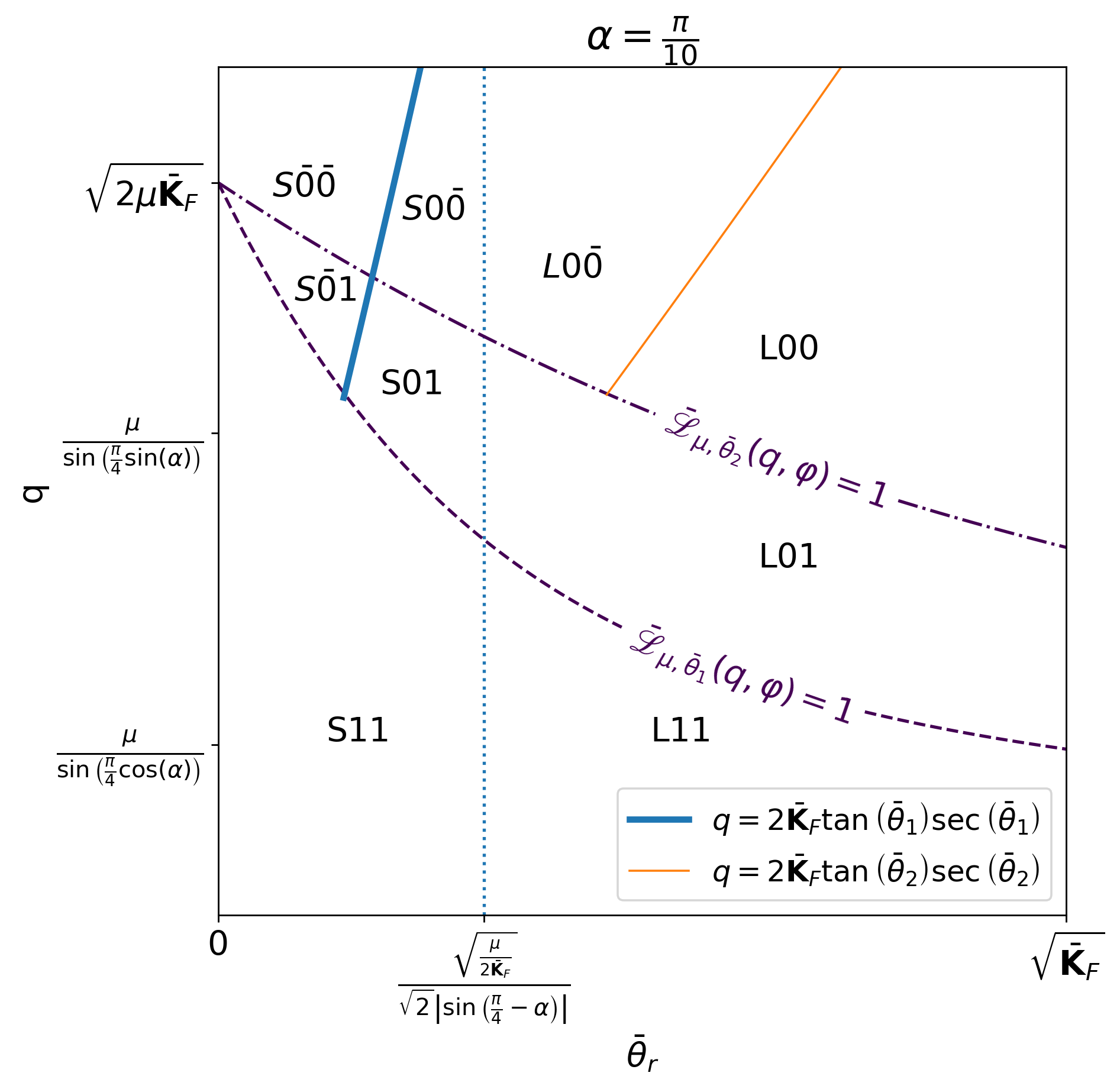}
  \caption{}
  \label{}
\end{subfigure}
\caption{All possible kinematic regions of the four fermion coupling for a circular Fermi surface with $\bar{\mathbf{K}}_F$ = 1, $\bar{v}_F$ = 1 and $\mu$ = 0.05. 
For the illustration purpose, 
we set $\beta_d^{-\frac{1}{3}}\sqrt{\frac{1}{Ng^{*}}} = 1$ 
for which 
$\bar{\theta}=\theta$.
The dashed and dash-dotted lines indicate the different values of $q$ and $\bar \theta_r$ for which dispersion of particle-hole pair with momentum $\vec{q}$ becomes equal to the energy scale $\mu$. The thick (thin) solid line indicates the different values of $q$ and $\bar \theta_r$ for which the linear term in dispersion becomes comparable to the quadratic term at angle $\bar{\theta}_1$ ($\bar{\theta}_2$).}
\label{eq:thetaq_r}
\end{figure}

At a fixed $\alpha$,
the crossover structure is shown in the plane of $\bar \theta_r$ and $q$ 
in \fig{eq:thetaq_r}.
Each region is marked by three letters
$P~V_1~V_2$, where 
$P=S$ or $L$,
and
$V_1,V_2=1,0$ or $\bar 0$.
The first letter is determined based on 
 whether $\bar \theta_r$ is smaller or larger than $\bar \theta_r^*$,
where $\bar \theta_r^*$, denoted as the vertical dashed line in the figures,
is determined from 
$l^{*}_{\bar{\theta}_1,\bar{\theta}_2}=l$.
In the region with $\bar \theta_r < \bar \theta_r^*$
 denoted with first letter $P=S$,
the separation between $\bar \theta_1$
and $\bar \theta_2$ is small (S) enough
that a fermion at $\bar \theta_1$
can be scattered to 
$\bar \theta_2$
by absorbing or emitting a critical boson with energy less than $\mu$.
In other words, two fermions are within one patch in this region.
On the other hand, 
the region with $\bar \theta_r > \bar \theta_r^*$
denoted as $P=L$ represents the region with a large (L) angular separation where $\bar \theta_1$ and $\bar \theta_2$ 
 can not be connected by a momentum of boson with energy $\mu$.
The second (third) symbol is determined by the presence or absence of the vertex correction created by virtual fermions located near angle $\bar \theta_1$ $\left(\bar \theta_2\right)$.
The crossovers caused by the vertex correction 
are denoted by the dashed (dash-dotted) curve : 
$l^{*}_{\bar \theta_1,\vec{q}} = l$
$\left(l^{*}_{\bar \theta_2,\vec{q}}=l\right)$.
In the region below the dashed (dash-dotted) curve denoted as $V_1=1$
($V_2=1$), the vertex correction from angle $\bar \theta_1$ $\left(\bar \theta_2\right)$ is fully `on'.
In the region above the dashed (dash-dotted) curve,
the vertex correction from angle $\bar \theta_1$ ($\bar \theta_2$) is suppressed.
Depending on how the vertex correction is suppressed, the region with suppression is further divided into two.
In the region with $V_1=0$ ($V_2=0$), $\vec q$ is not tangential to the patch at $\bar \theta_1$ ($\bar \theta_2$) and the four-fermion coupling is suppressed by $\mu/q$,
while
in the region with $V_1=\bar 0$ ($V_2=\bar 0$), $\vec q$ is tangential to the patch and it is only suppressed by 
$\mu \bar{\mathbf{K}}_{F,\bar \theta_1}/q^2$
($\mu \bar{\mathbf{K}}_{F,\bar \theta_2}/q^2$).
The boundaries between $V_i=0$ and $\bar 0$ are denoted as the straight solid lines.